\documentclass[11pt,letterpaper,twoside,openright]{report}

\usepackage{amsmath}
\usepackage[latin2]{inputenx}
\usepackage{graphicx}
\usepackage{subfigure}
\usepackage{textcomp}
\usepackage{amssymb}
\usepackage[top=3cm,left=2.5cm,right=2.5cm,bottom=3cm]{geometry}
\usepackage{cite}

\usepackage{epsfig}
\usepackage{hyperref}
\hypersetup{
	colorlinks=true,
	linkcolor=blue,
	citecolor=red,
}

\usepackage{tcolorbox}
\tcbuselibrary{theorems}

\DeclareMathOperator\arctanh{arctanh}
\renewcommand{\Im}{\text{Im}\,}
\renewcommand{\Re}{\text{Re}\,}

\author{Daniel Mata Pacheco}

\begin{document}

\thispagestyle{empty}
\begin{minipage}{0.15\textwidth}
	\includegraphics[width=\textwidth]{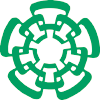}
\end{minipage}
\begin{minipage}{0.85\textwidth}
	\Large
	\begin{center}
		\bf Center for Research and Advanced Studies of the \\ National  Polytechnic Institute
	\end{center}
\end{minipage}

\vspace{1cm}
\centerline{\Large PHYSICS DEPARMENT}

\vspace*{1.5cm}
\begin{center}
	\huge \bf
	Cosmological implications from some proposals in Quantum Gravity
\end{center}
\vspace*{1.5cm}

\vspace{0.5cm}
\centerline{\large  by}
\vspace{1cm}
\centerline{\huge \bf Daniel Mata Pacheco}
\vspace{0.5cm}
\centerline{\large In order to obtain the}
\vspace{0.5cm}
\centerline{\Large \bf Doctor of Science}
\vspace{0.5cm}
\centerline{\Large \bf degree, speciality in}
\vspace{0.5cm}
\centerline{\Large \bf Physics}
\vspace{1.2cm}
\centerline{\Large Advisor: \hspace{1cm}\Large \bf Ph. D. H\'ector Hugo Garc\'ia Compe\'an}

\vspace{1.5cm}
{\large \bf Mexico City \hfill February, 2024}











\chapter*{Abstract}
\addcontentsline{toc}{section}{Abstract}

In this thesis we present the cosmological applications of some proposals from Quantum Gravity. Namely, we will explore classical and quantum cosmological implications of the Generalized Uncertainty Principle (GUP), the Ho\v{r}ava-Lifshitz (HL) theory of gravity and the Swampland Conjectures. Furthermore, we will also present a detailed analysis of Lorentzian Vacuum Transitions in various contexts.

First of all, we will present a study of the classical implications of a general form of the GUP to the standard inflationary scenario driven by a scalar field. We will describe an approximation procedure to obtain inflation similar to the slow roll criteria. Furthermore, we will obtain an analytical result that generalizes the standard cosmological constant solution in the form of a scalar field with its corresponding potential. With this result we will show that the only ingredient needed to obtain an exponentially expanding universe is a GUP with a correct form.

We will then present the study of the compatibility between the dS swampland conjecture and the Ho\v{r}ava-Lifshitz $F(\bar{R})$ theories. We will show that the conjecture implies constraints on the parameters of the theory. For an expanding universe of a power law form we will obtain that in some cases the constraint is in agreement with the expected UV behaviour of the theory, furthermore we can also study a limit to recover the standard $f(R)$ theories where the conjecture is fulfilled in a very restrictive form. In addition, we will show that for a constant Hubble parameter the theory is consistent with the conjecture whereas the standard $f(R)$ theories and General Relativity (GR) are not. Thus our results support the idea that the conjectures may be right and may be applicable to more scenarios, in particular with explicit breaking of Lorentz invariance.

Then we will study the effects of considering a GUP in the variables of the superspace in a model of the Wheeler-DeWitt (WDW) equation in HL gravity. We will show two possible equivalent ways to obtain a relevant deformation of the WDW equation that contains the information of the GUP. Furthermore, we will explore the infrared (IR) as well as the ultraviolet (UV) limit. In the IR we will obtain consistency with the results found in General Relativity. On the other hand, we will find analytical solutions in the UV that can exhibit oscillatory behaviour.

Finally, we will present a general study of the Lorentzian vacuum transitions. We will present a general method to compute the transition probabilities between two minima of a scalar field potential for any model of the superspace that leads to a generic form of the Hamiltonian constraint, by solving the WDW equation with a semiclassical expansion. We will only consider up to first order in this expansion but the method provides enough information to compute the probabilities up to any desired order. We will then apply the method in different scenarios. First we will use it to study the transitions in General Relativity with isotropic and anisotropic metrics, thus we will be able to explore explicitly the effect of anisotropy. Then, we will apply it to the HL theory employing inhomogeneous and homogeneous scalar fields. Finally, we will apply the method to a scenario where the variables of superspace describing the WDW equation obey a GUP, thus we will study the effect of the GUP in the transition probabilities as well. We will give a general interpretation of the results as probability distributions of creating universes with a given size. In this way, the GR results will always predict a singularity at the beginning of the universe. On the other hand, the HL theory  with an inhomogeneous scalar field predicts that the universe starts with a small but finite value, and the GUP solutions enhance the probability of creating universes with a non-zero size.

With these studies we will approach keys aspects of cosmology that are hoped to be solved by quantum gravity, that is we will explore (at a toy model level): the origin of inflation, the viability of the swampland conjectures and a plausible model to avoid the initial singularity. This thesis encompasses six research articles where these results were presented originally.


\tableofcontents

\chapter{Introduction}\label{Introduction}

The pursue of a quantum theory of gravity represented a logical step forward to the application of the quantum ideas that revolutionized physics on the 20th century. However, this task was found to be far more complicated and it remains as an open problem today, for some discussion in general of the subject see for instance \cite{Oriti:2009,Armas:2021,Kiefer:2004xyv}. There are many reasons why such a theory presents important conceptual problems. To begin with, the current classical theory of gravitation is General Relativity (GR) where the structure of spacetime is a dynamical object and thus a quantum theory of gravity should contain a quantum description of spacetime itself, something that is not required in the other quantum field theories. At present this is a still an unresolved problem. On the other hand, a quantum description of gravity is expected to be relevant at distances of the order of the Planck length, thus much smaller than ordinary quantum mechanics. This represents a formidable difficulty on trying to conceive laboratory experiments to explore such regimes with direct measurements. However, this does not mean that a quantum theory of gravity is not relevant on events that are accessible to us at present, since there are places of nature that provides natural scenarios where this type of theory is necessary. For example, the regions where gravity is very strong and restricted to a small region in spacetime such as a singularity within a black hole, or the physics of the very early universe. It is known that in the very early universe a period of inflation, that is of exponential expansion, must be present, therefore this expansion can grow signatures of a quantum behaviour of gravity to macroscopical size, thus it can provide indirect measurements of these quantum features. Therefore, cosmological scenarios represent a realistic setup to look for quantum signatures of gravity. Furthermore, it is expected that a quantum theory of gravity can provide the necessary completion to correctly answer the remaining questions of the very early universe, such as the big bang singularity or the origin of inflation. In this thesis we will treat these subjects in different scenarios derived or inspired by quantum gravitational proposals.

Despite the fundamental problems of a quantum theory of gravity mentioned earlier, there have been some approaches to describe gravity in a quantum regime with different levels of success and  conceptual grounds. One of the first approaches to pursue this goal consists on applying the standard canonical quantization procedure to General Relativity. In order to correctly apply the procedure, the ADM formalism is used, in which the spacetime is foliated by spacelike hypersurfaces and the dynamical object is the spatial three metric \cite{Arnowitt:1962hi}. In this formalism the system is described by constraints that are imposed on the quantum theory. These are the momentum and the Hamiltonian constrains, the latter ensures that the total Hamiltonian vanishes. Following the standard quantization procedure to these constraints, the Hamiltonian constrains leads to the Wheleer-DeWitt (WDW) equation with the idea that the universe as a hole can be described as a single entity with a wave functional defined on the superspace \cite{Wheeler,DeWitt}. In quantum cosmology such functional is usually called the wave function of the universe (for some general reviews see \cite{Halliwell:2009,VargasMoniz:2010zz,Bojowald:2020nwa}). The superspace is defined by the degrees of freedom of the three-metric as well as any matter content that is considered. However, for concrete applications, cosmological  metrics are used, which only have a few degrees of freedom. In this case, it is said that an approximation to the minisuperspace is performed. Although this quantization scheme turned out to be troublesome and the interpretation of the wave functional have been subjected to debate up to date (for example see \cite{Valentini:2021izg} for a recent discussion), there are some scenarios where this framework is relevant as an s-wave approximation to quantum behaviour. For example, it is believed that a vacuum transition can be the beginning of the universe, this kind of processes can be described by using an interpretation of the ratio of two solutions of the WDW equation as a probability to obtain a transition between both configurations. These ideas were developed in \cite{FMP1,FMP2} for transition between regions of different cosmological constants and recently they have been extended in \cite{deAlwis:2019dkc,Cespedes:2020xpn} where transitions between minima of a scalar field potential can be studied. These results were compared to the results obtained by using a path integral approach relying on a Euclidean computation and some important differences were outlined. 

The WDW approach to quantum gravity takes the classical description of gravity in General Relativity and try to apply a quantization procedure as it is performed with standard quantum mechanics. This approach is natural but it is limited to a semiclassical description. On the other hand, the best candidate for a full quantum theory of gravity is the framework of string theory. These approaches modify in a profound way the description of gravity by proposing models of 1-dimensional objects that can reproduce gravitation when the spectrum of closed strings is closely investigated. With these modifications the classical behaviour can be recovered at low energies by an effective action but they always carry additional fields that alter the effective theory profoundly. There is, however, one proposal for a quantum theory of gravity that can be thought of as to live in the middle of these approaches. That is, we can modify General Relativity at the classical level in order to have a theory that has a better quantum behaviour without the need to incorporate more radical changes. With this spirit in mind, based on the Lifhstiz scaling proposed on solid state physics, Petr Ho\v{r}ava proposed a modification of General Relativity that breaks the Lorentz invariance on the ultraviolet but render the theory power counting renormalizable as a result \cite{Horava:2009uw} (for some
recent reviews, see \cite{Weinfurtner:2010hz,Sotiriou:2010wn,Wang:2017brl,Mukohyama:2010xz} and references therein). There are various versions of such Ho\v{r}ava-Lifshtiz (HL) theory for gravitation but they all rely on such anisotropic scaling. The cosmological scenarios in this kind of theories have been studied extensively, finding new behaviours such as cosmological bounce scenarios \cite{Calcagni:2009ar,Brandenberger:2009yt,Czuchry:2009hz}   or the appearance of dark matter as an integration constant \cite{Mukohyama:2009mz}. The canonical procedure described earlier can also be pursued with this theory. In this way, the WDW equation is expected to have a better UV behaviour since it is derived from a theory that improves GR in the UV. Such WDW equations have also been considered extensively (for some examples see references \cite{Bertolami:2011ka,Christodoulakis:2011np,Pitelli:2012sj,Vakili:2013wc,Obregon:2012bt,Benedetti:2014dra,Cordero:2017egl}.). On the other hand, these theories can be generalized in the same way in which $f(R)$ theories generalize General Relativity leading to what has been called $F(\bar{R})$ Ho\v{r}ava-Lifshitz theories \cite{Chaichian:2010yi,Elizalde:2010ep}. These theories have some cosmological interesting features such as the presence of two regions of exponential expansion. Therefore, we can study the Ho\v{r}ava-Lifshitz theory of gravity, in any of its forms and generalizations, in order to explore quantum behaviour with very similar techniques as the ones employed in General Relativity.

Adopting a phenomenological point of view, we can avoid the complications involved by considering proposals of quantum theories of gravity, but still obtain some insights into what we can expect from them, by modifying the quantum theories that are well understood with some properties that we expect are going to arrive when gravity takes a quantum description. For example, it is believed that any theory of quantum gravity should have a minimum measurable length, this is of course present on string theory \cite{STLength,STLength2} and Loop Quantum Gravity but it can be justified in general by thought experiments on black hole physics as well \cite{Maggiore:1993rv,Scardigli:1999jh} (see for example \cite{Garay:1994en} for a more thorough discussion). We can incorporate such feature to quantum mechanics by modifying the uncertainty principle through a modification of the Heisenberg algebra. This kind of generalization have been called the Generalized Uncertainty Principle (GUP) and have been studied extensively in the literature. From the modifications that they imply to standard quantum mechanics and the changes on the mathematical description such as in \cite{Kempf:1994su,Shababi:2019ysa,Pedram:2011gw} or looking to constrain their defining parameters by experimental measurements \cite{Scardigli:2016ubl,Scardigli:2016pjs,Lambiase:2017adh,Vagenas:2017fwa,Bosso:2017hoq,Demir:2018akw,Bosso:2018ckz,Fu:2021zrd,Bushev:2019zvw}. Furthermore, modifications to the Heisenberg Uncertainty Principle (HUP) have also been found necessary when the quantization procedure takes place on a de Sitter or anti de Sitter space \cite{Park:2007az,Mignemi:2009ji,Gine:2020izd}, or by the theory of Doubly Special Relativity \cite{Magueijo:2001cr,Cortes:2004qn,Magueijo:2004vv,Ali:2009zq}. Thus, there are many motivations to generalize the HUP with different forms and implications, not only the search of a minimum measurable length. Since this generalization modify the quantum description in a profound way we can look to the modifications implied in the superspace through the  WDW equation, this has been done in a variety of ways, for example see \cite{Garcia-Compean:2001jxk,Vakili:2007yz,Vakili:2008tt,Kober:2011uj,Zeynali:2012tw,Faizal:2014rha,Faizal:2015kqa,Garattini:2015aca,Gusson:2020pgh,GUP}. Furthermore the GUP modifications are not restricted to the quantum theory, there are some ways in which we can derive modifications to the classical theory as well. In this sense implications regarding classical cosmology have also been studied, in particular the implications to the CMBR spectrum were investigated in \cite{Kempf:2000ac,Ashoorioon:2004vm,Ashoorioon:2004wd,Ashoorioon:2005ep}  as well as the changes that it implied to the Friedmann equations using different techniques in \cite{Cai:2005ra,Zhu:2008cg,Giardino:2020myz,Paliathanasis:2015cza,Giacomini:2020zmv,Battisti:2008du,Ali:2014hma,Moumni:2020uki,Atazadeh:2016yeh}. Therefore, considering a GUP allows to find modifications to the classical and quantum theories inspired by quantum gravity and other scenarios.

The must robust proposal of a quantum theory of gravity is the framework of string theory which encompass different theories whose mathematical description have been expanded for decades. This setup have many deep implications to physics  such as treating the dimension of spacetime as a quantity constrained by the consistency of the theory, among others. The phenomenological implications of such proposals have been studied extensively. For example, the look for a description of scalar field inflation derived from string theory have been an important area of research since there are many candidates for such a field, see for example \cite{Baumann:2014nda}.  However, there are important issues regarding the phenomenological aspects of these theories that have prevented this framework to be completely accepted. One of such complications arises because all the physical constants in these theories are obtained as expectation values of scalar fields in the vacuum, but there is not a single vacuum of the theory, there is instead a huge number of possibilities, this is the so called Landscape of string theory. However, at present a mechanism to choose dynamically such vacuum does not exists. There are many scenarios that can be formulated on these theories that will lead to effective low energy theories that can then be used to describe cosmological scenarios among other aspects. All the effective theories constructed in this way will be compatible to a completion to the full quantum theory, thus they will belong to the landscape. On the other hand, if we have a low energy effective theory that cannot be completed in the ultraviolet to string theory (or in general to any quantum theory of gravity), it is said that this theory belongs to the swampland. Therefore, exploring a possible way to distinguish an effective theory that can be completed to a correct quantum theory of gravity in the ultraviolet from those which cannot is an important subject that have attracted attention in the past years. With this classification in mind, conditions on an effective field theory (EFT) containing gravity to identify if it belongs to the landscape or to the swampland have been proposed  recently \cite{Vafa:2005ui,Brennan:2017rbf,Ooguri:2006in,Palti:2019pca,vanBeest:2021lhn,Grana:2021zvf}, they are called the swampland conjectures. Their goal is to identify general aspects of string theoretic realizations such that they may serve as  guidelines to characterize the effective theory.  The application of some of these conjectures such as the de Sitter conjecture (that claims that a dS background cannot be found in string theory \cite{Obied:2018sgi,Ooguri:2018wrx,Andriot:2018wzk,Agrawal:2018own,Roupec:2018mbn}) to cosmological scenarios have attracted a lot of attention leading to some insightful results. For example, the standard slow roll criteria for the inflation potential is not consisted with the dS conjecture \cite{Garg:2018reu,Ben-Dayan:2018mhe,Kinney:2018nny}. This can be resolved by considering multifield inflation, modifying the theory of gravity, or breaking Lorentz invariance among other approaches \cite{Motaharfar:2018zyb,Achucarro:2018vey,Artymowski:2019vfy,Benetti:2019smr,Elizalde:2022oej,Denef:2018etk,Trivedi:2020wxf,Yi:2018dhl,Brahma:2019kch,Trivedi:2021nss}. Therefore the conjectures can be used to explore if any given cosmological setup can present a well behaved quantum behaviour by providing information of its correctly completion to a full quantum regime.

In this thesis we will present the study of the different theories and proposals that we have described on relevant cosmological scenarios. We will present results in the classical and quantum regimes that will allow us to explore important conceptual problems such as the viability of the swampland conjectures, the initial singularity and the origin of inflation. The outline of this thesis is as follows:

In chapter \ref{CH-Preliminaries} we will present the mathematical and conceptual preliminaries to the thesis. Firstly, we will present a brief introduction to classical cosmology, describing the relevant cosmological metrics as well as the standard inflationary scenario driven by a scalar field. Then we will present the Hamiltonian formalism of General Relativity, in order to pursue the canonical quantum gravity procedure to obtain the Wheeler-DeWitt equation. Later on, we will discuss some aspects about the interpretation of the solutions and limitations of this procedure. Then we will present the action that describes the Ho\v{r}ava-Lifshitz theory of gravity and discuss the different versions of the theory, the way in which the scalar field can be coupled to such theories and the generalization to $F(\bar{R})$ theories. Next, we will describe the different forms that have been proposed of the Generalized Uncertainty Principle and describe some cosmological scenarios where its effects have been studied. Finally, we will present the swampland conjectures, in particular the dS conjecture, and discuss the way in which it can be used as a tool to test different cosmological scenarios.

We will begin to present our results by starting with a classical scenario in chapter \ref{CH-GUPInflation}. There we will study the modifications implied by a GUP on the classical inflationary scenario driven by a scalar field. Since there are many proposals for the GUP, we will not use a particular form, we will instead consider a general modification of the Heisenberg algebra. In order to find classical implications of the GUP, we will perform a classical limit consisting in modifying a Poisson bracket whenever the corresponding quantum commutator is modified. In some cases we will find that inflation can be obtained approximately by Hubble slow-roll parameters that depend on the standard slow roll parameters in a very different form that in the usual case, thus opening the possibility to more forms for the potential. Furthermore, we will show that there is an analytical solution describing an exponentially expanding universe that generalizes the standard cosmological constant solution by restricting the form of the GUP. It is found that even if the cosmological constant is set to vanish when the GUP vanishes, the expansion of the universe is present when the GUP is relevant. Thus we can relate the origin of inflation to a GUP with an appropriate form, there is no need of any other source. The results of this chapter were first presented in \cite{Garcia-Compean:2023aaj}.

Since we know that the Ho\v{r}ava-Lifshitz theory of gravity has a better ultraviolet behaviour than GR, we will study in chapter \ref{CH-HLFRSwampland}  the compatibility between the Ho\v{r}ava-Lifshitz $F(\bar{R})$ theories and the de Sitter swampland conjecture. We will perform such test using a flat Friedmann-Lema\^itre-Robertson-Walker (FLRW) metric which is of cosmological interest. We will find appropriate forms of the $F(\bar{R})$ function that allows us to use the conjecture to constrain the parameters of the theory. First of all, we will show that in standard $f(R)$ theories the conjecture can be fulfilled but in a very restrictive way. On the other hand, for the $F(\bar{R})$ theories we will firstly present the analysis with an ansatz of a power law form for the scale factor. There is one particular case in which the theory is described by only one parameter, whereas in the general case the theory is described by two parameters. In the particular case the conjecture leads to a region  on the  parameters space where it is expected ultraviolet behaviour, thus we find an agreement between the conjecture and the ultraviolet region of the theory. In the general case, we can obtain consistency with such region as well but we can also perform the $f(R)$ limit and obtain consistent results. Finally, we study the case of a constant Hubble parameter, the dS conjecture can also be fulfilled by constraining the parameters of the theory, however the constrictions make this compatibility exclusive to these theories. These results can be used to support the idea that the conjectures may be right and that they can also be applied to more scenarios than originally thought, in particular, in this case we obtain consistency with a theory that breaks Lorentz invariance in the ultraviolet. The results of this chapter were first presented in \cite{Garcia-Compean:2023fjl}.

We will then move on to a quantum scenario employing the canonical quantum procedure in the form of the Wheeler-DeWitt equation. In chapter \ref{CH-GUPWDW} we will study a way in which the GUP can be incorporated to deform the WDW equation obtained in the Ho\v{r}ava-Lifshtiz theory, using in particular a Kantowski-Sachs metric. Firstly, we will explore the WDW equation with standard uncertainty principle, and explore analytical solutions in the infrared limit as well as a convenient ultraviolet limit. From this we will show that one particular version of the theory derived by using detailed balanced can be obtained from the most general form of the theory by an appropriate choice of the parameters. We will then consider a GUP in the variables of minisuperspace. We will describe two possible ways to obtain a modification of the WDW equation with only up to second power terms in the momenta. One of them is useful to present an equation for the most general form of the theory. However, the second option allows to obtain analytical solutions in the limiting cases considered before. In the infrared limit we will obtain consistency with General Relativity. Moreover, in the ultraviolet limit we will obtain an analytical solution that displays oscillatory behaviour, but we will find that this can not be achieved in general and it depends on the parameters involved. These results are relevant since by introducing a GUP by hand to the WDW equation, we are improving its ability to correctly describe quantum behaviour. The results of this chapter were first presented in \cite{Garcia-Compean:2021kqi}. 

We will then use all the ingredients explored before to study a process which is purely quantum in a general form in chapter \ref{CH-Transitions}. We will study the vacuum transition probabilities between two minima of a scalar field potential by using the semiclassical Wentzel-Kramers-Brillouin (WKB) approximation to the WDW equation in a Lorentzian formalism. First of all we will present a general method to compute such probabilities for any model on superspace by considering a general form of the Hamiltonian constraint described by quadratic as well as linear terms in the momenta. In section \ref{S-TransitionsGR} we will use this method to compute probabilities in General Relativity. We will study first the FLRW  metric and obtain the same results as the ones obtained by using an Euclidean approach with path integral methods. However, our is method is not restricted to a specific form of the metric. Thus we will use it to study the transition probabilities with some homogeneous but anisotropic metrics such as the Kantowski-Sachs, biaxial Bianchi IX and the Bianchi III metric, and study what are the implications of anisotropy. Since our method is not restricted to General Relativity we will use it in section \ref{S-TransitionsHL} to study the transition probabilities in the Ho\v{r}ava-Lifshitz theory of gravity. We will consider and FLRW metric and homogeneous fields as usual but we will also study a field that can depend on the spatial variables as well. In this latter case we will find that the probabilities are modified in the UV region leading to the contrary behaviour than in GR, in fact vanishing when the scale factor vanishes. By giving an appropriate interpretation to the probabilities this represents a plausible resolution of the initial singularity, namely if the universe is created by a vacuum transition in this form the universe should be created without a singularity. Finally, in section \ref{S-TransitionsGUP} we will use the most general form of our method to study the implications of considering a GUP. To achieve this we will use the techniques developed in chapter \ref{CH-GUPWDW}. We will study such transitions with a FLRW metric as well as with anisotropic metrics and study the differences with the standard case. We will find that the probability of creating a universe with a nonzero size increases as the GUP parameter increases. Therefore, by studying vacuum transitions and interpreting the solutions in a proper manner, we will explore some relevant features of the very early universe, in particular, a possible way to avoid the initial singularity. The results of this chapter were first presented in \cite{Garcia-Compean:2021syl,Garcia-Compean:2021vcy,Garcia-Compean:2022ysy}. 

Finally, in chapter \ref{Ch-FinalRemarks} we will present our final remarks.

\chapter{Preliminaries}\label{CH-Preliminaries}

In this chapter we will present a brief conceptual and mathematical introduction to the subjects that we will be using in the rest of this thesis. Our intention is not to be exhaustive, instead we will focus only on the aspects that will be necessary for the development of the following chapters with the intention for this thesis to be as self-contained as possible. However, we will try to provide sufficient references so the reader that is interested in any of the subjects presented can explore further.  

\section*{Notes about units} 
In the majority of this work we will use natural units in which $c=\hbar=\kappa=M_{pl}=1$ where $\kappa=8\pi G$ with $G$ Newton's constant and $M_{pl}$ is the Planck mass. However, when dealing with classical cosmology we will maintain the $\kappa$ term in the equations so the results can be compared to the standard classical procedures. Furthermore, in chapter \ref{CH-Transitions} a semiclassical expansion on  $\hbar$ will be performed, thus in that chapter $\hbar\neq1$ and the units will be chosen appropriately for every example considered. Furthermore, when dealing with HL gravity we will explicitly write $M_{pl}$ when needed.

\section{Cosmology}\label{S-Cosmology}
Let us begin with a brief description of some aspects of classical cosmology. We will discuss the form of the metrics mostly used on cosmological scenarios, the standard model of inflation driven by a scalar field and the Hamiltonian formalism of General Relativity.

\subsection{Cosmological metrics}\label{SS-Metrics}
Most of the cosmological studies are based on the cosmological principle, which states that the universe at great scales described as one entity is homogeneous and isotropic. The metrics that describe the universe with these properties have the maximum number of symmetries, that is the maximum number of Killing vectors, and are categorized only by the sign of the constant spatial curvature. They are the Friedmann-Lema\^{\i}tre-Robertson-Walker metrics and can be written in general as \cite{Wald}
	\begin{equation}\label{DefFLRW}
		ds^2=-dt^2+a^2(t)
		\begin{cases}
			dr^2+\sin^2(r)d\Omega^2_{2} \\
			dr^2+r^2d\Omega^2_{2} \\
			dr^2+\sinh^2(r)d\Omega^2_{2}
		\end{cases}
	\end{equation}
where the first case corresponds to positive spatial curvature (closed spatial surface), the second to the zero curvature (flat and open spatial surface),  and the third one to the negative curvature case (open spatial surface). Furthermore, $d\Omega^2_{2}=d\theta^2+\sin^2\theta d\phi^2$ stands for the metric of the 2-sphere. In the latter expression $a(t)$ is called the scale factor which is responsible for the description of the expansion of the universe. For example, we can note that for the positive curvature case in (\ref{DefFLRW}) the metric of the spatial part is just the metric of the 3-sphere times a function that changes with time, thus it is an sphere whose size evolves with time dictated by the scale factor. Most of the discussion on cosmological models is performed in the context of these types of metrics, particular attention is put in the Hubble parameter $H$ defined by
	\begin{equation}\label{DefHubbleParameter}
		H=\frac{\dot{a}}{a} ,
	\end{equation}
where $\dot{a}=\frac{da}{dt}$. At present, there is a tension between the experimentally measured values of this parameter coming from the early universe, and the values arriving from measurements in the late time universe. This tension is a subject of intense debate, for recent reviews on this subject see for example \cite{DiValentino:2021izs,Dainotti:2023yrk,Hu:2023jqc} and references therein.

On the other hand, there has been a recent debate regarding the level of anisotropy in the universe. Observations from type Ia supernovae \cite{Colin:2018ghy} of the cosmic acceleration or  based-space x-ray measurements \cite{Migkas:2020fza} seem to imply certain degree of large scale anisotropy. However, analysis from the Cosmic Microwave Background (CMB) measures of temperature and polarization from Planck observatory \cite{Saadeh:2016sak}, or using supernovae \cite{Soltis:2019ryf} seems to strongly imply that the deviations from isotropy are disfavoured. The question of the level of anisotropy in our universe is fundamental for the cosmological models, for example implications of the anisotropy in galaxy clusters have been analyzed in \cite{Migkas:2021zdo}. More recently, it has been proposed that anisotropies seem to be compatible with the tension in the value of the Hubble constant, suggesting that modification of the current cosmological  paradigm is needed \cite{Krishnan:2021dyb,Krishnan:2021jmh}. Thus, the debate about the possible existence of a certain degree of anisotropy in our universe is still not conclusive. All these results show that there is a great interest from the observational point of view to study homogeneous but anisotropic metrics. These kind of anisotropic metrics have been considered and studied extensively from the theoretical point of view as well. Employing a construction based on the group of isometries they can be classified following the classification of 3-dimensional Lie algebras made by Bianchi \cite{Bianchi} and they correspond to 9 metrics named after him \cite{Estabrook:1968,Kamenshchik:2018ggn}. They can be written in general as \cite{Wiltshire:1995vk}
	\begin{equation}\label{DefBianchiMetrics}
		ds^2=-dt^2+g_{ij}(t)\omega^{i}\otimes\omega^{j} ,
	\end{equation}	
where $\omega^{i}$ are the invariant 1-forms associated with the isometry groups. Furthermore, a different homogeneous anisotropic metric can also be defined, namely the Kantowski-Sachs metric \cite{Kantowski:1966te} in the form
	\begin{equation}\label{DefKantowskiSachs}
		ds^2=-dt^2++a^2(t)dr^2+b^2(t)d\Omega^2_{2} .
	\end{equation}
This metric is useful since it can be used to study not only anisotropic cosmological models, but also the interior of a Schwarzschild black hole as well. For a thorough discussion about the classification and behaviours encountered with anisotropic metrics see for example \cite{Kamenshchik:2018ggn,Ryan:1975jw} and references therein.

\subsection{Scalar field inflation}\label{SS-Inflation}
The current paradigm of the cosmological history of the universe, states that in the very early universe, there should be a period where the expansion in the scale factor has an exponential form. This period is called inflation, and it is commonly described by a homogeneous scalar field, called the inflaton, coupled minimally to General Relativity with the action
	\begin{equation}\label{DefActionGRPSF}
		S=\int\sqrt{-g}\left[\frac{R}{2\kappa}+\frac{\dot{\phi}^2}{2}-V(\phi)\right] ,
	\end{equation}
where $\phi$ is the inflaton and $V(\phi)$ is its corresponding potential. In this context, the flat FLRW metric is used, which can be written in cartesian coordinates as
	\begin{equation}\label{DefFlatFLRW}
		ds^2=-dt^2+a^2(t)\left(dx^2+dy^2+dz^2\right) .
	\end{equation}

Varying the action with respect to the scalar field we can obtain the scalar field equation of motion in the form 
	\begin{equation}\label{FieldEoMStandard}
		\ddot{\phi}+3H\dot{\phi}+V'=0 ,
	\end{equation}
The Einstein equations are obtained after variation with respect to the metric, from the 00 component we obtain the Friedmann equation
	\begin{equation}\label{FriedmannStandardC}
		H^2=\frac{\kappa}{3}\left(\frac{\dot{\phi}^2}{2}+V\right) .
	\end{equation}
On the other hand, the Einstein equation coming from the spatial indices leads to the second Friedmann equation which can be derived from (\ref{FieldEoMStandard}) and (\ref{FriedmannStandardC}), thus it is not independent. It is known that the only analytical solution that can describe a period of accelerating expansion (i.e. a constant Hubble parameter) is when the scalar field vanishes and its potential is a constant, the cosmological constant. However, when the scalar field is present we can obtain this behaviour by two approximations, called the slow-roll approximation,  in the following way; first we ignore the kinetic term with respect to the potential in (\ref{FriedmannStandardC}), second we ignore the acceleration term with respect to the friction one (that is, the second term) in (\ref{FieldEoMStandard}). These two approximations are measure by the slow-roll Hubble parameters
	\begin{equation}\label{DefStandardHubbleSLowRoll}
		\epsilon_{H}=-\frac{\dot{H}}{H^2} , \hspace{1cm} \eta_{H}=-\frac{\ddot{\phi}}{H\dot{\phi}} .
	\end{equation}
Thus, the approximations are well justified if we demand that $\epsilon_{H},\eta_{H}\ll1$. However, we can write these two conditions in terms of the potential by using the standard slow-roll parameters defined by
	\begin{equation}\label{DefSlowRollP}
		\epsilon_{V}=\frac{1}{2\kappa}\left(\frac{V'}{V}\right)^2 , \hspace{1cm} \eta_{V}=\frac{1}{\kappa}\left(\frac{V''}{V}\right) ,
	\end{equation}
then, in the light of these approximations we can write
	\begin{equation}\label{RelationParameters} 
		\epsilon_{H}\thickapprox\frac{3\dot{\phi^2}}{2V}\thickapprox\epsilon_{V} , \hspace{1cm} \eta_{H}\thickapprox\eta_{V}-\epsilon_{V} .
	\end{equation}	
Therefore, when the standard slow-roll parameters are small, the slow-roll Hubble parameters are small and the approximations are justified. We note from the definition (\ref{DefSlowRollP}) that if we ask $\epsilon_{V}\ll1$, we obtain that $V$ is approximately constant and thus from the Friedmann equation (\ref{FriedmannStandardC}) we obtain an approximately constant Hubble parameter and thus a period of inflation.

\subsection{The ADM formalism}\label{SS-ADM}
As is well known, General Relativity can be formulated by considering the Einstein-Hilbert action and obtain the Einstein equations with a variational procedure with respect to the metric, as was performed in the last subsection. However, a Hamiltonian formalism for General Relativity can not be constructed directly from this action, since it has a covariant form and the time variable is treated in the same form as the spatial variables. In order to correctly perform a Hamiltonian formulation we need to employ the Arnowitt-Deser-Misner (ADM) formalism \cite{Arnowitt:1962hi}. Let us briefly review this formulation. (For a detailed introduction see \cite{Corichi:1991qqo,Kiefer:2004xyv} and for a thorough description of the application to cosmology see \cite{Ryan:1972Hm,Kolb:1990vq}).

In order to split the time variables from the covariant description, this formalism employs a foliation of spacetime with spatial hypersurfaces of constant time. That is, the four dimensional spacetime is decomposed into three dimensional hypersurfaces of constant time that fills the whole manifold and do not intersect.  In this form, the metric is written as 
	\begin{equation}\label{DefADMMetric}
		ds^2=-N^2dt^2+h_{ij}(dx^{i}+N^{i}dt)(dx^{j}+N^{j}dt) ,
	\end{equation}
where the spatial index $i$ runs from $1$ to $3$, $N$ is the lapse function that gives information of how a point moves in the time direction between hypersurfaces,  $N^{i}$ are the shift functions which give information of how a point can be moved in the spatial variables when moving through the hypersurfaces, and $h_{ij}$ is the three three-metric describing the induced metric on each spatial surface. The lapse function, the shift vector (composed of the three shift functions) and the three-metric represents the ADM variables in which we can express the whole equivalent formalism. With these variables the Einstein-Hilbert action can be written as
	\begin{equation}\label{EinsteinHilbertActionADM}
		S=\frac{1}{2\kappa}\int\int dtd^3x\sqrt{h}N\left[K_{ij}K^{ij}-K^2+R\right] ,
	\end{equation}
where $K_{ij}$ is the extrinsic curvature on the spatial hypersurface, $K$ its contraction, that is $K=K^{i}_{i}$, $h$ is the determinant of the spatial three-metric induced on the hypersurface and $R$ its corresponding Ricci scalar. With this form we can construct the Hamiltonian in the usual way in terms of the ADM variables. The Lagrangian $\mathcal{L}$ does not depend on the time derivatives of the lapse or shift functions, making them non dynamical variables. Thus their canonical momenta must vanish, that is
	\begin{equation}\label{DefPrimaryConstraints}
		\pi_{N}=\frac{\delta\mathcal{L}}{\delta \dot{N}}=0 , \hspace{1cm} \pi_{i}=\frac{\delta\mathcal{L}}{\delta \dot{N}_{i}}=0 ,
	\end{equation}
these are primary constraints. Therefore, the Legendre transformation is performed only with the variables of the three-metric. The corresponding canonical momenta are defined by
	\begin{equation}
		\pi_{ij}=\frac{\delta \mathcal{L}}{\delta \dot{h}^{(3)}_{ij}} ,
	\end{equation}
where a functional derivative is used, since in general, we will have an infinite number of degrees of freedom. Then the total Hamiltonian is written as
		\begin{equation}\label{DefTotalHamiltonian}
			\mathbb{H}=\int d^3x\left[N\mathcal{H}+N^{i}\mathcal{H}_{i}\right] ,
		\end{equation}
where	
	\begin{equation}
		\mathcal{H}=N\sqrt{h}\left[K_{ij}K^{ij}-K^2-R^{(3)}\right] , \hspace{1cm} \mathcal{H}_{i}=2\nabla_{j}\pi^{j}_{i} ,
	\end{equation}
and $\nabla_{i}$ is the covariant derivative with respect to the three-metric. As a consequence of the primary constraints (\ref{DefPrimaryConstraints}) it is obtained that the $N$ and $N_{i}$ functions act as Lagrange multipliers, and thus they are in principle arbitrary functions. Varying wih respect to those variables the secondary constraints are obtained
	\begin{equation}\label{ConstraintsADM}
		\mathcal{H}=0 , \hspace{1cm} \mathcal{H}_{i}=0 .
	\end{equation} 
The first of these equations is called the Hamiltonian constraint whereas the second set of equations are the momentum constraints. As stated earlier, the dynamical variables are the components of the three-metric $h_{ij}$ and they evolve in such a way that these constraints are preserved. In practice, the Hamiltonian constraint is equivalent to the Einstein equation with both temporal indices, whereas the momentum constraints are equivalent to the equations with one temporal and one spatial indices. The remaining Einstein equations are obtained by variation of the action with respect to the three-metric or as Hamilton equations with the total Hamiltonian.

\subsection{Friedmann equation in the Hamiltonian formalism}\label{SS-FriedmannHamiltonian}
Let us note that the FLRW metric (\ref{DefFLRW}), in particular the flat FLRW version (\ref{DefFlatFLRW}) is already in the ADM form (\ref{DefADMMetric}) with a unit lapse function and vanishing shift vector. Therefore the only degree of freedom  coming from the spatial metric is the scale factor. Thus we can incorporate the scalar field in the action (\ref{DefActionGRPSF}) and perform a Hamiltonian treatment to obtain the same equation as in the last section. Let us show how this procedure works. The canonical momenta of the scale factor and scalar field have the form
	\begin{equation}\label{DefMomentaGRFLRW}
		\pi_{a}=-\frac{6a\dot{a}}{\kappa} , \hspace{1cm} \pi_{\phi}=a^3\dot{\phi} .
	\end{equation}
In this case the momentum constraints are automatically satisfied and the Hamiltonian constraint is the vanishing of the total Hamiltonian constructed in the usual way, in terms of the scale factor and the scalar field. Thus, in this case the Hamiltonian constraint is
	\begin{equation}\label{SistStandardHCI}
		\mathcal{H}=-\frac{\kappa}{12a}\pi^2_{a}+\frac{1}{2a^3}\pi^2_{\phi}+a^3V(\phi)\simeq0 .
	\end{equation}
We also obtain the four Hamilton equations
	\begin{equation}\label{SistStandard1}
		\dot{a}=\{a,\mathcal{H}\}=\frac{\partial \mathcal{H}}{\partial \pi_{a}} , \hspace{1cm} \dot{\phi}=\{\phi,\mathcal{H}\}=\frac{\partial \mathcal{H}}{\partial \pi_{\phi}} ,
	\end{equation}
	\begin{equation}\label{SistStandard2}
		\dot{\pi}_{a}=\{\pi_{a},\mathcal{H}\}=-\frac{\partial \mathcal{H}}{\partial a} , \hspace{1cm} \dot{\pi}_{\phi}=\{\pi_{\phi},\mathcal{H}\}=-\frac{\partial \mathcal{H}}{\partial \phi} .
	\end{equation}
In this case the two equations in (\ref{SistStandard1}) are just the definitions of the momenta (\ref{DefMomentaGRFLRW}) as usual. The first equation in (\ref{SistStandard2}) leads to the equation coming from the variation of the action (\ref{DefActionGRPSF}) with respect to the scale factor, that is, the Einstein equation with spatial indices that will lead to the second Friedmann equation. On the other hand, the second equation in (\ref{SistStandard2}) leads to the scalar field equation (\ref{FieldEoMStandard}). Finally, the Friedmann equation (\ref{FriedmannStandardC}) is obtained after substituting the canonical momenta in the Hamiltonian constraint (\ref{SistStandardHCI}). Thus, the scalar field inflation scenario can be studied in the same form as in the last section using the Hamiltonian formalism.

\section{Canonical Quantum Gravity}\label{S-CQG}
The ADM formalism described in the last subsection is useful because with this setup the Hamiltonian description is written in terms of the true degrees of freedom, that is, with the variables describing the spatial three-metric. For this reason, sometimes this formalism is called geometrodynamics. Having obtained this form we can proceed to perform a canonical quantization procedure seeking a quantum description for gravity. Let us review how this procedure works and discuss its interpretations and limitations, for an extensive review on the subject see \cite{Kiefer:2004xyv}.

\subsection{The Wheeler-De Witt equation}\label{SS-WDW}
Once we have identified the true degrees of freedom, we can proceed to quantize the theory with the techniques developed by Dirac for a constrained system \cite{Dirac:1969}. In this way, in the coordinate representation the momenta will be promoted to functional derivative operators, whereas the constraints (\ref{ConstraintsADM}) will have to be imposed at the quantum level. Thus we have that the wave functional $\Psi$ will obey\footnote{Throughout this work we will use the same notation for the classical functions and their corresponding quantum operators. The context will provide the information of the correct interpretation.}
	\begin{equation}\label{DefWDW}
		\mathcal{H}\Psi=0 ,
	\end{equation}
	\begin{equation}\label{DefMomentumConstraintsQ}
		\mathcal{H}_{i}\Psi=0 .
	\end{equation}
This wave functional will depend on the true degrees of freedom, that is, on the variables of the three-metric as well as any other matter content that is considered. Therefore, it will be defined on a space of parameters where the coordinates are the degrees of freedom just mentioned. This is called the superspace\footnote{Although it has the same name, this space has nothing to do with the space of supersymmetric theories.}  and is in principle infinite dimensional (for a mathematical discussion of such space see \cite{Giulini:2009np}). However, in most practical applications of this formalism, one usually employs cosmological homogeneous metrics which can be described by only a few number of degrees of freedom, for example, the FLRW metrics are just described by the scale factor and the Bianchi metrics just by 3 functions at most. When such metrics are used, it is said that a minisuperspace approximation is performed, and then the functional derivatives can be exchanged by standard partial derivatives. In this context, the theory is usually just called quantum cosmology. In this quantization scheme, equation (\ref{DefWDW}) is called the Wheeler-DeWitt (WDW) equation \cite{Wheeler:1968,DeWitt:1967} and it has the form of a Schr\"odinger equation independent of time. The wave functional $\Psi$ appearing in the WDW equation is sometimes called the wave function of the universe with the idea that the universe as a hole can be described as a single entity obeying this set of equations. Furthermore, the momentum constraints (\ref{DefMomentumConstraintsQ}) imply the invariance of the theory under three-dimensional diffeomorphisms, then in most homogeneous metrics they are satisfied identically. Thus all the quantum treatment can be performed with the WDW equation alone. In the following we will briefly discuss some aspects of quantum cosmology, for a more thorough introductory discussion see for example \cite{Wiltshire:1995vk,Halliwell:2009,Carlip:2001wq,Kiefer:2008sw,Isham:1992ms} and references therein.

\subsection{Interpretation of the solutions}\label{SS-WDWI}
This proposal for a quantum theory of gravity was one of the first to be explored since it naturally follows from the Hamiltonian description of GR. However, the interpretation of the wave functional can not be performed as easily as in standard quantum mechanics, thus the correct interpretation is still a subject to rigorous debate. Let us review some general aspects on this regard.
	\begin{itemize}
		\item First of all, in most cases the wave functional does not lead to quadratic integrable functions and thus the Bohm interpretation of $|\Psi|^2$ as a probability measure can not be used  (see \cite{Valentini:2021izg} for a recent discussion), among other conceptual problems \cite{Kuchar:1991qf}.
		\item In general, the WDW equation will be of the  Klein-Gordon type on superspace, thus we can define a conserved current 
			\begin{equation}
				J=\frac{i}{2}\left(\Psi^{*}\nabla\Psi-\Psi\nabla\Psi^{*}\right) ,
			\end{equation} 
		that satisfies $\nabla\cdot J=0$. However, as in the Klein-Gordon case, the probability measure constructed from this current can display negative values and thus it can not be interpreted as a correct probability measure. For a thorough discussion of this subject see \cite{Kuchar:1991qf}
		\item Usually on quantum cosmology, the WKB approximation to the WDW equation is used to perform a semiclassical treatment. In this sense the wave functionals can be of the form $\Psi\sim e^{iS}$ or $\Psi\sim e^{-S}$. The oscillatory form represents classical behaviour since the solutions are going to be peaked around classical Lorentzian solutions, and will imply correlations between the coordinates and their canonical momenta. On the other hand, the second form is a solution belonging to a pure quantum regime prior to the classical transition. 
		\item Even if the solutions of the WDW equation cannot define a probability measure by themselves, it is possible to use them to construct a conditional probability that can be helpful to interpret results. In this sense, the ratio of absolute squares of the wave functionals for different configurations can be interpreted as relative probabilities between two states in which the system performs a transition \cite{DeAlwis:2019rxg,Cespedes:2020xpn}. 
		\item A more recent interpretation employs the Born-Oppenhaimer approximation in the presence of matter fields. In this regard, the heavy degrees of freedom coming from the gravitational part defines the background whereas the light degrees of freedom of the matter fields propagate in such background. In this way, it is possible to obtain quantum corrections to the scalar field modes that leads to modification to some parameters that can be measured. For more information on this interpretation see for instance \cite{Alberghi:2006th,Bini:2013fea,Kamenshchik:2013msa,Kamenshchik:2017kfs}.
	\end{itemize}

Therefore, even if the probabilistic interpretation of solutions of the WDW equation is troublesome and there is still an ongoing debate, we can focus on some aspects of such solutions to extract information from the results. In this sense, when dealing with the WDW equation in chapter \ref{CH-GUPWDW} we will look for oscillatory behaviour with a classical description in mind. Furthermore, we will use the conditional probability proposal to study the vacuum transitions in chapter \ref{CH-Transitions}.

\subsection{Boundary conditions}\label{SS-WDWBC}
Now that we have addressed the question of interpreting the solutions of the WDW equation we move on to the question of how to obtain such solutions in the first place. Let us remark that we have a differential equation that presumably captures some quantum behaviour of the universe treated as a single entity, however in order to obtain solutions to this equation we need to impose boundary conditions. In standard quantum mechanics such conditions are imposed externally to the system. However, if we are dealing with the universe as a whole, the nature of such conditions represents a new law by itself. In this regard, there are two proposals that have defined the standard uses of quantum cosmology;
	\begin{itemize}
		\item The Hartle-Hawking no boundary proposal \cite{Hawking:1983}: As the name suggest, this proposal states that the universe has no boundary at the beginning and thus the wave functional takes the form of an Euclidean path integral over compact 4-geometries bounded by the corresponding three-geometry. 
		\item The tunnelling proposal \cite{Vilenkin:1988}: This is a proposal directly in the superspace and states that at singular boundaries the wave function of the universe must consist only of outgoing modes. It can also be defined analogously to the no boundary proposal but using a Lorentzian path integral.
		\item Linde proposed \cite{Linde:1983cm,Linde:1983mx} that the Wick rotation employed by Hartle and Hawking should be done on the negative time instead of the positive one. However, for simple models this proposal leads to the same results as the Vilenkin proposal, thus it is usually stated under the tunnelling proposal as well.
	\end{itemize}
Both proposals lead to the paradigm that the universe was created from ``nothing'' (a state with zero geometry) by a tunnelling event and have different conceptual and operational problems. However, they have been studied extensively since their first appearance and have dictated the usual framework of quantum cosmology. For an early discussion of the differences between both proposals see \cite{Vilenkin:1994rn,Vilenkin:1998rp}. Furthermore, for some recent developments see \cite{Vilenkin:2018dch,Vilenkin:2018oja,Brahma:2020cpy,Matsui:2020tyd,Jonas:2021ucu,deAlwis:2018sec}.

In analogy to the Lorentzian formalism of the WDW equation just described, a formalism using the path integral approach can also be studied. On a formal level this approach is equivalent to the WDW equation although in most cases one uses a Wick rotation as in standard quantum field theories. In this formalism the problem of initial conditions is encountered in the form of the definitions of the contours of integration. For some recent discussion in this regard see \cite{deAlwis:2018sec,Feldbrugge:2017kzv,Feldbrugge:2017mbc,Feldbrugge:2018gin,Kaya:2019piu,Matsui:2021oio}.

\subsection{The problem of time}\label{SS-WDWTime}
One particular problem that arises in any quantization approach to gravity is the so called problem of time. In the quantum cosmology scenario it appears in the following form: We note that the  WDW equation (\ref{DefWDW}) is defined in the superspace, and thus there is not a time variable to properly define evolution. This is not at all surprising, since this behaviour is common to parametrized theories such as General Relativity. Let us review some solutions to this problem (for a formal discussion see \cite{Kuchar:1991qf})
\begin{itemize}
	\item Internal time: It can be proposed that before quantization one can interpret one of the variables as time, in this way we can obtain a functional Schr\"odinger equation instead of the WDW equation. In this proposal we can use for example the volume of the three-metric as such parameter or the mean extrinsic curvature which leads to the York time \cite{York:1972sj,Roser:2014foa}.
	\item Reference fluid: It can be proposed that a fluids fills the universe, the particles of the fluid identifies the space points and their clocks measure time \cite{Kucha:1991}. This procedure can also be performed with dust \cite{Brown:1994py}.
	\item Matter clocks: instead of using a perfect fluid, we can also use a scalar field as a clock. For a recent work regarding this subject see for example \cite{Rotondo:2019grb}.
\end{itemize}
All of the approaches to define a time parameter to the WDW equation have issues, at present there is not a general consensus on which can be the solution to this problem. However, even if we do not attempt to consider a time parameter to incorporate an evolution to the WDW solutions, we can still obtain physically relevant results by studying the oscillatory regions of such solutions, or by employing the conditional probability stated in the last subsection.

\subsection{Other applications}\label{SS-WDWA}
Nowadays it is believed that the paradigm of the WDW equation represents an approximation to the quantum behaviour of gravity in the same way as an s-wave approximation. Therefore it is expected that we can gain information about the quantum regime at least in a semiclassical approximation by using this formalism. The quantum cosmology methods presented have been extensively used in standard inflationary scenarios. However, the method is not restricted to General Relativity, it can be used on different theories of modified gravity as well, for example in the Ho\v{r}ava-Lifshitz theory as we will see in the next section. Furthermore, since we can derive low energy actions arriving from string theories, we can recover some quantum features by applying this procedure to those actions, for some examples on this line of work see \cite{Enqvist:1987uh,Enqvist:1987hx,Birmingham:1987,Luckock:1988,Pollock:1989uj,Wang_1992,Cavaglia:1999ka,Grigorian:2006yva,Gusin:2008zz,Gasperini:2021eri}. Moreover, the ideas of the tunnelling from nothing scenario can also be applied to study the swampland of string theories \cite{Kobakhidze:2004gm,Brustein:2005yn}. Therefore, at present, the canonical approach contained in the WDW equation continues to represent an important proposal to look for quantum behaviour of gravity.

\section{Ho\v{r}ava-Lifshitz theories of gravity}\label{S-HLGravity}
It is well known that General Relativity is not a renormalizable theory. Therefore, the look for quantum behaviour by employing the canonical formalism of the last section is necessarily limited. Let us review some aspects of a theory that improves the UV behaviour of GR and represents a better candidate to explore gravity at the quantum level.

\subsection{Action}\label{SS-HLAction}
The Lifshitz scaling from condensed matter physics was applied to modify the theory of General Relativity by Petr Ho\v{r}ava leading to the Ho\v{r}ava-Lifshitz theory of gravity \cite{Horava:2009uw} (for some recent reviews, see \cite{Weinfurtner:2010hz,Sotiriou:2010wn,Wang:2017brl,Mukohyama:2010xz} and references therein). The theory employs an anisotropic scaling of space and time variables described by
	\begin{equation}\label{DefScallingHL}
		x^{i}\rightarrow bx^{i} , \hspace{1cm} t\rightarrow b^{z}t ,
	\end{equation}
where $z$ is the dynamical critical exponent. In the ultraviolet $z=3$ and the theory becomes power counting renormalizable. We note that since the space and time variables scale differently, this theory is no longer Lorentz invariant at high energies. Since such scaling must be preserved under any change of variables, the theory is no longer invariant to the full four dimensional diffeomorphism, the symmetry is reduced to the so called foliation preserving diffeomorphisms. The action of this theory can be written as
\cite{Bertolami:2011ka,Sotiriou:2009gy,Sotiriou:2009bx}
	\begin{equation}\label{DefActionHL}
		S_{HL}=\frac{1}{2\kappa}\int d^4xN\sqrt{h}\left[K^{ij}K_{ij}-\lambda K^2+\mathcal{L}(h_{ij})\right] ,
	\end{equation}
where the metric is written in the ADM form (\ref{DefADMMetric}), $\lambda$ is a dynamical coupling constant susceptible to quantum corrections, that defines the restricted foliation compatible with the Lifshitz scaling, and $\mathcal{L}(h_{ij})$ is a potential term for gravity that is constructed only by higher order derivative terms in the spatial components of the curvature, giving rise to a ghost-free theory. There are two versions of this theory regarding the choice for such potential term:
	\begin{itemize}
		\item Detailed balance condition: Originally it was proposed that such term can be derived from a geometrical action, thus adding a new symmetry to the theory. In this way the potential term can be written as
			\begin{equation}
				\mathcal{L}(h_{ij})\sim E^{ij}G_{ijkl}E^{kl} ,
			\end{equation}	
		where $G_{ijkl}$ is the metric in superspace and $E_{ij}$ is derived from
			\begin{equation}
				\sqrt{h}E^{ij}=\frac{\delta W[g_{kl}]}{\delta g_{ij}} ,
			\end{equation}
		for some action $W$, this is called the detailed balance condition. It is useful since the number of constants needed in the potential term is reduced, but it leads to consistency problems.
		\item General case: If we abandon such condition, in the general scenario, the potential term contains the scalar curvature of the three-metric $R^{(3)}$  and seven constants accompanying higher spatial derivative terms written in terms of the Ricci scalar of the spatial three-metric. 
	\end{itemize}
The most general form of the action without using detailed balanced is then expressed as
	\begin{multline}\label{ActionHL}
		S_{HL}=\frac{M^2_{pl}}{2}\int_{M}d^3xdtN\sqrt{h}\left[K_{ij}K^{ij}-\lambda K^2-\Lambda M^2_{pl}+R^{(3)}-\frac{1}{M^2_{pl}}\left(g_{2}(R^{(3)})^{2}+g_{3}R^{(3)}_{ij}R^{(3) ij}\right)\right. \\ \left. -\frac{1}{M^4_{pl}}\left(g_{4}(R^{(3)})^{3}+g_{5}R^{(3)}R^{(3) i}_{j}R^{(3) j}_{i}+g_{6}R^{(3) i}_{j}R^{(3) j}_{k}R^{(3) k}_{i}+g_{7}R^{(3)}\nabla^{2}R^{(3)}+g_{8}\nabla_{i}R^{(3)}_{jk}\nabla^{i}R^{(3) jk}\right)\right]\\ + M^2_{pl}\int_{\partial M}d^3x\sqrt{h}K,
	\end{multline}
where  we have added a cosmological constant term $\Lambda$ and a boundary term that for most cases can be omitted, we have also denoted the Ricci tensor and curvature scalar in three dimensions as $R^{(3)}_{ij}$ and $R^{(3)}$ respectively. Furthermore, all $g_{n}$ ($n=2,...,8$) are positive dimensionless running coupling constants. In addition to the two versions of the theory defined by the choice of the gravitational potential term, the lapse function appearing in the action can lead to different versions of the theory as well, classified by the variables in which it may depend on:
	\begin{itemize}
		\item Projectability condition: In this case the lapse function only depends on the time variable, and thus the Hamiltonian constraint is a global constrain. Therefore, we can not use it to construct an equation of motion, as it was used for example to obtain the Friedmann equation in the last part of subsection \ref{SS-ADM}.
		\item Without projectability: In this case, the lapse function is allowed to depend also on the spatial variables. Thus the Hamiltonian constraint is local, that is, valid on every spatial point.
	\end{itemize}
In any of the two versions there is enough freedom to choose convenient values for the lapse function, but in general, the global nature of the Hamiltonian constraint leads to an integration constant, and thus a new term in the equations of motion \cite{Sotiriou:2010wn,Mukohyama:2009mz}. Therefore, even though in both cases we can choose the lapse function to be one, the equations of motion will be different.

This theory has been studied extensively in the classical regime, in particular it has led to new relevant cosmological behaviours such as bouncing universes \cite{Calcagni:2009ar,Brandenberger:2009yt,Czuchry:2009hz} or the appearance of a dark matter term as a constant of integration \cite{Mukohyama:2009mz}, to name a few. The compatibility with observational data have also been studied \cite{Dutta:2009jn,Dutta:2010jh,Harko:2009qr,Nilsson:2018knn,Nilsson:2019bxv,Nilsson:2021ute,Escamilla-Rivera:2020llu}.

\subsection{The infrared limit}\label{SS-HLInfraredLimit}
As can be easily seen from (\ref{ActionHL}), we can try to recover standard General Relativity by performing an infrared limit consisting of $z\to1$ (the Lorentz invariance is recovered) accompanied by $\lambda\to1$ (the full diffeomorphism symmetry is recovered, consequently the foliation takes the form as the usual in  the ADM formalism). In addition, the higher order derivative terms in the action have to be neglected in order to get the correct limit, that is $g_{n}\to0$ for all $n$.  In this case, it is obtained that $\mathcal{L}(h_{ij})\to R^{(3)}$. However, this limit was found to be troublesome, since it remains an additional degree of freedom  which leads to a perturbative IR instability \cite{Sotiriou:2010wn,Mukohyama:2010xz,Izumi:2011eh,Gumrukcuoglu:2011ef}. The non-projectable version  has the possibility to remove this unphysical degree of freedom.

In order to improve the issues in the IR limit found in the original theory, the Ho\v{r}ava-Lifshitz theory of gravity has been generalized in a number of ways. For example, in \cite{Zhu:2011xe,Zhu:2011yu,Huang:2012ep}, an extra $U(1)$ field was added.

\subsection{Coupling matter}\label{SS-HLMatter}
In order to couple a scalar field to this theory $\phi(t,x^{i})$,  we need to consider actions that are invariant to the symmetries of the theory, that is that take into account  the anisotropic scaling symmetry and UV renormalizability. The general scalar action in HL gravity  is found to consist of spatial derivatives up to order 6. 
For simplicity, we do not consider terms of products of derivatives, then this action is written in the form \cite{Kiritsis:2009sh}
	\begin{equation}\label{ActionMatter}
		S_{m}=\frac{1}{2}\int dtd^3x \sqrt{h}N\left[\frac{(3\lambda-1)}{2N^2}\left(\dot{\phi}-N^i\partial_{i}\phi\right)^2+F(\phi)\right] ,
	\end{equation}
where the function $F(\phi)$ is given by
	\begin{equation}\label{DefinitionF}
		F(\phi)=\phi\left(c_{1}\Delta\phi-c_{2}\Delta^2\phi+c_{3}\Delta^3\phi\right)-V(\phi) ,
	\end{equation}
where $\Delta$ denotes the three-metric Laplacian, whereas $V(\phi)$ is the potential for the scalar field. The constant $c_{1}$ is the velocity of light in the IR limit, the two other constants
are related to the energy scale $M$ (which can be taken as the Planck mass $M_{pl}$) as
	\begin{equation}
		c_{2}=\frac{1}{M^2} , \hspace{0.5cm} c_{3}=\frac{1}{M^4} .
	\end{equation} 

We note that for homogeneous scalar fields, that is with a time dependence only, the action will be very similar to that of standard General Relativity, the difference will only be a constant regarding $\lambda$ in the kinetic term. Therefore, since the higher spatial derivative terms in the action (\ref{ActionHL}) are abscent for an FLRW flat metric, we can see that most of the cosmological analysis with this metric will be as in standard GR, in particular the inflationary scenario performed in \ref{SS-Inflation} can be carried out in the same manner, there will only be an extra term with the $\lambda$ parameter in the expressions. On the other hand, for inhomogeneous fields the actions will be different and the results are expected to deviate from GR results significantly.

\subsection{Quantum cosmology in HL gravity}
Since the Ho\v{r}ava-Lifhitz theory represents an improvement over GR in the high energy regime, it is natural that quantum gravity aspects of the theory are of great interest. Furthermore, the form of the action is naturally written in the ADM form. Thus, the canonical quantization techniques of the last section can be used. These approaches  have been extensively studied. For example, some of the works regarding quantum cosmology in Ho\v{r}ava-Lifshitz gravity are \cite{Bertolami:2011ka,Christodoulakis:2011np,Pitelli:2012sj,Vakili:2013wc,Obregon:2012bt,Benedetti:2014dra,Cordero:2017egl}. 

\subsection{$F(\bar{R})$ Ho\v{r}ava-Lifshitz theories}\label{SS-FRHL}
There are many ways to generalize GR by modifying the Einstein-Hilbert action. One of the most important theories in this regard is the $f(R)$ theories \cite{Artymowski:2019vfy}. The gravitational action for these proposals in the Jordan frame is written as
\begin{equation}\label{ActionfR}
	S=\frac{1}{2\kappa}\int d^4x\sqrt{-g}f(R) ,
\end{equation} 
where $f(R)$ is some well-behaved function of the curvature scalar $R$. In this scenario, it is thought that the action of GR is just the first term on a Taylor expansion of the $f$ function around a small curvature. These theories are of great interest since they lead to many new interesting cosmological behaviours, see for instance \cite{DeFelice:2010aj}. 

As we can see from the form of action (\ref{ActionHL}), it is natural to generalize HL theories in the same way as the generalization leading to $f(R)$ theories just described. In this case, the action for the generalization is written as
\begin{equation}\label{ActionFBarR}
	S_{F(\bar{R})}=\int d^4x\sqrt{h}NF(\bar{R}) ,
\end{equation}
where $\bar{R}$ can be seen as a generalization of $R$ which includes the new terms of spatial derivatives abscent in the GR scenario, and is proposed to have the form \cite{Chaichian:2010yi,Elizalde:2010ep}
\begin{equation}\label{DefBarR}
	\bar{R}=K^{ij}K_{ij}-\lambda K^2+2\mu\nabla_{\rho}(n^{\rho}\nabla_{\nu}n^{\nu}-n^{\nu}\nabla_{\nu}n^{\rho})+\mathcal{L}(h_{ij}) ,
\end{equation}
where $\mu$ is a constant. The term with the $\mu$ parameter is usually omitted in the standard $f(R)$ proposal since in that case it is just a total derivative term. However, for these new theories such term is necessary. As we can see from (\ref{ActionFBarR}) and (\ref{DefBarR}), the limit $\lambda\to1$, $\mu\to1$ of this theory leads to the standard $f(R)$ theory (\ref{ActionfR}). This theory was originally proposed employing the detailed balanced condition in a series of papers \cite{Chaichian:2010yi,Elizalde:2010ep}. They turned out to be relevant in cosmology since the solutions with the flat FLRW metric describe new and interesting behaviours such as two periods of accelerating expansion. These theories were later studied further, for example the Ekpyrotic scenario was presented in \cite{Lopez-Revelles:2012xal}. Pursuing a more general version, the detailed balanced condition was abandoned in \cite{Carloni:2010nx}. A Hamiltonian analysis for these theories was performed in \cite{Chaichian:2010zn}, and the relation to general scalar tensor HL theories was investigated in \cite{Kluson:2011ff}. Furthermore, an extension to incorporate a $U(1)$ field was performed in \cite{Kluson:2010za}. Thus, these theories represent an interesting generalization to the HL theory of gravity, in particular to cosmological scenarios. For this reason, in chapter \ref{CH-HLFRSwampland} we will study a new feature for these theories in the form of their compatibility with the dS swampland conjecture.

\section{Generalized Uncertainty Principle}\label{S-GUP}
The Heisenberg Uncertainty Principle is a key result of standard quantum mechanics, it represented one of the first features that changed the paradigm of classical theories. However, as the quantum ideas were pursued in different contexts, a generalization of this principle was found to be necessary motivated within different scenarios. Let us review some of these scenarios and their implications to cosmology at the classical, as well as at the quantum level.

\subsection{GUP proposals}\label{SS-GUPProposals}
We can categorize the motivations from such generalization as follows:
\begin{itemize}
	\item Minimum measurable length: At the present time it is believed that a quantum theory of gravity, whatever its form, must contain a minimum measurable length,  since it is expected that the standard classical view of spacetime as a smooth manifold breaks down and a more fundamental structure should appear \cite{STLength,STLength2,Maggiore:1993rv,Scardigli:1999jh,Garay:1994en}, which is incompatible with the HUP. Thus, we can modify the uncertainty principle, by modifying the Heisenberg algebra, in the well established quantum theories, in order to obtain some insight to the modifications that we expect are going to arrive with the full quantum gravity theory. 
	\item Quantization in (anti) de Sitter space: It has been found that when the quantization procedure takes place in a dS or AdS space the uncertainty principle has to be modified in order to incorporate a maximum distance given by the dS (or AdS) radius, this has been called the Extended Uncertainty Principle (EUP) \cite{Park:2007az,Mignemi:2009ji,Gine:2020izd}.
	\item Modifications in the coordinates and momenta: Modifying the uncertainty principle by incorporating a dependence on the momentum as well as the position have also been performed, this proposal is called the Generalized Extended Uncertainty Principle (GEUP) in \cite{Park:2007az,Mignemi:2009ji}. 
	\item Doubly Special Relativity (DSR): On the other hand, in the context of the Doubly Special Relativity theory it has also been found that a modification of the HUP is needed \cite{Magueijo:2001cr,Cortes:2004qn} which leads to a form that is compatible with string theory and black hole physics as well \cite{Magueijo:2004vv,Ali:2009zq}. 
	\item Non-commutativity: Furthermore, considering non-commutativity on the variables of spacetime can also lead to a modification of the HUP which are of great interest as well. For some reviews, see for instance \cite{Douglas:2001ba,Szabo:2001kg}. 
\end{itemize}

Strictly speaking, only the modifications that seek a minimum measurable length are called the Generalized Uncertainty Principle. However, in this work we will refer to any modification of the uncertainty principle in the form of a modification to the Heisenberg algebra as a GUP for simplicity of notation. 

The easiest way to incorporate a minimum measurable length to the quantum theory is by considering that the GUP takes the form \cite{Kempf:1994su}
	\begin{equation}\label{SimplestGUP}
		[x,p]=i(1+\gamma^2p^2) ,
	\end{equation}
where $\gamma$ is a small parameter which will play the role of the minimal measurable length. In general it can be shown that proposing a GUP in the form\cite{Shababi:2019ysa}
	\begin{equation}
		[x,p]=iF(p^2) ,
	\end{equation}
a minimal measurable length can be incorporated to the theory if  the function $F(x^2)/x$ has a minimum, provided that the states are chosen such that $\langle p\rangle=0$. Therefore, a wide variety of GUPs have been proposed in the literature. Some examples are the following:
	\begin{itemize} 
		\item It can be proposed a GUP with a function that coincides with (\ref{SimplestGUP}) on a Taylor series at first order, for example the GUP inspired by non-commutative Snyder space \cite{Snyder:1946qz,Battisti:2008du} in the form
		\begin{equation}\label{GUPSR}
			[x,p]=i\sqrt{1-\gamma p^2} .
		\end{equation}
		\item In the same spirit, an exponential proposal have also been studied \cite{Miao:2013wua}
		\begin{equation}
			[x,p]=ie^{\gamma p^2} .
		\end{equation}
		\item A function that generalizes  (\ref{SimplestGUP}) as a geometric series have  been studied  as well in \cite{Pedram:2011gw}
		\begin{equation}\label{GUPSeries}
			[x,p]=\frac{i}{1-\gamma^2p^2}.
		\end{equation}
		This form is also compatible with Doubly Special Relativity and loop quantum cosmology.  
	\end{itemize}
These kind of proposals have been extensively studied in the literature. From the purely theoretical point of view see for example \cite{Kempf:1994su,Anacleto:2015mma,Anacleto:2015rlz,Ali:2015ola,Pramanik:2014mma,Faizal:2015Dir,Faizal:2016zlo,Masood:2016wma}. For studies that look to constrain the GUP parameters by established measurements from the classical as well as the quantum point of view see  for example \cite{Scardigli:2016ubl,Scardigli:2016pjs,Lambiase:2017adh,Vagenas:2017fwa,Bosso:2017hoq,Demir:2018akw,Bosso:2018ckz,Fu:2021zrd}. For a proposal of direct measurement in experiment see \cite{Bushev:2019zvw}.

\subsection{Cosmological applications}\label{SS-GUPCosmolApp}
Any of the GUP proposals have a significant impact on the quantum theory since it completely modifies the involved operators \cite{Kempf:1994su}. Furthermore, a cosmological setup represents a natural interesting scenario to study the implications of these generalizations since as we stated in the Introduction, the early universe provides a natural framework to study quantum gravity phenomenology. The different forms of cosmological applications derived from a GUP that have been studied previously are:
	\begin{itemize}
		\item Modifications to the WDW equation: The GUP can be implemented in the variables of superspace, leading in this way to a deformation of the  Wheeler-DeWitt equation, this have been performed for example in \cite{Garcia-Compean:2001jxk,Vakili:2007yz,Vakili:2008tt,Kober:2011uj,Zeynali:2012tw,Faizal:2014rha,Faizal:2015kqa,Garattini:2015aca,Gusson:2020pgh,GUP,Garcia-Compean:2021wgv}
		\item Modifications to the scalar field modes: An important feature of standard inflationary scenarios is the quantization of the inflaton modes that are responsible for the CMBR spectrum. A GUP can also be applied to the modes in this procedure, this has been carried out in \cite{Kempf:2000ac,Ashoorioon:2004vm,Ashoorioon:2004wd,Ashoorioon:2005ep}, where it was shown that there appears an ambiguity regarding the action for the scalar modes.
		\item A modification to the Friedmann equation for a perfect fluid or for a scalar field (\ref{FriedmannStandardC}) can also be pursued by some classical limit starting from the modifications derived from the GUP at the quantum level. This has been performed in different ways:
			\begin{itemize}
				\item It can be obtained using a thermodynamical approach, this has been carried out in \cite{Cai:2005ra,Zhu:2008cg,Giardino:2020myz} where it was found that the results are identical to the behaviour encountered with the standard HUP case, at least for the forms of the GUP considered.
				\item An scalar field can be incorporated by reconstructing the effects of the GUP in the form of a new action for the scalar field, this was carried out  in \cite{Paliathanasis:2015cza,Giacomini:2020zmv}, leading to compatibility with an exponentially expanding universe. However, we note that in this case the GUP is only in the scalar field, not in the scale factor.
				\item A procedure using the noncommutative approach have also been performed in \cite{Bina:2007wj,Leon:2022oeo,Perez-Payan:2011cvf,Toghrai:2021nuz}.
				\item A classical limit can be implemented through the modification of the Poisson brackets in the same form as the GUP modifies the corresponding quantum commutators. In this way, a universe filled with a perfect fluid was considered in \cite{Battisti:2008du,Ali:2014hma,Moumni:2020uki,Atazadeh:2016yeh}. It was found that the resulting solutions describe a cosmological bounce, or an static universe depending on the specific form of the GUP.
			\end{itemize}
		\item  A similar treatment exploring and comparing, the classical and the quantum implications of a GUP and polymer quantum mechanics, in the context of cosmology have been performed in \cite{Barca:2019ane,Barca:2021epy,Barca:2023epu}.   
	\end{itemize}

Therefore, the GUP proposal have many applications in cosmology ranging from the classical to the quantum point of view. We can use it in the coordinates of space to infer non-trivial modifications to the quantum theories that we know and in that way look for signatures to quantum gravity. Moreover, we can also use such proposals on the canonical formalism of section \ref{S-CQG}. In that case, the GUP is an ingredient that incorporates to the WDW equation an extra quantum gravitational feature. In this way we expect to improve the ability of such formalism to describe the quantum behaviour. In the present work we will be interested in both applications of the GUP, at the classical and at the quantum level.

\section{Swampland conjectures}\label{S-SConjectures}
The first string theory was proposed many decades ago. The area has grown vastly and it now represents a framework where different realizations can be explored. As we stated in the Introduction, the  swampland conjectures \cite{Vafa:2005ui,Brennan:2017rbf,Ooguri:2006in,vanBeest:2021lhn,Grana:2021zvf} try to constrain an effective theory so we can discover if it can be completed to string theory in the ultraviolet, therefore belonging to the landscape, for an extensive review see \cite{Palti:2019pca}. These conjectures try to extract  key elements of a quantum gravity theory from generic behaviours found on string theoretic realizations. Although features derived from string theory models are used as guidelines, the conjectures are expected to be valid in a general sense, even if string theory fails to be the complete form of quantum gravity. 

The cosmological scenario have naturally attracted much attention in string theory, seeking to find a suitable construction within this framework that can correctly describe the physics involved compatible with observations. In particular, from the inflationary point of view, many candidates for the inflaton field have been proposed from string theory (see, for example, \cite{Baumann:2014nda}). Therefore, we can expect that a cosmological setup is a natural scenario to test the viability of the conjectures, or assuming their validity, use them as a new ingredient to constrain the models, once we identify them as effective theories. Let us briefly summarize some of the conjectures that have direct consequences on inflationary models:
	\begin{itemize}
		\item The de Sitter conjecture: Despite many different efforts that have been pursued, a dS background  has been impossible to obtain starting with a complete proposal on any higher dimensional string theory. This apparent impossibility have led to the dS conjecture that now claims that it is not possible to obtain such background \cite{Obied:2018sgi,Andriot:2018wzk,Roupec:2018mbn}. This conjecture implies a direct conflict with the standard slow roll inflationary scenario of General Relativity as we will see in the next subsection.
		\item The Distance conjecture: This conjecture constrain the distance on field space that the scalar fields can travel, in order to not spoil the effective field theory validity. In other words, it implies a maximum limit for $\Delta\phi$, where $\phi$ is any scalar field of the EFT. This can have immediate consequences for the duration of inflation for example. For relations to the de Sitter conjecture see \cite{Ooguri:2018wrx}.
		\item The Trans-planckian Censorship conjecture: In the usual inflationary model, it is widely accepted that quantum fluctuations of the inflaton are the seeds for the late-time structure formation, since such fluctuations grow because of the exponential expansion. However, it is known that a trans-planckian problem exists in this scenario, that is if going back in time we are lead to a perturbation that is smaller than the Planck length, since in such scenario the viability of the effective theory breaks down. In order to avoid this problem, this conjecture states that any effective field theory consistent with quantum gravity can not generate such trans-planckian problem, in this way setting bounds for the Hubble parameter \cite{Bedroya:2019snp}. A relation with the distance conjecture can be found in \cite{Andriot:2020lea,Brahma:2019vpl}.
	\end{itemize}

Although the three conjectures have direct implication to cosmological  models, in the following we will focus only on the dS conjecture. For some recent references regarding the other conjectures see for example \cite{Agrawal:2018own,Scalisi:2018eaz,Lin:2019pmj,Tenkanen:2020xbb}.

\subsection{The dS conjecture}\label{SS-dSConjecture}
We consider a theory of gravity (not necessarily GR) coupled  to a single scalar field, as standard in inflationary models. The first proposal of the dS conjecture was written originally in the following form  \cite{Obied:2018sgi,Andriot:2018wzk,Roupec:2018mbn,Ooguri:2018wrx,Agrawal:2018own}:
	\begin{equation}\label{dSConjecture}
		|V_{\phi}|\geq cV ,
	\end{equation}
where $V$ is the scalar field potential, $V_{\phi}$ is the derivative of the potential with respect to the scalar field, and $c$ is an order $1$ constant. Later on, this conjecture was refined to incorporate another possibility, which is written as 
	\begin{equation}\label{dSConjecture2}
		V_{\phi\phi}<-\widetilde{c}V ,
	\end{equation}
where $\widetilde{c}$ is another order $1$ constant. Now, the dS conjectures states that either one of the two inequalities, the first dS conjecture (\ref{dSConjecture}), or the second dS conjecture (\ref{dSConjecture2}), must be satisfied. We can easily see that (\ref{dSConjecture}) and (\ref{dSConjecture2}) are incompatible with the standard slow-roll conditions that require the parameters in (\ref{DefSlowRollP}) to be small. Therefore,  the dS conjecture is not compatible with the standard cosmological picture of an inflationary universe driven by a scalar field that obeys the slow-roll conditions on General Relativity \cite{Garg:2018reu,Ben-Dayan:2018mhe,Kinney:2018nny}. However, there are many forms to obtain an inflationary scenario and consistency with this conjecture. For example considering warm inflation \cite{Motaharfar:2018zyb}, multi-field inflation \cite{Achucarro:2018vey}, or generalizations of GR in the form of $f(R)$ and Brans-Dicke theories \cite{Artymowski:2019vfy,Benetti:2019smr,Elizalde:2022oej}  among other proposals \cite{Denef:2018etk,Trivedi:2020wxf,Yi:2018dhl,Brahma:2019kch,Trivedi:2021nss}.

The swampland conjectures are a young area that have attracted much attention and can have potential deep implications on cosmological models, as well as effective theories that tries to capture some aspects of the quantum gravity behaviour, for example the implication of these conjectures to the wave function of the universe have been studied in \cite{Brahma:2020cpy,Matsui:2020tyd}. Although this program relies on conjectures, and for that reason it can comprehensively be seen with scepticism, it represents a new angle to study cosmological models. With this idea in mind, in chapter \ref{CH-HLFRSwampland} we will use the dS conjecture to explore its compatibility with the $F(\bar{R})$ Ho\v{r}ava-Lifshtz theories. We will se how this procedure can lead to an argument that supports the conjecture a posteriori, as well as a possible indication that the conjecture may be applicable to more scenarios than originally thought.
	\chapter{Scalar field inflation driven by a classical limit of the GUP}\label{CH-GUPInflation}
	
	In this chapter we present the study of the modifications induced on scalar field inflation, by considering a general form of the GUP through a modification of the Heisenberg algebra. 
	
	As we pointed out in subsection \ref{SS-GUPCosmolApp} there are various ways to find the modifications on the Friedmann equation as a consequence of a GUP. In this chapter we will use a classical limit in which a modification of the quantum commutator will correspond to a modification of the corresponding Poisson bracket on the classical theory. This method has been explored before when the only source of the gravitational field is a perfect fluid \cite{Battisti:2008du,Ali:2014hma,Moumni:2020uki,Atazadeh:2016yeh}, in section \ref{S-NoScalarField} we will review these works. Then, in section \ref{S-ScalarField}  we will present the generalization of this treatment by taking into account a scalar field with an arbitrary potential. However, as we pointed in subsection \ref{SS-GUPProposals}, there are many scenarios that motivates a modification of the Heisenberg algebra and can take many forms and imply different behaviours. Therefore, the analysis will not be restricted to a specific form of the GUP, instead we will constrain the functions defining the GUP by looking for an appropriate cosmological behaviour. Thus, we will employ the GUP by general proposals which will be categorized in four cases that will be treated separately in subsections \ref{S-Case1}-\ref{S-Case4}. We will encounter that for some proposals the way in which inflation can be obtained is by approximations to the equations of motion in the same line as the ones used in subsection \ref{SS-Inflation}, but in this case three slow roll Hubble parameters will be needed, instead of the standard two. Furthermore, they will depend on the standard slow roll parameters in a very different form than in the usual case, in some cases they can be less restrictive to the form of the potential, therefore allowing for more ways to describe inflation. Furthermore, we will show that for all the general forms of the GUP considered there is an analytical solution which describes a constant Hubble parameter, that is an exponentially expanding universe, which will be found to be a generalization to the cosmological constant solution. It will be found that even if the modifications implied by the GUP can be neglected in some limit at the same time that the cosmological constant is set to vanish, the inflationary solution is present when the GUP is relevant. Therefore, an appropriate modification of the Heisenberg algebra is sufficient to produce an exponentially expanding universe driven by a scalar field, there is no need of any other source. The results of this chapter were first presented in \cite{Garcia-Compean:2023aaj}.
	
	\section{Modifications to the cosmology of a perfect fluid by a GUP}
	\label{S-NoScalarField}
	Let us begin with a brief discussion of how the modification from the Heisenberg algebra derived from the GUP have been studied in a classical limit, in particular  how the Friedmann equations are modified at the classical level when the only gravitational source  is a perfect isotropic fluid with energy density $\rho$ and pressure $p$. In this chapter we will consider General Relativity, thus the gravitational action will be given by the Einstein-Hilbert action, furthermore since we want to explore the Friedmann equation, we will consider the FLRW flat metric (\ref{DefFlatFLRW}). The Hamiltonian constraint that follows from the standard formulation is written as
	\begin{equation}\label{SistNSF1}
		\mathcal{H}=-\frac{\kappa}{12a}P^2_{a}+a^3\rho\simeq0 ,
	\end{equation}
	where $P_{a}$ is the canonical momentum associated to the scale factor $a$. Following the standard procedure, the Hamilton equations of motion will be given by
	\begin{equation}\label{SistNSF2}
		\dot{a}=\{a,\mathcal{H}\}=\frac{\partial \mathcal{H}}{\partial P_{a}} , \hspace{1cm} \dot{P_{a}}=\{P_{a},\mathcal{H}\}=-\frac{\partial\mathcal{H}}{\partial a}.
	\end{equation}
	Furthermore, the fluid obeys the standard conservation law
	\begin{equation}\label{SistNSF3}
		\rho+3H(\rho+p)=0 ,
	\end{equation}
	where $H$ is the Hubble parameter (\ref{DefHubbleParameter}). As in the standard scenario, the first equation in (\ref{SistNSF2}) is just the definition of $P_{a}$. The Friedmann equation is obtained as in subsection (\ref{SS-FriedmannHamiltonian}), that is after substituting the momentum $P_{a}$ into the Hamiltonian constraint (\ref{SistNSF1}), in this case this leads to
	\begin{equation}\label{FriedmannStandard}
		H^2=\frac{\kappa}{3}\rho .
	\end{equation}
	Once again, the second equation in (\ref{SistNSF2}) leads to the second Friedmann equation after using (\ref{SistNSF3}) and can be derived from (\ref{FriedmannStandard}). Therefore the scenario is the same as the one presented in subsection \ref{SS-FriedmannHamiltonian} with the only difference that the fluid is the only gravitational source.
	
	When the canonical quantization procedure is used, an equivalence between the classical Poisson brackets of any two variables and the quantum commutator of their corresponding operators is established, this takes the form
	\begin{equation}\label{PoissionConm}
		\{\cdot,\cdot\}\leftrightarrow i[\cdot,\cdot] .
	\end{equation}
	 where $\{\cdot,\cdot\}$ denotes the classical Poisson bracket and  $[\cdot,\cdot]$ the quantum commutator. Thus, by employing this relation we can obtain modifications of the classical equations of motion by modifying the Poisson brackets whenever the commutator is modified by the GUP. In particular, with this procedure  the modifications to the Friedmann equations can be obtained. Let us consider a general form of a GUP 
	\begin{equation}
		[a,P_{a}]=iF(P^2_{a}) ,
	\end{equation}
	where $F$ is any function that must depend on a parameter $\gamma$ such that $\lim_{\gamma\to0}F=1$ (as in all the cases shown in subsection \ref{SS-GUPProposals}), this proposal can lead to a minimum value for the position if the function is chosen properly. Therefore, the Poisson brackets will be modified as
	\begin{equation}\label{NSFConmu}
		\{a,P_{a}\}\rightarrow\{a,P_{a}\}F .
	\end{equation}
	
	Considering a Hamiltonian system with coordinates $x_{i}$ and momenta $p_{i}$, in general we can express the Poisson bracket of any two functions $A$ and $B$ in the form
	\begin{equation}\label{PoissonGeneral}
		\{A,B\}=\left(\frac{\partial A}{\partial x_{i}}\frac{\partial B}{\partial p_{j}}-\frac{\partial A}{\partial p_{i}}\frac{\partial B}{\partial x_{j}}\right)\{x_{i},p_{j}\} .
	\end{equation}
	Therefore, in the case at hand, we obtain from (\ref{NSFConmu}) and (\ref{PoissonGeneral}) the following equations of motion
	\begin{equation}\label{SistNSFGUP1}
		\dot{a}=\frac{\partial H}{\partial P_{a}}F(P^2_{a}) ,\hspace{1cm}
		\dot{P}_{a}=-\frac{\partial H}{\partial a}F(P^2_{a}) .
	\end{equation}
	We remark that the form of the Hamiltonian constraint (\ref{SistNSF1}) is not modified by the GUP. However, we can interpret the first equation in (\ref{SistNSFGUP1}) as a modification to the definition of the momentum $P_{a}$. Therefore, even if the form of the Hamiltonian constraint is unaltered, the Friedmann equation will be modified in a particular form for every function $F$. We note that in order to obtain the Friedmann equation, the first equation in (\ref{SistNSFGUP1})  must be solved for the momentum, and then substitute the solution back into the Hamiltonian constraint. Thus, this procedure can not be carried out in general, we must know the specific form of the $F$ function to proceed. Let us review  some particular functions that have been studied previously:
	\begin{itemize}
		\item Firstly, taking the GUP defined by (\ref{GUPSR}), the Poisson brackets will be modified as
		\begin{equation}
			\{a,P_{a}\}=\sqrt{1-\gamma P^2_{a}} ,
		\end{equation}
		then, the first equation in (\ref{SistNSFGUP1}) takes the form
		\begin{equation}\label{GUP1EOM}
			\dot{a}=-\frac{\kappa N}{6a}P_{a}\sqrt{1-\gamma P^2_{a}} ,
		\end{equation}
		which is actually a quartic equation for $P_{a}$. Substituting the solution back into (\ref{SistNSF1}), we obtain the modified Friedmann equation \cite{Battisti:2008du} 
		\begin{equation}\label{ModifiedFriedmann1}
			\left(\frac{\dot{a}}{a}\right)^2=\frac{\rho \kappa}{3}\left[1-\frac{12\gamma\rho}{\kappa}a^4\right] .
		\end{equation}
		\item Next, we consider a GUP that is compatible with DSR. For a single coordinate it takes the form
		\begin{equation}
			\{a,P_{a}\}=1-2\gamma P_{a} ,
		\end{equation}
		the equation that defines the momentum leads to a quadratic equation, with the solutions we obtain the modified Friedman equation \cite{Ali:2014hma}
		\begin{equation}\label{ModifiedFriedmann2}
			\left(\frac{\dot{a}}{a}\right)^2=\frac{\kappa\rho}{3}\left[1-8\gamma\sqrt{\frac{3\rho}{\kappa}}a^2+\frac{48\gamma^2\rho}{\kappa}a^4\right] .
		\end{equation}
		\item Taking now the GUP defined by the geometric series (\ref{GUPSeries}), we have
		\begin{equation}
			\{a,P_{a}\}=\frac{1}{1-\gamma P^{2}_{a}} .
		\end{equation}
		which leads to quadratic equation for $P_{a}$ and to the modified Friedmann equation \cite{Moumni:2020uki}
		\begin{equation}\label{ModifiedFriedmann3}
			\left(\frac{\dot{a}}{a}\right)^2=\frac{\kappa\rho}{3}\left[\frac{12\rho\gamma}{\kappa}a^4-1\right]^{-2} .
		\end{equation}
	\end{itemize}
	These results show, that as we pointed out, the modified Friedmann equation is determined by the specific form of the $F$ function considered. However, all the equations obtained describe a cosmological bounce, i.e. there is a value of the scale factor different from zero for which $\dot{a}=0$. Thus, the cosmology contained in these equations is not related to the standard inflationary paradigm. This may seem as a generic feature when the only gravitational source is a perfect fluid, however things can change significantly if we consider instead a scalar field equipped with a potential. This scenario was considered by a noncommutative approach (similar to a GUP) in \cite{Bina:2007wj,Toghrai:2021nuz}. Moreover, considering a GUP and a scalar field but without a potential was performed in \cite{Battisti:2008du}, however the analysis was simplified by a proper choice for the lapse function. Furthermore, a study with a GUP and a potential using a numerical analysis was presented in \cite{Paliathanasis:2015cza,Giacomini:2020zmv}, these results points towards an agreement between the GUP modified scenario and considering a scalar field with the standard inflationary behaviour.  Therefore, in the following section we will generalize the analysis presented in this section to take into account a scalar field with a general potential and look for consistency with the standard inflationary paradigm by using the classical limit just employed to obtain analytical solutions.
	
	\section{Modifications to the cosmology of a scalar field by a GUP}
	\label{S-ScalarField}
	The standard Hamiltonian treatment when the gravitational source is a scalar field $\phi$ with its corresponding potential $V$ was presented in subsection \ref{SS-FriedmannHamiltonian}. In particular, we showed that the Hamiltonian constrain takes the form
		\begin{equation}\label{SistStandardHC}
			\mathcal{H}=-\frac{\kappa}{12a}P^2_{a}+\frac{1}{2a^3}P^2_{\phi}+a^3V(\phi)\simeq0 .
		\end{equation}
	We now proceed as before, that is, by modifying the Poisson brackets of the classical theory. As we saw in the last section, the Friedmann equation depends on the specific form of the GUP considered, however we intend to present an analysis as general as possible. Therefore we will consider a general form of the GUP with arbitrary functions that can depend on the coordinates and the momenta and we will restrict its form only by seeking a consistency with the standard cosmological behaviour. However, as was shown in the last section, the variables in which the function can depend is important since the GUP will modify the equations that  defines the momenta. Thus we will categorize the possible general forms of the GUP and treat each case separately. Such classification is defined as follows.
	
	\begin{itemize}
		\item \textbf{Case I}. The simplest form of the GUP is described by the diagonal case in which
		\begin{equation}\label{DefCase1}
			[a,P_{a}]=[\phi, P_{\phi}]=iF(a,\phi) ,
		\end{equation}
		where $F(a,\phi)$ is considered to be a function only of the coordinates $a$ and $\phi$.
		\item \textbf{Case II}. Moving on to consider the simplest way to incorporate a dependence on the momenta, we propose a GUP described by two different functions, one for each coordinate in the following form
		\begin{equation}\label{DefCase2}
			[a,P_{a}]=iF_{1}(a,P_{a}) , \hspace{1cm} [\phi,P_{\phi}]=iF_{2}(\phi,P_{\phi}) .
		\end{equation}
		\item \textbf{Case III}. The next scenario considers a similar form as in case I, but now the function can depend on the coordinates as well as the momenta, it is a more complicated form but still has a diagonal form 
		\begin{equation}\label{DefCase3}
			[a,P_{a}]=[\phi, P_{\phi}]=iF(a,\phi,P_{a},P_{\phi}) .
		\end{equation}
		\item \textbf{Case IV}. Finally, the most general GUP proposal is obtained in the non-diagonal case, that is when there are four functions modifying the commutators in the following form
		\begin{equation}\label{DefCase4}
			[a,P_{a}]=iF_{1} , \hspace{1cm} [\phi,P_{\phi}]=iF_{2} , \hspace{1cm} [a,P_{\phi}]=iG_{1} , \hspace{1cm} [\phi,P_{a}]=iG_{2},
		\end{equation}
		where all functions $F_{1,2}$ and $G_{1,2}$ can depend on the momenta as well as both coordinates.
	\end{itemize}
	In the following subsections we will study the modifications in the classical equations of motion for each case separately, paying particular attention to the implications on the inflationary scenario.
	
	\subsection{Case I}
	\label{S-Case1}
\begin{figure}[h]
		\centering
		\includegraphics[width=0.6\textwidth]{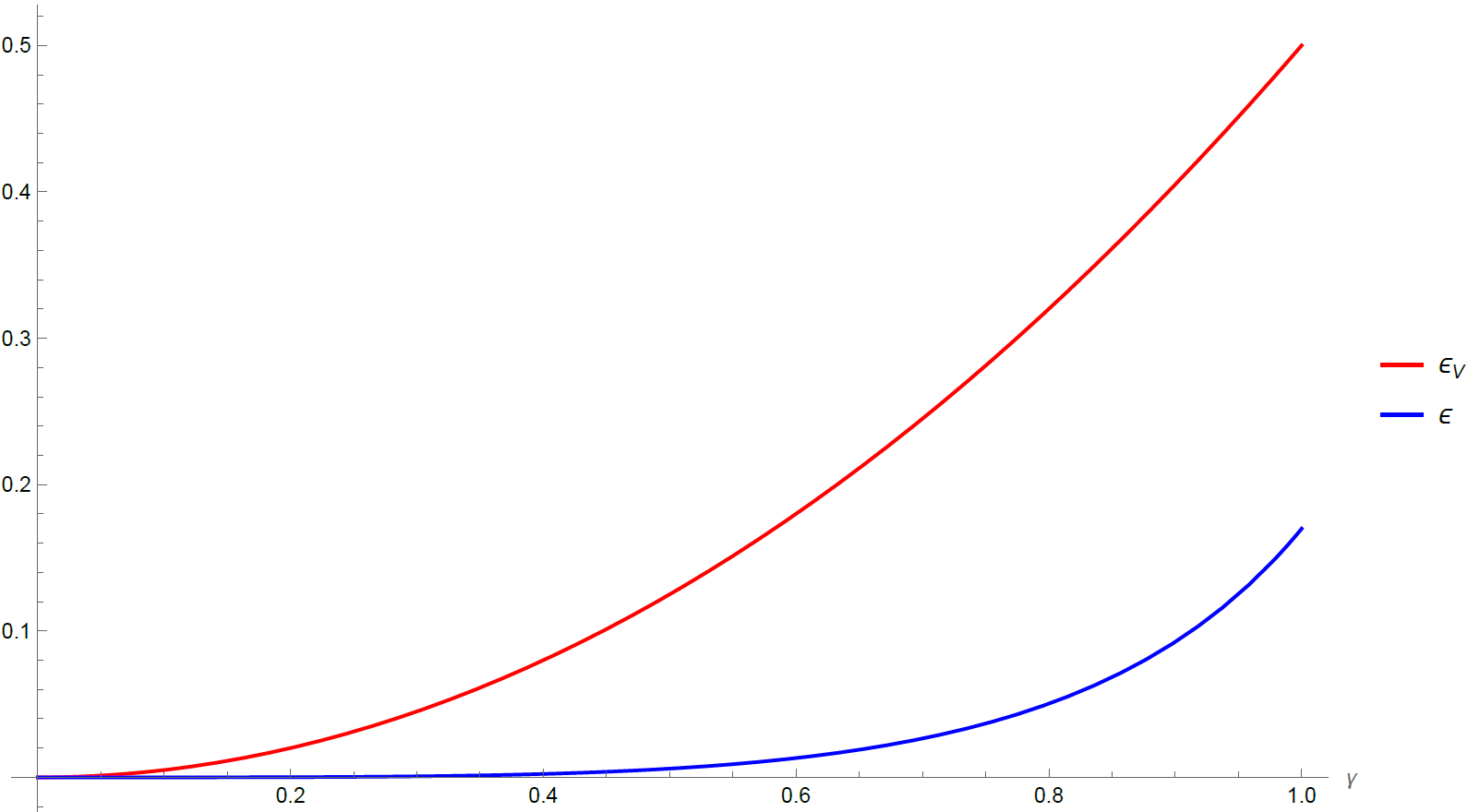}
		\caption{Slow roll Hubble parameters that guarantee the inflationary behaviour for an exponential potential. The standard slow roll parameter is shown in the red curve, whereas the Hubble parameter in the GUP scenario is shown in the blue curve. We find that $\epsilon<\epsilon_{V}$ in all the range. } \label{FigParameters}
	\end{figure}

	Let us start by considering the simplest GUP proposal, in the diagonal form with no dependence on the momenta, defined by (\ref{DefCase1}). The equations of motion are modified as
	\begin{equation}\label{SistCase1-1}
		\dot{a}=\frac{\partial \mathcal{H}}{\partial P_{a}}F=-\frac{\kappa P_{a}}{6a}F , \hspace{1cm} \dot{\phi}=\frac{\partial\mathcal{H}}{\partial P_{a}}F=\frac{P_{\phi}}{a^3}F ,
	\end{equation}
	\begin{equation}\label{SistCase1-2}
		\dot{P}_{a}=-\frac{\partial\mathcal{H}}{\partial a}F , \hspace{1cm} \dot{P}_{\phi}=-\frac{\partial\mathcal{H}}{\partial\phi}F .
	\end{equation}
	Let us remark that this is the simplest case because the redefinition of the momenta from (\ref{SistCase1-1}) is trivial, the equation can be solved in general in terms of the coordinates as
	\begin{equation}\label{Case1RedMomenta}
		P_{a}=-\frac{6a\dot{a}}{\kappa F} , \hspace{1cm} P_{\phi}=\frac{a^3\dot{\phi}}{F} .
	\end{equation}
	then, substituting back into the Hamiltonian constraint (\ref{SistStandardHC})  we obtain the modified Friedmann equation
	\begin{equation}\label{Case1Friedmann}
		H^2=\frac{\kappa}{3}\left[\frac{\dot{\phi}^2}{2}+F^2(a,\phi)V\right].
	\end{equation}
	The modified equation of motion for the scalar field is obtained from the second equation in (\ref{SistCase1-2}) leading to
	\begin{equation}\label{Case1ScalarEq}
		\ddot{\phi}+(3H-H_{F})\dot{\phi}+F^2V'=0 ,
	\end{equation}
	where we have defined $H_{F}=\frac{\dot{F}}{F}$. With these set of equations we can explore the modifications in the standard picture of inflation without the need of specifying a particular form for $F$. As we presented in subsection \ref{SS-Inflation}, in the standard scenario, scalar field inflation is realized after making two approximations, in the Friedmann equation (\ref{FriedmannStandardC}) it is assumed that $\frac{\dot{\phi}^2}{2}\ll V$ and in the scalar field equation (\ref{FieldEoMStandard}) it is imposed that $\ddot{\phi}\ll H\dot{\phi}$.  These approximations are guaranteed when the two slow roll Hubble parameters (\ref{DefStandardHubbleSLowRoll}) are small, which is satisfied when the standard slow roll parameters  (\ref{DefSlowRollP}) are small. Let us proceed in an analogous manner, that is, in the Friedmann equation (\ref{Case1Friedmann}) we are going to neglect the kinetic term $\frac{\dot{\phi}^2}{2}$ with respect to $F^2V$ and in the field equation (\ref{Case1ScalarEq}) we are going to neglect $\ddot{\phi}$ with respect to the friction term $(3H-H_{F})\dot{\phi}$. By doing these approximations we hope to obtain an exponential expansion as in the standard case. Thus let us define the slow roll Hubble parameters as
	\begin{equation}\label{Case1DefHubbParameters}
		\epsilon=-\frac{\dot{H}}{H^2} , \hspace{1cm} \eta=\frac{\ddot{\phi}}{(3H-H_{F})\dot{\phi}} .
	\end{equation}
	With these definitions we obtain
	\begin{equation}
		\epsilon=\frac{3\dot{\phi}^2}{2F^{2}V}-\frac{3HH_{F}}{kF^2V} ,
	\end{equation}
	then in this case assuming $\epsilon\ll1$ does not guarantee that the kinetic term can be neglected with respect to the potential. We can guarantee this fact by defining a third Hubble parameter as
	\begin{equation}\label{Case1DefHubbParameters2}
		\omega=\frac{\dot{\phi^2}}{2F^2V} .
	\end{equation}
	With these parameters, the approximations will be justified with an approximately constant $H$ by demanding $\epsilon,\eta,\omega\ll1$. Furthermore, these Hubble parameters can be related to the standard slow roll parameters by
	\begin{equation}\label{Case1Epsilon}
		\epsilon=\epsilon_{V}\left[\frac{1+\frac{H_{F}}{3\epsilon_{V} H^2}\left(2H_{F}-3H-\frac{H^2_{F}}{3H}\right)}{1+\frac{H_{F}}{9H^2}\left(H_{F}-6H\right)}\right] ,
	\end{equation}
	\begin{equation}\label{Case1Eta}
		\eta=\frac{\eta_{V}}{3\left[1+\frac{H_{F}}{9H^2}(H_{F}-6H)\right]}-\frac{3H^2}{(3H-H_{F})^2}\epsilon-\frac{2H_{F}}{3H-H_{F}}-\frac{\dot{H}_{F}}{(3H-H_{F})^2} .
	\end{equation}
	Since we only require that $\epsilon,\eta,\omega\ll1$, we note from the above expression that this can be achieved even if $\epsilon_{V}$ or $\eta_{V}$ are of order 1, the constrictions can be put in the form of the $F$ function. This implies a bigger set of possibilities since the constrictions on the form of the potential can be relaxed. However, the approximations used only guarantee that we can neglect the kinetic term in the Friedmann equation (\ref{Case1Friedmann}), thus we obtain
	\begin{equation}\label{Case1AproxH}
		H^2\thickapprox \frac{\kappa}{3}F^2V .
	\end{equation} 
	Therefore, in order to obtain a solution describing exponential expansion, the $F$ function must be chosen appropriately. Assuming the standard slow roll criteria $\epsilon_{V},\eta_{V}\ll1$ we obtain that $V$ is approximately constant, therefore if we choose $F$ to be a function only of the scalar field potential, that is taking $F=F(V(\phi))$, the above analysis guarantees that there will be exponential expansion if $F$ leads to  $\epsilon,\eta,\omega\ll1$. On the other hand, if we do not restrict $V$ to be a constant (contrary to the standard slow roll criteria), we see from (\ref{Case1AproxH}) that a constant Hubble parameter can still be present if $F$ is chosen in such a way that the term $F^2V$ is a constant. Let us consider a particular example to illustrate this possibility. In order to have a constant $\epsilon_{V}$ parameter  we choose $V=V_{0}e^{\gamma\phi}$, where $\gamma$ is the constant related to the GUP and $V_{0}$ is an independent constant, then an appropriate form for the GUP function is
	\begin{equation}
		F=\sqrt{\frac{A}{V_{0}}}e^{-\frac{\gamma}{2}\phi} , 
	\end{equation}
	where $H\thickapprox\frac{\kappa A}{3}$ with $A$ a positive constant, then we obtain
	\begin{equation}
		\epsilon\thickapprox\frac{\gamma^2}{2\kappa}\left[\frac{1-\sqrt{1-\frac{2\gamma^2}{3\kappa}}}{1-\frac{\gamma^2}{3\kappa}+\sqrt{1-\frac{2\gamma^2}{3\kappa}}}\right] ,
	\end{equation}
	\begin{equation}
		\eta\thickapprox\frac{\frac{2\gamma^2}{3\kappa}}{1-\frac{\gamma^2}{3\kappa}+\sqrt{1-\frac{2\gamma^2}{3\kappa}}}-\frac{\frac{2}{3}\epsilon}{1-\frac{\gamma^2}{3\kappa}+\sqrt{1-\frac{2\gamma^2}{3\kappa}}}-\frac{2\left(1-\sqrt{1-\frac{2\gamma^2}{3\kappa}}\right)}{1+\sqrt{1-\frac{2\gamma^2}{3\kappa}}} ,
	\end{equation} 
	\begin{equation}
		\omega\thickapprox\frac{3\kappa}{\gamma^2}\left[1-\frac{\gamma^2}{3\kappa}-\sqrt{1-\frac{2\gamma^2}{3\kappa}}\right].
	\end{equation}
	In this case the standard slow roll parameters are given by $\epsilon_{V}=\frac{\gamma^2}{2\kappa}$ and $\eta_{V}=\frac{\gamma^2}{\kappa}$. From the above we find that the absolute value of the three parameters increase when $\gamma$ increases, thus the three are small when $\gamma$ is small. In Figure \ref{FigParameters} we show the behaviour of the GUP Hubble parameter $\epsilon$ and the slow roll parameter $\epsilon_{V}$ (that coincides with the standard Hubble parameter for small $\gamma^2$). We show that $\epsilon<\epsilon_{V}$, therefore the scenario modified by the GUP is less restrictive than the standard slow roll scenario. For example, choosing $\gamma=0.5$ (and $\kappa=1$) we obtain $\epsilon\thickapprox0.00595$, $\eta\thickapprox-0.00217$, $\omega\thickapprox0.0455$, thus the approximations are well justified. On the other hand $\epsilon_{V}=0.125$ which is two orders of magnitude bigger than $\epsilon$.
	
	We have shown that the standard inflationary scenario can be obtained by approximations as in the standard case. However, using the system of equations modified by the GUP, it is possible to perform a full analytical analysis. The set of equations describing the cosmological setup are the Hamiltonian constraint (\ref{SistStandardHC}) and the modified equations of motion (\ref{SistCase1-1}) and (\ref{SistCase1-2}). We can express the latter in the form
	\begin{equation}\label{Case1SisE2}
		P_{a}=-\frac{6a\dot{a}}{\kappa F}  , \hspace{1cm}
		P_{\phi}=\frac{a^3\dot{\phi}}{F} ,
	\end{equation}
	\begin{equation}\label{Case1SisE4}
		\dot{P}_{a}=-\left[\frac{\kappa}{12a^2}P^2_{a}-\frac{3}{2a^4}P^2_{\phi}+3a^2V\right]F , \hspace{1cm}
		\hspace{1cm} \dot{P}_{\phi}=-a^3V'F .
	\end{equation}
	As we stated earlier, $F$ has a dependence on a parameter $\gamma$ such that $\lim_{\gamma\to0}F=1$ in order to recover the standard scenario in such limit. We then propose an exponentially expanding universe ansatz,  that is $a(t)=a_{0}e^{Ht}$ with $a_{0}$ and $H$ constants. It can be shown that the analytical solution of this system of five equations is
	\begin{equation}\label{Case1Solutions1}
		P_{a}=-\frac{6a^2_{0}H}{\kappa}\frac{e^{2Ht}}{F} , \hspace{1cm} P_{\phi}=\pm a^3_{0}e^{3Ht}\sqrt{\frac{2H\dot{F}}{\kappa F^3}} ,
	\end{equation}
	\begin{equation}\label{Case1Solutions2}
		\dot{\phi}=\pm\sqrt{\frac{2H}{\kappa}}\sqrt{\frac{\dot{F}}{F}} , \hspace{1cm}
		V=\frac{H}{\kappa F^3}\left(3HF-\dot{F}\right) .
	\end{equation}
	From consistency of the above equations we impose that $\frac{\dot{F}}{F}\geq0$. However, from the last equation we can write the Hubble parameter as
	\begin{equation}
		H=\frac{\dot{F}}{6F}\left[1\pm\sqrt{1+\frac{12\kappa F^4V}{\dot{F}^2}}\right] ,
	\end{equation}
	thus, since we are seeking a constant $H$, we can impose the condition $\dot{H}=0$, then the last expression leads to 
	\begin{equation}
		\dot{F}^2-F\ddot{F}=0 .
	\end{equation}
	The general solution of this equation is $F=Ae^{\alpha t}$ with $A$ and $\alpha$ constants. However, in order to obtain the appropriate limit when $\gamma\to0$, and for simplicity, we choose $A=1$ and $\alpha=\alpha(\gamma)$, such that $\lim_{\gamma\to0}\alpha=0$. With this form, we obtain from (\ref{Case1Solutions1}) and (\ref{Case1Solutions2})
	\begin{equation}\label{Case1SolutionsFinal}
		P_{a}=-\frac{6a^2_{0}H}{\kappa}e^{(2H-\alpha)t} , \hspace{1cm} P_{\phi}=\pm\sqrt{\frac{2H\alpha}{\kappa}}a^3_{0}e^{(3H-\alpha)t} , \hspace{1cm} \dot{\phi}=\pm\sqrt{\frac{2H\alpha}{\kappa}} ,
	\end{equation}
	furthermore, the scalar field potential takes the form	
	\begin{equation}
		V=\frac{H}{\kappa}(3H-\alpha)e^{-2\alpha t} .
	\end{equation}
	However, consistently with the latter expression, we can write the potential in general as
	\begin{equation}\label{Case1Potential}
		V=Be^{-2\alpha t} ,
	\end{equation}
	where $B$ is an independent constant. Therefore, the resulting Hubble constant will be described by two parameters, namely $\alpha$ and $B$, taking the form
	\begin{equation}\label{Case1Hubble}
		\tcboxmath[colback=red!10!white,colframe=red]
		{
		H=\frac{\alpha}{6}\left(1+\sqrt{1+\frac{12\kappa B}{\alpha^2}}\right) .
		}
	\end{equation}
	Thus, by choosing an exponential form in the time variable for the  $F$ function  we obtain a solution that describes an exponentially expanding universe with a constant Hubble parameter, without any approximation. The scalar field is linear in time and the potential has an exponential form. We chose $\alpha$ to be determined by the constant $\gamma$ in a proper way to fulfil $\lim_{\gamma\to0}\alpha=0$, then if $B$ is independent of $\gamma$ we obtain
	\begin{equation}
		\lim_{\gamma\to0}H=\sqrt{\frac{\kappa B}{3}}.
	\end{equation}
	We also note that in this limit $\lim_{\gamma\to0}\dot{\phi}=0$ and $\lim_{\gamma\to0}V=B$. Therefore, we can identify $B$ with the cosmological constant, since in this limit the scalar field will vanish and the potential will only be a constant. Thus, the solution obtained is actually a generalization of the well known positive cosmological constant solution, but it is generalized through a  scalar field with a corresponding potential. Since $\alpha\geq0$, we note from (\ref{Case1Hubble}) that the Hubble parameter is bigger than in the standard cosmological constant solution, thus the expansion in the presence of the GUP is faster. On the other hand, we can choose $B$ to be a function of $\gamma$, in a way in which $\lim_{\gamma\to0}B=0$ is fulfilled, then we obtain $\lim_{\gamma\to0}V=0$ and $\lim_{\gamma\to0}H=0$, that is we recover the static universe. Therefore, even if we start with a static universe, the appropriate form of the GUP will create an exponentially expanding universe by creating the appropriate scalar field and its potential. Thus the GUP is capable of creating and exponentially expanding universe in the standard form of an inflaton, without the need of a cosmological constant or any other extra source.
	
	Let us discuss the form of the scalar field potential obtained. From (\ref{Case1Solutions1}), neglecting an integration constant we obtain
	\begin{equation}\label{Case1ScalarField}
		\phi=\pm\sqrt{\frac{2H\alpha}{\kappa}}t , \hspace{1cm} V(\phi)=Be^{\mp\sqrt{\frac{2\kappa \alpha}{H}}\phi} .
	\end{equation}
	Let us choose the minus sign for the field, then 
	\begin{equation}
		\frac{V'}{V}=\sqrt{\frac{12\kappa}{1+\sqrt{1+\frac{12B\kappa}{\alpha^2}}}} .
	\end{equation}
	Considering $B$ as the cosmological constant (independent of $\gamma$) we obtain from the above expression that $\lim_{\gamma\to0}\frac{V'}{V}=0$, thus in this case the standard slow roll criteria in the $\epsilon_{V}$ slow roll parameter (\ref{DefSlowRollP}) is satisfied when $\gamma\ll1$. On the other hand, in the static universe case, $B$ has a dependence on $\gamma$, therefore if we choose $B$ such that $\lim_{\gamma\to0}\frac{B}{\alpha^2}=0$  (in units of $\kappa=1$), we obtain $\lim_{\gamma\to0}\frac{V'}{V}=\sqrt{6}$ which is an order $1$ constant. Thus, in this case the inflationary procedure does not fulfil the slow roll criteria when $\gamma\to0$. In contrast, the dS swampland Conjecture (\ref{dSConjecture}) can be fulfilled. In order to obtain consistency with the conjecture in general we are lead to the condition 
	\begin{equation}
		B\leq\frac{\alpha^2}{12}\left[\left(\frac{12}{c^2}-1\right)^2-1\right] ,
	\end{equation}
	where  $c\in[1,\sqrt{6}]$. In particular, choosing $B=\alpha^2$ this simplifies to 	
	\begin{equation}
		1\leq\frac{1}{12}\left[\left(\frac{12}{c^2}-1\right)^2-1\right] ,
	\end{equation}
	which can be fulfilled if $c\in[1,\sqrt{\sqrt{13}-1}]\thickapprox[1,1.614]$. Thus in the static universe case, we have an inflationary scenario consistent with the dS conjecture for different values of the constant $c$. This is a relevant result, since as we explained in section \ref{S-SConjectures} the conjecture is consistent for theories that have a correct UV completion, thus it is in agreement with the fact that the modification of the Heisenberg algebra that we are proposing comes from a correct quantum gravity behaviour.
	
	The condition to have a constant Hubble parameter in this case is that $F$ has an exponential form with the time variable. However, we defined this case by asking $F$ to be a function  only of the coordinates. Furthermore, from the ansatz of the scale factor and (\ref{Case1ScalarField}), we note that $F$ can be written as a power term of the scale factor or an exponential function of the scalar field, that is it can be written as 
	\begin{equation}\label{Case1F}
		F=\left(\frac{a}{a_{0}}\right)^{c_{1}}=e^{c_{2}\phi} ,
	\end{equation}
	where
	\begin{equation}
		c_{1}=\frac{\alpha}{H}=\frac{6}{1+\sqrt{1+\frac{12\kappa B}{\alpha^2}}} ,
	\end{equation}
	\begin{equation}
		c_{2}=\alpha\sqrt{\frac{\kappa}{2H\alpha}}=\sqrt{\frac{3\kappa}{1+\sqrt{1+\frac{12\kappa B}{\alpha^2}}}} .
	\end{equation}
	When $B$ is independent of $\gamma$, we have that $\lim_{\gamma\to0}c_{1}=\lim_{\gamma\to0}c_{2}=0$, therefore $c_{1},c_{2}\ll1$ for $\gamma\ll1$. Thus, using the general relation
	\begin{equation}
		\Delta x\Delta p\geq \frac{1}{2}|\langle[x,p]\rangle| ,
	\end{equation}
	we obtain that the modified uncertainty principle is written as
	\begin{equation}\label{Case1UPa}
		\Delta a\Delta P_{a} \geq \frac{1}{2a^{c_{1}}_{0}}|\langle a^{c_{1}}\rangle| ,
	\end{equation}
	\begin{equation}\label{Case1UPfi}
		\Delta\phi\Delta P_{\phi}\geq\frac{1}{2}|\langle e^{-c_{2}\phi}\rangle| .
	\end{equation}
	However, since $c_{2}$ is small we can expand the exponential in the scalar field, keeping only quadratic terms. Furthermore, we choose states in which 
	$\langle\phi\rangle=0$, thus we have that
	\begin{equation}
		\langle e^{-c_{2}\phi}\rangle \thickapprox \langle1-c_{2}\phi+\frac{c^2_{2}}{2}\phi^2 \rangle=1+\frac{c^2_{2}}{2}\langle\phi^2\rangle=1+\frac{c^2_{2}}{2}(\Delta\phi)^2.
	\end{equation}
	Therefore the uncertainty relation in the scalar field is expressed as
	\begin{equation}
		\Delta P_{\phi}\geq \frac{2+c^2_{2}(\Delta\phi)^2}{4\Delta\phi} .
	\end{equation}
	We note that the right hand side function has a minimum value in $\Delta\phi_{min}=\frac{\sqrt{2}}{c_{2}}$, which implies a minimum measurable value for the scalar field momentum given by $(\Delta P_{\phi})_{min}=\frac{c_{2}}{\sqrt{2}}$. This conclusion is valid for the static universe case as long as $\lim_{\gamma\to0}\frac{\alpha^2}{B}=0$ so it is guaranteed that $c_{2}\ll1$. Unfortunately, in this case we cannot reduced the uncertainty principle in the scale factor  (\ref{Case1UPa}), since the right hand side cannot be related to $\Delta a$ when $c_{1}\ll1$. The only scenario in which we can relate these two quantities is by taking $c_{1}=2n$ with $n$ a natural number. This can be obtained by proposing
	\begin{equation}
		B=\frac{\alpha^2}{12\kappa}\left[\left(\frac{3}{n}-1\right)^2-1\right] ,
	\end{equation}
	it follows that $H=\frac{\alpha}{2n}$ and $c_{2}=\sqrt{\kappa n}$. Then, we note that $\lim_{\gamma\to0}B=0$, thus it is a particular example of the static universe case. Employing the property $\langle a^{2n}\rangle\geq(\langle a^{2}\rangle)^{n}$ and choosing states in which $\langle a \rangle=0$ we obtain from (\ref{Case1UPa}) that
	\begin{equation}\label{Case1Uncertainty}
		\Delta P_{a}\geq \frac{(\Delta a)^{2n-1}}{2a_{0}^{2n}} .
	\end{equation}
	In contrast, considering the easiest form of the GUP in the scale factor (\ref{SimplestGUP}), that is taking $F(a)=1+\gamma^2a^2$, the uncertainty principle is written as
	\begin{equation}\label{Case1UncertaintyG}
		\Delta P_{a}\geq \frac{1+\gamma^2(\Delta a)^2}{\Delta a} .
	\end{equation}
	Let us remark that (\ref{Case1Uncertainty}) has the same behaviour as (\ref{Case1UncertaintyG}) when $n=1$ for big values of $\Delta a$, that is the minimum value for $\Delta P_{a}$ grows linearly with $\Delta a$. However, for the simplest case (\ref{Case1UncertaintyG}) there is a minimum value for $\Delta P_{a}$ obtained for some $\Delta a\neq0$, if we consider smaller values of $\Delta a$ beyond this point $\Delta P_{a}$ increases as in the standard scenario without a GUP. In contrast, in our solution the minimum value gets displaced to the origin, therefore our result leads to the unexpected behaviour that $\Delta P_{a}\to0$ implies $\Delta a\to0$, loosing the region of standard behaviour. In this case however  $c_{2}$ is not a small number and we cannot say more about the uncertainty principle for the scalar field.
	
	\subsection*{A realistic scenario}
	In the presented analysis we have only considered a scalar field to be the source of the gravitational field, the scalar field is defined by the $\gamma$ parameter describing the GUP. Although this is a toy model for the universe, we can ask ourselves how to conceive a more realistic scenario for this solution, it can be done as follows: We remark that by definition $\gamma$ is a parameter that measures how great the modification from the GUP in the Heisenberg algebra is, then it should be related to the scale in which the effects of a quantum theory of gravity is expected to be relevant. Therefore, a more realistic scenario will be to consider this parameter as a function of the size of the universe, or depending inversely with its energy density, in such a way that in the very early universe we expect a non zero value for $\gamma$ (when the effects of the GUP are expected to be relevant), as the expansion proceeds $\gamma$ should decease, then for big values for the scale factor, that correspond to classical values, this parameter decreases until it can be neglected, therefore in this moment, the inflaton field vanishes as well as its corresponding potential (in the static universe case), or the cosmological constant remains (in the other case). Therefore, at this point the standard cosmological scenario can follow with the other gravitational sources of the universe, the scalar field does not decay, it just vanishes because the $\gamma$ parameter can be neglected.
	
	\subsection{Case II}
	\label{S-Case2}
	Let us consider now a dependence on the momenta for the GUP proposal. First of all, we will consider the simplest case where a  GUP is proposed for each coordinate independently (\ref{DefCase2}). In this case the Hamiltonian constraint is the same as before (\ref{SistStandardHC}), the remaining equations in the system are the equations of motion
	\begin{equation}\label{Case2SisE2}
		P_{a}=-\frac{6a\dot{a}}{\kappa F_{1}}  ,\hspace{1cm}
		P_{\phi}=\frac{a^3\dot{\phi}}{F_{2}} ,
	\end{equation}
	\begin{equation}\label{Case2SisE4}
		\dot{P}_{a}=-\left[\frac{\kappa}{12a^2}P^2_{a}-\frac{3}{2a^4}P^2_{\phi}+3a^2V\right]F_{1} , \hspace{1cm}
		\hspace{1cm} \dot{P}_{\phi}=-a^3V'F_{2} .
	\end{equation}
	with $F_{1}=F_{1}(a,P_{a})$ and $F_{2}=F_{2}(\phi,P_{\phi})$. As we pointed out earlier, in order to obtain the Friedmann equation we need to specify the form in which the $F$ functions depend on the momenta such that we can solve for $P_{a}$ and $P_{\phi}$ in (\ref{Case2SisE2}) and then substitute these results back in (\ref{SistStandardHC}). However, as we haven seen in section \ref{S-NoScalarField} this procedure will modify completely the form of the Friedmann equation. For example, taking the simplest possibility where the modification by a GUP is implemented only in the scale factor, that is taking $F_{2}=1$, the resulting modified Friedmann equations will have the form in (\ref{ModifiedFriedmann1})-(\ref{ModifiedFriedmann3}) when the appropriate form for the $F_{1}$ function is chosen, the only difference will be that the energy density $\rho$ is changed to $\frac{\dot{\phi}^2}{2}+V$ as expected. However, when $F_{2}$ is taken into account we can obtain more forms of the Friedmann equations, in particular we are interested in obtaining consistency with the standard cosmological behaviour. Therefore, instead of proposing a particular ansatz for the $F$ functions, we will perform the analytical treatment of the system of equations analogous to the one presented in the last section. In this case, the system of five equations can also be solved  with the exponential ansatz $a=a_{0}e^{Ht}$, the solutions are given by
	\begin{equation}
		P_{a}=-\frac{6a^2_{0}H}{\kappa}\frac{e^{2Ht}}{F_{1}} , \hspace{1cm} P_{\phi}=\pm a^3_{0}e^{3Ht}\sqrt{\frac{2H\dot{F}_{1}}{\kappa F^3_{1}}} ,
	\end{equation}
	\begin{equation}\label{SolGeneralCampo}
		\frac{\dot{\phi}}{F_{2}}=\pm\sqrt{\frac{2H}{\kappa}}\sqrt{\frac{\dot{F}_{1}}{F^3_{1}}} , \hspace{1cm}
		V=\frac{H}{\kappa F^3_{1}}\left(3HF_{1}-\dot{F}_{1}\right) .
	\end{equation}  
	We note that as a consequence of considering a different function for each coordinate, $F_{2}$ is relevant only in the equation involving $\dot{\phi}$. Therefore, the potential has the same form as in case I, but in terms of $F_{1}$. Thus following the same procedure, we impose the condition $\dot{H}=0$ which leads to $F_{1}=e^{\alpha t}$. Using this form we can obtain the appropriate form for the potential, thus we propose again $V=Be^{-2\alpha t}$. The only difference with case I is that in this case the form of the scalar field is not uniquely determined, since $F_{2}$ is not restricted by the system. Thus we have the freedom to choose the $F_{2}$ function as we prefer and it will not alter the expansion of the universe, it will only alter the form of the scalar field and thus the form of the potential to obtain this solution. Therefore we obtain
	\begin{equation}\label{Case2Soluciones}
		P_{a}=-\frac{6a^2_{0}H}{\kappa}e^{(2H-\alpha)t} , \hspace{1cm} P_{\phi}=\pm\sqrt{\frac{2H\alpha}{\kappa}}a^3_{0}e^{(3H-\alpha)t} , \hspace{1cm} \frac{\dot{\phi}}{F_{2}}=\pm\sqrt{\frac{2H\alpha}{\kappa}}e^{-\alpha t} ,
	\end{equation}
	where the Hubble parameter will be given by
	\begin{equation}
		H=\frac{\alpha}{6}\left(1+\sqrt{1+\frac{12\kappa B}{\alpha^2}}\right) .
	\end{equation}
	Making a general ansatz of the form $F_{2}=f_{2}(P_{\phi})g_{2}(\phi)$, we obtain from the last expression in (\ref{Case2Soluciones}) that the scalar field obeys the equation
	\begin{equation}\label{Case2EcField}
		\int\frac{d\phi}{g_{2}(\phi)}=\pm\sqrt{\frac{2H\alpha}{\kappa}}\int f_{2}\left(\pm\sqrt{\frac{2H\alpha}{\kappa}}a^3_{0}e^{(3H-\alpha)t}\right)e^{-\alpha t}dt .
	\end{equation}
	
	In the solutions for this case the Hubble parameter and the potential (in terms of the time variable) have the same form as in case I. Thus, the analysis presented in case I for the Hubble parameter is applicable here. In particular, $B$ can be identified with the cosmological constant, or it can be considered in terms of $\gamma$ which leads us to the static universe scenario. What is different than in case I for the scale factor is that the $F_{1}$ function can also be written as powers of the scale factor as in (\ref{Case1F}), but from (\ref{Case2Soluciones}) we see that we can also write it as powers of the scale factor momentum in the form
	\begin{equation}\label{Case2F1}
		F_{1}=\left(c_{3}P_{a}\right)^{c_{4}} ,
	\end{equation}
	where
	\begin{equation}
		c_{3}=-\frac{\kappa}{6a^2_{0}H}=-\frac{\kappa}{a^2_{0}\alpha\left(1+\sqrt{1+\frac{12\kappa B}{\alpha^2}}\right)} ,
	\end{equation}
	\begin{equation}
		c_{4}=\frac{\alpha}{2H-\alpha}=\frac{6}{\sqrt{1+\frac{
					12\kappa B}{\alpha^2}}-5} .
	\end{equation}
	Taking $B$ as the cosmological constant, we have that $c_{4}\ll1$ and  $c_{3}\to-\frac{1}{a^2_{0}}\sqrt{\frac{\kappa}{12B}}$ when $\gamma\to0$. Then, the uncertainty principle in the scale factor in this case is written as
	\begin{equation}
		\Delta a\Delta P_{a}\geq\frac{|c_{3}^{c_{4}}|}{2}|\langle P_{a}^{c_{4}}\rangle| ,
	\end{equation}
	where $c_{4}$ is a small number, for this reason we cannot relate the right hand side to $\Delta P_{a}$. The only example in which we can relate both of them is if we propose that $c_{4}=2n$ with $n$ a natural number, this can be obtained by choosing
	\begin{equation}
		B=\frac{\alpha^2}{12\kappa}\left[\left(\frac{3}{n}+5\right)^2-1\right] ,
	\end{equation}
	then $c_{3}=-\frac{\kappa}{a^2_{0}\alpha\left(6+\frac{3}{n}\right)}$. Thus the uncertainty principle is written as
	\begin{equation}
		\Delta a\geq\frac{1}{2}\left(\frac{\kappa}{a^2_{0}\alpha\left(6+\frac{3}{n}\right)}\right)^{2n}(\Delta P_{a})^{2n-1} ,
	\end{equation}
	which describes the same unexpected behaviour as in case I, that is $\Delta P_{a}\to0$ implies $\Delta a\to0$. However for $\Delta P_{a}\neq0$, we note that since the coefficient depends inversely  with $\alpha$, for small values of $\alpha$ we always have a much bigger uncertainty in the position than in the momentum.
	
	On the other hand, since the solutions do not imply a specific form for the $F_{2}$ function, we can obtain the standard implications of the GUP by choosing the appropriate form for this function. For example, if we propose that it is a function only of $P^2_{\phi}$, or a function only of $\phi^2$, such that $F(x^2)/x$ has a minimum value, we obtain a minimal measurable value for the scalar field or for the momentum respectively. The only requirement to write the potential in terms of the momentum in an analytical form, is that we can solve (\ref{Case2EcField}) for the time variable $t$, since the form of $V$ with respect to time is restricted to be an exponential function. Let us consider two particular examples:
	\begin{itemize}
		\item Choosing $F_{2}=f_{2}(P_{\phi})$ we obtain from (\ref{Case2EcField}) that the scalar field will be given in general by
		\begin{equation}
			\phi=\mp\sqrt{\frac{2H}{\kappa\alpha}}\int f_{2}\left(\pm\sqrt{\frac{2H\alpha}{\kappa}}a^3_{0}x^{(1-3H/\alpha)}\right)dx ,
		\end{equation}
		where $x=e^{-2\alpha t}$ and the potential is always given by $V=Bx^2$. Thus in order to write an analytical expression for the potential $V=V(\phi)$, we need proposals for the $f_{2}$ functions such that we can solve for $x$ in the latter expression.
		\item Choosing $F_{2}=g_{2}(\phi)$ we can always solve for the time variable in terms of the scalar field, then the potential will be given in general by
		\begin{equation}
			V(\phi)=\frac{3\kappa B}{1+\sqrt{1+\frac{12\kappa B}{\alpha^2}}}\left[\int\frac{d\phi}{g_{2}(\phi)}\right]^2 .
		\end{equation}
	\end{itemize}
	
	Finally, let us consider a particular example of the general solution found that has some interesting features. We note from (\ref{Case2Soluciones}) that by taking $\alpha=2H$, the scale factor momentum will be a constant. This is helpful because we can then propose $F_{1}(a,P_{a})=f_{1}(P_{a})g_{1}(a)=f_{1}(P_{a})e^{2Ht}$. Then $f_{1}(P_{a})$ will be constant, independently of the particular form of $f_{1}$. Thus we look for a solution to the system of equations (\ref{SistStandardHC}), (\ref{Case2SisE2}) and (\ref{Case2SisE4}) employing the ansatz $V=Be^{-2\alpha t}=Be^{-4Ht}$. We obtain that the solutions are
	\begin{equation}
		P_{a}=\pm6a^2_{0}\sqrt{\frac{B}{\kappa}} , \hspace{1cm} P_{\phi}=\pm2a^3_{0}e^{Ht}\sqrt{B} ,
	\end{equation}
	thus the scalar field obeys the equation
	\begin{equation}
		\frac{\dot{\phi}}{F_{2}}=\pm2\sqrt{B}e^{-2Ht} ,
	\end{equation}
	and the Hubble parameter is written in terms of the $f_{1}$ function as
	\begin{equation}
		H=\mp\sqrt{\kappa B}f_{1}\left(\pm6a^2_{0}\sqrt{\frac{B}{\kappa}}\right) .
	\end{equation}
	Thus we have the freedom to propose forms of the $f_{1}$ function as in the standard cases. In particular with the proposal $f_{1}(P_{a})=\frac{1}{1-\gamma P^2_{a}}$ we can obtain big values for $H$ in the vicinity of the value that makes $f_{1}$ divergent, by choosing $B$ in a proper manner. We note that in this particular example $\lim_{\gamma\to0}H=\lim_{\gamma\to0}\alpha=0$, then $\lim_{\gamma\to0}B=0$, thus this solution is a particular example of the static universe case.
	
	\subsection{Case III}
	\label{S-Case3}
	Let us consider now the GUP diagonal proposal but with a dependence on the momenta defined by (\ref{DefCase3}). We have the same system of equations than in case I, but now $F$ can depend on the coordinates and in the momenta as well. Thus we have the Hamiltonian constraint (\ref{SistStandardHC}) and the equations of motion
	\begin{equation}\label{Case3SisE2}
		P_{a}=-\frac{6a\dot{a}}{\kappa F(a,\phi,P_{a},P_{\phi})}  , \hspace{1cm}
		P_{\phi}=\frac{a^3\dot{\phi}}{F(a,\phi,P_{a},P_{\phi})} ,
	\end{equation}
	\begin{equation}\label{Case3SisE4}
		\dot{P}_{a}=-\left[\frac{\kappa}{12a^2}P^2_{a}-\frac{3}{2a^4}P^2_{\phi}+3a^2V\right]F(a,\phi,P_{a},P_{\phi}) , 
		\hspace{0.5cm} \dot{P}_{\phi}=-a^3V'F(a,\phi,P_{a},P_{\phi}) .
	\end{equation}
	Let us consider firstly the simpler case in which the dependence is only in the total momentum. However, we note from the Hamiltonian constraint (\ref{SistStandardHC}) that the minisuperspace is not defined by a flat euclidean metric, in contrast it takes the form
	\begin{equation}
		G^{\mu\nu}=\begin{pmatrix}
			G^{aa} & G^{a\phi} \\
			G^{\phi a} & G^{\phi\phi}
		\end{pmatrix}=\begin{pmatrix}
			-\frac{\kappa}{12a} & 0 \\
			0 & \frac{1}{2a^3} 
		\end{pmatrix}.
	\end{equation}
	Therefore, the total momentum have to be expressed by considering this metric. Thus it must be defined as the kinetic part of the Hamiltonian constraint, that is 
	\begin{equation}\label{DefTotalMomentum}
		\mathcal{P}^2=-\frac{\kappa}{12a}P^2_{a}+\frac{1}{2a^3}P^2_{\phi}.
	\end{equation}
	However, from the Hamiltonian constraint it follows that $\mathcal{P}^2=-a^3V$. Therefore, if $F=F(\mathcal{P}^2)$ we actually have $F=F(-a^3V(\phi))$. Thus this particular case is effectively the same scenario as in case I, then all that was derived in subsection (\ref{S-Case1}) is applicable to this case as well. In particular, the Friedmann equation can be expressed in general as in (\ref{Case1Friedmann}), that is
	\begin{equation}
		H^2=\frac{\kappa}{3}\left[\frac{\dot{\phi}^2}{2}+F^2(-a^3V(\phi))V(\phi)\right] .
	\end{equation}
	The procedure to obtain inflation through approximations can also be done in the same form, as well as the exact solution describing the exponentially expanding universe. The only difference is that now the condition to obtain such solution is 
	\begin{equation}
		F(\mathcal{P}^2)=e^{\alpha t},
	\end{equation} 
	but it was found in that case that $a=a_{0}e^{Ht}$ and $V(\phi)=Be^{-2\alpha t}$, thus for this solution the total momentum is
	\begin{equation}
		\mathcal{P}^2=-a^3V=-a^3_{0}Be^{(3H-2\alpha)t} ,
	\end{equation}
	then we can write $F$ as a power term of the total momentum as well
	\begin{equation}
		F=\left(-\frac{\mathcal{P}^2}{a^3_{0}B}\right)^{\frac{\alpha}{3H-2\alpha}}=\left(-\frac{\mathcal{P}^2}{a^3_{0}B}\right)^{\frac{2}{\sqrt{1+\frac{12\kappa B}{\alpha^2}}-3}} .
	\end{equation}
	However, this form it is not particularly helpful since it is not possible to extract general consequences to the uncertainty relations of the coordinates.
	
	Let us abandon now the constriction of a dependence only through the total momentum, thus we consider the most general dependence on $F$ in the form $F=F(a,\phi,P_{a},P_{\phi})$. In this case in order to obtain a general form of the Friedmann equation, we need to solve for the momenta in the coupled system of equations in (\ref{Case3SisE2}), thus we need a particular form for the $F$ function. Therefore, once we consider a dependence on the momenta we cannot obtain a general Friedmann equation. However the system of equations have the same exact solution as in case I with the same condition, that is we have to impose $F=e^{\alpha t}$, then the analysis of case I follows in the same way. In particular, the solution takes the form (\ref{Case1SolutionsFinal}) with the potential (\ref{Case1Potential}) and the Hubble parameter  (\ref{Case1Hubble}). Then, the scalar field is linear with time, the potential is exponential with the scalar field and both momenta are exponentials with time. Thus, we can write the $F$ function as in (\ref{Case1F}) in terms of the coordinates, or as a function of the scale factor momentum as in (\ref{Case2F1}). Furthermore, in this case we can also write it as a function of the scalar field momentum as
	\begin{equation}\label{Case3F}
		F=(c_{5}P_{\phi})^{c_{6}} ,
	\end{equation}
	where
	\begin{equation}
		c_{5}=\pm\frac{1}{a^3_{0}}\sqrt{\frac{\kappa}{2H\alpha}}=\pm\frac{1}{\alpha a^3_{0}}\sqrt{\frac{3\kappa}{1+\sqrt{1+\frac{12B\kappa}{\alpha^2}}}} , \hspace{1cm} c_{6}=\frac{\alpha}{3H-\alpha}=\frac{2}{\sqrt{1+\frac{12B\kappa}{\alpha^2}}-1},
	\end{equation}
	which leads to the uncertainty relation
	\begin{equation}
		\Delta\phi\Delta P_{\phi}\geq \frac{|c_{5}^{c_{6}}|}{2}|\langle P_{\phi}^{c_{6}}\rangle| ,
	\end{equation}
	we note that  $c_{6}$ is a small number when $\alpha\ll1$ but $c_{5}\gg1$ in the same limit. Once again, we can reduce this expression in terms of uncertainties by proposing $c_{6}=2n$ with $n$ a positive integer, which can be realized by choosing
	\begin{equation}
		B=\frac{\alpha^2}{12\kappa n}\left(2+\frac{1}{n}\right) ,	
	\end{equation}
	which will be a particular example of static universe scenario when $\alpha\to0$. With this choice, the uncertainty principle leads to
	\begin{equation}
		\Delta\phi\geq\frac{1}{2}\left(\frac{3\kappa n}{\alpha^2a^6_{0}(1+2n)}\right)^{n}(\Delta P_{\phi})^{2n-1} .
	\end{equation}
	Therefore,  we obtain the unexpected behaviour that $\Delta \phi\to0$ implies that $\Delta P_{a}\to0$ once again, but we also have that for $\Delta\phi\neq0$, the uncertainty in the scalar field is much greater than the uncertainty in its momentum since $\alpha$ is a small number.
	
	\subsection{Case IV}
	\label{S-Case4}
	
	Finally, let us consider the most general scenario, where the form of the GUP  is non-diagonal (\ref{DefCase4}). For simplicity we will first consider the case in which $F_1=F_2$ and $G_1=G_2$, this will be called case A in the following. We will study this case by analyzing the system of equations with the hope of obtaining an analytical solution describing inflation as in the latter cases. We will then consider the most general case, case B, in which $F_1\not=F_2$ and $G_1\not=G_2$ and proceed in the same form.
	
	\subsection*{Case A}	
	Let us start with the simplest non-diagonal form of the GUP defined by $F_{1}=F_{2}=F$ and $G_{1}=G_{2}=G$, where both $F$ and $G$ are functions of the coordinates and any of the momenta. As we have established by now, both functions must contain a parameter $\gamma$ in order to recover the standard commutator relations in the limit $\gamma\to0$. Therefore we require that such functions obey
	\begin{equation}\label{Case4SLimits}
		\lim_{\gamma\to0}F=1 , \hspace{1cm} \lim_{\gamma\to0}G=0 .
	\end{equation}
	The system of equations in this case takes the form 
	\begin{equation}\label{Case4SSist-1}
		\mathcal{H}=-\frac{\kappa P^2_{a}}{12a}+\frac{P^2_{\phi}}{2a^3}+a^3V=0 ,
	\end{equation} 
	\begin{equation}\label{Case4SSist-2}
		\dot{a}=-\frac{\kappa P_{a}}{6a}F+\frac{P_{\phi}}{a^3}G , \hspace{1cm}
		\dot{\phi}=-\frac{\kappa P_{a}}{6a}G+\frac{P_{\phi}}{a^3}F ,
	\end{equation}
	\begin{equation}\label{Case4SSist-4}
		\dot{P}_{a}=\left[-\frac{\kappa P^2_{a}}{12a^2}+\frac{3P^2_{\phi}}{2a^4}-3a^2V\right]F-a^3V'G ,
	\end{equation}
	\begin{equation}\label{Case4SSist-5}
		\dot{P}_{\phi}=\left[-\frac{\kappa P^2_{a}}{12a^2}+\frac{3P^2_{\phi}}{2a^4}-3a^2V\right]G-a^3V'F .
	\end{equation}
	Let us remark that in cases I and III the general solution of the system of equations gave us the form of the momenta, the scalar field  and its potential  in terms of the function $F$ (\ref{Case1Solutions1}), (\ref{Case1Solutions2}). Then, the appropriate form for this function was obtained  by imposing the condition of a constant $H$, in this form a unique solution was found for the  complete system. However, in this case the number of equations defining the system is the same but the number of variables has increased by one, thus  we expect that the system will not constrain both $F$ and $G$, so we will have to rely on giving appropriate ansatz for these functions in order to obtain a constant Hubble parameter. Let us consider the exponentially expanding universe ansatz $a=a_{0}e^{Ht}$, then from (\ref{Case4SSist-1}) and (\ref{Case4SSist-2}) we can write everything in terms of one of the momenta, choosing $P_{\phi}$ as the independent variable we obtain
	\begin{equation}\label{Case4SSolMomenta}
		P_{a}=\frac{6}{\kappa F}\left[\frac{P_{\phi}G}{a^2}-Ha^2\right] , \hspace{0.5cm}
		V=\left[\frac{3G^2}{\kappa F^2a^2}-\frac{1}{2}\right]\frac{P^2_{\phi}}{a^6}+\frac{1}{\kappa F^2}\left[3H^2-\frac{6HP_{\phi}G}{a^4}\right],
	\end{equation}
	\begin{equation}\label{Case4SSolMomentaF}
		\dot{\phi}=\frac{P_{\phi}}{a^3}\left(\frac{F^2-G^2}{F}\right)+\frac{HGa}{F} .
	\end{equation}
	Furthermore, from (\ref{Case4SSist-4}) and (\ref{Case4SSist-5}) we obtain 
	\begin{equation}\label{Case4SDiffEq}
		-\frac{\kappa P^2_{a}}{3a^2}+\frac{3P^2_{\phi}}{a^4}=\frac{\dot{P}_{a}F-\dot{P}_{\phi}G}{F^2-G^2} , \hspace{1cm}
		V'=\frac{\dot{V}}{\dot{\phi}}=\frac{\dot{P}_{a}G-\dot{P}_{\phi}F}{a^3(F^2-G^2)} ,
	\end{equation}
	which represents a system of differential equations for $P_{\phi}$. So the analytical solution for the whole system will be obtained after solving these differential equations. However, in the particular case we are considering, both equations in (\ref{Case4SDiffEq}) lead to 
	\begin{multline}\label{Case4SMomentumEq}
		\frac{3(F^2-G^2)}{a^4}\left(1-\frac{4G^2}{\kappa F^2a^2}\right)P^2_{\phi}+\left(1-\frac{6}{\kappa a^2}\right)G\dot{P}_{\phi}+\frac{6G}{\kappa a^2}\left[6H-\frac{4HG^2}{F^2}+\frac{\dot{F}}{F}-\frac{\dot{G}}{G}\right]P_{\phi}\\+\frac{6Ha^2}{\kappa}\left[\frac{2HG^2}{F^2}-\frac{\dot{F}}{F}\right]=0 .
	\end{multline}
	This differential equation is hard to solve in general, therefore continuing the procedure to obtain a general solution and then impose $\dot{H}=0$ is really hard to follow. However, instead of looking for a general constraint for the GUP functions $F$ and $G$, we can propose a convenient ansatz to the system of equations in order to obtain a constant $H$. We note from (\ref{Case4SSist-2}) that the Hubble parameter can  be written as
	\begin{equation}
		H=-\frac{\kappa P_{a}}{6a^2}F+\frac{P_{\phi}G}{a^4} ,
	\end{equation}	
	thus we can guarantee that $H$ is a constant by proposing
	\begin{equation}\label{Case4SAnsatzMomenta}
		P_{a}=-\frac{6Aa^2}{\kappa F} , \hspace{1cm} P_{\phi}=\frac{Ba^4}{G} .
	\end{equation}
	where $A$ and $B$ are constants. Then (\ref{Case4SMomentumEq}) simplifies to
	\begin{multline}\label{Case4SMomentumEq2}
		\left[3B^2\left(1-\frac{F^2}{G^2}\right)+B\left(\frac{\dot{G}}{G}-4H\right)\right]a^2_{0}e^{2Ht}=\frac{12B^2}{\kappa}\left(\frac{G^2}{F^2}-1\right)\\+\frac{6}{\kappa}\left[2HB+\frac{2HG^2}{F^2}(H-2B)+\frac{\dot{F}}{F}(B-H)\right] .
	\end{multline}
	From this expression we can see that an exponential ansatz for both $F$ and $G$ will greatly simplify the equation, furthermore as we encountered in the last cases an exponential form for the $F$ function was indeed the only compatible form. Taking into account the correct limits (\ref{Case4SLimits}) we propose
	\begin{equation}
		F=e^{\alpha t} , \hspace{1cm} G(t)=\delta e^{\beta t} ,
	\end{equation}
	where $\alpha,\beta$ and $\delta$ are constants that depend on $\gamma$ and fulfil $\lim_{\gamma\to0}\alpha,\beta,\delta=0$. Then (\ref{Case4SMomentumEq2}) simplifies to
	\begin{multline}
		\left[3B+\beta-4H\right]Ba^2_{0}e^{2Ht}-\frac{3B^2}{\delta^2}a^2_{0}e^{2(H+\alpha-\beta)t}-\frac{12\delta^2}{\kappa}\left[B^2+H(H-2B)\right]e^{2(\beta-\alpha)t}\\+\frac{6}{\kappa}\left[2B^2-2HB+\alpha(H-B)\right]=0 .
	\end{multline}
	The consistent solution is obtained with $\beta=\alpha$ and leads to the following system of equations
	\begin{equation}\label{Case4SConstSys1}
		3B=[3B+\alpha-4H]\delta^2 ,
	\end{equation}
	\begin{equation}\label{Case4SConstSys2}
		2B^2-2HB+\alpha(H-B)=2\delta^2[B^2+H(H-2B)] .
	\end{equation}
	In this case, the potential (\ref{Case4SSolMomenta}) takes the form
	\begin{equation}
		V=\frac{3(H-B)^2}{\kappa}e^{-2\alpha t}-\frac{B^2a^2_{0}}{2\delta^2}e^{2(H-\alpha)t} .
	\end{equation}
	However, as we have seen in the last sections, it is convenient to write the potential in terms of an independent constant, since it will have a particular physical meaning, thus we propose that the potential is of the form
	\begin{equation}
		V=De^{-2\alpha t}+Ee^{2(H-\alpha)t} ,
	\end{equation}
	where $D$ will be an independent constant, but $E$ will be expressed in terms of the other constants. From this we have $D=\frac{3(H-B)^2}{\kappa}$, then taking into account  (\ref{Case4SConstSys1}) and (\ref{Case4SConstSys2}), we can obtain the solution in terms of just two  variables. It is convenient to choose $D$ and $\delta$ as the two independent constants, then the solution takes the form
	\begin{equation}\label{Case4ConsSolutions1}
		B=2\sqrt{\frac{\kappa D}{3}}\left[\frac{\delta^2(\delta^2-2)}{3-\delta^2}\right] , \hspace{1cm} \alpha=2\sqrt{\frac{\kappa D}{3}}\left[\frac{\delta^2(\delta^2-1)}{3-\delta^2}\right] ,
	\end{equation}
	\begin{equation}\label{Case4ConsSolutions2}
		\tcboxmath[colback=red!10!white,colframe=red]
		{
		H=\sqrt{\frac{\kappa D}{3}}\left[\frac{2\delta^4-5\delta^2+3}{3-\delta^2}\right] .
		}
	\end{equation}
	Furthermore, since $\lim_{\gamma\to0}\delta=0$ we have $\lim_{\gamma\to0}\alpha=0$ as expected, but we also have $\lim_{\gamma\to0}B=0$ and $\lim_{\gamma\to0}H=\sqrt{\frac{\kappa D}{3}}$. Therefore, if $D$ is independent of $\gamma$, we can identify it once again with the cosmological constant. On the other hand if $D$ depends on $\gamma$ and obeys $\lim_{\gamma\to0}D=0$, then we will have the static universe case. Therefore, in this particular non-diagonal case we have obtained a solution describing the same physics as in the previous sections.
	
	Since we want to describe an exponentially expanding universe, we seek a positive value for the Hubble constant, then from (\ref{Case4ConsSolutions2}) we see that $\delta$ must be restricted to the regions $\delta^2<1$ or $\frac{3}{2}<\delta^2<3$. Furthermore, if we want a faster expansion than in the cosmological constant case, $\delta$ must be constrained to the region $2<\delta^2<3$. Moreover, solving (\ref{Case4SSolMomentaF}) and setting the integration constant to vanish, the scalar field takes the form
	\begin{equation}\label{Case4SScalarField}
		\phi=\frac{\delta(\delta^2-1)a_{0}}{2\delta^4-5\delta^2+3}e^{Ht} ,
	\end{equation}
	therefore in the limit $\delta\to0$ the kinetic term vanishes as in the previous cases. With this solution we note that the potential is written as
	\begin{equation}\label{Case4SPotential}
		V=D\left[\left(\frac{2\delta^4-5\delta^2+3}{\delta(\delta^2-1)a^{2}_{0}}\phi\right)^{\frac{4\delta^2(1-\delta^2)}{2\delta^4-5\delta^2+3}}-\frac{2a^2_{0}\kappa \delta^2(\delta^2-2)^2}{3(3-\delta^2)^2}\left(\frac{2\delta^4-5\delta^2+3}{\delta(\delta^2-1)a^{2}_{0}}\phi\right)^{\frac{6(1-\delta^2)}{2\delta^4-5\delta^2+3}}\right] ,
	\end{equation}
	and obeys $\lim_{\gamma\to0}V=D$ as expected. Furthermore, from (\ref{Case4SScalarField}) and (\ref{Case4SAnsatzMomenta}) we find that in this case both momenta, the scale factor and the scalar field are exponential functions on the time variable. Thus the $F$ and $G$ functions can be written as in the following equivalent forms
	\begin{equation}
		F=\left(\frac{a}{a_{0}}\right)^{d_{2}}=(d_{1}\phi)^{d_{2}}=(d_{3}P_{a})^{d_{4}}=(d_{5}P_{\phi})^{d_{6}} ,
	\end{equation}
	where
	\begin{equation}
		d_{1}=\frac{2\delta^4-5\delta^2+3}{\delta(\delta^2-1)a_{0}} , \hspace{1cm}  d_{2}=\frac{2\delta^2(\delta^2-1)}{2\delta^4-5\delta^2+3}, \hspace{1cm} 
		d_{3}=-\frac{1}{2a^2_{0}}\sqrt{\frac{\kappa}{3D}} , \hspace{1cm}
	\end{equation}
	\begin{equation} d_{4}=\frac{\delta^2}{\delta^2-3} , \hspace{1cm}
		d_{5}=\sqrt{\frac{3}{\kappa D}}\frac{3-\delta^2}{2a^4_{0}\delta(\delta^2-2)} , \hspace{1cm} d_{6}=\frac{\delta^2}{3(\delta^2-2)}
	\end{equation}
	and $G=\delta F$. Thus the uncertainty relations lead to
	\begin{equation}
		\Delta a\Delta P_{a}\geq \frac{1}{2a^{d_{1}}_{0}}|\langle a^{d_{1}}\rangle|=\frac{|d_{3}^{d_{4}}|}{2}|\langle P_{a}^{d_{4}}\rangle|,
	\end{equation}
	\begin{equation}
		\Delta\phi\Delta P_{\phi}\geq\frac{|d_{2}^{d_{1}}|}{2}|\langle\phi^{d_{1}}\rangle|=\frac{|d_{5}^{d_{6}}|}{2}|\langle P_{\phi}^{d_{6}}\rangle | ,
	\end{equation}
	\begin{equation}
		\Delta a\Delta P_{\phi}\geq\frac{\delta}{2a^{d_{1}}_{0}}|\langle a^{d_{1}}\rangle |=\frac{\delta|d_{5}^{d_{6}}|}{2}|\langle P_{\phi}^{d_{6}}\rangle | ,
	\end{equation}
	\begin{equation}
		\Delta\phi\Delta P_{a}\geq\frac{\delta|d_{2}^{d_{1}}|}{2}|\langle \phi^{d_{1}}\rangle |=\frac{\delta|d_{3}^{d_{4}}|}{2}|\langle P_{a}^{d_{4}}\rangle| .
	\end{equation}
	Since $\lim_{\gamma\to0}\delta=0$ we expect that $\delta\ll1$. Then, $d_{2}$ is small, whereas $d_{1}$ and $d_{5}$ are big numbers. We can not simplify the above expression in terms of uncertainties since $d_{2}, d_{4}$ or $d_{6}$ can be written as odd integers only in the region $\frac{3}{2}<\delta^2<2$. Thus, in this particular region we would obtain similar results as in the previous cases, that is, we would obtain that all the uncertainties on the coordinates and the momenta will vanish at the same time. Let us remark that the solution obtained in this case was found through a convenient ansatz on the momenta that guarantee a constant Hubble parameter. In this case, we did not find a uniquely general constraint in $F$ and $G$ for this kind of solution to appear,  thus we can not assure that it is the only way to obtain a constant Hubble parameter.
	
	\subsection*{Case B}
	
	Finally, let us study the most general case defined by (\ref{DefCase4}) with  $F_{1}\neq F_{2}$ and $G_{1}\neq G_{2}$. As in the previous case we note that these functions must depend on $\gamma$ and obey
	\begin{equation}\label{Case4Limits}
		\lim_{\gamma\to0}F_{1}=\lim_{\gamma\to0}F_{2}=1 , \hspace{1cm} \lim_{\gamma\to0}G_{1}=\lim_{\gamma\to0}G_{2}=0.
	\end{equation}
	In this case the system of equations is
	\begin{equation}\label{Case4Sist1}
		\mathcal{H}=-\frac{\kappa P^2_{a}}{12a}+\frac{P^2_{\phi}}{2a^3}+a^3V=0 ,
	\end{equation} 
	\begin{equation}\label{Case4Sist2}
		\dot{a}=-\frac{\kappa P_{a}}{6a}F_{1}+\frac{P_{\phi}}{a^3}G_{1} ,\hspace{1cm}
		\dot{\phi}=-\frac{\kappa P_{a}}{6a}G_{2}+\frac{P_{\phi}}{a^3}F_{2},
	\end{equation}
	\begin{equation}\label{Case4Sist4}
		\dot{P}_{a}=\left[-\frac{\kappa P^2_{a}}{12a^2}+\frac{3P^2_{\phi}}{2a^4}-3a^2V\right]F_{1}-a^3V'G_{1} ,
	\end{equation}
	\begin{equation}\label{Case4Sist5}
		\dot{P}_{\phi}=\left[-\frac{\kappa P^2_{a}}{12a^2}+\frac{3P^2_{\phi}}{2a^4}-3a^2V\right]G_{2}-a^3V'F_{2} .
	\end{equation}	
	Let us proceed as in the particular case studied before, that is, we use (\ref{Case4Sist1}) and (\ref{Case4Sist2}) to write everything in terms of $P_{\phi}$, then we obtain
	\begin{equation}
		P_{a}=\frac{6}{\kappa F_{1}}\left[\frac{P_{\phi}G_{1}}{a^2}-a^2H\right] , \hspace{1cm} 
		\dot{\phi}=\frac{P_{\phi}}{F_{1}a^3}\left(F_{1}F_{2}-G_{1}G_{2}\right)+\frac{aHG_{2}}{F_{1}} ,
	\end{equation}
	\begin{equation}
		V=\left[\frac{3G^2_{1}}{\kappa F^2_{1}a^2}-\frac{1}{2}\right]\frac{P^2_{\phi}}{a^6}+\frac{3}{\kappa F^2_{1}}\left[H^2-\frac{2HG_{1}P_{\phi}}{a^4}\right] .
	\end{equation}
	Thus, from (\ref{Case4Sist4}) and (\ref{Case4Sist5}) we obtain
	\begin{equation}\label{Case4EcMomentum1}
		\left(1-\frac{4G^2_{1}}{\kappa F^2_{1}a^2}\right)\frac{3P^2_{\phi}}{a^4}-\frac{12}{\kappa F^2_{1}}\left(H^2a^2-\frac{2HG_{1}}{a^2}P_{\phi}\right)=\frac{\dot{P}_{a}F_{2}-\dot{P}_{\phi}G_{1}}{F_{1}F_{2}-G_{1}G_{2}} ,
	\end{equation}
	\begin{equation}\label{Case4EcMomentum2}
		V'=\frac{\dot{P}_{a}G_{2}-\dot{P}_{\phi}F_{1}}{a^3(F_{1}F_{2}-G_{1}G_{2})} .
	\end{equation}
	In the particular case A, the two expressions led to the same differential equations for $P_{\phi}$. However in this case this will no longer hold, since from (\ref{Case4EcMomentum1}) we obtain
	\begin{multline}\label{Case4EcMomentumC1}
		\frac{3(F_{1}F_{2}-G_{1}G_{2})}{a^4}\left(1-\frac{4G_{1}^2}{\kappa F^2_{1}a^2}\right)P^2_{\phi}+\left(1-\frac{6F_{2}}{\kappa a^2F_{1}}\right)G_{1}\dot{P}_{\phi}\\+\frac{6F_{2}}{\kappa a^2F_{1}}\left[G_{1}\left(6H-\frac{4HG_{1}G_{2}}{F_{1}F_{2}}+\frac{\dot{F_{1}}}{F_{1}}\right)-\dot{G}_{1}\right]P_{\phi}+\frac{6Ha^2}{\kappa F^2_{1}}\left[2HG_{1}G_{2}-\dot{F_{1}}F_{2}\right]=0 ,
	\end{multline}
	whereas from (\ref{Case4EcMomentum2}) we obtain
	\begin{multline}\label{Case4EcMomentumC2}
		\frac{6(F_{1}F_{2}-G_{1}G_{2})}{Ha^4}\left[\frac{G_{1}-G_{2}}{\kappa F^2_{1}a^2}\left(\dot{G}_{1}-\frac{G_{1}\dot{F}_{1}}{F_{1}}\right)+\frac{2HG_{1}}{\kappa F^2_{1}a^2}(G_{2}-2G_{1})+\frac{H}{2}\right]P^2_{\phi}\\+\frac{6G_{1}(G_{1}-G_{2})}{\kappa HF^2_{1}a^6}(F_{1}F_{2}-G_{1}G_{2})P_{\phi}\dot{P}_{\phi}+G_{2}\left[1+\frac{6}{\kappa a^2}\left(\frac{G_{1}}{F^2_{1}}(G_{1}-G_{2})-\frac{F_{2}G_{1}}{F_{1}G_{2}}\right)\right]\dot{P}_{\phi}\\+\frac{6}{\kappa F^2_{1}a^2}\left[\dot{F}_{1}\left(F_{2}(2G_{1}-G_{2})+\frac{2G_{1}G_{2}}{F_{1}}(G_{2}-G_{1})\right)-\dot{G}_{1}\left(F_{1}F_{2}+G_{2}(G_{2}-G_{1})\right)\right.\\\left.+2H\left(F_{1}F_{2}(2G_{1}+G_{2})-2G^2_{1}G_{2}\right)\right]P_{\phi}-\frac{6Ha^2}{\kappa F^2_{1}}\left[\frac{\dot{F}_{1}}{F_{1}}G_{2}(G_{2}-G_{1})+F_{2}\dot{F}_{1}-2HG^2_{2}\right]=0 .
	\end{multline}
	However, in this general case we note from (\ref{Case4Sist2}) that the Hubble parameter can also be written as
	\begin{equation}
		H=-\frac{\kappa P_{a}}{6a^2}F_{1}+\frac{P_{\phi}}{a^4}G_{1}.
	\end{equation}
	Thus an appropriate ansatz that guarantee a constant $H$ is
	\begin{equation}
		P_{a}=-\frac{6Aa^2}{\kappa F_{1}} , \hspace{1cm} P_{\phi}=\frac{Ba^4}{G_{1}}.
	\end{equation}
	Consequently, the system of equations (\ref{Case4EcMomentumC1}) and (\ref{Case4EcMomentumC2}) simplifies to
	\begin{multline}
		B\left[3B\left(\frac{F_{1}F_{2}}{G^2_{1}}-\frac{G_{2}}{G_{1}}\right)+4H-\frac{\dot{G}_{1}}{G_{1}}\right]a^2=\frac{6H}{\kappa}\left(\frac{F_{2}\dot{F}_{1}}{F^2_{1}}-\frac{2HG_{1}G_{2}}{F^2_{1}}\right)\\+\frac{12B^2}{\kappa}\left(\frac{F_{2}}{F_{1}}-\frac{G_{1}G_{2}}{F^2_{1}}\right)+\frac{6BF_{2}}{\kappa F_{1}}\left(\frac{4HG_{1}G_{2}}{F_{1}F_{2}}-2H-\frac{\dot{F}_{1}}{F_{1}}\right) ,
	\end{multline}
	\begin{multline}
		\frac{B}{G_{1}}\left[3B\left(\frac{F_{1}F_{2}}{G_{1}}-G_{2}\right)+G_{2}\left(4H-\frac{\dot{G}_{1}}{G_{1}}\right)\right]a^2\\=\frac{6B^2}{\kappa H}\left[\left(1-\frac{G_{2}}{G_{1}}\right)\left(\frac{G_{1}G_{2}}{F^2_{1}}-\frac{F_{2}}{F_{1}}\right)\left(4H-\frac{\dot{G}_{1}}{G_{1}}\right)\right.\\\left.+ \left(\frac{G_{2}}{F^2_{1}G_{1}}-\frac{F_{2}}{F_{1}G^2_{1}}\right)\left((G_{1}-G_{2})\left(\dot{G}_{1}-\frac{G_{1}\dot{F}_{1}}{F_{1}}\right)+2HG_{2}(G_{2}-2G_{1})\right)\right]\\+\frac{6B}{\kappa}\left\{G_{2}\left(\frac{F_{2}}{F_{1}G_{2}}-\frac{G_{1}-G_{2}}{F^2_{1}}\right)\left(4H-\frac{\dot{G}_{1}}{G_{1}}\right)-\frac{1}{G_{1}F^2_{1}}\left[-\dot{G}_{1}(F_{1}F_{2}+G_{2}(G_{2}-G_{1}))\right.\right.\\\left.\left.+\dot{F}_{1}\left(F_{2}(2G_{1}-G_{2})+\frac{2G_{1}G_{2}}{F_{1}}(G_{2}-G_{1})\right)+2H(F_{1}F_{2}(2G_{1}+G_{2})-2G^2_{1}G_{2})\right]\right\}\\+\frac{6H}{\kappa F^2_{1}}\left[\frac{\dot{F}_{1}}{F_{1}}G_{2}(G_{2}-G_{1})+F_{2}\dot{F}_{1}-2HG^2_{2}\right] .
	\end{multline}
	From this system of equations we note that a convenient ansatz is to propose an exponential form for the four functions. Considering the limiting behaviours (\ref{Case4Limits}) we propose
	\begin{equation}
		F_{1}=e^{\alpha t} , \hspace{1cm} F_{2}=Ce^{\alpha t} , \hspace{1cm} G_{1}=h_{1}e^{\alpha t}, \hspace{1cm} G_{2}=h_{2}e^{\alpha t} ,
	\end{equation}
	where $\lim_{\gamma\to0}\alpha,h_{1,2}=0$ and $\lim_{\gamma\to0}C=1$. However, the above system of equations imply $h_{1}=h_{2}=\eta$ as well as
	\begin{equation}\label{AuxAux1}
		\frac{3B}{\eta}\left(\frac{C}{\eta}-\eta\right)+4H-\alpha=0 ,
	\end{equation}
	\begin{equation}
		2B^2-2HB+\alpha(H-B)=\frac{2\eta^2}{C}\left[B^2-H(H-2B)\right] .
	\end{equation}
	Thus the potential is constrained to be two terms of exponentials of the time variables once again, thus we write it as
	\begin{equation}\label{Case4Potential}
		V=De^{-2\alpha t}+Ee^{2(H-\alpha)t} ,
	\end{equation}
	where $D$ is an independent constant related to $B$ and $H$ by $D=\frac{3(H-B)^2}{\kappa}$ and $E$ will depend on the other constants. The solution of the system of equations is
	\begin{equation}
		B=2\sqrt{\frac{\kappa D}{3}}\frac{\eta^2(\eta^2-2C)}{C(3C-\eta^2)} , \hspace{1cm} \alpha=2\sqrt{\frac{\kappa D}{3}}\frac{\eta^2(\eta^2-C)}{C(3C-\eta^2)} ,
	\end{equation}
	\begin{equation}
		\tcboxmath[colback=red!10!white,colframe=red]{
		H=\sqrt{\frac{\kappa D}{3}}\left[\frac{2\eta^4-5\eta^2C+3C^2}{C(3C-\eta^2)}\right] ,
		}
	\end{equation}
	it also follows that
	\begin{equation}\label{Case4KineticField}
		\dot{\phi}=a_{0}\sqrt{\frac{\kappa D}{3}}\frac{\eta(\eta^2-C)}{3C-\eta^2}e^{Ht} .
	\end{equation}
	We know that $\lim_{\gamma\to0}\eta=0$ , then $\lim_{\gamma\to0}\alpha=\lim_{\gamma\to0}B=\lim_{\gamma\to0}\dot{\phi}=0$ and $\lim_{\gamma\to0}H=\sqrt{\frac{\kappa D}{3}}$ with $\lim_{\gamma\to0}V=D$. Therefore, we can identify $D$ with the cosmological constant once again if it is independent of $\gamma$, but even if it depends on $\gamma$ and vanishes in that limit, leading to a static universe, we still have an exponential expansion thanks to the GUP as in the previous cases. By imposing the condition $H\geq0$, the allowed values of $\eta$ and $C$ will be constrained, however,  in this case such condition will be easier to fulfil than in the particular case, because of the presence of $C$. We also constrain these parameters to obtain a faster expansion than in the cosmological constant case, this leads to $2C<\eta^2<3C$, which can be less restrictive to $\eta$ than in the particular case by choosing appropriate values of $C$. Furthermore, solving (\ref{Case4KineticField}) and setting the integration constant to zero, we obtain the scalar field
	\begin{equation}
		\phi=\frac{a_{0}C\eta(\eta^2-C)}{2\eta^4-5\eta^2C+3C^2}e^{Ht} ,
	\end{equation}
	thus the potential takes the same form as before, that is we obtain
	\begin{multline}
		V=D\left[\left(\frac{2\eta^4-5\eta^2C+3C^2}{a_{0}C\eta(\eta^2-C)}\phi\right)^{-\frac{4\eta^2(\eta^2-C)}{2\eta^4-5\eta^2C+3C^2}}\right.\\\left.-\frac{2\kappa a^2_{0}\eta^2(\eta^2-2C)^2}{3C^2(3C-\eta^2)^2}\left(\frac{2\eta^4-5\eta^2C+3C^2}{a_{0}C\eta(\eta^2-C)}\phi\right)^{-\frac{6C(\eta^2-C)}{2\eta^4-5\eta^2C+3C^2}}\right] .
	\end{multline}
	Finally, in this case both the momenta and both coordinates are exponentials in time, thus we can write the $F$'s and $G$'s functions as powers of any of those variables in the form
	\begin{equation}
		F_{1}=\left(\frac{a}{a_{0}}\right)^{k_{2}}=(k_{1}\phi)^{k_{2}}=(k_{3}P_{a})^{k_{4}}=(k_{5}P_{\phi})^{k_{6}} ,
	\end{equation}
	where
	\begin{equation}
		k_{1}=\frac{2\eta^4-5\eta^2C+3C^2}{\eta(\eta^2-C)a_{0}C} , \hspace{1cm}  k_{2}=\frac{2\eta^2(\eta^2-C)}{2\eta^4-5\eta^2C+3C^2} , \hspace{1cm} 
		k_{3}=-\frac{1}{2a^2_{0}}\sqrt{\frac{\kappa}{3D}} , \hspace{1cm}
	\end{equation}
	\begin{equation} 
		k_{4}=\frac{\eta^2}{\eta^2-3C} , \hspace{1cm}
		k_{5}=\sqrt{\frac{3}{\kappa D}}\frac{C(3C-\eta^2)}{2a^4_{0}\eta(\eta^2-2C)} , \hspace{1cm} k_{6}=\frac{\eta^2}{3(\eta^2-2C)} ,
	\end{equation}
	and $F_{2}=CF_{1}$, $G_{1}=G_{2}=\eta F_{1}$. Thus the uncertainty relations lead to
	\begin{equation}
		\Delta a\Delta P_{a}\geq \frac{1}{2a^{k_{1}}_{0}}|\langle a^{k_{1}}\rangle |=\frac{|k_{3}^{k_{4}}|}{2}|\langle P_{a}^{k_{4}}\rangle |,
	\end{equation}
	\begin{equation}
		\Delta\phi\Delta P_{\phi}\geq\frac{|Ck_{2}^{k_{1}}|}{2}|\langle \phi^{k_{1}} \rangle |=\frac{|Ck_{5}^{k_{6}}|}{2}|\langle P_{\phi}^{k_{6}}\rangle | ,
	\end{equation}
	\begin{equation}
		\Delta a\Delta P_{\phi}\geq\frac{h}{2a^{k_{1}}_{0}}|\langle a^{k_{1}} \rangle |=\frac{h|k_{5}^{k_{6}}|}{2}|\langle P_{\phi}^{k_{6}}\rangle | ,
	\end{equation}
	\begin{equation}
		\Delta\phi\Delta P_{a}\geq\frac{h|k_{2}^{k_{1}}|}{2}|\langle \phi^{k_{1}} \rangle |=\frac{h|k_{3}^{k_{4}}|}{2}|\langle P_{a}^{k_{4}}\rangle | .
	\end{equation}
	These expression can not be reduced further since we can not make $k_{2}, k_{4}$ or $k_{6}$ to be an odd integer consistently with the limits of $\eta$ and $C$ when $\gamma\to0$. Such ansatz can only be proposed for non small values of $\gamma$, in that case it will lead to the same unexpected behaviour as in the other cases.
	
	\section{Discussion}\label{S-D-GUPInflation}
	
	Finally, let us summarize and give a discussion of the results presented in this chapter.
	
	We have presented a general study of the implications of a classical limit of the GUP, on the classical inflationary scenario driven by a scalar field.  Such limit was implemented through modifications of the Poisson brackets of the classical theory in the same way as the GUP modifies the corresponding quantum commutators. This study represents the generalization of the treatment in \cite{Battisti:2008du,Ali:2014hma,Moumni:2020uki,Atazadeh:2016yeh} by considering explicitly a scalar field and its corresponding potential. We note that in these works they only considered a perfect fluid, the cosmological solutions describe bouncing universes or a static solution. On the other hand, we have shown that when the scalar field is incorporated an agreement with the standard inflationary behaviour is found. In fact, we showed that we can obtain an exponentially expanding universe through an approximation method with slow roll parameters as in the standard case, but we also obtained analytical solutions. We remark that different approaches to incorporate the effects of a GUP at the classical level have found agreement with the standard inflationary cosmology before \cite{Cai:2005ra,Zhu:2008cg,Giardino:2020myz,Paliathanasis:2015cza,Giacomini:2020zmv} but not employing a general analytic treatment of the equations of motion through the classical limit employed in this chapter.
	
	Instead of considering a particular form for the GUP we considered a general modification of the Heisenberg algebra and classified the different forms of the GUP only by the variables in which it may depend. In this way we identified four distinct cases and we treat each of them separately in subsections \ref{S-Case1} to \ref{S-Case4}.
	
	When the GUP is implemented in a diagonal form in  case I and a particular example of case III with a dependence only in the total momentum, we obtained the general form of the Friedmann and scalar field equations. With these equations we obtained an inflationary solution by implementing an approximation procedure in the same line as the standard slow roll criteria. We showed that inflation can be obtained with three slow roll Hubble parameters that depend on the standard potential slow roll parameters in a very different form than in the usual case. Therefore, it is possible to obtain an exponentially expanding universe even if the standard slow roll parameters are not small. We used a particular form of the potential as an exponential function to show an explicit example in which the conditions found can be less restrictive than the standard scenario.
	
	We also showed that for every form of the GUP, that is for all four cases, we can obtain an analytical solution describing an exponentially expanding universes with an explicit form of the scalar field  and its potential. This solution can be seen as a generalization to the cosmological constant solution (which is the only analytical solution describing inflation in the standard scenario). However we showed that even if the cosmological constant is set to vanish when the GUP can be neglected, we still have the constant Hubble solution when the GUP is relevant, thus we can identify this solution as being defined exclusively by the GUP. Then we can propose a more realistic scenario by considering the GUP parameter $\gamma$ as varying with the size of the universe, then for the  very early universe $\gamma\neq0$ and it will not be necessary to incorporate any other matter content, the universe will expand in an exponential form because of the GUP. Then, as the universe expands $\gamma$ decreases, in a such way that for a big enough universe $\gamma$ can be neglected, then the scalar field will vanish and it will only remain a cosmological constant or nothing derived from the GUP. Thus at this time standard classical cosmology can follow. We remark that the only requirement to obtain this solution was to constrain the functions defining the GUP to be exponentials with the time variable. However, the form of the scalar field and its potential is different for every case.
	
	For cases I and III the scalar field has a linear dependence in time and the potential is an exponential function of the scalar field. In these cases we explicitly showed that we can have a scenario where the potential obeys the standard slow roll criteria, however we can also obtain that the potential obeys the dS conjecture at the same time that it describes an exponentially expanding universe. Furthermore, in this case the constriction on the form of the GUP lead the function to be written as powers of the scale factor, any of the two momenta or an exponential in the scalar field. With this form a minimal measurable value for the scalar field momentum is encountered when $\gamma$ is small. However, for the scale factor we cannot derived a conclusive statement since the form of the GUP is not written in terms of only uncertainties. We can only simplify this relation by choosing a very particular form for the constants involved. However, in this case we obtain that for non small values of the uncertainties we can have the same behaviour as the one encountered with the simplest form of the GUP, but for very small uncertainties we loose the standard behaviour, that is, instead of a divergence we obtain the unexpected behaviour that both uncertainties vanish at the same time. We remark that this only occurs by selecting a very specific form of the GUP and is not a general feature.
	
	In case II the GUP function for the scale factor take the same form. But the function defining the GUP in the scalar field was not restricted. Therefore, the scalar field and its potential (written in terms of the field) are not restricted. Then, in this case we can consider standard GUP proposals and obtain the corresponding phenomenological features without spoiling the inflationary solution, the only thing that is changed is the form of the potential.
	
	Let us remark that in cases I-III the constriction found on the GUP function was found to be the only way to produce a constant Hubble parameter. However, case IV was more difficult, in particular we did not find a general constriction to the form the GUP. Instead, we proposed a convenient ansatz and find a solution describing exponentially expanding universes with the same phenomenological features of the other cases. The only difference originated in the form of the scalar field, since for this case it was an exponential function with time that lead to potentials defined by two power terms of the scalar field. Then , the uncertainty relations were written as powers of all the coordinates and momenta and in general they cannot be reduced to a relation of the uncertainties only. They can only be reduced in that way by choosing the parameters in a very specific form leading to the same unexpected behaviour as for the other cases, however this cannot be done for small values of $\gamma$.
	
	Finally, we remark that we have found analytical solutions describing exponentially expanding universes that generalize the standard cosmological constant case. It is very interesting that the form in which the standard cosmological constant solution is generalized to incorporate a GUP is precisely by the appearance of a scalar field with and an adequate potential. Unfortunately, the forms of the GUP that allow this solution have not the standard forms of the GUP that have been studied previously (the only exception will be in cases I and III that we in fact encountered a GUP that leads to a minimum measurable scalar field momentum). Thus, the results presented in this chapter can be complemented by studying the general implications of the GUPs that we have found. Furthermore, it will be also interesting to study in a more formal ground what are the implications that the modification of the Poisson brackets imply to the classical theory in order to have a deeper understanding of the classical implications of a GUP.

	\chapter{Consistency of the Ho\v{r}ava-Lifshitz $F(\bar{R}$) theories and the Swampland conjectures}\label{CH-HLFRSwampland}
	 
	 In this chapter we will study the consistency between the dS swampland conjecture and the Ho\v{r}ava-Lifshitz $F(\bar{R})$ theories, using a flat FLRW metric.
	 
	 First of all, in section \ref{S-SWFR}, we will summarize the way in which standard $f(R)$ theories are studied in the light of the dS conjecture, we will consider the scenario where the conjecture can be made independent of $R$, and thus it can be used to constraint the parameters of the theory. We will show that such theories can be compatible with the conjecture but with a very restrictive form. Then, in section \ref{S-SWFRHL} we will perform the full analysis using the HL $F(\bar{R})$ theories. We will construct the relevant forms of the $F(\bar{R})$ functions starting from relevant proposals for the scale factor of cosmological interest, then we will show that such forms will lead to constrictions on the parameters of the theories. In some cases the regimes of compatibility will be identified with an ultraviolet behaviour of the theories, signalling consistency with the conjecture and expected behaviour. In other cases, it will be shown that these theories are compatible with the conjecture in the scenario where General Relativity and standard $f(R)$ theories are not compatible. Therefore, the results of this chapter can be used as an argument to support the idea that the dS conjecture may indeed capture key features of quantum gravity. The results of this chapter were first presented in \cite{Garcia-Compean:2023fjl}.
	
	\section{dS Swampland conjecture in $f(R)$ theories}
	\label{S-SWFR}
	
	Let us start by giving a brief review of how can we study the dS swampland conjecture in the light of the standard $f(R)$ theories. The action for such theories in the Jordan frame is of the general form (\ref{ActionfR}). However, we need to transform this action to a form suitable to study its compatibility with the dS conjecture, let us then consider the equivalent action \cite{DeFelice:2010aj,Faraoni:2010pgm}
	\begin{equation}\label{EqActionfR}
		S=\frac{1}{2}\int d^4x\sqrt{-g}[\psi R-U],
	\end{equation}
	where $\psi=\frac{df}{dR}=f_{R}$ is an auxiliary field and we have used $U=R\psi-f(R)$. Then, we can perform a conformal transformation in order to describe gravity in the Einstein frame, such transformation is written as
	\begin{equation}\label{ConfTransfR}
		\widetilde{g}_{\mu\nu}=e^{2\varphi}g_{\mu\nu} ,
	\end{equation}
	where $\varphi=\frac{1}{2}\ln f_{R}$. Then, in the Einstein frame we obtain the action
	\begin{equation}\label{ActionfRF}
		S=\int d^4x\sqrt{-\widetilde{g}}\left[\frac{\widetilde{R}}{2}-\frac{1}{2}\widetilde{g}^{\mu\nu}\widetilde{\nabla}_{\mu}\phi\widetilde{\nabla}_{\nu}\phi-V(\phi)\right] ,
	\end{equation}
	which is express in the form of an action of gravity plus a scalar field $\phi$ defined by
	\begin{equation}\label{ScalarFieldfR}
		\phi=\sqrt{\frac{3}{2}}\ln f_{R} ,
	\end{equation}
	furthermore, the scalar field potential is
	\begin{equation}\label{ScalarPotentialfR}
		V(\phi)=\frac{Rf_{R}-f}{2f^2_{R}} . 
	\end{equation}
	Let us remark that we have started in the Jordan by considering only a gravitational action, we did not consider any external scalar field. The canonically coupled scalar field $\phi$ has appeared in the process of reaching the Einstein frame through a conformal transformation, therefore it is only geometry that defines the scalar field and its potential. By doing this procedure we will see that the conjecture can be used to constrain the parameters of the theory and thus provide a way to explore if the theory itself can  be compatible with the conjecture. We can se from (\ref{dSConjecture}) that the conjecture is satisfied immediately if the potential is negative. Therefore, we will always be interested in the scenario where the potential is positive. Therefore, we note from the scalar field potential (\ref{ScalarPotentialfR}), that we must impose the condition
	\begin{equation}\label{ConditionfR}
		Rf_{R}-f>0 .
	\end{equation}
	Furthermore, from (\ref{ScalarFieldfR}) we also note that in order to have an scalar field properly defined we demand that $f_{R}>0$. Therefore, the derivative of the potential with respect to the scalar field can be written as
	\begin{equation}\label{fRDerivPotential}
		\frac{dV}{d\phi}=\frac{dV}{dR}\frac{dR}{df_{R}}\frac{df_{R}}{d\phi} .
	\end{equation}
	Using the above expression, the first dS conjecture (\ref{dSConjecture}) takes the general form 
	\begin{equation}\label{dSConjfR}
		|2f-Rf_{R}|\geq\sqrt{\frac{3}{2}}c(Rf_{R}-f) ,
	\end{equation}
	whereas the second dS conjecture (\ref{dSConjecture2}) is written as
	\begin{equation}\label{dSConjfR2}
		\frac{f^2_{R}+f_{RR}(Rf_{R}-4f)}{3f_{RR}(Rf_{R}-f)}<-\widetilde{c} .
	\end{equation}
	Using these expressions, the conjectures were studied in a general way in \cite{Artymowski:2019vfy}. They look for the constraints on the  $f$ function in order to make it compatible with the conjecture. Furthermore, in \cite{Benetti:2019smr}, a function of the form $f(R)=R^{1+\epsilon}$, where $\epsilon$ is a small number, was studied. It was found that the conjecture is compatible by restricting the $\epsilon$ parameter to a region that coincides with the one of phenomenological interest. Finally, in \cite{Elizalde:2022oej}, the first dS conjecture was found to be in good agreement with these theories and with numerical data. 
	
	Let us remark that in general  (\ref{dSConjfR}) will lead to an inequality involving $R$. However, we can use the conjecture to constrain the parameters of the theory by seeking a possible way in which such expression can be made independent of $R$, in that way making it valid for all values of $R$. This scenario can be achieved by imposing 
	\begin{equation}\label{CondionfRConstant}
		\frac{|2f-Rf_{R}|}{Rf_{R}-f}=A ,
	\end{equation}
	where $A$ is a constant, it has to be positive so we can fulfil (\ref{ConditionfR}). Let us remark that this represents an ansatz for the $f(R)$ function, therefore it is not the only way to fulfil (\ref{dSConjfR}), it is however the only way to make this expression independent of $R$. Since we have an absolute value in the left hand side, we  have two possibilities. First of all if such term is positive $2f>Rf_{R}$, then (\ref{CondionfRConstant}) leads to 
	\begin{equation}\label{fRConsForm}
		f(R)=\beta R^{\gamma} ,
	\end{equation}
	where $\gamma=\frac{A+2}{A+1}$ with $1<\gamma<2$. In order to have a well defined scalar field and not ghostly gravitons, we demand that $f_{R}>0$. Furthermore, in order to have positive mass of the curvature fluctuations, we also require $f_{RR}>0$ \cite{DeFelice:2010aj}. The expression in (\ref{fRConsForm}) fulfils both conditions for $\beta>0$. Substituting this ansatz in the first dS conjecture (\ref{dSConjfR}), we obtain that $\gamma$ is constrained to the region
	\begin{equation}\label{fRRestrictGamma}
		\gamma\leq\frac{2+\sqrt{\frac{3}{2}}c}{1+\sqrt{\frac{3}{2}}c} .
	\end{equation}
	In particular, for $c=1$, this leads to $\gamma\lesssim1.45$. Thus, the conjecture implies constraints on the exponent of the $f$ function in such a way that the GR value is not allowed, the exponent must be bigger, but its allowed region is very small.
	
	On the other hand, if the term within the absolute value is negative, that is if $2f<Rf_{R}$, we obtain the same form (\ref{fRConsForm}), but with $\gamma=\frac{A-2}{A-1}$, therefore $\gamma>2$. We obtain that in this case the conditions $f_{R}>0$ and $f_{RR}>0$ can be fulfilled for $\beta>0$. However, the first dS conjecture cannot be fulfilled for these values of $\gamma$.
	
	Moreover, using the form of the $f(R)$ function given by (\ref{fRConsForm}), we determine from (\ref{dSConjecture2}) that the second dS conjecture is written as
	\begin{equation}
		\frac{(\gamma-2)^2}{3(\gamma-1)^2}<-\widetilde{c} ,	
	\end{equation}
	which is independent of $R$, but this expression cannot be fulfilled for any value of $\gamma$. Therefore, the form for the $f$ function obtained to make the dS conjecture independent of $R$, is incompatible with the second dS conjecture. This is not a problem since we only need to fulfil one of the two versions of the conjecture, and we have shown that the first version is indeed fulfilled.
	
	Therefore, we can conclude that the only way to fulfil the first dS conjecture independently of $R$ for $f(R)$ theories is to consider the $f$ function as a power term in $R$. The conjecture constrain the exponent of this function to be close to the GR value, furthermore the second dS conjecture is never valid. In the following section, we will pursue a similar procedure to look for the viability of the dS conjecture with HL $F(\bar{R})$ theories.

	\section{dS Swampland conjecture in HL $F(\bar{R})$ theories}
	\label{S-SWFRHL}
	
	In the remaining of this chapter we will consider a flat FLRW metric (\ref{DefFlatFLRW}), which as we have seen is of great cosmological interest. Furthermore, as we saw in chapter \ref{CH-Preliminaries}, the general action of Ho\v{r}ava-Lifshitz gravity takes the form (\ref{ActionHL}). Then, for the flat FLRW metric the gravitational potential term vanishes, because for this metric the spatial part is just the scale factor times the three-dimensional euclidean flat metric (\ref{DefFlatFLRW}). Thus, in the following, we will not be concerned with the explicit form for this term, since at the level of equations of motion that we will employ, we can treat any of the two version in the same form. 
		
	Moreover, as we saw in section \ref{S-SConjectures}, General Relativity is not consistent with the dS conjecture with the flat FLRW metric in  the context of the slow roll scenario, then it is interesting to study if the Ho\v{r}ava-Lifshitz theory of gravity can be consistent with this metric in particular. In that context, in \cite{Wu:2019xtv} the swampland conjectures were studied when a non-linear dispersion relation is present. Such feature comes naturally from HL gravity and all its possible extensions (such as \cite{Zhu:2011yu}). However, as we have pointed out before, for an FLRW flat metric, the equations of motion are almost identical in all types of HL theories to the ones obtained in GR, the only difference are constants. Therefore, the inflationary scenario must be carried out in the same form as in GR, that is by employing the slow roll criteria, thus the incompatibility with the dS conjecture will be present in this theory as well. However, we can consider a different proposal to describe an expanding universe in these theories without the need of an extra scalar field. This can be achieved by considering the HL $F(\bar{R})$ theories presented in subsection \ref{SS-FRHL}. Considering only the flat FLRW metric at the level of equations of motion, all the versions proposed for the $F(\bar{R})$ generalization will lead to the same results. Therefore, the following analysis is valid for all these type of theories.
	
	With the flat FLRW metric, we obtain from (\ref{DefBarR}) that we can write
	\begin{equation}\label{BarRFLRW}
		\bar{R}=(3-9\lambda+18\mu)H^2+6\mu\frac{d}{dt}\left(H\right) ,
	\end{equation}
	where $H$ is the Hubble parameter. In contrast, the standard curvature scalar for this metric is
	\begin{equation}\label{RFLRW}
		R=12H^2+6\frac{d}{dt}\left(H\right) .
	\end{equation}
	Therefore, for this metric the difference between the two variables are just the appearance of the two parameters $\mu$ and $\lambda$, then we can provide an interpretation of $\bar{R}$ as to be related to the curvature, at least for this metric. 
	
	Let us consider in the Jordan frame only gravity described by the action (\ref{ActionFBarR}), then the hamiltonian constraint provides the first equation of motion in the form \cite{Chaichian:2010yi}
	\begin{equation}\label{EoM1}
		F(\bar{R})-6\left[(1-3\lambda+3\mu)H^2+\mu\dot{H}\right]F'(\bar{R})+6\mu H\frac{dF'(\bar{R})}{dt}-\frac{C}{a^3}=0 ,
	\end{equation}
	where $\dot{H}=\frac{dH}{dt}$,  $F'=\frac{dF}{d\bar{R}}$ and $C\neq0$ is an integration constant in the projectable version of the theory (in standard HL theory this constant can be considered to represent dark matter for the way in which it appears in the equations of motion when it takes positive values \cite{Mukohyama:2009mz}). In the non-projectable version of the theory, we have $C=0$. This is the only difference between both versions of the theory that we will be considering. The second equation of motion is obtained by varying with respect to the three-metric, which leads to \cite{Chaichian:2010yi}	  
	\begin{equation}\label{EoM2}
		F(\bar{R})-2(1-3\lambda+3\mu)(\dot{H}+3H^2)F'(\bar{R})+2(3\lambda-1)H\frac{dF'(\bar{R})}{dt}+2\mu\frac{d^2F'(\bar{R})}{dt^2}=0 .
	\end{equation}
	Employing a constant Hubble parameter ansatz $H=H_{0}$ in the above equations, in addition to considering a polynomial form for the $F(\bar{R})$ function of second or third order, the solutions obtained from this set of equations describe a universe with two periods of exponential expansion \cite{Chaichian:2010yi,Elizalde:2010ep}. This behaviour is exclusive of these theories since it is removed in the limit $\mu,\lambda\to1$ which is the $f(R)$ limit. This result shows the importance of these kinds of theories.
	
	Let us discuss an important point before we enter to the main analysis of this chapter. The swampland conjectures were proposed originally in the context of theories with Lorentz invariance. However, as we discussed in section \ref{S-HLGravity}, HL theories and all their generalizations, break explicitly the Lorentz invariance in the UV. Thus, it is natural to worry about the applicability of the conjectures to these kinds of theories. In this regard, we assume an \textit{a posteriori} position. That is, if we apply the conjectures and find that they are incompatible with these theories, we may conclude that Lorentz invariance is a key requirement of the conjectures. On the other hand, if the conjectures are found to be compatible with these theories, and as we will show in the next section they actually lead to some expected behaviours, then the results can support the idea that the conjectures may be applicable to a context without Lorentz invariance. Actually, in \cite{Trivedi:2021nss}, the compatibility of the conjectures with a cosmological context without Lorentz invariance was studied with positive results as well.

	Let us study now the dS conjecture with these type of theories. Since in the Jordan frame we only have the action of gravity, we need to perform a conformal transformation to the Einstein frame, as it was done for the $f(R)$ theories of the last section. In this case we perform a conformal transformation in the three-metric of the HL $F(\bar{R})$ theories, in order to study the compatibility with the dS conjecture in the Einstein frame. We consider an equivalent action of (\ref{ActionFBarR}) for any metric with an auxiliary field in the form
	\begin{equation}\label{AuxActionBarF}
		S=\frac{1}{2}\int d^4x \sqrt{h}N\left[\bar{R}F'(\bar{R})-U\right] ,
	\end{equation}
	where $U=\bar{R}F'(\bar{R})-F(\bar{R})$, We then employ a conformal transformation, but in this case only on the three-metric, given by
	\begin{equation}\label{ConformalTrans3}
		h_{ij}=e^{-\bar{\phi}}\widetilde{h}_{ij} ,
	\end{equation}
	then by considering metrics with lapse function $N=1$ and shift functions $N^{i}=0$, as for the flat FLRW metric, we obtain in the Einstein frame the gravitational action  \cite{Elizalde:2010ep,Carloni:2010nx}
	\begin{multline}\label{FbarRActionAux}
		S=\frac{1}{2}\int d^4x\sqrt{\widetilde{h}}\left[\widetilde{K}^{ij}\widetilde{K}_{ij}-\lambda \widetilde{K}^2+\left(-\frac{1}{2}+\frac{3\lambda}{2}-\frac{3\mu}{2}\right)\widetilde{h}^{ij}\dot{\widetilde{h}}_{ij}\dot{\bar{\phi}}\right. \\ \left.+\left(\frac{3}{4}-\frac{9\lambda}{4}+\frac{9\mu}{2}\right)\dot{\bar{\phi}}^2 -\mathcal{L}(e^{-\bar{\phi}}h_{ij})-2V(\bar{\phi})\right],
	\end{multline}
	where 
	\begin{equation}\label{AuxScalarField}
		\bar{\phi}=\frac{2}{3}\ln F'(\bar{R}) , \hspace{0.5cm} V(\bar{\phi})=\frac{\bar{R}F'-F}{2F'}. 
	\end{equation}
	We would like that the action in the Einstein frame takes the form of an action of gravity plus an action for a canonically coupled scalar field (that is, that it appears as in standard scalar theories), however from the above we note that we have an extra term that combines the metric with the scalar field. Therefore, in order to eliminate such term we choose
	\begin{equation}\label{ElectioLambda}
		\mu=\lambda-\frac{1}{3}.
	\end{equation}
	Furthermore, in order to obtain a canonically coupled scalar field we also perform the redefinition 
	\begin{equation}\label{RedScalarField}
		\phi=\alpha\bar{\phi}=\frac{2\alpha}{3}\ln F'(\bar{R}),
	\end{equation}
	where 
	\begin{equation}\label{DefAlpha}
		\alpha=\sqrt{\frac{3}{4}-\frac{9\lambda}{4}+\frac{9\mu}{2}}=\frac{\sqrt{3(3\lambda-1)}}{2} .
	\end{equation}
	Thus, we finally obtain in the Einstein frame the action
	\begin{equation}\label{ActionBarFRE}
		S=\int d^4x\sqrt{\widetilde{h}}\left[\frac{\widetilde{K}^{ij}\widetilde{K}_{ij}-\lambda \widetilde{K}^2-\mathcal{L}(e^{-\bar{\phi}}h_{ij})}{2}+\frac{\dot{\phi}^2}{2}-V(\phi)\right] .
	\end{equation}
	However, we have not obtained the HL theory of gravity in the Einstein frame, since the action for the scalar field does not contain the higher derivative terms in (\ref{ActionMatter}). Nevertheless, since the general form is described by an action for gravity plus an action for a scalar field, we can use it to investigate the compatibility with the dS conjecture. We remark that we only require an action that describes gravity, not necessarily GR. For example, in \cite{Yi:2018dhl}, the conjectures have been studied with a Gauss-Bonnet extra term in the action. Let us remark that since we made the choice (\ref{ElectioLambda}), we can no longer recover the $f(R)$ theories in any limit (since we can not choose $\mu\to1$ and $\lambda\to1$ at the same time). We will study firstly this case since this simplification allows a complete analytic analysis. The general action where this choice is not required will be studied in subsections \ref{SS-G} and \ref{SS-CH}.
	
	With the simplification of the action in the Einstein frame, we are going to use (\ref{fRDerivPotential}) expressed in terms of $\bar{R}$. Then the first dS conjecture (\ref{dSConjecture}) takes the general form
	\begin{equation}\label{dSConjectureBarFRAux}
		\frac{|V_{\phi}|}{V}=\frac{3}{2\alpha}\frac{|F|}{\bar{R}F'-F}\geq c ,
	\end{equation}
	which can be written as
	\begin{equation}\label{dSConjectureBarFR}
		\tcboxmath[colback=red!10!white,colframe=red]
		{
		|F|\geq\frac{2\alpha c}{3}(\bar{R}F'-F) .
		}
	\end{equation}
	A condition for the scalar field to be positive will also be required for these theories, thus we impose that
	\begin{equation}\label{ConditionBarFR}
		\tcboxmath[colback=red!10!white,colframe=red]
		{
		\bar{R}F'-F>0.
	}
	\end{equation}
	On the other hand, the second dS conjecture takes the general form
	\begin{equation}\label{dSSecondConjectureBarFR}
		\tcboxmath[colback=red!10!white,colframe=red]	
		{
		\frac{V_{\phi\phi}}{V}=\frac{9}{4\alpha^2}\frac{(F')^2-F''F}{F''(RF'-F)}<-\widetilde{c} .
	 	}
	\end{equation}
	
	We note from these results that the only way in which the conjecture can be independent of $\bar{R}$ is by considering $F(\bar{R})$ as some power of $\bar{R}$, as it happened with the standard $f(R)$ theories. However, the HL $F(\bar{R})$ theories have not been studied as extensively as the $f(R)$ theories, then we will proceed in a different way. Instead of proposing a convenient ansatz for the $F$ function, we will propose a relevant cosmological ansatz for the Hubble parameter and use the system of equations of motion in the Jordan frame to construct the appropriate $F(\bar{R})$ function. Then, we will analyze if such function can fulfil the conjecture at the same time that leads to cosmologically interesting solutions. 
	
	With the particular choice (\ref{ElectioLambda}), the system of equations of motion (\ref{EoM1}) and (\ref{EoM2}) takes the form
	\begin{equation}\label{System1}
		F(\bar{R})+2(3\lambda-1)H\frac{dF'(\bar{R})}{dt}+\frac{2}{3}(3\lambda-1)\frac{d^2F'(\bar{R})}{dt^2}=0 ,
	\end{equation}
	\begin{equation}\label{System2}
		F(\bar{R})-2(3\lambda-1)\dot{H}F'(\bar{R})+2(3\lambda-1)H\frac{dF'(\bar{R})}{dt}-\frac{C}{a^3}=0 .
	\end{equation}
	We also obtain in this case
	\begin{equation}\label{System3}
		\bar{R}=(3\lambda-1)(3H^2+2\dot{H}) .
	\end{equation}
	Furthermore, from (\ref{System1}) and (\ref{System2}) we can obtain the differential equation
	\begin{equation}\label{System4}
		\frac{d^2}{dt^2}F'(\bar{R})+3\dot{H}F'(\bar{R})+\frac{3C}{2(3\lambda-1)a^3}=0 .
	\end{equation}
	These equations represent the system that we will solve in order to construct the appropriate $F(\bar{R})$ function by starting with a relevant cosmological ansatz. We note however that $C$ change the set of equations, consequently the possible solutions. Thus, we will consider  separately the two different scenarios, namely the non-projectable case in which we set such constant to zero, and the projectable case in which it has to be different from zero.
	
	\subsection{Non-Projectable version}
	\label{SS-NP}
	Let us  begin with the non-projectable case in which $C=0$. The most relevant cosmological ansatz for the Hubble parameter is a constant value, that describes an inflationary period where the universe has an exponential expansion. With this proposal a solution with two different periods of accelerated expansion was found in \cite{Chaichian:2010yi}. However, with the particular choice implemented by choosing the value of $\mu$ in (\ref{ElectioLambda}), this ansatz leads to a solution with a vanishing $F$ function, therefore it is inconsistent. Thus,  we propose instead an ansatz describing an expanding universe but in the form of a power law with time, that is we propose for the scale factor in the Jordan frame the form
	\begin{equation}\label{ScaleFactorJordan}
		a(t)=t^n ,
	\end{equation}
	where $n$ is a positive constant. Thus, the Hubble parameter for this ansatz is $H=n/t$. Then, from (\ref{System3}) we obtain that $\bar{R}$ is written in terms of the time variable as
	\begin{equation}\label{BRTime}
		\bar{R}=\frac{(3\lambda-1)n(3n-2)}{t^2} .
	\end{equation}
	Furthermore, from (\ref{RedScalarField}) we can see that in order to have a properly defined scalar field we must impose the condition $3\lambda>1$. Since we want rapidly expanding universes, we will be generally interested in values of $n$ that are not small, therefore considering positive values for $\bar{R}$, we obtain from (\ref{BRTime}) the condition $3n-2>0$. Furthermore,  we obtain from this result that $\bar{R}$ grows inversely with time, a behaviour that could be expected of the curvature.  With this result, the differential equation in (\ref{System4}) is written as
	\begin{equation}\label{NPFEq}
		t^2\frac{d^2}{dt^2}F'(\bar{R})-3nF'(\bar{R})=0 ,
	\end{equation}
	which has as the general solution
	\begin{equation}\label{NPFAux}
		F'(\bar{R})=c_{1}t^{\alpha_{+}}+c_{2}t^{\alpha_{-}} ,
	\end{equation}
	where $c_{1}$ and $c_{2}$ are integration constants and
	\begin{equation}\label{NPAlphas}
		\alpha_{\pm}=\frac{1}{2}\left[1\pm\sqrt{1+12n}\right] .
	\end{equation}
	The remaining equations of the system are satisfied identically. Therefore, using (\ref{BRTime}), we can write the $F$ function in terms of $\bar{R}$, thus we obtain
	\begin{equation}\label{NPFBR}
		F(\bar{R})=-\frac{A_{1}}{\beta_{+}\bar{R}^{\beta_{+}}}+\frac{A_{2}}{\beta_{-}}\bar{R}^{\beta_{-}} ,
	\end{equation}
	where the positive constants $\beta_{\pm}$ are defined by
	\begin{equation}\label{NPBetas}
		\beta_{+}=\frac{\alpha_{+}}{2}-1=\frac{1}{4}\left[\sqrt{1+12n}-3\right] \geq0 , \hspace{0.5cm} \beta_{-}=1-\frac{\alpha_{-}}{2}=\frac{1}{4}\left[\sqrt{1+12n}+3\right]>\frac{3}{2}.
	\end{equation}
	The last inequality follows from the condition $3n-2>0$ in both cases, we have also defined
	\begin{equation}
		A_{1}=c_{1}\left[(3\lambda-1)n(3n-2)\right]^{\beta_{+}+1} , \hspace{0.5cm} A_{2}=\frac{c_{2}}{\left[(3\lambda-1)n(3n-2)\right]^{\beta_{-}-1}} .
	\end{equation}
	
	We note that the general form of the solution contains two powers terms of $\bar{R}$, then we expect that the dS conjecture can not be fulfilled independently of $\bar{R}$ with the two terms at the same time, it can only be fulfilled in this way when each term is taken independently. Moreover, the positivity condition of the scalar field potential in the Einstein frame (\ref{ConditionBarFR}) is written as
	\begin{equation}\label{NPCondition}
		\bar{R}F'-F=\frac{A_{1}}{\bar{R}^{\beta_{+}}}\left(1+\frac{1}{\beta_{+}}\right)+A_{2}\bar{R}^{\beta_{-}}\left(1-\frac{1}{\beta_{-}}\right)>0.
	\end{equation}
	Furthermore, from (\ref{NPBetas}) we obtain that 
	\begin{equation}\label{BFBetasP}
		1+\frac{1}{\beta_{+}}=\frac{\sqrt{1+12n}+1}{\sqrt{1+12n}-3}>0 , \hspace{0.5cm} 1-\frac{1}{\beta_{-}}=\frac{\sqrt{1+12n}-1}{\sqrt{1+12n}+3}>0  .
	\end{equation}
	Thus taking positive values for the integration constants $c_{1}$ and $c_{2}$ we can easily fulfil (\ref{NPCondition}). In this case, we obtain
	\begin{equation}\label{NPAbsF}
		|F|=\bigg\rvert-\frac{A_{1}}{\beta_{+}\bar{R}^{\beta_{+}}}+\frac{A_{2}}{\beta_{-}}\bar{R}^{\beta_{-}}\bigg\rvert .
	\end{equation}
	We note that the $F$ function cannot have a definite sign for all values of $\bar{R}$ because there is a minus sign in the first term. Thus considering both terms at the same time, we cannot fulfill the dS conjecture for all values of $\bar{R}$, as anticipated. Let us then consider each term separately. 
	
	Firstly we choose  $c_{1}=0$ and $c_{2}>0$, that is we consider only the positive power term in $F(\bar{R})$. Then, the first dS conjecture (\ref{dSConjectureBarFR}) leads to
	\begin{equation}\label{dSConjecturNPPositive}
		\tcboxmath[colback=red!10!white,colframe=red]
		{
		\frac{1}{3}<\lambda\leq\frac{1}{3}+\frac{16}{c^2(\sqrt{1+12n}-1)^2} .
		}
	\end{equation}
	On the other hand, choosing $c_{1}>0$ and $c_{2}=0$, that is considering only the negative power term in $F(\bar{R})$, the first dS conjecture leads to
	\begin{equation}\label{dSConjecturNPNegative}
		\tcboxmath[colback=red!10!white,colframe=red]{
		\frac{1}{3}<\lambda\leq\frac{1}{3}+\frac{16}{c^2(\sqrt{1+12n}+1)^2}.}
	\end{equation}
	
	In both cases the positivity of the potential condition (\ref{ConditionBarFR}) is automatically satisfied. Thus, for both cases the dS conjecture leads  to a constraint for the HL parameter $\lambda$. We note that as the value of $n$ increases, that is as the expansion is faster, the upper bound from $\lambda$ decreases and it gets closer to the limiting value $\lambda=1/3$. Thus, in order to have a fast expansion and fulfil the conjecture independently of $\bar{R}$ at the same time, we conclude that $\lambda$ must be bigger but close to $1/3$. Therefore, $\lambda$ is constrained to a region away for its IR limit value ($\lambda\to1$). Furthermore, since neither $\beta_{+}$ nor $\beta_{-}$ in (\ref{NPFBR}) depend on $\lambda$,  the exponent of the $F(\bar{R})$ function is not constrained by the conjecture. Thus, we have the freedom to choose any positive values for these terms, contrary to what happened in the standard $f(R)$ case.
	
	On the other hand, in this case the second dS conjecture (\ref{dSSecondConjectureBarFR}) is written in general as
	\begin{equation}\label{FBRdS2General}
		\begin{array}{l}
			\dfrac{V_{\phi\phi}}{V}\vspace{3pt}\\
			=\dfrac{9}{4\alpha^{2}}\dfrac{\left[\dfrac{A_{1}}{\bar{R}^{\beta_{+}+1}}+A_{2}\bar{R}^{\beta_{-}-1}\right]^2-\left[-\dfrac{A_{1}}{\beta_{+}\bar{R}^{\beta_{+}}}+\dfrac{A_{2}}{\beta_{-}}\bar{R}^{\beta_{-}}\right]\left[-\dfrac{A_{1}(\beta_{+}+1)}{\bar{R}^{\beta_{+}+2}}+A_{2}(\beta_{-}-1)\bar{R}^{\beta_{-}-2}\right]}{\left[-\dfrac{A_{1}(\beta_{+}+1)}{\bar{R}^{\beta_{+}+2}}+A_{2}(\beta_{-}-1)\bar{R}^{\beta_{-}-2}\right]\left[\dfrac{A_{1}}{\bar{R}^{\beta_{+}}}\left(1+\dfrac{1}{\beta_{+}}\right)+A_{2}\bar{R}^{\beta_{-}}\left(1-\frac{1}{\beta_{-}}\right)\right]} \vspace{3pt}\\<-\widetilde{c}.
		\end{array}
	\end{equation}
	Taking  $c_{1}=0$, that is the positive powers of $\bar{R}$, we obtain 
	\begin{equation}\label{DBRdS2P}
		\frac{9}{4\alpha^{2}}\frac{1}{(\beta_{-}-1)^2}<-\widetilde{c} ,
	\end{equation}
	however the term on the left hand side is always positive, thus this inequality can never be fulfilled. On the other hand, taking $c_{2}=0$, that is considering only the negative power terms in $\bar{R}$, we obtain 
	\begin{equation}\label{DBRdS2N}
		\frac{9}{4\alpha^{2}}\frac{1}{(\beta_{+}+1)^2}<-\widetilde{c} ,
	\end{equation}
	which is also never fulfilled. Thus, we can conclude that the second dS conjecture is never compatible with the obtained forms for the $F(\bar{R})$ function.
	
	In summary, we have shown that in HL $F(\bar{R})$ theories we can  obtain solutions that describe expanding universes with a power law scale factor in the Jordan frame. Through the equations of motion these solutions lead naturally to power law terms for the $F(\bar{R})$ function. After performing the conformal transformation, we can obtain in the Einstein frame a scalar field and its potential defined by geometry, then we can apply the dS conjecture. The first version of the dS conjecture can be studied independently of $\bar{R}$ taking each term on the solution of $F$ separately. In both cases it leads to an inequality for the $\lambda$ parameter  which gets constrained to the UV limit $\lambda\to1/3$ for fast expanding universes. This result is in agreement with the difficulties of achieving this scenario in GR. Moreover, we showed that the second dS conjecture can never be fulfilled.
	
	Finally, we note that in the Jordan frame we have an expanding universe with a power law form, let us explore what behaviour is encountered in the Einstein frame, where there is a scalar field coupled to gravity. In order to obtain such behaviour, we note that both frames are related by the conformal transformation (\ref{ConformalTrans3}) only on the three-metric, thus in the Einstein frame we also have a flat FLRW metric, with a scale factor given by
	\begin{equation}\label{NPScaleFactorEF}
		\widetilde{a}(t)=t^n\left[c_{1}t^{\frac{1}{2}\left(1+\sqrt{1+12n}\right)}+c_{2}t^{\frac{1}{2}\left(1-\sqrt{1+12n}\right)}\right]^{1/3}.
	\end{equation}
	Let us consider each term separately, as we have considered for the dS conjecture. The negative powers of $\bar{R}$ are obtained with the choice $c_{1}=1$ and $c_{2}=0$. Then, the scale factor takes the form
	\begin{equation}\label{ScaleFactorEFN}
		\widetilde{a}(t)=t^{\frac{1}{6}\left[6n+1+\sqrt{1+12n}\right]}=t^{\theta_{+}} .
	\end{equation}
	We note that $\theta_{+}$ does not depend on $\lambda$, therefore the constraint derived from the dS conjecture in (\ref{dSConjecturNPNegative}) does not constrain this parameter. Furthermore, we note that such exponent is always positive, therefore in this case the universe is always expanding as well. 
	
\begin{figure}[h!]
		\centering
		\includegraphics[width=0.65\textwidth]{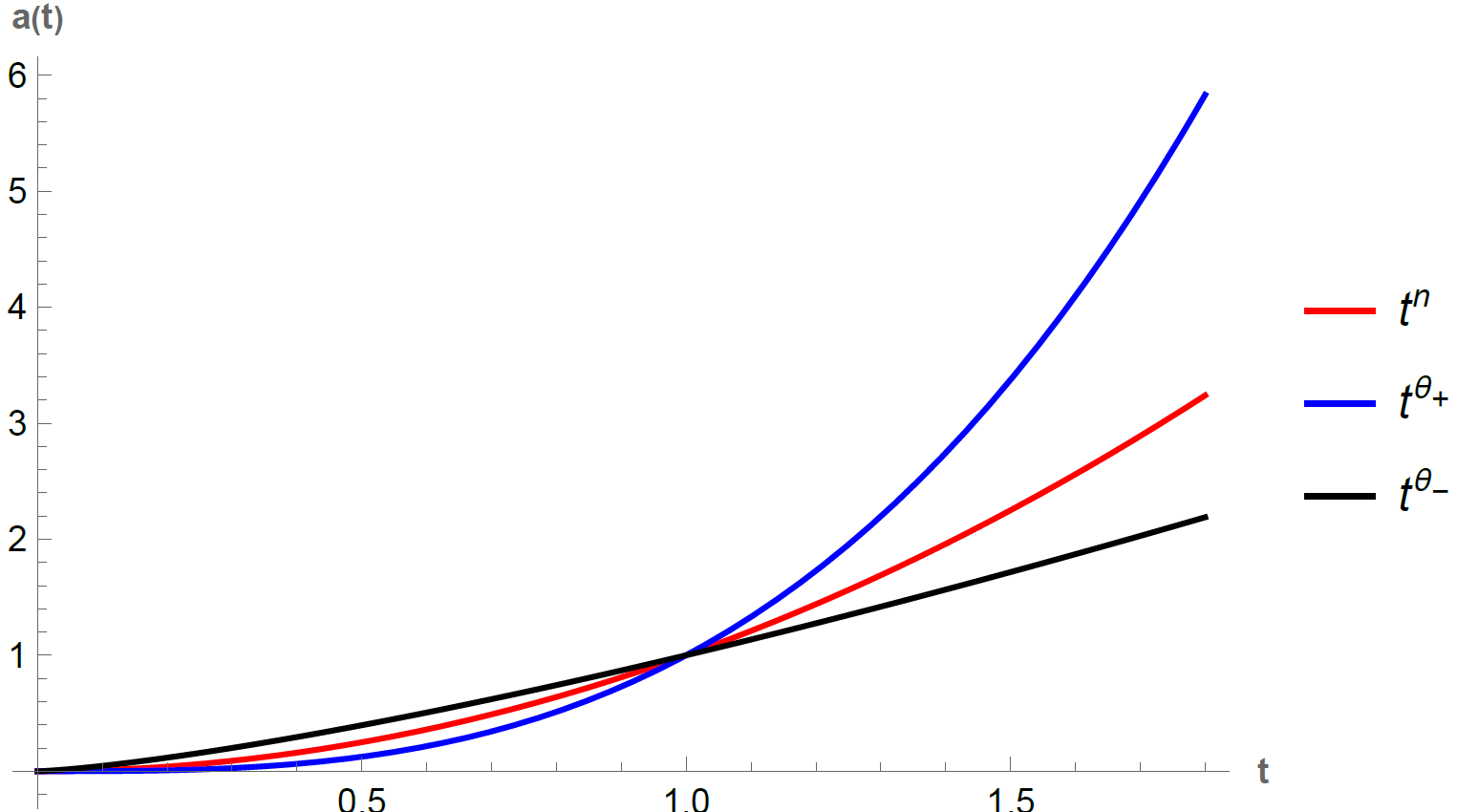}
		\caption{Scale factor with $n=2$ in the Jordan frame (Red curve) and in the Einstein frame with negative power terms of $\bar{R}$ in $F$ (Blue curve) and with positive power terms (Black curve).}
		\label{ScaleFactorPC}
	\end{figure}

	On the other hand, choosing $c_{1}=0$ and $c_{2}=1$ we get the positive power term in $\bar{R}$. For this term the scale factor is given by
	\begin{equation}\label{ScaleFactorEFP}
		\widetilde{a}(t)=t^{\frac{1}{6}\left[6n+1-\sqrt{1+12n}\right]}=t^{\theta_{-}} .
	\end{equation}
	We note again that $\theta_{-}$ is independent of $\lambda$ and always takes positive values, thus it is not constrained and it always describes expanding universes in the Einstein frame. Moreover, we can show that in general,
	\begin{equation}\label{DiffThetas}
		\theta_{-}<n<\theta_{+} .
	\end{equation}
	Thus, for negative power terms of $\bar{R}$ in $F$, we have in the Einstein frame, a more rapidly expanding universe than in the Jordan frame for $t>1$. On the contrary, for positive powers, we have a slower expanding universe. However, for small time values $t<1$, this behaviour is reversed, and the fastest expanding universe corresponds to positive powers of $\bar{R}$. We show the behaviour just described  for $n=2$ in Figure \ref{ScaleFactorPC}.
	
	\subsection{Projectable version}
	\label{SS-P}
	We now study the projectable case, that is we consider $C\neq0$ in the system of equations (\ref{System1})-(\ref{System4}). We employ the same ansatz as in the last subsection, that is, an expanding universe in the form of a power law of the scale factor in the Jordan frame. Thus, the form for the $\bar{R}$ parameter (\ref{BRTime}) is still valid. In this case, the differential equation in (\ref{System4}) is written as
	\begin{equation}\label{EqAuxP}
		\frac{d^2}{dt^2}F'(\bar{R})-\frac{3n}{t^2}F'(\bar{R})+\frac{3C}{2(3\lambda-1)t^{3n}}=0 .
	\end{equation}
	The general solution of this equation is composed of the general homogeneous solution (the same as in the non-projectable case with $C=0$) and a particular solution. In this form, we obtain that the general solution is written as
	\begin{equation}\label{BFTimeP}
		F'(\bar{R})=c_{1}t^{\alpha_{+}}+c_{2}t^{\alpha_{-}}-\frac{3C}{2(3\lambda-1)(9n^2-12n+2)}t^{2-3n},
	\end{equation}
	which is valid avoiding the roots of the polynomial $9n^2-12+2$, which are $n\approx0.1952,1.138$. In this case, the $F(\bar{R})$ function is written as
	\begin{equation}\label{BarRFN}
		F(\bar{R})=-\frac{A_{1}}{\beta_{+}\bar{R}^{\beta_{+}}}+\frac{A_{2}}{\beta_{-}}\bar{R}^{\beta_{-}}-A_{C}\bar{R}^{3n/2} ,
	\end{equation}
	where we have defined
	\begin{equation}
		A_{C}=\frac{C}{(3\lambda-1)n(9n^2-12n+2)\left[2(3n-2)(3\lambda-1)\right]^{3n/2-1}}.
	\end{equation}
	
	Furthermore, in this case the condition for the positivity of the potential (\ref{ConditionBarFR}) is written as
	\begin{equation}\label{PotentialPositiveProyectable}
		\bar{R}F'-F=\frac{A_{1}}{\bar{R}^{\beta_{+}}}\left(1+\frac{1}{\beta_{+}}\right)+A_{2}\bar{R}^{\beta_{-}}\left(1-\frac{1}{\beta_{-}}\right)-A_{C}\bar{R}^{3n/2}\left(\frac{3n-2}{2}\right)>0 .
	\end{equation}
	Since $\beta_{\pm}$ fulfils (\ref{BFBetasP}), we obtain that this condition is fulfilled for all values of $\bar{R}$ when $A_{C}<0$, which implies $\frac{C}{9n^2-12n+2}<0$. Therefore, we have two possibilities:
	\begin{itemize}
		\item $C>0$ and $\frac{2}{3}<n<\frac{2+\sqrt{2}}{3}$.
		\item $C<0$ and $n>\frac{2+\sqrt{2}}{3}$ .
	\end{itemize}
	Moreover, in this case we have
	\begin{equation}\label{PAbsF}
		|F(\bar{R})|=\bigg\rvert-\frac{A_{1}}{\beta_{+}\bar{R}^{\beta_{+}}}+\frac{A_{2}}{\beta_{-}}\bar{R}^{\beta_{-}}-A_{C}\bar{R}^{3n/2}\bigg\rvert .
	\end{equation}
	However, since in this case the last term will always be present (since $C\neq0$), the function $F$ will not have a definite sign for all values of $\bar{R}$ taking $c_{2}=0$, then in this case we will not find constraints from the dS conjecture valid for all values of $\bar{R}$. Therefore, we will take $c_{1}=0$ and consider only the positive power term. In this case, the dS conjecture can be studied for all positive values of $\bar{R}$, but not independently of it as we stated before. Thus with this choice, the first dS conjecture leads to
	\begin{equation}\label{dSConjectureProyectable}
		A_{2}\bar{R}^{\beta_{-}}\left[\frac{1}{\beta_{-}}-\frac{2\alpha c}{3}\left(1-\frac{1}{\beta_{-}}\right)\right]-A_{C}\bar{R}^{3n/2}\left[1-\frac{2\alpha c(3n-2)}{6}\right]\geq0 .
	\end{equation}
	We note that in order to fulfil this inequality for all values of $\bar{R}$, we must impose that each term within square brackets is positive. Thus, these two constraints lead to
	\begin{equation}\label{PLambda1}
		\lambda\leq\frac{1}{3}+\frac{16}{c^2(\sqrt{1+12n}-1)^2} ,
	\end{equation}
	\begin{equation}\label{PLambda2}
		\tcboxmath[colback=red!10!white,colframe=red]{
		\lambda\leq\frac{1}{3}+\frac{4}{c^2(3n-2)^2} .}
	\end{equation} 
	We note that the first constraint (\ref{PLambda1}) is the same as the one obtained in the non-projectable case (\ref{dSConjecturNPPositive}). However, since we have obtained two constraints in the same parameter, we only need to impose the stronger one. It turns out that (\ref{PLambda1}) is more restrictive than (\ref{PLambda2}) only if $n<\frac{2+\sqrt{2}}{3}\approx1.138$. Thus, for most of the values of $n$, the dS conjecture is satisfied by (\ref{PLambda2}) for every value of $\bar{R}$. Then, taking $C\neq0$, we are lead to a stronger condition for the $\lambda$ parameter for most cases.
	
	On the other hand, from the general form of the $F$ function (\ref{BarRFN}), taking $c_{1}=0$ we obtain in this case
	\begin{equation}
		F'=A_{2}\bar{R}^{\beta_{-}-1}-\frac{3n}{2}A_{C}\bar{R}^{\frac{3n}{2}-1} ,
	\end{equation}
	\begin{equation}
		F''=A_{2}(\beta_{-}-1)\bar{R}^{\beta_{-}-2}-\frac{3n}{4}(3n-2)A_{C}\bar{R}^{\frac{3n}{2}-2} .
	\end{equation}
	Then since we have imposed that $-A_{C}>0$, we have $F''>0$ for all $\bar{R}>0$. However, in this case we obtain
	\begin{multline}
		(F')^2-F''F=\frac{A_{2}^2}{\beta_{-}}\bar{R}^{2\beta_{-}-2}+\frac{3n}{2}A^{2}_{C}\bar{R}^{3n-2}\\+A_{2}A_{C}\left[-3n+\frac{3n}{2\beta_{-}}\left(\frac{3n-2}{2}\right)+(\beta_{-}-1)\right]\bar{R}^{3n/2+\beta_{-}-2} ,
	\end{multline} 
	thus, $(F')^2-F''F$ does not have a definite sign for all values of $\bar{R}$. Therefore, we can not study the second dS conjecture (\ref{dSSecondConjectureBarFR}) in the projectable version for all values of $\bar{R}$ in general.
	
	In summary, in the projectable case, the first dS conjecture can not be fulfilled independently of $\bar{R}$, instead it is fulfilled for all positive values of $\bar{R}$ and only for positive power terms in the $F(\bar{R})$ function. The conjecture also constrain the $\lambda$ parameter near the UV region, actually for most values of $n$, this new condition is more restrictive than in the non-projectable case. 
	
\begin{figure}[h!]
		\centering
		\includegraphics[width=0.65\textwidth]{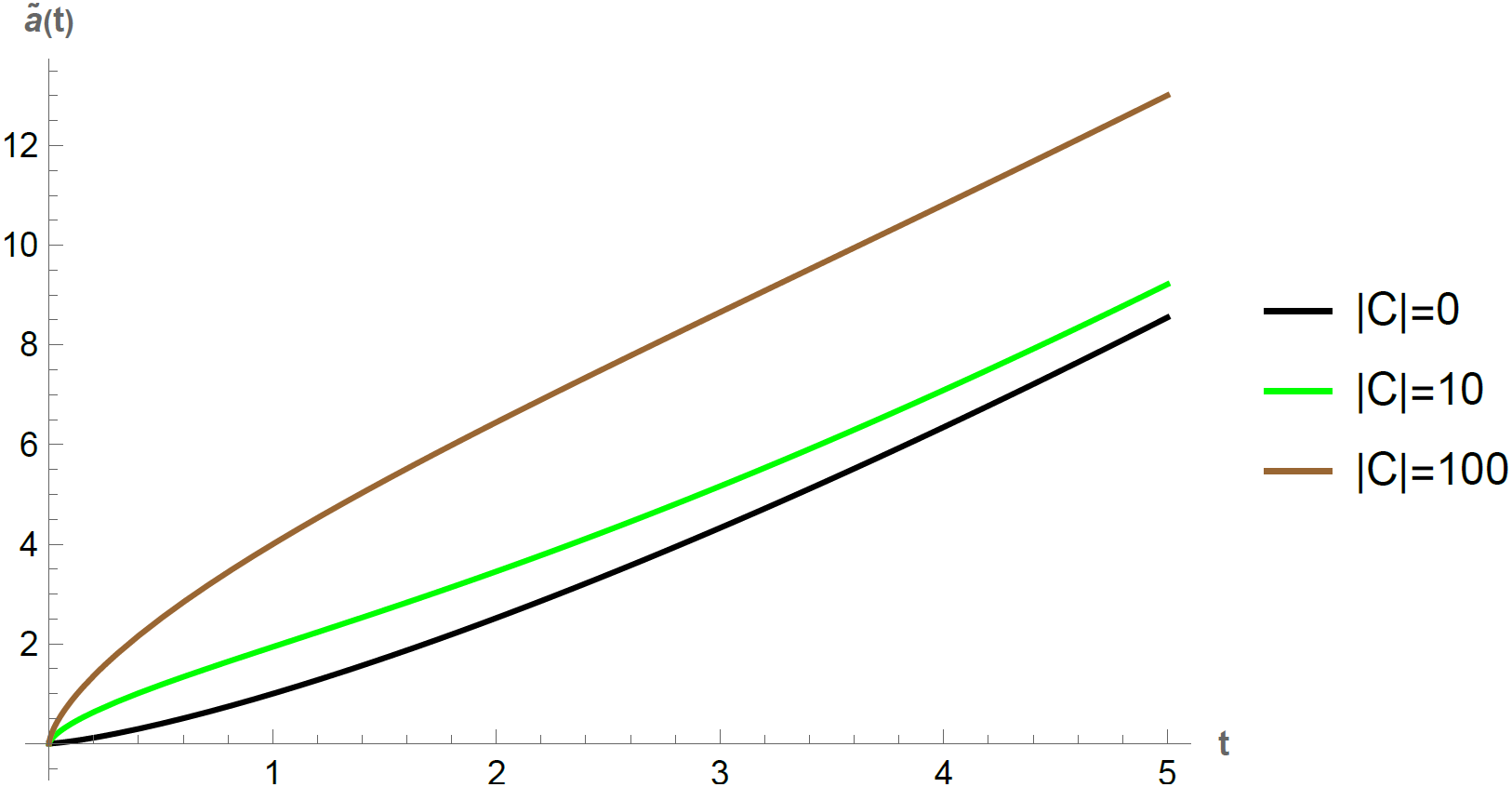}
		\caption{Scale factor in the Einstein frame for the projectable case with $\lambda=0.39$, $c=1$, $n=2$ and choosing $C=-|C|$ for $|C|=0$ (Black curve) (Non-projectable case), $|C|=10$ (Green curve) and $|C|=100$ (Brown curve). In all cases, we have expanding universes.}
		\label{ScaleFactorPr}
	\end{figure}

	In the projectable case we also have a flat FLRW metric in the Einstein frame. Taking $c_{2}=1$, the scale factor is 
	\begin{equation}\label{PScaleFactorEins}
		\widetilde{a}=t^n\left[t^{\frac{1}{2}(1-\sqrt{1+12n})}-\frac{3C}{2(3\lambda-1)(9n^2-12n+2)}t^{2-3n}\right]^{\frac{1}{3}} .
	\end{equation}
	Thus, in this case, the scale factor depends on $\lambda$, then it gets constrained by the dS conjecture, it creates a lower bound for it. Furthermore, we note that both terms are positive, thus the  universe is always expanding in this case as well. In Figure \ref{ScaleFactorPr}, we plot this scale factor choosing $\lambda=0.39$, $c=1$ and $n=2$ for different values of $C$. Since we choose $n=2$, in order to fulfil the condition $A_{C}<0$, we need to take negative values for $C$. Thus, we write $C=-|C|$. From this figure we see that the effect of projectability, ($|C|\neq0$), is that as $|C|$ increases, the scale factor increases, making the expansion faster.
	
	Finally, considering $c_{1}=0$ and $c_{2}=1$ in (\ref{BarRFN}) we obtain an $F(\bar{R})$ function with two terms of positive powers of $\bar{R}$. Thus, we can look for a possible way to write this function in the approximation form of an  Einstein-Hilbert term $\bar{R}$ plus corrections. From the dS conjecture constraint (\ref{PLambda1}) we obtain that if $n$ is close to its limiting value $2/3$, we can have access to the IR limit $\lambda\to1$. Thus, we will study the form that the $F(\bar{R})$ takes for values of $n$ close to $2/3$. Let us write
	\begin{equation}\label{Approxn}
		n=\frac{2}{3}(1+\epsilon) ,
	\end{equation}
	with $\epsilon\ll1$. In this scenario we can take $C>0$, then the most restrictive inequality is (\ref{PLambda1}). Then, taking  $c_{1}=0$, $c_{2}=1$ in (\ref{BarRFN}), we obtain
	\begin{multline}\label{BFREpsilon}
		F(\bar{R})=\frac{3C}{4(1+\epsilon)|1-2\epsilon^2|(4\epsilon)^{\epsilon}(3\lambda-1)^{\epsilon+1}}\bar{R}^{\epsilon+1}\\+\frac{4}{[\sqrt{9+8\epsilon}+3]\left[\frac{4}{3}\epsilon(\epsilon+1)(3\lambda-1)\right]^{(\sqrt{9+8\epsilon}-1)/4}}\bar{R}^{(\sqrt{9+8\epsilon}+3)/4} .
	\end{multline}
	For $\epsilon<<1$, we can make the approximation \cite{Benetti:2019smr}
	\begin{equation}\label{Approx}
		\bar{R}^{1+\epsilon}=\sum_{\nu=0}^{\infty}\frac{\epsilon^{\nu}\bar{R}\ln^{\nu}(\bar{R})}{\nu!}\simeq\bar{R}+\epsilon\bar{R}\ln(\bar{R}) .
	\end{equation}
	Thus, the latter is written as
	\begin{equation}\label{BFREpsilonF}
		F(\bar{R})\approx\frac{3C}{4(1+\epsilon)|1-2\epsilon^2|(4\epsilon)^{\epsilon}(3\lambda-1)^{\epsilon+1}}\bar{R}+F_{c}(\bar{R}),
	\end{equation}
	with the correction term
	\begin{multline}\label{BFRCorrec}
		F_{c}(\bar{R})=\frac{3C\epsilon}{4(1+\epsilon)|1-2\epsilon^2|(4\epsilon)^{\epsilon}(3\lambda-1)^{\epsilon+1}}\bar{R}\ln\bar{R}\\ +\frac{4}{[\sqrt{9+8\epsilon}+3]\left[\frac{4}{3}\epsilon(\epsilon+1)(3\lambda-1)\right]^{(\sqrt{9+8\epsilon}-1)/4}}\bar{R}^{(\sqrt{9+8\epsilon}+3)/4} .
	\end{multline}
	Furthermore, the scale factor in the Einstein frame (\ref{PScaleFactorEins}) takes the form
	\begin{equation}
		\bar{a}(t)=t^{\frac{2}{3}(1+\epsilon)}\left[\frac{1}{t^{\frac{1}{2}(\sqrt{9+8\epsilon}-1)}}+\frac{3C}{4(3\lambda-1)|1-2\epsilon^2|}\frac{1}{t^{2\epsilon}}\right]^{\frac{1}{3}} .
	\end{equation}
	Therefore, for these values of $n$, we can take values of $\lambda$ greater than one, that is a form of an IR limit, and the resulting universes are always expanding. However, from (\ref{System3}) we obtain in this limit
	\begin{equation}
		\lim_{\lambda\to1}\bar{R}=6H^2+4\dot{H} .
	\end{equation}
	We note that this result does not coincide with the $R$ value in GR (\ref{RFLRW}). Thus, since we have chosen (\ref{ElectioLambda}) in this case, the $\lambda\to1$ IR limit does not correspond to GR. We will be able to perform a correct GR limit in subsection \ref{SS-G}, where we will avoid this particular choice and consider the conjectures in the general case.
	
	\subsection{Constraints on $\lambda$}
	\label{SS-Lamdas}
	So far, we have used the particular case defined by $\mu=\lambda-1/3$, and we have shown that the first dS conjecture leads to constraints on the $\lambda$ parameter, for both the projectable and the non-projectable versions of the theory.

\begin{figure}[h!]
		\centering
		\includegraphics[width=0.7\textwidth]{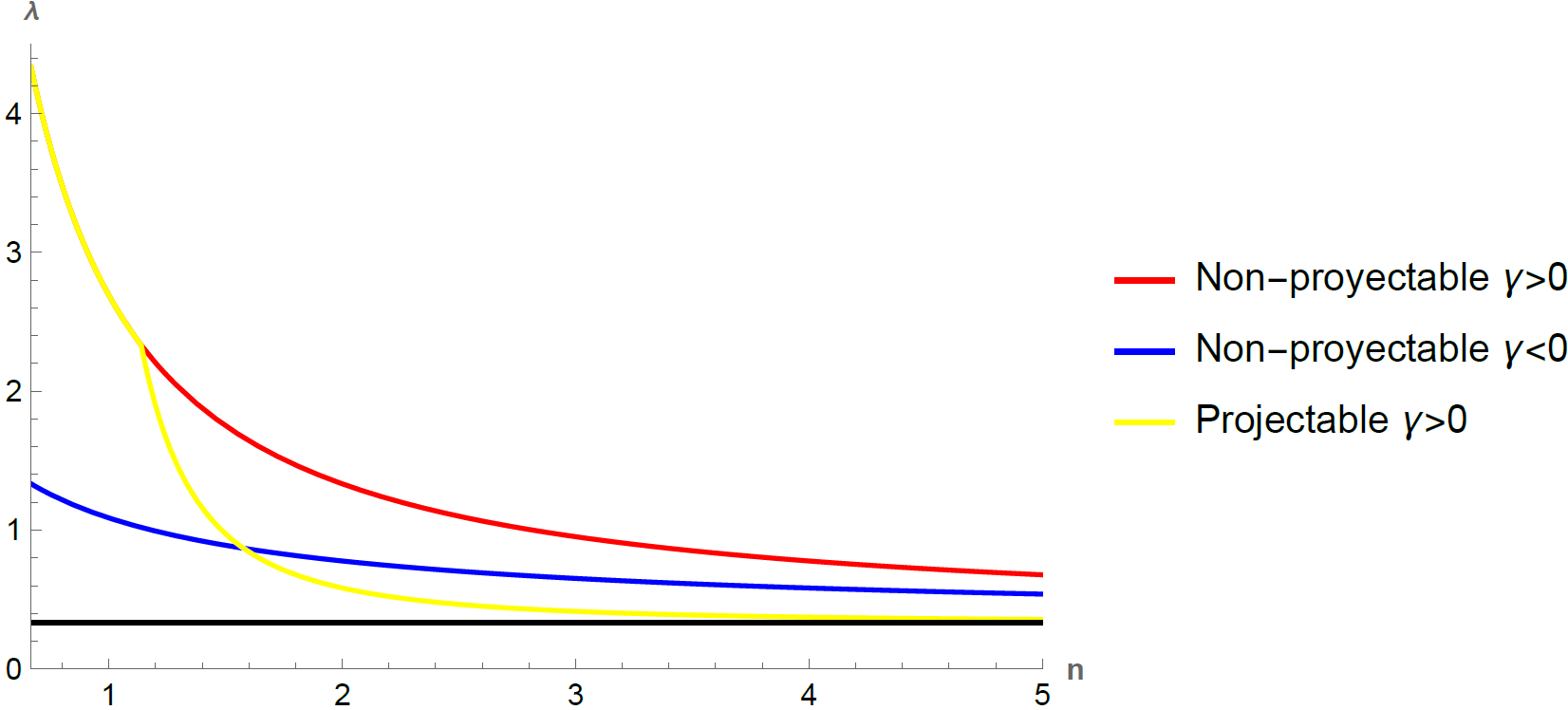}
		\caption{Allowed region for the $\lambda$ parameter consistent with the dS conjecture with $c=1$. The lower bound $1/3$ is indicated with the horizontal black line. For the non-projectable case, we have the generic form $F(\bar{R})\sim\bar{R}^{\gamma}$. The upper bounds for the non-projectable case are shown for $\gamma<0$ (Blue curve), and for $\gamma>0$ (Red curve). For the projectable case, we have the additional term in $F(\bar{R})$ with $C\neq0$ and $\gamma>0$. The corresponding upper bound is shown as well (Yellow curve). }
		\label{Lambdas}
	\end{figure}

	We want to obtain a visual representation of these constraints. Thus in Figure \ref{Lambdas} we present a plot of such constraints choosing in all cases $c=1$. For the non-projectable version, we plot the upper values  (\ref{dSConjecturNPPositive}) and (\ref{dSConjecturNPNegative}) varying the $n$ parameter. It is shown that the region where $\lambda>1$ is allowed only for small values of $n$,  furthermore the allowed region is reduced as $n$ grows in both cases. It is shown that in general, the allowed region for positive power terms in the $F(\bar{R})$ function is bigger than for the negative power terms. On the other hand, for the projectable case, the behaviour of the corresponding upper bound in (\ref{PLambda1}) and (\ref{PLambda2}) is also shown. We only plot the most restrictive constraint as we vary $n$. We can see that that the constraint in this case is the same as the one obtained in the non-projectable case for small values of $n$. However, when $n$ takes big enough values, the constraint coming from the projectable case is more restrictive, thus the allowed range of $\lambda$ is reduced and decreases inversely with $n$. 
	
	\subsection{The General Case}
	\label{SS-G}
	Let us move on to the general scenario, in this case $\mu$ and $\lambda$ in  (\ref{FbarRActionAux}) will be considered as independent parameters. The scalar field is defined in the same form by (\ref{RedScalarField}). We note that in this case the action does not consist only of a gravitational action and the action for a canonically coupled scalar field, we have an extra term, the third one in (\ref{FbarRActionAux}), that combines the metric with the scalar field. However, we still have a correct definition for the scalar field with the appropriate kinetic term in the action as in the particular case, we also have a concrete definition of the corresponding potential. Thus, adopting the \textit{a posteriori} position indicated earlier, we will propose that the dS conjecture can still be applicable for this action. This kind of argument of applying the swampland conjectures in scenarios where the scalar field is not canonically coupled have been studied before, for example in \cite{Brahma:2019kch}.

	From the definition of the scalar field in (\ref{RedScalarField}) we note that in order to have a proper definition we need the parameter $\alpha$ defined in (\ref{DefAlpha}) to be a real number, this fact  leads to the condition
	\begin{equation}\label{D1}
		\lambda<\frac{1}{3}+2\mu.
	\end{equation}
	
	From the system of equations (\ref{EoM1}) and (\ref{EoM2}), we can derive the following differential equation
	\begin{equation}\label{GEoMEx}
		\frac{d^2F'}{dt^2}+\frac{3\lambda-1-3\mu}{\mu}H\frac{dF'}{dt}+\frac{3\lambda-1}{\mu}\dot{H}F'+\frac{C}{2\mu a^3}=0 .
	\end{equation}
	In the following, we will consider $\mu,\lambda>0$. Furthermore, we are also going to use the ansatz for the scale factor in the Jordan frame as in the particular case, that is we will employ $a(t)=t^n$. Then, in the non-projectable case ($C=0$) the latter expression leads to
	\begin{equation}\label{GEoM}
		t^2\frac{d^2F'}{dt^2}+\frac{3\lambda-1-3\mu}{\mu}nt\frac{dF'}{dt}-\frac{(3\lambda-1)n}{\mu}F'=0 .
	\end{equation}
	
	Furthermore, in this case we obtain from (\ref{BarRFLRW}) that
	\begin{equation}\label{GRbart}
		\bar{R}=\frac{D}{t^2} ,
	\end{equation}
	where $D=3n[n(6\mu+1-3\lambda)-2\mu]$. Therefore, in the general case we still can think of $\bar{R}$ as to be related to the curvature. In the particular case studied before, we considered positive values for $\bar{R}$, this fact led us to a lower bound for $n$ which was compatible with expanding universes. For consistency and simplicity, in the general case we will also take positive values of $\bar{R}$ only. Thus $D>0$, which leads to the constraint
	\begin{equation}\label{D2}
		\lambda<\frac{1}{3}+\frac{2\mu(3n-1)}{3n}  ,
	\end{equation}
	from which it can de deduced that $n>1/3$. Thus the condition for a correct definition of the scalar field and the constriction to take only positive values for $\bar{R}$ have led to upper bounds for $\lambda$. Now, we proceed as in the particular case, that is we will find the solution of (\ref{GEoM}) and construct the $F(\bar{R})$ function. The general solution of (\ref{GEoM}) is
	\begin{equation}\label{GF'}
		F'(\bar{R})=c_{1}t^{\delta_{+}}+c_{2}t^{\delta_{-}} ,
	\end{equation}
	with
	\begin{equation}\label{DefDeltas}
		\delta_{\pm}=\frac{1}{2}\left[3n+1-B\pm\sqrt{(B+1-3n)^2+12n}\right] ,
	\end{equation}
	where we have defined
	\begin{equation}\label{DefB}
		B=\frac{3\lambda-1}{\mu}n .
	\end{equation}
	Using the relation (\ref{GRbart}), we can integrate the latter expression to obtain
	\begin{equation}\label{GF}
		F(\bar{R})=-\frac{c_{1}D^{\delta_{+}/2}}{\gamma_{+}\bar{R}^{\gamma_{+}}}+\frac{c_{2}D^{\delta_{-}/2}}{\gamma_{-}}\bar{R}^{\gamma_{-}} ,
	\end{equation}
	where $c_{1,2}$ are constants of integration and 
	\begin{equation}\label{DefGammas}
		\gamma_{+}= -1 + \frac{\delta_{+}}{2} , \hspace{0.5cm} \gamma_{-}=1-\frac{\delta_{-}}{2}  .
	\end{equation}
	In this case, the condition of the positivity of the scalar field potential is written as
	\begin{equation}\label{GCondtionPotential}
		\bar{R}F'-F=\frac{c_{1}D^{\delta_{+}/2}}{\bar{R}^{\gamma_{+}}}\left(1+\frac{1}{\gamma_{+}}\right)+c_{2}D^{\delta_{-}/2}\bar{R}^{\gamma_{-}}\left(1-\frac{1}{\gamma_{-}}\right)>0 ,
	\end{equation}
	which has the same form as (\ref{NPCondition}) with $\beta_{\pm}$ substituted by $\gamma_{\pm}$. Thus, the condition can be fulfilled for all values of $\bar{R}$, by considering each term to be positive. Choosing $c_{1}>0$ and $c_{2}>0$, we obtain the conditions $\delta_{+}>2$ and $\delta_{-}<0$. The first condition is automatically satisfied with (\ref{D2}), however the second condition leads to 
	\begin{equation}\label{D3}
		\lambda>\frac{1}{3} .
	\end{equation}
	Furthermore, we can show from (\ref{D2}) and (\ref{D3}), that  $\gamma_{\pm}>0$ always, therefore in the general case the $F(\bar{R})$ function  (\ref{GF}) is written in the form of two power terms in $\bar{R}$, one with positive exponent and one with negative exponent. Furthermore, the negative power term always has a negative coefficient, whereas the positive power term always has a positive coefficient, in agreement with the particular case studied before.
	
	Therefore in the general case before the dS conjecture, we have three inequalities that constrain the  $\lambda$  parameter, namely expressions (\ref{D1}), (\ref{D2}) and (\ref{D3}). Since the three constraints are imposed on the same parameter, it is sufficient to consider only the most restrictive one. Therefore we obtain the constraint
	\begin{equation}\label{D3C}
		\tcboxmath[colback=red!10!white,colframe=red]{
		\frac{1}{3}<\lambda<\frac{1}{3}+\frac{2\mu(3n-1)}{3n} .}
	\end{equation}
	This constraint implies that as  $\mu$ goes to zero, $\lambda$ becomes closer to its limiting UV value $1/3$. Furthermore, we also obtain that $\mu\neq0$ in order to correctly fulfill the inequalities. Thus, the simpler versions of $F(\bar{R})$ theories that do not take into account $\mu$ are inconsistent with this conformal transformation. Thus, considering both parameters is needed as explained in \cite{Chaichian:2010yi}.
	
	In the general case the first dS conjecture is still of the form (\ref{dSConjectureBarFR}) but now with the $F(\bar{R})$ function (\ref{GF}). As performed in the particular case we will consider each term separately, so we can fulfil the conjecture independently of $\bar{R}$. Taking $c_{2}=0$ we consider the negative power term, then the conjecture leads to
	\begin{equation}\label{D4}
		\tcboxmath[colback=red!10!white,colframe=red]{
		\sqrt{\frac{3\mu(6n-B)}{n}}\left[\sqrt{(B+1-3n)^2+12n}+3n+1-B\right]\leq\frac{12}{c} .}
	\end{equation}
	On the other hand, taking $c_{1}=0$ for the positive power term, the conjecture leads to
	\begin{equation}\label{D5}
		\tcboxmath[colback=red!10!white,colframe=red]{
		\sqrt{\frac{3\mu(6n-B)}{n}}\left[\sqrt{(B+1-3n)^2+12n}+B-3n-1\right]\leq\frac{12}{c} .}
	\end{equation}
	We note that in general, both constraints (\ref{D4}) and (\ref{D5}) will lead to a region of validity for $\lambda$ and $\mu$. Thus we must intersect the resulting region  with the results of (\ref{D3C}) in order to obtain a consistent model compatible with the dS conjecture. Unfortunately, the expressions are complicated to solve analytically, thus we will rely on a numerical analysis.
	
\begin{figure}[h!]
		\centering
		\includegraphics[width=0.65\textwidth]{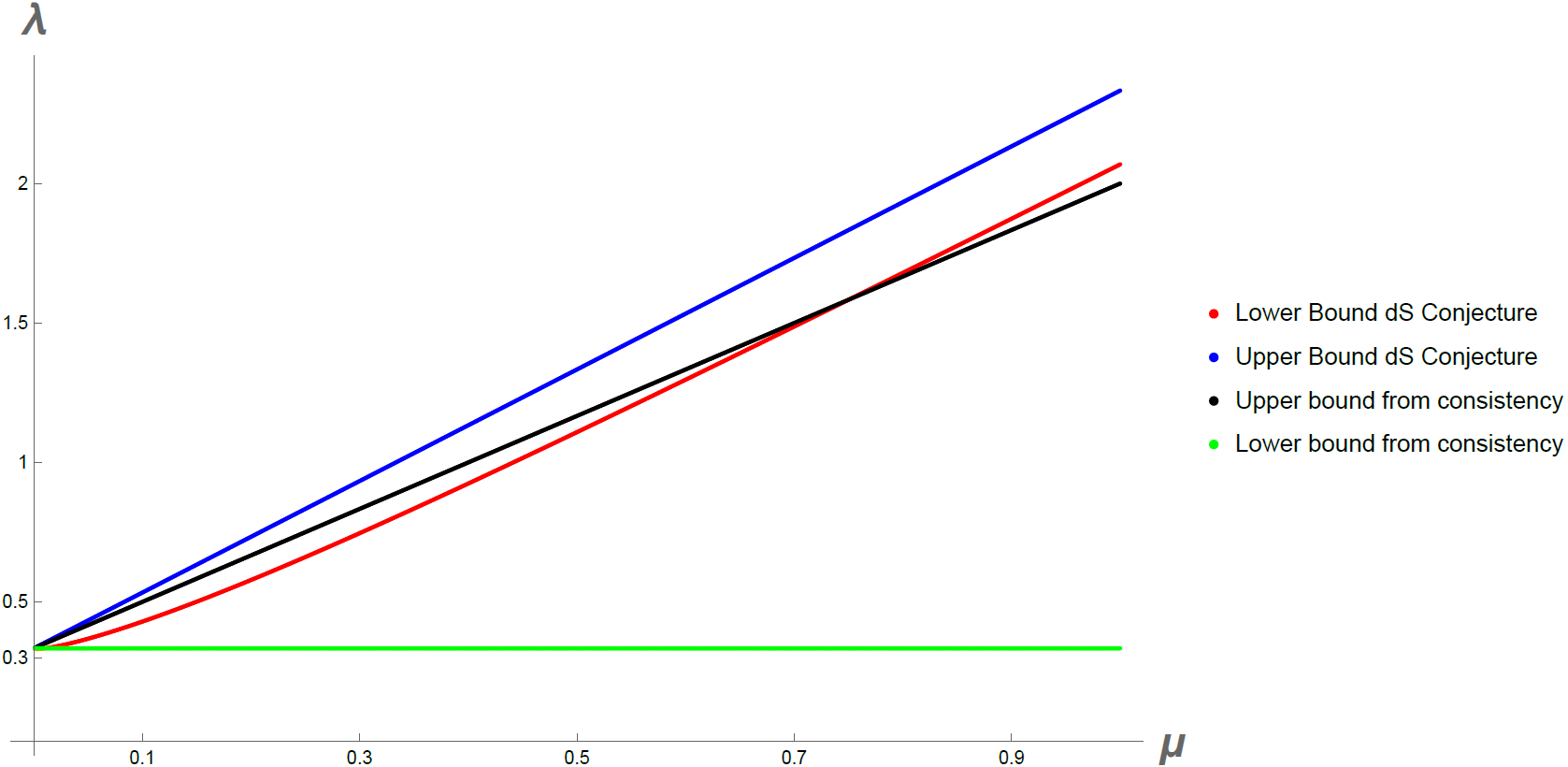}
		\caption{Allowed region of validity with a consistent theory and the dS conjecture for $\lambda$, in the general scenario without projectability, considering only the negative power term in $F(\bar{R})$ for $n=2$ and $c=2$. The lower and upper bounds from the dS conjecture are shown with the red and blue curves, respectively. The lower ($\lambda=1/3$) and upper bounds coming from (\ref{D3C}) are also shown with the green and black curves, respectively. The allowed values of $\mu$ are constrained to the region where the red curve shows smaller values than the black one.}
		\label{GeneralLambdasN}
	\end{figure}

	We consider first the negative power term in the $F(\bar{R})$ function, then upper and lower bounds from (\ref{D4}) are found numerically. In Figure \ref{GeneralLambdasN} these bounds are shown for $n=2$, $c=2$ and varying the value of the $\mu$ parameter. Furthermore, in this figure we also plot the upper and lower bounds coming from (\ref{D3C}). Therefore, the final allowed region must be the one in which the upper bound from (\ref{D3C}) is bigger than the lower bound from (\ref{D4}). Therefore an upper bound for $\mu$ appears. Then, the allowed region for $\lambda$ increases as $\mu$ increases, but there exists a maximum value. Considering smaller values of $n$ or bigger values of $c$, the constrain on $\mu$ is more restrictive, thus $\lambda$ gets constrained closer to $1/3$.
	
\begin{figure}[h!]
		\centering
		\includegraphics[width=0.65\textwidth]{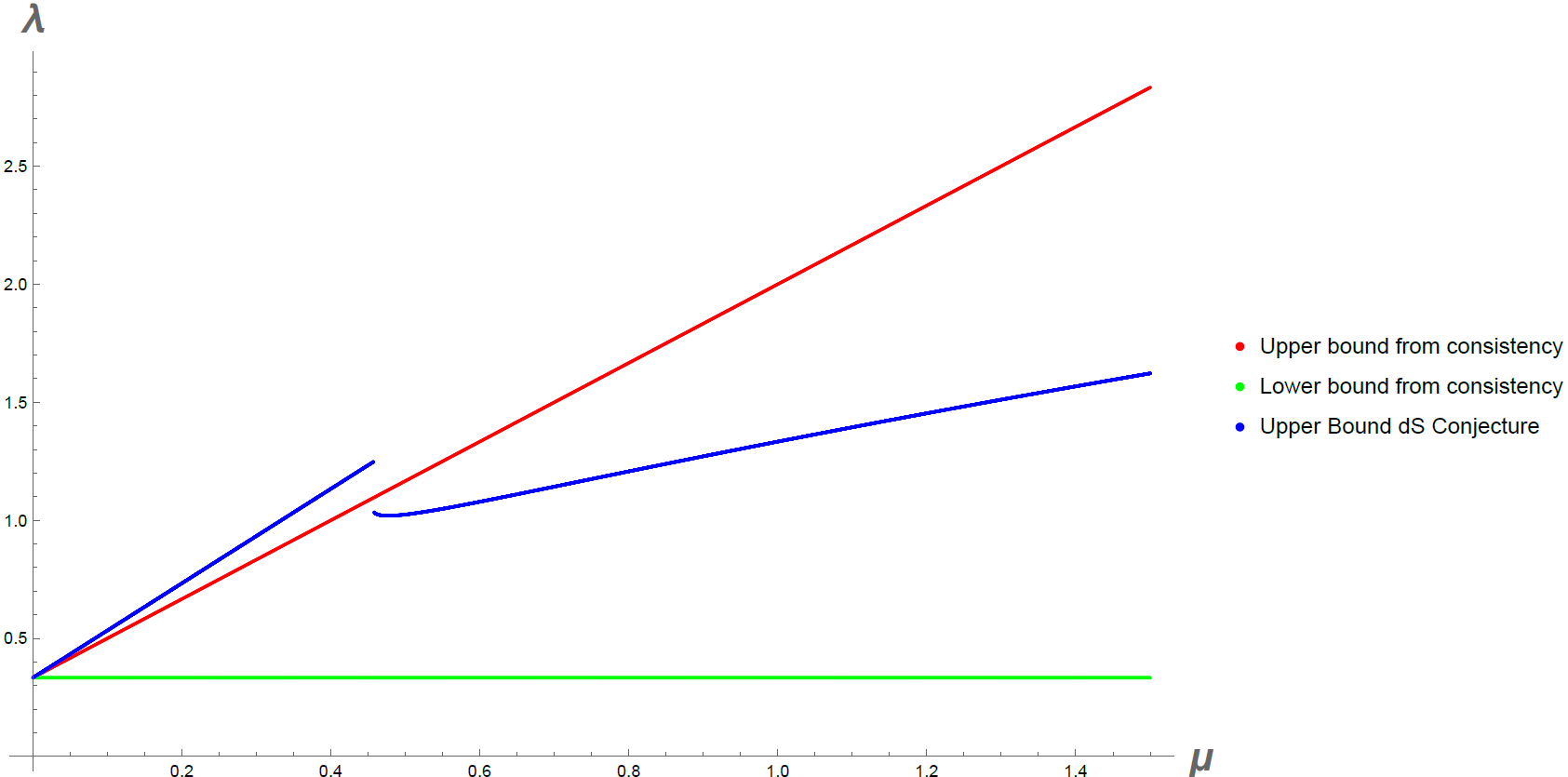}
		\caption{Allowed region of validity with a consistent theory and the dS conjecture for $\lambda$, in the non-projectable general scenario for the positive power term in $F(\bar{R})$ for $n=2$ and $c=1$. The lower ($\lambda=1/3$) and upper bounds from the consistency constrain (\ref{D3C}) are shown with the green and red curves respectively. The upper bound from the dS conjecture is shown with the blue curve. For large enough values of $\mu$, there are two allowed regions. We show only the lowest upper bound since the other region is inconsistent with (\ref{D3C}).}
		\label{GeneralLambdasP}
	\end{figure}

	On the other hand, for the positive power term in $F(\bar{R})$, we find performing a numerical analysis, that for small values of $\mu$ (\ref{D5}) leads to an upper bound for $\lambda$ that is bigger than the one coming from (\ref{D3C}), therefore the conjecture is automatically satisfied as a consequence of the consistency condition (\ref{D3C}) in this case. However, when $\mu$ takes values large enough, (\ref{D5}) leads to two regions of validity: the first is an upper bound which is smaller than (\ref{D3C}), the second one is a region with an upper and lower bounds bigger than (\ref{D3C}), therefore the second region is inconsistent. Thus, for large values of $\mu$, the dS conjecture is more restrictive than the consistency constraint (\ref{D3C}). Furthermore, both upper bounds grow with $\mu$, then in this case we can have access to larger values of $\lambda$, it is constrained to a region near $1/3$ if $\mu$ is small enough. In Figure \ref{GeneralLambdasP} we show this behaviour for $n=2$ and $c=1$. Taking smaller values of $n$, or larger values of $c$, the region where the dS conjecture is more restrictive than (\ref{D3C}) is found for smaller values of $\mu$.

	On the other hand, the second dS conjecture takes the same form as in (\ref{FBRdS2General}) with $\gamma_{\pm}$ instead of $\beta_{\pm}$. Therefore, for the positive power term, we have (\ref{DBRdS2P}) with $\gamma_{-}$, and for the negative power term we have (\ref{DBRdS2N}) with $\gamma_{+}$. Therefore, the second dS conjecture is never fulfilled in the general scenario either.
	
	In the general case, the metric in the Einstein frame is also of the flat FLRW form, the scale factor will be given by
	\begin{equation}\label{GSFAux}
		\widetilde{a}(t)=t^n\left[c_{1}t^{\delta_{+}}+c_{2}t^{\delta_{-}}\right]^{1/3} .
	\end{equation}
	Thus, choosing the negative power term ($c_{1}=1$ and $c_{2}=0$), the scale factor takes the form 
	\begin{equation}\label{GSFN}
		\widetilde{a}(t)=t^{\theta_{+}} ,
	\end{equation}
	where
	\begin{equation}\label{DefTheta+}
		\theta_{+}=\frac{1}{6}\left[9n+1-B+\sqrt{(B+1-3n)^2+12n}\right] .
	\end{equation}
	In the same way, for the positive power term ($c_{1}=0$ and $c_{2}=1$), the scale factor takes the form
	\begin{equation}\label{GSFP}
		\widetilde{a}(t)=t^{\theta_{-}} ,
	\end{equation}
	with
	\begin{equation}\label{DefTheta-}
		\theta_{-}=\frac{1}{6}\left[9n+1-B-\sqrt{(B+1-3n)^2+12n}\right] .
	\end{equation}
	Thus, we see that in the general scenario, the scale factor depends on $\lambda$ and $\mu$. Therefore the consistency constraint (\ref{D3C}) and the allowed region compatible with the first dS conjecture constrain the possible values. Moreover, we can show that $\theta_{\pm}>0$ is satisfied, thus we always have expanding universes in the Einstein frame as well.
	
	In summary, the dS conjecture in the general scenario leads to constraints on the parameters of the theory as in the particular case. Here we obtain two constraints, one coming from consistency of the theory and the other one arriving from the dS conjecture, the allowed region of validity restricts $\lambda$ in terms of $\mu$ around $1/3$. Furthermore, the resulting description in the Einstein frame are expanding universes. However, one important aspect to remark is that in the general scenario we can have access to larger values of $\lambda$ by considering appropriate values for $\mu$. This opens the possibility to consider the relevant limit to the $f(R)$ theories $\lambda,\mu\to1$. In this case, we have also shown that the second dS conjecture is never fulfilled. 
	
	As we have stated earlier, the general scenario allows us to explore correctly the limit $\mu,\lambda\to1$ which, as previously discussed, should correspond to standard $f(R)$ theories. In this limit, (\ref{D1}) and (\ref{D3}) are automatically satisfied, whereas (\ref{D2}) leads simply to $n>1/2$. Then, the first dS conjecture will lead simply to upper bounds for $n$. 
	
	For the negative power term, the dS constraint (\ref{D4}) leads to $c\lesssim0.87$, which is inconsistent with $c$ being an order $1$ constant, thus in this case the dS conjecture can not be consistent in this limit. On the other hand, for the positive power term, the dS constraint (\ref{D5}) leads to
	\begin{equation}\label{GAVn}
		\frac{1}{2}<n\leq\frac{\sqrt{3}}{c}\left(\frac{c+\sqrt{3}}{2c-\sqrt{3}}\right) , \hspace{0.5cm}\text{with}\hspace{0.5cm} 1\leq c\lesssim3.46 .
	\end{equation}
	The largest upper bound is obtained where $c=1$ and corresponds to approximately $17.66$. With these bounds, we obtain the allowed region for the exponent in the $F$ function from (\ref{DefDeltas}) which is given by
	\begin{equation}\label{GAVdelta}
		\gamma_{-}\big\rvert_{n=1/2}=1.25<\gamma_{-}\leq\gamma_{-}\big\rvert_{n=17.66}\simeq1.866 .
	\end{equation}
	The obtained region for the dS conjecture consistency (\ref{GAVdelta}) must be compared to the one obtained in the standard $f(R)$ case given by (\ref{fRRestrictGamma}). We note that the lower bound from (\ref{GAVdelta}) comes from the dependence of $\delta_{-}$ on $n$ and the condition $n>1/2$ coming from (\ref{D2}). We know that the system of equations (\ref{EoM1}) and (\ref{EoM2}) reduces to the standard $f(R)$ case in the limit $\mu,\lambda\to1$, $C\to0$. Therefore if we propose $a(t)=t^n$, the same form of the $f$ function (\ref{GF}) can be constructed, thus after imposing the positivity of $R$ , that is $D>0$ in the standard $f(R)$ theories, we can obtain the same lower bound, those we find consistency for the lower bound. However, the upper bound is actually different. This occurs because although in the Jordan frame the $F(\bar{R})$ theory recovers the standard $f(R)$ in the mentioned limit, the transformation to the Einstein frame employed for the $F(\bar{R})$ theories (\ref{ConformalTrans3}) was performed only on the three-metric. In contrast, in the standard $f(R)$ theories, the conformal transformation used (\ref{ConfTransfR}) is in the complete four-metric. Thus, the resulting scalar field and scalar potential are different for both theories, therefore the first dS conjecture has in general a different form. However, let us remark that although the analytical forms of both upper bounds are different, their maximum values are not too far apart. In the standard $f(R)$ theories, we obtain approximately the maximum upper bound $1.45$, whereas in the $F(\bar{R})$ theories we obtain approximately the maximum upper bound $1.866$. Thus, the $F(\bar{R})$ theories imply a bigger region of validity for the exponent, although it is still not too far from one. Therefore, in the limit $\mu,\lambda\to1$, the $F(\bar{R})$ theories lead to a consistent result with the standard $f(R)$ theories. That is, we obtain $\bar{R}\to R$, then the exponent on the $F(\bar{R})$ functions gets constrained to be bigger than one (actually bigger than $1.25$), but not too big (smaller than approximately $1.866$). Furthermore, using (\ref{DefTheta-}), we can obtain that in this limit the scale factor has the same form as in (\ref{GSFP})  with 
	\begin{equation}
		\frac{1}{3}<\theta_{-}\lesssim 17.083 .
	\end{equation}
	Thus, in the Einstein frame the velocity for the expansion of the universe can be quite large.
	
\begin{figure}[h!]
		\centering
		\includegraphics[width=0.65\textwidth]{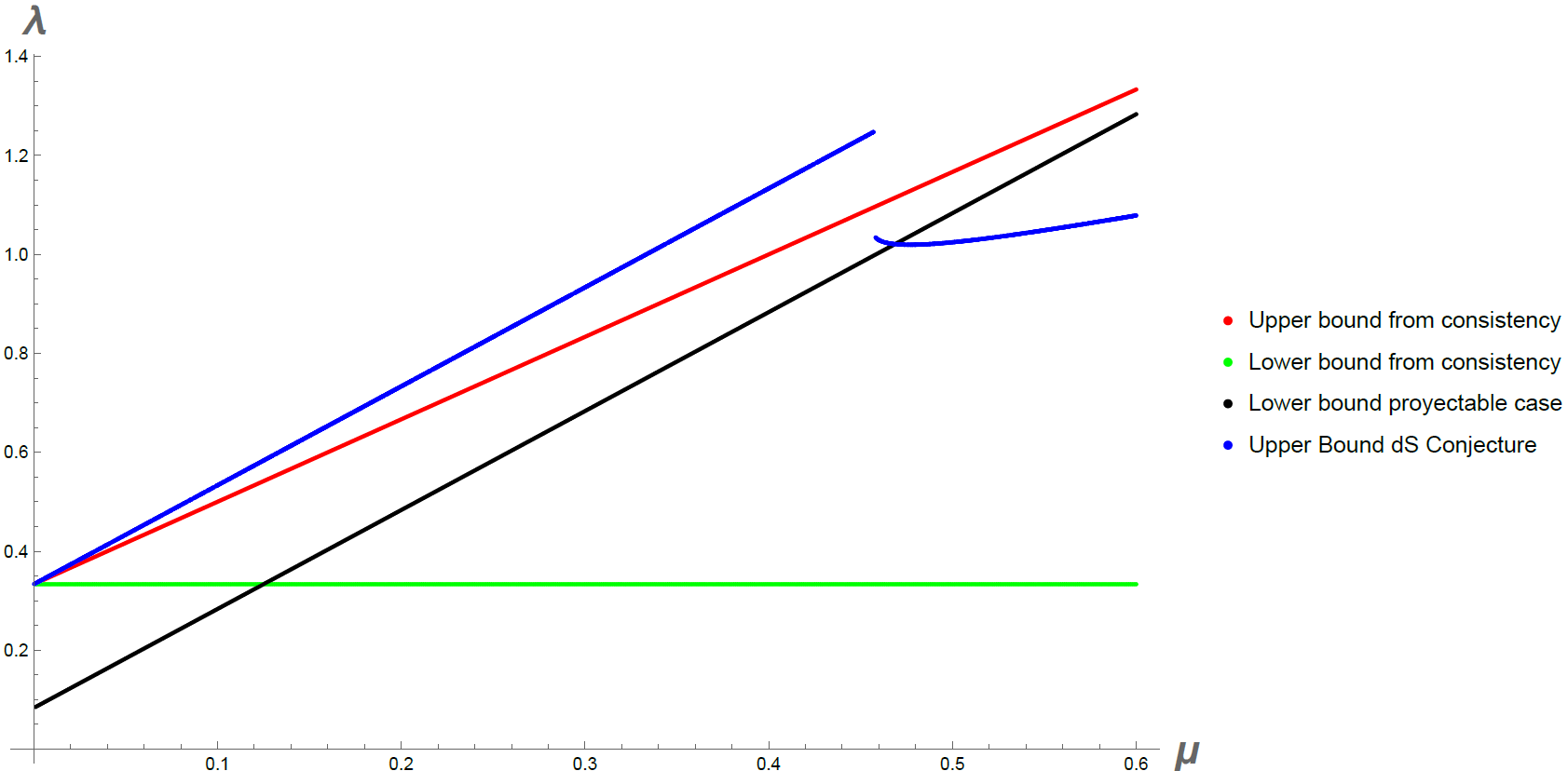}
		\caption{Allowed region of validity with a consistent theory and the dS conjecture for $\lambda$ in the general scenario with projectability for $n=2$ and $c=1$. The lower ($\lambda=1/3$) and upper bounds from the consistency constraints (\ref{D3C}) are shown with the green and red curves, respectively. The upper bound from the dS conjecture is shown with the blue curve. The lower bound from (\ref{DProy}) is shown with the black curve. The allowed region for $\mu$ is then constrain to the region where the black curve displays smaller values than the blue or red curves.}
		\label{GeneralLambdasPPr}
	\end{figure}

	Finally, let us study the projectable case in the general setup. We note that in the general case, the $F$ function has the same form as the one obtained in the particular case, the only difference is the specific form of the exponents. Thus, the projectable case can be treated in the same form as in subsection \ref{SS-P}. We obtain in this case that the $F(\bar{R})$ function takes the form
	\begin{equation}\label{FProjectableGeneral}
		F(\bar{R})=-\frac{c_{1}D^{\delta_{+}/2}}{\gamma_{+}\bar{R}^{\gamma_{+}}}+\frac{c_{2}D^{\delta_{-}/2}}{\gamma_{-}}\bar{R}^{\gamma_{-}}-\frac{C}{3\mu n(18n^2-15n+B-3nB+2)D^{3n/2-1}}\bar{R}^{3n/2} .
	\end{equation}
	The condition for the positivity of the potential is written in the same form as in (\ref{PotentialPositiveProyectable}), thus it can also be fulfilled for all values of $\bar{R}$ by taking the coefficient of the new term to be positive and imposing $n>2/3$. We choose $c_{1}=0$, then the first dS conjecture leads to a constrain written with two terms as in (\ref{dSConjectureProyectable}), therefore we can ask for the two terms to be positive in order to fulfil the conjecture for all values of $\bar{R}$. The first resulting constraint in this case is the same as the one obtained in (\ref{D5}), on the other hand the second constraint leads to
	\begin{equation}\label{DProy}
		\tcboxmath[colback=red!10!white,colframe=red]{
		\lambda\geq\frac{1}{3}+2\mu-\frac{4}{(3n-2)^2c^2} .}
	\end{equation}
	Therefore, in the projectable case we obtain an extra constraint, in addition to (\ref{D3C}) and (\ref{D5}). We note that this new inequality is actually a lower bound for $\lambda$. In Figure \ref{GeneralLambdasPPr} we show the bounds encounter earlier for the constraints (\ref{D3C}) and (\ref{D5})  for $n=2$ and $c=1$, we now add to the plot the constraint coming from (\ref{DProy}). We note that in order to fulfil the complete system of inequalities, an upper bound for $\mu$ appears, thus $\lambda$ becomes constrained to be close to the UV limit $1/3$ once again. 
	
	Furthermore, in the $f(R)$ limit, the new constraint (\ref{DProy}) takes the form
	\begin{equation}
		n\leq\frac{1}{\sqrt{3}c}+\frac{2}{3} .
	\end{equation}
	We note that the maximum value for the upper bound appears when $c=1$ and gives approximately $1.244$. Therefore the upper bound from the projectable case on $n$ is more restrictive than in the non-projectable case. Furthermore, with these constraints we obtain
	\begin{equation}\label{GRGProyectable}
		\gamma_{-}\big\rvert_{n=2/3}=1.295<\gamma_{-}\leq\gamma_{-}\big\rvert_{n=1.244}\simeq1.407 .
	\end{equation}
	Thus, the exponent of the $F(\bar{R})$ functions is constrained with bounds that are even more restrictive than in the standard $f(R)$ theories. However, the form of the $F$ function has an extra power term with an exponent of $3n/2$. Thus, taking small values of $n$ around $2/3$ as in (\ref{Approxn}), we can write this function as GR plus correction terms in the same form of (\ref{BFREpsilonF}) (the only difference will be that the coefficients of each term and the exponent of the second term in (\ref{BFRCorrec}) will have a different form). However in this case the infrared limit will actually correspond to GR since in this case we can obtain consistently $\bar{R}\to R$. 
	
	\subsection{Constant Hubble Parameter}
	\label{SS-CH}
	Let us remark that in the particular case considered in subsections \ref{SS-NP}-\ref{SS-Lamdas},  a solution with a constant Hubble parameter was not consistent with the choice of $\mu$ in terms of $\lambda$. It was for this reason that  we used the ansatz (\ref{ScaleFactorJordan}). Furthermore, in the general case, we have only considered the same ansatz, and use it to constrain the parameters of the theory. However, in the general case, there should not be an inconsistency problem for a constant Hubble parameter, actually this ansatz was used in reference \cite{Chaichian:2010yi} and explicit solutions were obtained for particular forms of the $F(\bar{R})$ function. Therefore, in this last subsection, we will employ this ansatz and analyze its compatibility with the dS conjecture as well. 
	
	If the Hubble parameter is a constant, we obtain from (\ref{BarRFLRW}) that
	\begin{equation}\label{ConstantBarR}
		\bar{R}=3(1-3\lambda+6\mu)H^2 ,
	\end{equation}
	therefore $\bar{R}$ is also a constant. Then the $F(\bar{R})$ function does not depend on time. Thus, both equations in the system (\ref{EoM1}) and (\ref{EoM2}), lead to 
	\begin{equation}\label{ConstantEoM}
		F(\bar{R})-6H^2(1-3\lambda+3\mu)F'(\bar{R})=0 ,
	\end{equation}
	where we have considered the non projectable case ($C=0$) for consistency of the equation. Now, since $\bar{R}$ is a constant, this equation does not lead to a unique solution. In particular, in  \cite{Chaichian:2010yi}, some $F(\bar{R})$ functions were proposed as polynomials. With these forms, they encountered two periods of accelerating expansion. However, as we have seen before from (\ref{dSConjectureBarFR}), if we want to fulfil the dS conjecture independently of $\bar{R}$, as it was performed on the previous sections, we must propose the form $F(\bar{R})=A\bar{R}^{\gamma}$ with $A$ and $\gamma$ constants.  Then,  we obtain  from (\ref{ConstantBarR}) and (\ref{ConstantEoM}) that the exponent is given by
	\begin{equation}
		\gamma=\frac{3\lambda-1-6\mu}{2\left(3\lambda-1-3\mu\right)} .
	\end{equation}
	Furthermore, the condition for a positive potential (\ref{ConditionBarFR}) can be fulfilled, in addition we can obtain a real value for $\alpha$ (\ref{DefAlpha}) by imposing the condition
	\begin{equation}\label{ConstantCondition1}
		\lambda<\frac{1}{3}+\mu ,
	\end{equation}
	then we obtain that $\bar{R}>0$ always. The first dS conjecture is written in the same form as in (\ref{dSConjectureBarFR}), in this case this leads to
	\begin{equation}\label{ConstantdSConjecture}
		\frac{(3\lambda-1)\sqrt{1-3\lambda+6\mu}}{2\sqrt{3}(3\mu-3\lambda+1)}\leq\frac{1}{c} .
	\end{equation}
	Let us point out that in the standard $f(R)$ limit $\mu,\lambda\to1$, the constraint (\ref{ConstantCondition1}) is satisfied, but the dS constraint (\ref{ConstantdSConjecture}) leads to $c<\frac{\sqrt{3}}{2}\simeq0.866$, which is not compatible with $c$ being an order 1 constant. Furthermore, the general solutions for the two constraints (\ref{ConstantCondition1}) and (\ref{ConstantdSConjecture}) are two validity regions given by
	\begin{equation}
		\tcboxmath[colback=red!10!white,colframe=red]{
		\lambda<\frac{1}{3} , \hspace{1cm} \mu\geq\frac{3\lambda-1}{6} ,}
	\end{equation}
	and
	\begin{equation}
		\tcboxmath[colback=red!10!white,colframe=red]{
		\lambda>\frac{1}{3} , \hspace{1cm} \mu\geq-\frac{1}{3}+\frac{c^2(3\lambda-1)^2}{36}+\lambda+\frac{1}{36}\sqrt{c^2(3\lambda-1)^3\left[12+c^2(3\lambda-1)\right]} .}
	\end{equation} 
	Then, the dS conjecture can be fulfilled in the region where $\lambda<1/3$ and $\mu$ takes any positive value or in the region where $\lambda>1/3$ and $\mu$ is bounded from below with a value that grows with $\lambda$ and $c$. We note that taking the limit $\mu,\lambda\to1$ is prohibited in any of the two regions. Furthermore, from (\ref{ConformalTrans3}) we note that the scale factor in the Einstein frame takes the form
	\begin{equation}
		\tilde{a}(t)=\left[\frac{3\lambda-1-6\mu}{2(2\lambda-1-3\mu)}\left[\left(3-9\lambda+18\mu\right)H^2\right]^{\frac{3\lambda-1}{2(3\mu-3\lambda+1)}}\right]e^{Ht} ,
	\end{equation}
	thus we have an exponentially expanding universe in the Einstein frame as well. In conclusion, a constant Hubble parameter is compatible with the dS conjecture leading to constraints over the parameters of the theory that forbids the infrared behaviour $\lambda,\mu\to1$, therefore this compatibility is exclusive to Ho\v{r}ava--Lifshitz $F(\bar{R})$ theories.
	
	\section{Discussion}\label{S-DHLFRSwampland}
	Finally, let us summarize and discuss the results presented in this chapter.
	
	We presented a study of the consistency between the dS swampland conjecture and the Ho\v{r}ava-Lifshitz $F(\bar{R})$ theories, in particular we used a FLRW metric. We did not consider any matter content in the Jordan frame, then we performed a conformal transformation to the Einstein frame, it was in this process that a scalar field appeared with a corresponding potential. Thus, in the Einstein frame we were able to apply the conjecture to obtain information of the theory since the scalar field and its potential were defined entirely by the geometry. We focused on scenarios where we can make the conjecture valid for all values of $\bar{R}$ in order to use it to constraint the parameters of the theories, in this way we were able to study the compatibility of the conjecture with the theory itself.
	
	We first considered the standard $f(R)$ functions that have been studied before in \cite{Artymowski:2019vfy}. We showed that the only way to fulfil the conjecture independently of $R$ is by taking $f$ as power terms in $R$. The exponent is constrained by the conjecture to be greater than $1$ (not GR) but less than around $1.45$, thus the conjecture can be satisfied in a very restrictive way.
	
	We then considered the HL $F(\bar{R})$ theories. Since we used a FLRW flat metric, the analysis is valid for every version of the theory. Furthermore, instead of proposing an ansatz for the $F$ function, we proposed an ansatz for the Hubble parameter and use the equations of motion to construct the corresponding $F$ function, then we analyzed if these functions could lead to interesting cosmological solutions at the same time that fulfil the dS conjecture. In the process of getting to the Einstein frame, we encountered that choosing a particular value for $\mu$ in terms of $\lambda$ simplifies the action by eliminating an extra term that combines the scalar field with the metric. Thus we first considered this particular case in subsections \ref{SS-NP}-\ref{SS-Lamdas}. We then considered the general case with two independent parameters in subsection  \ref{SS-G}.
	
	For the particular case we proposed an ansatz that describes an expanding universe with a power law form. We studied the projectable and non-projectable cases separately. The differences between both versions is that in the projectable case the conjecture can not be fulfilled independently of $\bar{R}$, but it can be studied for all positive values of $\bar{R}$, thus we can still use it to constraint the parameters of the theory. We found that for both cases the dS conjecture implies constraints on the $\lambda$ parameter, it gets restricted around the limiting value $1/3$. We showed such constraints in Figure \ref{Lambdas}. Furthermore, we showed that the faster the expansion of the universe, the closer $\lambda$ gets to the UV limit $1/3$. Thus we found a consistency with the complications obtained in GR, since in order to fulfil the conjecture the $\lambda$ parameter is restricted to a region that does not allow the infrared limit. We studied the scale factor in the Einstein frame as well and showed that it always describes expanding universes. It is only on the projectable case where the scale factor is constrained by the conjecture, we showed the behaviour of these scale factors in Figures \ref{ScaleFactorPC} and \ref{ScaleFactorPr}.
	
	Then we studied the general version where the two parameters were considered as independent. We proposed that the conjecture can still be applied to this case since we still can identify an action of gravity and a properly defined scalar field with its potential. In this case we also studied the projectable and non-projectable cases with the same differences as before. We found that the dS conjecture leads to a constraint in both parameters, furthermore the consistency of the theory also implies a constraint of the two parameters, then the allowed values for the parameters are the region that fulfils both constraints. We studied each scenario with a numerical analysis. In the non-projectable case for negative power terms in the $F$ function we obtained Figure \ref{GeneralLambdasN}, where we showed that $\mu$ is bounded from above and consequently $\lambda$ gets restricted around $1/3$. For positive power terms we obtained  Figure  \ref{GeneralLambdasP} where we showed that the upper bound for $\lambda$ can increase with $\mu$, thus in this scenario we could performed the $f(R)$ limit $\mu,\lambda\to1$. We obtain consistency with the standard $f(R)$ theories since the conjecture led to a constraint on the exponent of the function which is constrained to be  greater than  $1.25$ and lesser than around $1.866$. The lower bound can be realized in standard $f(R)$ theories as well whereas the upper bound is different since in HL $F(\bar{R})$ theories the conformal transformation is performed only on the three-metric, however we can say that it is in agreement since the numerical bounds are close. On the other hand, for the projectable case we obtained Figure \ref{GeneralLambdasPPr} which provided a new constraint that implied a constriction of $\lambda$ around $1/3$ once again. However, we could still consider the $f(R)$ limit and we obtain in this case a more restrictive constraint.
	
	In the general case we were also able to study the ansatz of an exponentially expanding universe, which was forbidden in the particular case. We assumed a power form for the $F$ function since the equations of motion do not provide a unique form. The dS conjecture and consistency of the theory led once again to constraints on the parameters of the theory. However the regions of validity forbids the $f(R)$ limit, Thus, a consistency with the dS conjecture and a constant Hubble parameter is exclusive to this theories.  Let us point out that as we discussed in subsection \ref{SS-dSConjecture}, GR is inconsistent with the dS conjecture when there is a scalar field whose potential obeys the slow roll conditions, that is when the Hubble parameter is approximately constant. Therefore, it is interesting to see that with a constant Hubble parameter that describes an inflationary scenario, GR and standard $f(R)$ theories are not consistent with the dS conjecture, whereas HL $F(\bar{R})$ theories are, in fact, consistent.
	
	Finally, we can conclude that our results support the idea that the dS conjecture may be encoding key aspects of quantum gravity since we have found agreement between the conjecture and a region of the parameters where we expect UV behaviour in one case, or exclusive to the HL $F(\bar{R})$ theories in the other. We also point out that these theories break Lorentz invariance in the UV, however we are still obtaining consistent results, thus we can also use these results to support the idea that the dS conjecture is applicable to theories without general covariance (as was also considered before in \cite{Trivedi:2021nss}), and thus they may be, in fact, more general than originally thought. We also remark that our results show that these kinds of theories are relevant to be studied further. Since they do not only lead to interesting classical cosmologies, they are indeed compatible with the dS conjecture, in particular in the case in which neither GR nor $f(R)$ theories are. Thus these theories may be a better candidate to study inflationary solutions with a correct quantum description. 

	Finally, let us mention a possible scenario to extend this work. As we saw in subsection \ref{SS-Inflation} the inflationary paradigm is described with the FLRW flat metric, and it is for this scenario that the dS conjecture is not fulfilled in GR as discussed in subsection \ref{SS-dSConjecture}. Those are the main reasons why we use it in this chapter. However, it would be interesting to use a closed or open FLRW metric or even other anisotropic metrics, since by doing so we will obtain different equations of motion between the different versions of the HL $F(\bar{R})$ theory, then we will obtain different scenarios regarding the dS conjecture. This could be used to explore what version of the theory is the most useful using the dS conjecture as a guideline.
	
\chapter{GUP effects in the WDW equation derived from Ho\v{r}ava-Lifshitz gravity}\label{CH-GUPWDW}
	
In this chapter we present the study of the modifications induced on the Wheeler-DeWitt equation in Ho\v{r}ava-Lifshitz gravity when thee coordinates of the minisuperspace obey a GUP, we will particularly focus on a Kantowski-Sachs metric. There is a possible connection within a GUP in the space coordinates and  Ho\v{r}ava-Lifshitz gravity as studied in \cite{Myung:2009ur,Myung:2009gv}. However, in the present chapter we will consider a GUP in the variables of the superspace, treating such modification as an extra independent ingredient to the WDW equation. Therefore, by considering the WDW equation in  HL  gravity and incorporating a GUP as an extra consideration, we are indeed incorporating  two contributions to the  WDW equation that are expected to be relevant in the ultraviolet, therefore, we expect that the ability of the WDW equation to describe quantum gravity behaviour is enhanced. 

First of all, in section \ref{S-KSNormal} we will study the standard WDW equation (that is with coordinates that obey the HUP) derived from HL gravity. We will obtain analytical solutions in the infrared, as well as in a particular ultraviolet limit of interest. In this framework we will show that the solution found with the version of the theory that employs projectability and detailed balanced, can be found as an approximation of the general theory without detailed balance. We then move on, in section \ref{S-KSGUP} to consider that the set of variables on minisuperspace obeys a GUP. We will present two equivalent methods to deduce  the modification of the WDW equation in this context. The first one is useful to present the general WDW equation. However, in section \ref{S-KSIRUV} we will consider the second option, which will be helpful to obtain analytical solutions in the limiting cases considered before. In the IR limit we obtain consistency with the results found in  GR. On the other hand, in the UV limit we will obtain a new analytical solution. For this solution we will explore the regions of oscillatory behaviour, since as we pointed out in subsection \ref{SS-WDWI}, such regions can be related to the classical behaviour.	The results of this chapter were first presented in \cite{Garcia-Compean:2021kqi}.

	\section{Standard Wheeler-DeWitt equation in HL gravity}
	\label{S-KSNormal} 	
	Let us begin by considering the standard case, where the variables of  minisuperspace obey the HUP. We will consider the most general version of HL gravity, that is without detailed balance, whose action is given by (\ref{ActionHL}). In the process to deduce the WDW equation for this theory the classical differences between the projectable and non-projectable versions of the theory are irrelevant, therefore in this chapter we will always use the projectable version. The Kantowski-Sachs metric defined in (\ref{DefKantowskiSachs}) can be written in a parametrization proposed by Misner \cite{Misner} in the form
	\begin{equation}\label{KSMetric}
		ds^2=-N^2(t)dt^2+e^{2\sqrt{3}\beta}dr^2+e^{-2\sqrt{3}(\beta+\Omega)}\left(d\theta^2+\sin^2\theta
		d\phi^2\right),
	\end{equation}
	where $\beta$ and $\Omega$ are real functions of the the time variable. We will not consider any matter content, thus the minisuperspace will be defined by $\beta$ and $\Omega$, which are related to the anisotropic nature of the metric. The action of HL gravity  (\ref{ActionHL}) in this case leads to
	\begin{multline}\label{ActionKSHL}
		S=V_{0}\int dt N\left\{\left[\frac{3}{N^2}(3-\lambda)\dot{\beta}^2+\frac{12}{N^2}(1-\lambda)\dot{\beta}\dot{\Omega}+\frac{6}{N^2}(1-2\lambda)\dot{\Omega}^2\right]e^{-\sqrt{3}(\beta+2\Omega)}\right. \\ \left. +2e^{\sqrt{3}\beta}\left(1-\Lambda
		e^{-2\sqrt{3}(\beta+\Omega)}\right)-g_{r}e^{\sqrt{3}(3\beta+2\Omega)}-g_{s}e^{\sqrt{3}(5\beta+4\Omega)}\right\},
	\end{multline}
	where we have defined
	\begin{equation}
		g_{r}=  2g_{2}+g_{3} , \hspace{1cm} g_{s}=2g_{4}+g_{5}+\frac{g_{6}}{2}.
	\end{equation}
	Furthermore, we have factorized the volume of the particular spatial slice, which is given by
	\begin{equation}
		V_{0}=\int\sin\theta d^3x,
	\end{equation}
	If we consider the whole space time described by the metric, the volume $V_{0}$ would diverge, it can only be finite if we consider a properly compactified spatial slice. However, since it appears only as a global multiplicative factor, we will ignore it hereafter. Following the standard procedure, the hamiltonian constraint leads in this case to
	\begin{multline}\label{HamiltonianConstraintO}
		H=\frac{N}{12}\frac{e^{\sqrt{3}(\beta+2\Omega)}}{3\lambda-1}\left[(2\lambda-1)P^2_{\beta}-2(\lambda-1)P_{\beta}P_{\Omega}+\frac{1}{2}(\lambda-3)P^2_{\Omega}\right. \\ \left. -24(3\lambda-1)e^{-2\sqrt{3}\Omega}\left(1-\Lambda e^{-2\sqrt{3}(\beta+\Omega)}\right)+12g_{r}(3\lambda-1)e^{2\sqrt{3}\beta}\right. \\ \left.
		+12g_{s}(3\lambda-1)e^{2\sqrt{3}(2\beta+\Omega)}\right]\simeq0 ,
	\end{multline}
	where $P_{\beta}$ and $P_{\Omega}$ are canonical momenta of the corresponding variables. This constraint is fulfilled globally since we have chosen the projectable version of the theory, thus the dependence on the spatial variables has only given a global multiplicative factor after integration. If we want to consider the non-projectable version of the theory, we would have obtained the same constraint but valid only locally. However both expressions are equivalent because we have chosen the fields on minisuperspace to depend only on the time variable. The first five terms of the hamiltonian (\ref{HamiltonianConstraintO}) corresponds to the constraint obtain in Ho\v{r}ava-Lifshitz gravity with	detailed balance condition for this metric in \cite{Obregon:2012bt}. The new term obtained from the general theory is a manifestation of the cubic terms in $R$ in the form $e^{4\sqrt{3}\beta +2 \sqrt{3}\Omega}$ which represents a higher-correction in the radius of the Kantowski-Sachs model. Therefore, as it was described in \cite{Bertolami:2011ka,Pitelli:2012sj,Sotiriou:2009gy,Sotiriou:2009bx} we have obtained that the general projectable version without detailed balanced that we have employed reduces consistently to the one with detailed balance. Let us perform the standard canonical quantization procedure on the hamiltonian constraint, that is, in the coordinate representation we promote the momenta to operators in the form
	\begin{equation}\label{MomentaOrig}
		P_{\beta}=-i\frac{\partial}{\partial\beta} , \hspace{1cm} P_{\Omega}=-i\frac{\partial}{\partial\Omega},
	\end{equation}
	considering the standard commutator relations
	\begin{equation}\label{UPO}
		\left[\Omega,P_{\Omega}\right]=i, \hspace{0.5cm} \left[\beta,P_{\Omega}\right]=0 , \hspace{0.5cm} \left[\Omega,P_{\beta}\right]=0 , \hspace{0.5cm} \left[\beta,P_{\beta}\right]=i.
	\end{equation}
	In this way, the general WDW equation in the standard HUP case is obtained in the form
	\begin{multline}\label{WDWO}
		\left[\frac{1}{2}(\lambda-3)\frac{\partial^2}{\partial\Omega^2}-2(\lambda-1)\frac{\partial}{\partial\Omega}\frac{\partial}{\partial\beta}+(2\lambda-1)\frac{\partial^2}{\partial\beta^2}\right. \\ \left. +24(3\lambda-1)e^{-2\sqrt{3}\Omega}\left(1-\Lambda e^{-2\sqrt{3}(\beta+\Omega)}-\frac{g_{r}}{2}e^{2\sqrt{3}(\beta+\Omega)}-\frac{g_{s}}{2}e^{4\sqrt{3}(\beta+\Omega)}\right)\right]\Psi(\beta,\Omega)=0.
	\end{multline}
	In order to obtain exact solutions for this equation we will focus in two limiting cases, namely an infrared and a particular ultraviolet limit.
	
	In the Misner parametrization for the Kantowski-Sachs metric we have that the time variable can be expressed as $t=e^{-\sqrt{3}(\beta+\Omega)}$. Furthermore, the IR limit is obtained through $t\gg1$ and $\lambda\to 1$. Therefore, in this limit we obtain a very small	curvature (very large curvature radius) with respect to the Planck length, that is $R_{c}\gg L_{Pl}$. Thus, we obtain for this limit that the relevant contributions will come from the exponentials with negative signs. Therefore, neglecting the cosmological constant we obtain in this limit that the WDW equation takes the form
	\begin{equation}\label{WDWOGR}
		\left[-\frac{\partial^2}{\partial\Omega^2}+\frac{\partial^2}{\partial\beta^2}+48e^{-2\sqrt{3}\Omega}\right]\Psi(\Omega,\beta)=0,
	\end{equation}
	which corresponds to the appropriate equation obtained in GR. The solutions of this equation take the form
	\begin{equation}\label{SolWDWOGR}
		\Psi^{\pm}_{\nu}(\Omega,\beta)=e^{\pm i\nu\sqrt{3}\beta}K_{i\nu}\left(4e^{-\sqrt{3}\Omega}\right),
	\end{equation}
	where $K_{i\nu}$ are the modified Bessel functions of the second kind.
	
\begin{figure}[h!]
		\centering
		\includegraphics[width=0.6\textwidth]{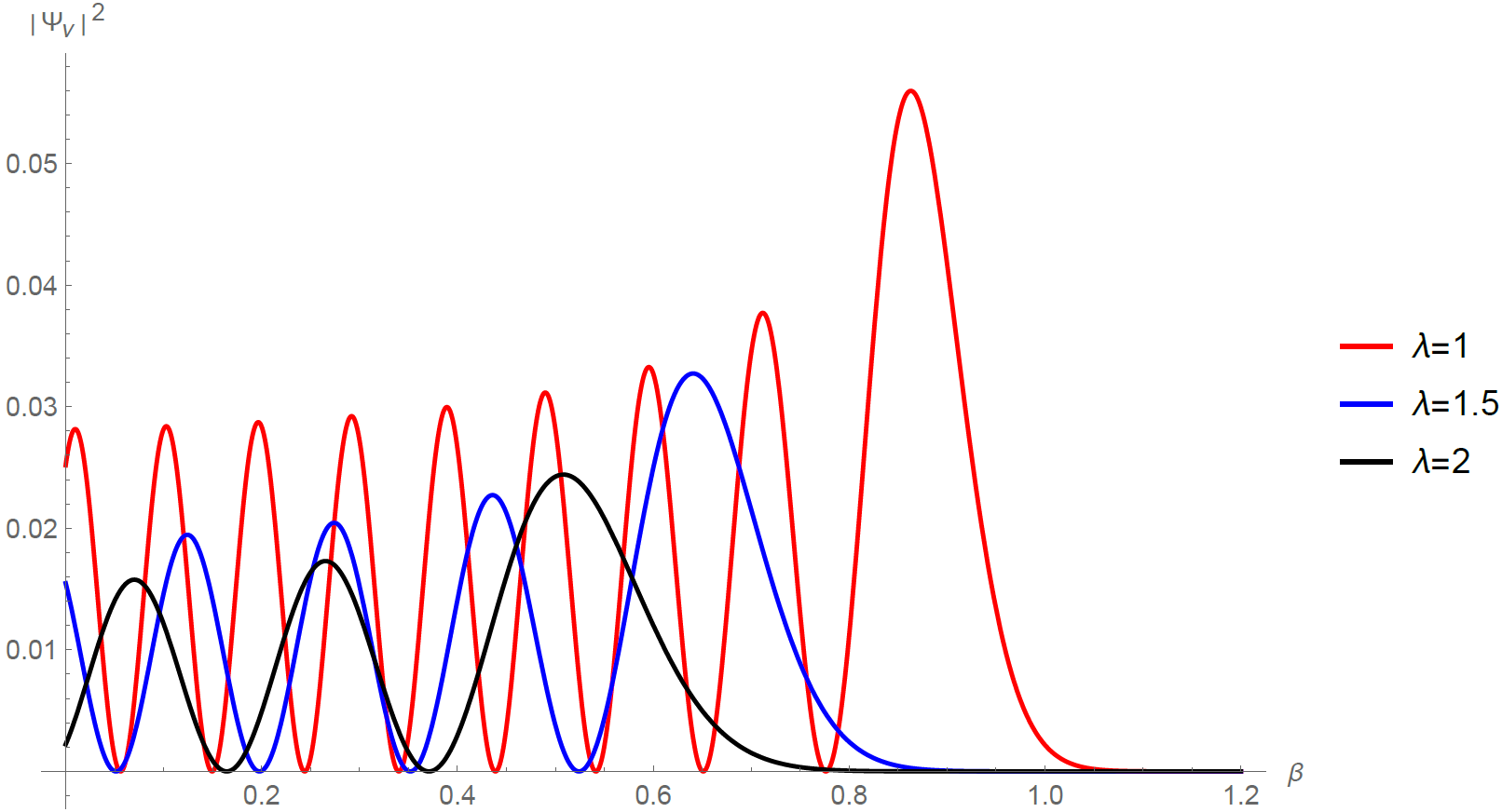}
		\caption{Absolute value squared of the wave functional $|\Psi_{\nu}|^2$ solution to the UV standard WDW equation, choosing $\nu=20i$ for $\lambda=1$ (red curve), $\lambda=1.5$ (blue curve) and $\lambda=2$ (black curve). }
		\label{WGUP}
	\end{figure}

	We now consider the UV limit achieved when $t\ll1$. In this case, this limit corresponds to a huge curvature (very small curvature radius) which may be of the same order of magnitude than $L_{Pl}$. Then the relevant contributions to the WDW equation are the exponentials with positive signs in (\ref{HamiltonianConstraintO}). However, for simplicity we will consider the particular case in which  $g_{r}=2$ and $g_{s}=0$, for this case the WDW  equation takes the form
	\begin{equation}\label{WDWOUV}
		\left[\frac{\lambda-3}{2}\frac{\partial^2}{\partial\Omega^2}-2(\lambda-1)\frac{\partial}{\partial\Omega}\frac{\partial}{\partial\beta}+(2\lambda-1)\frac{\partial^2}{\partial\beta^2}-3\mu^2(2\lambda-1)e^{2\sqrt{3}\beta}\right]\Psi(\Omega,\beta)=0,
	\end{equation}
	where $\mu^2=8\frac{3\lambda-1}{2\lambda-1}$.  This equation was obtained in \cite{Obregon:2012bt} for the version of  HL gravity with projectability and with detailed balance. However in our analysis we have obtained (\ref{WDWOUV}) as a particular case of the theory without detailed balance after a correct choice of parameters. The normalized analytical solutions for this equation are written as
	\begin{multline}\label{SolWDWUV}
		\Psi_{\nu}(\beta,\Omega)=\frac{3^{1/4}}{\pi}\frac{\sqrt{3\lambda-1}}{\sqrt{2}(2\lambda-1)}\sqrt{\sinh\left(\frac{\pi\sqrt{3\lambda-1}|\nu|}{\sqrt{2}(2\lambda-1)}\right)}e^{\pm \sqrt{3}\nu\Omega}\left[\mu e^{\sqrt{3}\beta}\right]^{\frac{\lambda-1}{2\lambda-1}\nu}\\ \times K_{\frac{\sqrt{3\lambda-1}}{\sqrt{2}(2\lambda-1)}\nu}\left(\mu e^{\sqrt{3}\beta}\right),
	\end{multline}
	where $\nu$ is constrained to be a purely imaginary number. Let us explore the behaviour of this solution in a graphical form, in Figure \ref{WGUP} we plot the squared absolute value of this wave functional and take different values for $\lambda$. Since $\nu$ is constrained to be a purely imaginary number, in this case after taking the absolute value squared of the wave functional, the dependence on $\Omega$  vanishes. Therefore we only present a one dimensional plot against $\beta$. We show that there is a region of oscillatory behaviour with increasing height that reaches a maximum peak, then the oscillations stop and the absolute value squared decreases. This behaviour is found generically for any value of $\lambda$ in the region of interest. The implications of varying $\lambda$ are found to be a decrease of the height of the oscillations and a reduction to the region of oscillatory behaviour as $\lambda$ increases.
	
	\section{Modified Wheeler-DeWitt equation by a GUP in HL gravity}
	\label{S-KSGUP}
	
	Let us move on to study what are the modification induced on the WDW equation by considering a set of variables on minisuperspace that obey a GUP. This can be performed in different ways, however we will employ the techniques developed in \cite{GUP} in the context of GR, and generalize them to the HL gravity theory, this will allow us to obtain differential equations up to order two in the momenta. We will employ a generalization to more variables of the simplest form of the GUP (\ref{SimplestGUP}), which in the simplest case leads to a non-zero minimum uncertainty in the position variable. That is, we will propose the following commutation relation in the variables on minisuperspace 
	\begin{equation}\label{RelationGUP}
		[q_{i},p_{j}]=i  \delta_{ij}\left(1+\gamma^2p^2\right),
	\end{equation}
	where $q_{1}=\Omega$, $q_{2}=\beta$, $p_{1}=P_{\Omega}$, $p_{\beta}=P_{\beta}$, $\gamma$ is a small constant with units of inverse momentum and the total momentum squared is $p^2=p^{k}p_{k}$, which  must be computed using the metric in superspace. Furthermore, we also introduce coordinates $q'_{i}$ that obey the usual HUP, that is they obey $[q'_{i},p_{j}]=i\delta_{ij}$. In the coordinate representation we can relate the variables of the two sets of coordinates in the form
	\begin{equation}\label{RelationsCoordinates}
		q_{i}=q'_{i}\left(1+\gamma^2p^2\right).
	\end{equation}
	Furthermore, we choose the following representation for the momentum operators
	\begin{equation}\label{MomOperatorsGUP}
		P_{\Omega}=-i \frac{\partial}{\partial\Omega'} , \hspace{1cm} P_{\beta}=-i \frac{\partial}{\partial\beta'}.
	\end{equation}
	Taking into consideration the metric on the minisuperspace in (\ref{WDWO}), we have in this case that
	\begin{equation}\label{PS}
		p^2=(2\lambda-1)P^2_{\beta}-2(\lambda-1)P_{\beta}P_{\Omega}+\frac{1}{2}(\lambda-3)P^2_{\Omega}.
	\end{equation}
	
	In order to obtain a modification of the  WDW equation derived from the GUP we have to rewrite each term in (\ref{HamiltonianConstraintO}) in terms of prime coordinates. Since we will deal with different exponential terms, we are going to use repeatedly the Zassenhaus formula (Baker-Campbell-Hausdorff formula, see references \cite{Casas:2012,Magnus:1954zz}) which states that for the operators $A$ and $B$ the following is fulfilled
	\begin{equation}\label{ZSFormula}
		e^{A+B}=e^{A}e^{B}e^{-\frac{1}{2}[A,B]}e^{\frac{1}{6}\left([A,[A,B]]+2[B,[A,B]]\right)}
		\cdots
	\end{equation}
	where $\cdots$ denotes terms involving higher order commutators.
	
	Let us consider all the terms which does not depend on the momenta in (\ref{WDWO}) as a gravitational potential. Thus, considering the first term of such potential and employing the relations (\ref{RelationsCoordinates}) we obtain that it can be written as
	\begin{equation}\label{ExpansionOmegaM1}
		\exp[-2\sqrt{3}\Omega]=\exp\left[-2\sqrt{3}\Omega'-2\sqrt{3}\gamma^2\Omega'p^2
		\right].
	\end{equation}
	Then, using the Zassenhaus formula (\ref{ZSFormula}) with $A=-2\sqrt{3}\Omega'$ and $B=-2\sqrt{3}\gamma^2\Omega'p^2$ we obtain
	\begin{multline}\label{ExpansionOmegaM}
		\exp[-2\sqrt{3}\Omega]\simeq \exp\left[-2\sqrt{3}\left\{1-2\gamma^2(\lambda-3)\right\}\Omega'\right]\exp\left[-2\sqrt{3}\gamma^2\Omega'p^2\right]
		\\ \times \exp\left[-6i\gamma^2\Omega'\left\{(\lambda-3)P_{\Omega}-2(\lambda-1)P_{\beta}\right\}\right],
	\end{multline}
	where we have only kept terms up to second order in $\gamma$ since it is a small parameter, thus the higher order commutators can be neglected. Following the same procedure in the remaining terms contributing to the potential we obtain 
	\begin{multline}\label{ExpansionBetaP}
		\exp\left[2\sqrt{3}\beta\right]\simeq\exp\left[2\sqrt{3}\left\{1-4\gamma^2(2\lambda-1)\right\}\beta'\right]\exp\left[2\sqrt{3}\gamma^2\beta'p^{2}\right]\\ \times \exp\left[-12i\gamma^2\beta'\left\{(2\lambda-1)P_{\beta}-(\lambda-1)P_{\Omega}\right\}\right],
	\end{multline}
	\begin{multline}\label{ExpansionBOM}
		\exp\left[-2\sqrt{3}(\beta+2\Omega)\right]\simeq \exp\left[-2\sqrt{3}(1+12\gamma^2)(\beta'+2\Omega')\right]\exp\left[-2\sqrt{3}\gamma^2(\beta'+2\Omega')p^2\right] \\  \times \exp\left[-12i\gamma^2(\beta'+2\Omega')(P_{\beta}-2P_{\Omega})\right],
	\end{multline}
	\begin{multline}\label{ExpansionBOP}
		\exp\left[2\sqrt{3}(2\beta+\Omega)\right] \simeq \exp\left[2\sqrt{3}\left\{1-6\gamma^2(3\lambda-1)\right\}(2\beta'+\Omega')\right]\exp\left[2\sqrt{3}\gamma^2(2\beta'+\Omega')p^2\right] \\  \times \exp\left[-6i\gamma^2(3\lambda-1)(2\beta'+\Omega')(2P_{\beta}-P_{\Omega})\right].
	\end{multline}
	
	Now that we have expressed all the factors describing the potential in terms of prime coordinates, we can study how are they applied to the wave functional. We begin with the first term in
	(\ref{ExpansionOmegaM}). We find that up to second order in $\gamma$ and the momenta we can write
	\begin{multline}\label{AppWF11}
		e^{-2\sqrt{3}\Omega}\Psi(\Omega,\beta)\simeq\exp\left[-2\sqrt{3}\left\{1-2\gamma^2(\lambda-3)\right\}\Omega'\right]\exp\left[-2\sqrt{3}\gamma^2\Omega'p^2\right]\\ \times \exp\left[-6i\gamma^2\Omega'\left\{(\lambda-3)P_{\Omega}-2(\lambda-1)P_{\beta}\right\}\right]\Psi(\Omega',\beta'),
	\end{multline}
	Then, the application of the exponential terms with linear momenta to the wave functional can be studied in two different ways. The first one consists in factorizing up to second order in $\gamma$ the exponential in the form
	\begin{multline}\label{Expansion}
		\exp\left[-6i\gamma^2\Omega'\left\{(\lambda-3)P_{\Omega}-2(\lambda-1)P_{\beta}\right\}\right]\\=\exp\left[12i\gamma^2(\lambda-1)\Omega'P_{\beta}\right]\exp\left[-6i\gamma^2(\lambda-3)\Omega'P_{\Omega}\right],
	\end{multline}
	then, defining $\Omega'=e^{y}$ we obtain from (\ref{MomOperatorsGUP}) that
	\begin{equation}
		\Omega'P_{\Omega}=-i \frac{\partial}{\partial y}.
	\end{equation}
	Therefore, the last term in (\ref{Expansion}) acts on the wave functional as a translation operator for $y$, which in turns represents an scaling of  $\Omega'$ in the wave functional. Thus we obtain
	\begin{equation}\label{ScalingDef}
		\exp\left[-6i\gamma^2(\lambda-3)\Omega'P_{\Omega}\right]\Psi(\Omega',\beta')=\Psi(e^{-6\gamma^2(\lambda-3)}\Omega',\beta').
	\end{equation}
	Moreover, the first exponential term of (\ref{Expansion}) can be expanded as a power series, then by keeping only up to second order in $\gamma$ we obtain 
	\begin{multline}\label{FirstOptionExp}
		\exp\bigg\{-6i\gamma^2\Omega'\left((\lambda-3)P_{\Omega}-2(\lambda-1)P_{\beta}\right)\bigg\}\Psi(\Omega',\beta') \\ \simeq \left[1+12i\gamma^2\Omega'(\lambda-1)P_{\beta}\right]\Psi(e^{-6\gamma^2(\lambda-3)}\Omega',\beta').
	\end{multline}
	Substituting back this expression into (\ref{AppWF11}) and expanding in the same form the second exponential  we finally obtain that the first option leads to
	\begin{multline} \label{AppWF1}
		e^{-2\sqrt{3}\Omega}\Psi(\Omega,\beta)\simeq
		e^{-2\sqrt{3}[1-2\gamma^2(\lambda-3)]\Omega'}
		\bigg\{1+12i\gamma^2\Omega'(\lambda-1)P_{\beta} \\
		-2\sqrt{3}\gamma^2\Omega'p^2 \bigg\}
		\Psi(e^{-6\gamma^2(\lambda-3)}\Omega',\beta').
	\end{multline}
	
	On the other hand, we have a second option, instead of interpreting the second term in (\ref{Expansion}) as an scaling on $\Omega'$,  we can expand this exponential in a power series as well and keep only up to second order in the parameter $\gamma$. Choosing this path we obtain that the second options leads to
	\begin{multline}\label{AppWF2}
		e^{-2\sqrt{3}\Omega}\Psi(\Omega,\beta)\simeq\exp^{-2\sqrt{3}\big[1-2\gamma^2(\lambda-3)\big]\Omega'}\bigg\{1-6i\gamma^2\Omega'\big[(\lambda-3)P_{\Omega}-2(\lambda-1)P_{\beta}\big]
		\\ -2\sqrt{3}\gamma^2\Omega'p^2\bigg\}\Psi(\Omega',\beta').
	\end{multline}
	Let us remark that  both options described are equivalent at the level of approximation employed, which corresponds to second order in $\gamma$ and momenta. Therefore, we have the freedom to choose any of the options, in the following we will find the most useful option in each particular case.
	
	We point out that the arguments used to obtain (\ref{ScalingDef}) are applicable to any term that contains a factor of a product of a coordinate times its momentum within an exponential. Therefore, any of the terms contributing to the potential (\ref{ExpansionBetaP})-(\ref{ExpansionBOP}) have two possible expansions in the same way as (\ref{AppWF1}) and (\ref{AppWF2}). Therefore, we will employ the most helpful option in each case.
	
	Now that we have explored the general way in which any of the terms	contributing to the potential in (\ref{WDWO}) are expressed in terms of prime coordinates and the two possible applications to the wave functional, we have all the information to obtain the modified WDW equation. For simplicity we take in all cases the second option of the form (\ref{AppWF2}), then the modified general WDW equation takes the form
	\begin{equation}\label{WDWGUPGeneral}
		\tcboxmath[colback=red!10!white,colframe=red]
		{
		\begin{split}
		\bigg\{\left[1+\gamma^2F(\Omega',\beta')\right]\left[(2\lambda-1)P^2_{\beta}-\right. & \left. 2(\lambda-1)P_{\beta}P_{\Omega}+\frac{1}{2}(\lambda-3)P^2_{\Omega}\right] \\   &+i\gamma^2G(\Omega',\beta',P_{\Omega},P_{\beta}) +\widetilde{V}(\Omega',\beta')\bigg\}\Psi(\Omega',\beta')=0,
		\end{split}
		}
	\end{equation}
	where we have defined
	\begin{multline}
		F(\Omega',\beta')=48\sqrt{3}(3\lambda-1)\Omega'e^{-2\sqrt{3}\left[1-2\gamma^2(\lambda-3)\right]\Omega'}\\-48\sqrt{3}(3\lambda-1)\Lambda(\beta'+2\Omega')e^{-2\sqrt{3}(1+12\gamma^2)(\beta'+2\Omega')} \\ +24\sqrt{3}g_{r}(3\lambda-1)\beta'e^{2\sqrt{3}\left[1-4\gamma^2(2\lambda-1)\right]\beta'}\\+24\sqrt{3}g_{s}(3\lambda-1)(2\beta'+\Omega')e^{2\sqrt{3}\left[1-6\gamma^2(3\lambda-1)\right](2\beta'+\Omega')},
	\end{multline}
	\begin{multline}
		G(\Omega',\beta',P_{\Omega},P_{\beta})=144(3\lambda-1)\Omega'\left[(\lambda-3)P_{\Omega}-2(\lambda-1)P_{\beta}\right]e^{-2\sqrt{3}\left[1-2\gamma^2(\lambda-3)\right]\Omega'}\\-288(3\lambda-1)\Lambda(\beta'+2\Omega')(P_{\beta}-2P_{\Omega})e^{-2\sqrt{3}(1+12\gamma^2)(\beta'+2\Omega')} \\-144g_{r}(3\lambda-1)\beta'\left[(2\lambda-1)P_{\beta}-(\lambda-1)P_{\Omega}\right]e^{2\sqrt{3}\left[1-4\gamma^2(2\lambda-1)\right]\beta'}\\-72g_{s}(3\lambda-1)^2(2\beta'+\Omega')(2P_{\beta}-P_{\Omega})e^{2\sqrt{3}\left[1-6\gamma^2(3\lambda-1)\right](2\beta'+\Omega')},
	\end{multline}
	\begin{multline}
		\widetilde{V}(\Omega',\beta')=-24(3\lambda-1)e^{-2\sqrt{3}\left[1-2\gamma^2(\lambda-3)\right]\Omega'}+24(3\lambda-1)\Lambda e^{-2\sqrt{3}(1+12\gamma^2)(\beta'+2\Omega')} \\ +12g_{r}(3\lambda-1)e^{2\sqrt{3}\left[1-4\gamma^2(2\lambda-1)\right]\beta'}+12g_{s}(3\lambda-1)e^{2\sqrt{3}\left[1-6\gamma^2(3\lambda-1)\right](2\beta'+\Omega')}.
	\end{multline}
	
	This equation is very complicated to solve in general. Thus, with the hope of obtaining analytical solutions, in the next section we will restrict ourselves to the limiting cases considered for the standard case in section \ref{S-KSNormal}.

	\section{Infrared and Ultraviolet limits}
	\label{S-KSIRUV}
	
	In order to correctly perform the limits of interest, we note that due to the expansions used in order to obtain (\ref{ExpansionOmegaM})-(\ref{ExpansionBOP}) we have changed the exponentials in all the terms of the potential in the following form
	\begin{itemize}
		\item $e^{-2\sqrt{3}\Omega}\to e^{-2\sqrt{3}\left[1-2\gamma^2(\lambda-3)\right]\Omega'}$
		\item $\Lambda e^{-2\sqrt{3}(\beta+2\Omega)}\to \Lambda e^{-2\sqrt{3}(1+12\gamma^2)(\beta'+2\Omega')}$
		\item $g_{r}e^{2\sqrt{3}\beta}\to g_{r}e^{2\sqrt{3}\left[1-4\gamma^2(2\lambda-1)\right]\beta'}$
		\item $g_{s}e^{2\sqrt{3}(2\beta+\Omega)}\to g_{s}e^{2\sqrt{3}\left[1-6\gamma^2(3\lambda-1)\right](2\beta+\Omega)}$.
	\end{itemize}
	Therefore, since the overall sign of the exponential determines in which limit they are relevant, we need to be careful not to change the signs of each term, such that the same terms contribute in each limiting case considered in section \ref{S-KSNormal}. Therefore, we demand that the new factors in all the exponential terms are positive, that is we propose the constraints
	\begin{equation}\label{ConditionGammaLambda}
		\frac{1}{\gamma^2}>\sup\left(6(3\lambda-1),4(2\lambda-1)\right)= \begin{cases}
			6(3\lambda-1) ,\hspace{0.5cm} \lambda>\frac{1}{5} , \\
			4(2\lambda-1) ,\hspace{0.5cm} \lambda<\frac{1}{5} .
		\end{cases}
	\end{equation}
	However, since $\gamma$ is a very small parameter, these constraints can be easily fulfilled. Thus, we can proceed as in the standard case.

	Then, in the IR limit (obtained with $t\gg1$ and $\lambda\to 1$) the relevant terms are the ones containing negative exponentials, as in section \ref{S-KSNormal}. Therefore, neglecting the cosmological constant and employing the first option in the form of (\ref{AppWF1}), we obtain in this infrared limit the WDW equation
	\begin{equation}\label{WDWGUPIR}
		\tcboxmath[colback=red!10!white,colframe=red]{
		\left[\left(1+96\sqrt{3}\gamma^2\Omega'e^{-2\sqrt{3}(1+4\gamma^2)\Omega'}\right)\left(P^2_{\beta}-P^2_{\Omega}\right)-48e^{-2\sqrt{3}(1+4\gamma^2)\Omega'}\right]\Psi\left(e^{12\gamma^2}\Omega',\beta'\right) \simeq0 .}
	\end{equation}
	Which corresponds to the result obtained in General Relativity in \cite{GUP} as expected.
	
	Let us study now the same UV limit as in section \ref{S-KSNormal}, since an analytical solution was found for this particular example in the standard case. Since we imposed the constraint (\ref{ConditionGammaLambda}), we know that the relevant terms in this limit are the $g_{r}$ and $g_{s}$ ones once again. Therefore, choosing $g_{r}=2$ and $g_{s}=0$, and employing the expansion of the first form in (\ref{AppWF11}) for the $g_{r}$ term, we obtain that the WDW equation in this limit takes the form
	\begin{multline}\label{WDWGUPUVO}
		\left\{\left[1+48\sqrt{3}(3\lambda-1)\gamma^2\beta'e^{2\sqrt{3}\left[1-4\gamma^2(2\lambda-1)\right]\beta'}\right]\left[(2\lambda-1)\frac{\partial^2}{\partial\beta'^2}-2(\lambda-1)\frac{\partial}{\partial\beta'}\frac{\partial}{\partial\Omega'}\right.\right. \\ \left.\left. +\frac{\lambda-3}{2}\frac{\partial^2}{\partial\Omega'^2}\right] -36(2\lambda-1)(\lambda-1)\mu^2\gamma^2\beta'e^{2\sqrt{3}\left[1-4\gamma^2(2\lambda-1)\right]\beta'}\frac{\partial}{\partial\Omega'} \right. \\ \left. - 3(2\lambda-1)\mu^2e^{2\sqrt{3}\left[1-4\gamma^2(2\lambda-1)\right]\beta'}\right\}\Psi(\Omega',e^{-12\gamma^2(2\lambda-1)}\beta')\simeq0 .
	\end{multline}
	Considering the form of the solution in the standard case given by (\ref{SolWDWUV}), we propose a general ansatz in the form
	\begin{equation}\label{UVAnsatz}
		\Psi\left(\Omega',e^{-12\gamma^2(2\lambda-1)}\beta'\right)=e^{\sqrt{3}\nu\Omega'}\left[e^{\sqrt{3}\beta'}\right]^{\frac{\lambda-1}{2\lambda-1}\nu}\chi(\beta') .
	\end{equation}
	Substituting back (\ref{UVAnsatz}) into (\ref{WDWGUPUVO}) we obtain  that the $\chi(\beta')$ function must obey
	\begin{equation}\label{GeneralChis}
		\left[\frac{d^2}{d\beta'^2}-V\right]\chi(\beta')=0,
	\end{equation}
	where we have defined the function
	\begin{equation}\label{Potentials}
		V=\frac{3\nu^2(3\lambda-1)}{2(2\lambda-1)^2}+\frac{1+12(\lambda-1)\sqrt{3}\nu\gamma^2\beta'}{1+48\sqrt{3}(3\lambda-1)\gamma^2\beta'e^{2\sqrt{3}\left[1-4\gamma^2(2\lambda-1)\right]\beta'}}\frac{24(3\lambda-1)}{(2\lambda-1)}e^{2\sqrt{3}\left[1-4\gamma^2(2\lambda-1)\right]\beta'}.
	\end{equation}
	Furthermore, in the UV limit considered we have that  $t=e^{-\sqrt{3}(\beta'+\Omega')}\ll1$. We note that this condition can be achieved by considering $\beta'\gg1$. Therefore, taking this condition for the $\beta'$ function, we can make an approximation for the denominator of (\ref{Potentials}) in the form
	\begin{equation}\label{Approx}
		1+48\sqrt{3}(3\lambda-1)\gamma^2\beta'e^{2\sqrt{3}\left[1-4\gamma^2(2\lambda-1)\right]\beta'}\simeq48\sqrt{3}(3\lambda-1)\gamma^2\beta'e^{2\sqrt{3}\left[1-4\gamma^2(2\lambda-1)\right]\beta'},
	\end{equation}
	thus, the  $V$ function can be approximated as
	\begin{equation}\label{DefPotentialsUV}
		V^{UV}=V\bigg\rvert_{\beta'\gg1}=A_{\nu,\lambda}+\frac{B_{\gamma,\lambda}}{\beta'},
	\end{equation}
	where
	\begin{equation}\label{DefsA&B}
		A_{\nu,\lambda}=\frac{3\nu}{2\lambda-1}\left[\frac{\nu(3\lambda-1)}{2(2\lambda-1)}+2(\lambda-1)\right] , \hspace{1cm}
		B_{\gamma,\lambda}=\frac{1}{2\sqrt{3}(2\lambda-1)\gamma^2}.
	\end{equation}
	
	Then, with the approximated form for the $V$ function (\ref{DefPotentialsUV}) we obtain that the general solution of (\ref{GeneralChis}) takes the form
	\begin{equation}\label{SolutionChis}
		\chi(\beta')=C M_{\rho,1/2}\left(2\sqrt{A_{\nu,\lambda}}\beta'\right)+D W_{\rho,1/2}\left(2\sqrt{A_{\nu,\lambda}}\beta'\right) ,
	\end{equation}
	where $C$ and $D$ are constants of integration, furthermore $M_{k,m}(z)$ and $W_{k,m}(z)$ denotes the Whittaker functions and
	\begin{equation}
		\rho=-\frac{B_{\gamma,\lambda}}{2\sqrt{A_{\nu,\lambda}}}.
	\end{equation}
	
\begin{figure}[h!]
		\centering
		\includegraphics[width=0.6\textwidth]{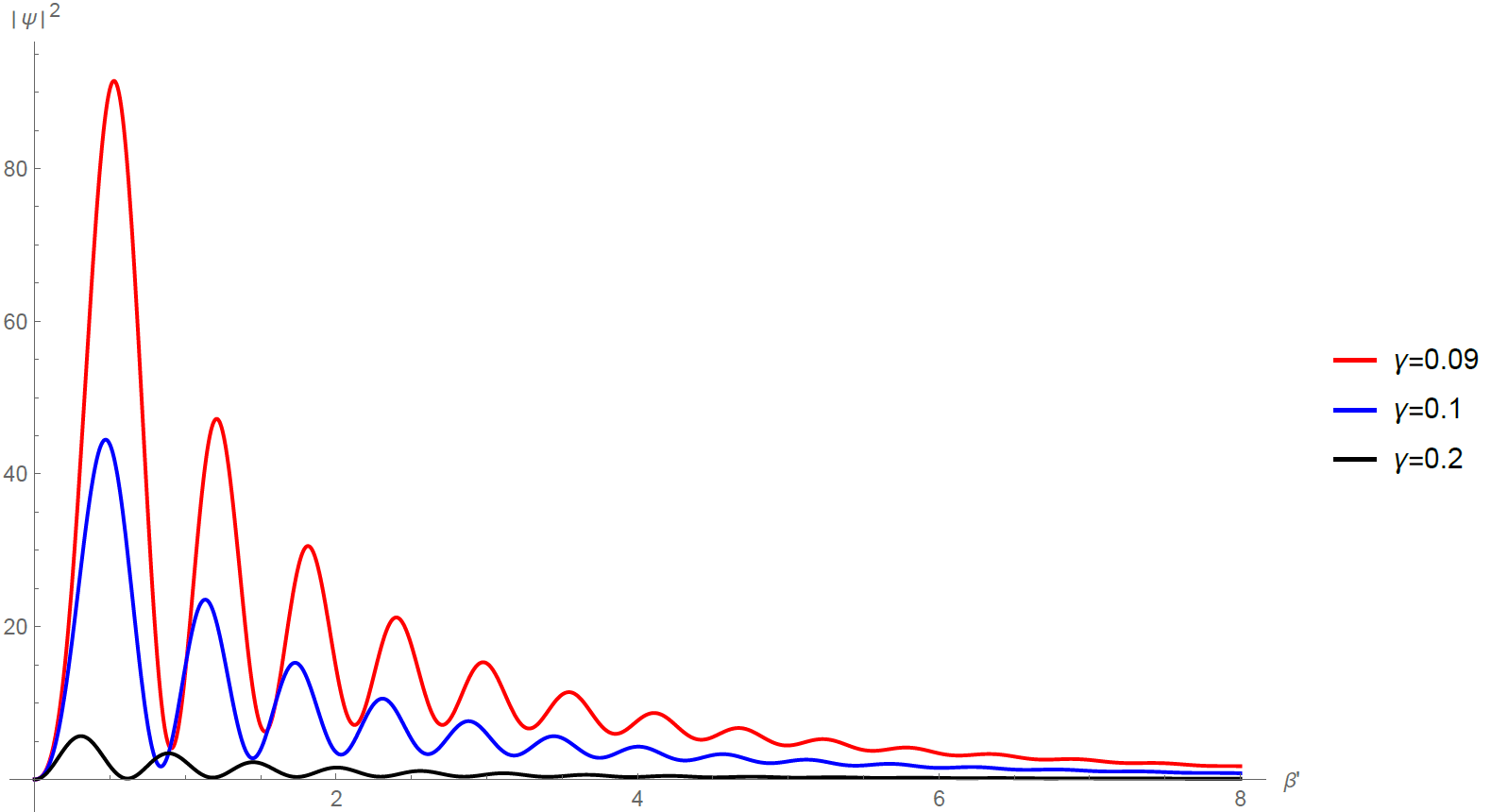}
		\caption{Projection of the absolute value squared of the wave functional $|\Psi|^2$ solution to the UV modified WDW equation by a GUP, choosing $|Ce^{\sqrt{3}\nu\Omega'}|^2=1$, $\lambda=1.5$ and $\nu=10i-1$ for $\gamma=0.09$ (red curve), $\gamma=0.1$ (blue curve) and $\gamma=0.2$ (black curve). }
		\label{WFL1.5NC}
	\end{figure}

\begin{figure}[h!]
		\centering
		\includegraphics[width=0.6\textwidth]{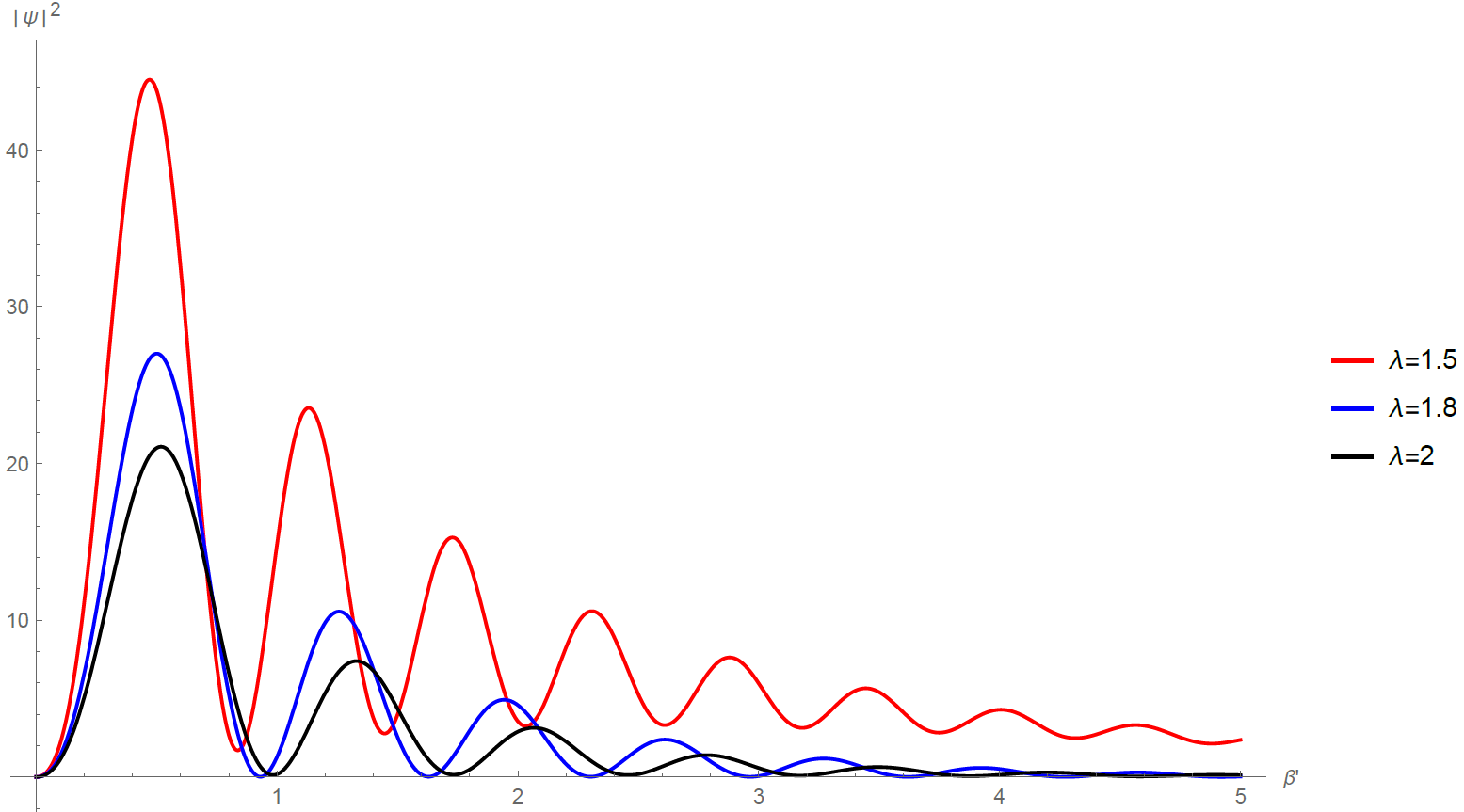}
		\caption{Projection of the absolute value squared of the wave functional $|\Psi|^2$ solution to the UV modified WDW equation by a GUP, choosing $|Ce^{\sqrt{3}\nu\Omega'}|^2=1$, $\gamma=0.1$ and $\nu=10i-1$ for $\lambda=1.5$ (red curve), $\lambda=1.8$  (blue curve) and $\lambda=2$  (black curve). }
		\label{WFLVNCF}
	\end{figure}

\begin{figure}[h!]
		\centering
		\includegraphics[width=0.6\textwidth]{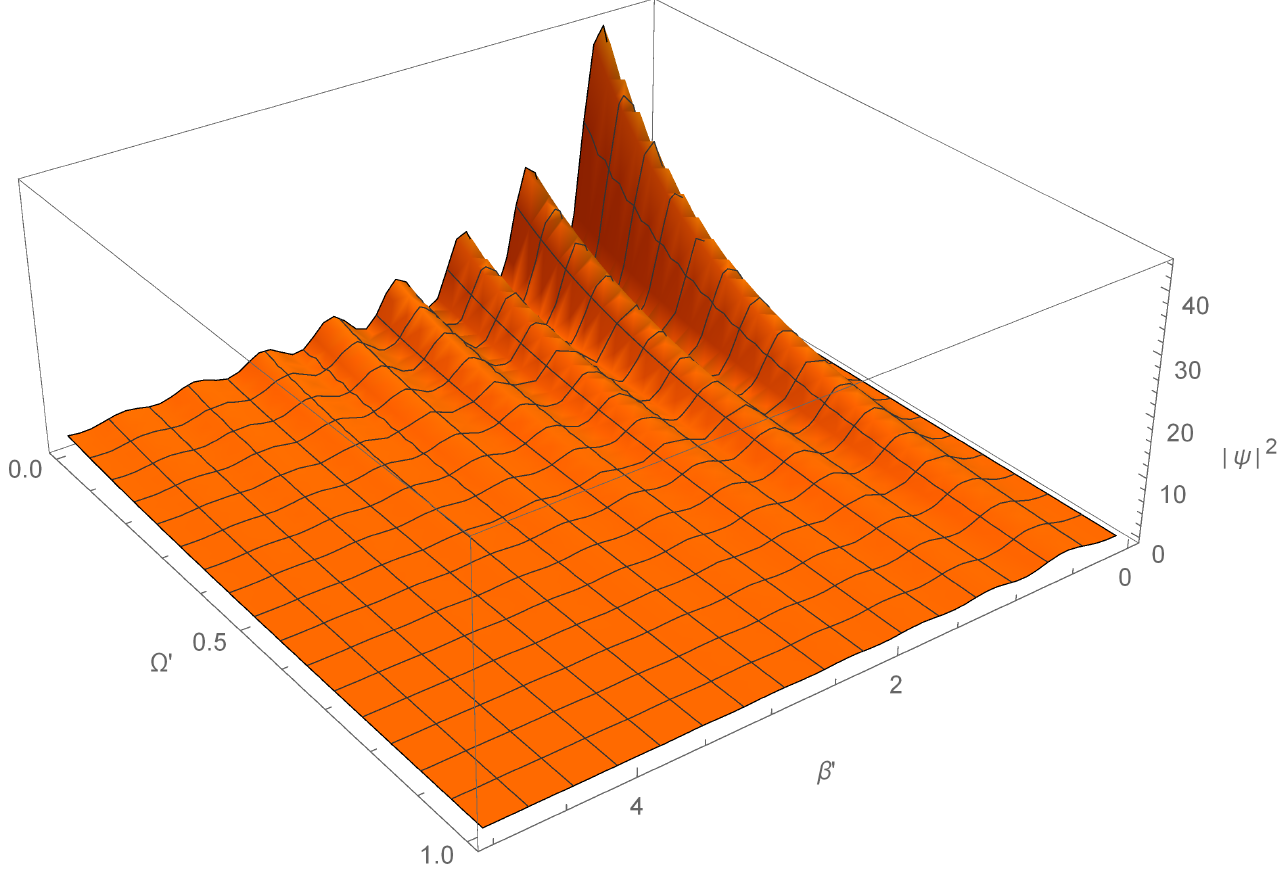}
		\caption{Absolute value squared of the complete wave functional $|\Psi|^2$ solution to the UV modified WDW equation by a GUP, choosing $|C|^2=1$, $\lambda=1.5$, $\gamma=0.1$ and $\nu=10i-1$. }
		\label{WF3D}
	\end{figure}

	In contrast to the standard situation studied in section \ref{S-KSNormal}, in the present solution the parameter $\nu$ is not restricted to take an specific form. Thus in the GUP scenario we will only constrain such parameter by seeking a well defined behaviour for the wave functionals, and can be a real, imaginary or even a complex number. However, as we discussed in subsection \ref{SS-WDWI}, we are looking for a region of oscillatory behaviour for the wave functional since such region can be interpreted as Lorentzian geometries \cite{Hawking,Wada:1985cp,Halliwell:2009}. However, it can be shown \cite{OscillationsN,OscillationsT} that the Whittaker equation that obeys the $\chi$  function does not have oscillatory solutions if $A_{\nu,\lambda}$ and  $B_{\gamma,\lambda}$ are real positive numbers, which is found when $\nu$ is a positive real number and $\lambda>1/2$ (which is a region of interest to us). Therefore, we will not propose  positive real values for the $\nu$ parameter. Furthermore, for generic values of the parameters, the function $|W_{\rho,1/2}|^2$ exhibits a monotonically decreasing behaviour, therefore in order to obtain oscillatory behaviour we will consider only the $M$ function in (\ref{SolutionChis}). Thus,  the solution for the wave functional will be given by
	\begin{equation}\label{SolWaveFunc}
		\tcboxmath[colback=red!10!white,colframe=red]{
		\Psi\left(\Omega',e^{-12\gamma^2(2\lambda-1)}\beta'\right)=Ce^{\sqrt{3}\nu\Omega'}\left[e^{\sqrt{3}\beta'}\right]^{\frac{\lambda-1}{2\lambda-1}\nu}M_{\rho,1/2}\left(2\sqrt{A_{\nu,\lambda}}\beta'\right).}
	\end{equation}
	We find that the squared absolute value of this wave functional $|\Psi|^2$ with respect to $\beta'$ does not exhibit oscillatory behaviour for all values of the parameters. In most cases it only presents a monotonically increasing behaviour, which is not of interest for us. Nevertheless, it is possible to obtain an oscillatory behaviour for some parameters. For example, for $\lambda>1$ such behaviour can be present by choosing negative real numbers for  $\nu$ (with $A_{\nu,\lambda}$ non-positive), or choosing such parameter as a complex number with negative real part. In Figures \ref{WFL1.5NC} and \ref{WFLVNCF} we show plots of the projection of $|\Psi|^2$ with respect to $\beta'$ choosing $|Ce^{\sqrt{3}\nu\Omega'}|^2=1$ for different values of $\gamma$ and $\lambda$  respectively. In particular, in Figure \ref{WFL1.5NC} we choose $\lambda=1.5$ and $\nu=10i-1$, we show curves for three different values of $\gamma$. We note that as $\gamma$ increases, the projection of $|\Psi|^2$ decreases. On the other hand, in Figure \ref{WFLVNCF} we choose  $\nu=10i-1$ and $\gamma=0.1$, in this case  we show curves for three different values of $\lambda$. We obtain analogously that as $\lambda$ increases, the projection of $|\Psi|^2$ decreases. In both cases we encountered a limited region for  $\gamma$ and $\lambda$, since we can spoil the oscillatory behaviour by varying too much one of these two parameter as well as varying $\nu$. We also note from this plots that the oscillatory behaviour is present for small values of $\beta'$, but it is reduced as $\beta'$ increases, eventually this behaviour disappears and we obtain a monotonically decreasing behaviour. Therefore, we need to choose the parameters of the solution in a proper form such that the the region of oscillatory behaviour is in agreement with the region where the UV approximation (\ref{Approx}) employed is valid. The same scenario is present when $\nu$ takes real negative values. Finally, in Figure \ref{WF3D} we show the complete squared absolute value of the wave functional $|\Psi|^2$, we choose $|C|^2=1$, $\nu=10i-1$, $\lambda=1.5$ and $\gamma=0.1$. We see that the dependence on $\Omega'$ is just a negative exponential as shown in (\ref{SolWaveFunc}). 
	

\begin{figure}[h!]
		\centering
		\includegraphics[width=0.6\textwidth]{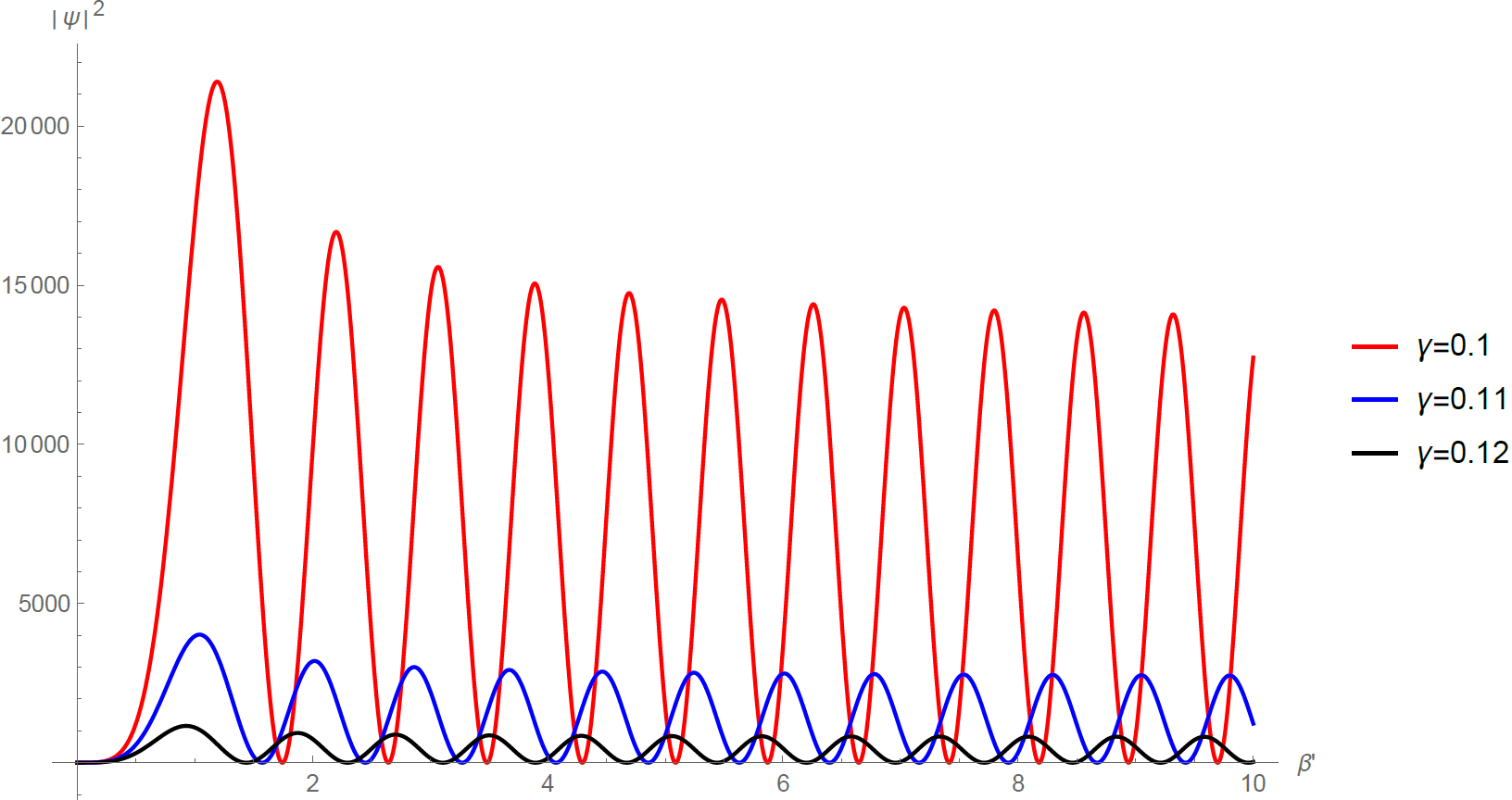}
		\caption{Projection of the absolute value squared of the wave functional $|\Psi|^2$ solution to the UV modified WDW equation by a GUP, in the particular case $\lambda=1$, choosing $|Ce^{\sqrt{3}\nu\Omega'}|^2=1$, and $\nu=5i$ for $\gamma=0.1$ (red curve), $\gamma=0.11$ (blue curve) and $\gamma=0.12$ (black curve). }
		\label{WFL1NI}
	\end{figure}

	We note from (\ref{DefsA&B}) that when $\lambda=1$, $A_{\nu,\lambda}$ only depends on $\nu^2$, thus we expect a different behaviour in this particular case. In this scenario a solution presenting an oscillatory behaviour can also be found by choosing $\nu$ as a pure imaginary number. However, in this case the oscillations do not decrease in height after a few peaks. In Figure \ref{WFL1NI} we show the plot of the projection of $|\Psi|^2$ with respect to $\beta'$ choosing $|Ce^{\sqrt{3}\nu\Omega'}|^2=1$ for this case. We choose $\nu=5i$ and show curves for three different values of $\gamma$. In this case we also obtain that the $|\Psi|^2$ increases inversely with $\gamma$.

	From these results we conclude that the solution of the WDW equation modified by the GUP in the particular UV limit considered  (\ref{UVAnsatz}) can be obtained in a similar form as the one obtained in the standard case (\ref{SolWDWUV}). We obtain that the special function involved is modified from a Bessel function in the HUP case to a Whitakker function in the GUP scenario. In both cases we can obtain a region of oscillatory behaviour for the squared absolute value of the wave functional. For $\lambda>1$ we obtained in both cases a finite region of oscillatory behaviour starting from the origin at $\beta=0$ ($\beta'=0$ in the GUP case). However, in the standard scenario this is achieved after constraining the $\nu$ parameter to be a purely imaginary number. Furthermore, the oscillations increase in height until they reach a maximum value. On the other hand, for the GUP case an oscillatory solution is found when $\nu$ is a real negative number or a complex number with negative real part. In this case the oscillations show the opposite behaviour, that is they decrease in height until the oscillations stop. Furthermore, we remark that in order to obtain the analytical solution in the GUP scenario we used the UV limit in the form $\beta'\gg1$, therefore the starting region of oscillatory behaviour is not accessible to this solution. Moreover, in both cases we obtained  that as $\lambda$ increases the squared wave functional decreases. Finally, in the GUP case, the choice $\lambda=1$ is an special scenario. In this case the oscillatory behaviour does not decrease in height. In contrast, for the standard case in (\ref{SolWDWUV}) this value of $\lambda$ is not special, it produces the same behaviour described earlier.

	\subsection*{The $\gamma\to0$ limit}
	Before we end up this chapter let us discuss a subtle issue regarding the $\gamma\to0$ limit. As we expect such limit should transform the GUP solution into the standard case solution. In fact, if we take such limit in (\ref{WDWGUPUVO}) we obtain the standard case result (\ref{WDWOUV}). We also note that in this limit the differential equation in the GUP case (\ref{GeneralChis}) takes the form of the equation obeyed by the modified Bessel function of second order with the appropriate change of variables. Therefore, in this limit the ansatz (\ref{UVAnsatz}) gives smoothly the same solution of the standard case (\ref{SolWDWUV}) as expected. However after taking the UV limit,  we note from (\ref{DefsA&B}) that  $B_{\gamma,\lambda}$ is divergent when $\gamma\to0$. Therefore the solution in the GUP case (\ref{SolWaveFunc}) does not simplifies to the solution of the standard case. The reason for this impediment is that in order to obtain analytical solutions, we employed the  ultraviolet limit that leads to (\ref{Approx}). We note that this approximation is no longer valid in the $\gamma\to0$ limit. Therefore, the region of validity of the solutions found in the GUP case in terms of the Wittaker functions is defined by appropriate values of $\beta'$ and $\gamma$ such that the approximation is valid.
	
	We can try to propose a method to recover the standard case solution (\ref{SolWDWUV}) from analytical solutions of the GUP case. We may propose to expand the denominator in the second term of (\ref{Potentials}), keeping only up to second order in the $\gamma$ parameter. However, in this case $\gamma$ is accompanied by a positive exponential term depending on $\beta'$. Therefore, we can not justify to keep an approximation up to second order on $\gamma$ in the UV limit, because the whole term is not small in general. It can only be small for small values of $\beta'$, which does not correspond to the UV limit that we are considering. Therefore, we believe that the UV approximation used in the denominator is a more compelling method even if we cannot longer recover the standard case solution.
	
	\section{Discussion}\label{S-DCH-GUPWDW} 
	Finally, let us summarize and discuss the results presented in this chapter.
	
	We presented the study ot the Wheeler-DeWitt equation for the Kantowski-Sachs metric in Ho\v{r}ava-Lifshitz gravity with coordinates of minisuperspace that obey a GUP using the procedure presented originally in \cite{GUP}.
	
	We first studied the WDW equation arriving from the standard commutator relations. We particularly focused on two limiting cases, the IR limit that displayed consistency with GR, and a particular UV limit after a proper choice of the parameters, that coincides with the result obtained from the version of HL theory with projectability and detailed balance, but now only as a particular example of the more general form derived from the theory without detailed balance. We presented analytical solutions for both limits, in particular the UV solution was show in Figure \ref{WGUP}. This solution exhibits a region of oscillatory behaviour with increasing height until a maximum peak is reached, then the oscillations stop leading to a region of exponential decay. Furthermore, it was found that the absolute value squared of the wave functional decreases with $\lambda$.
	
	We then considered the generalization to more variables of the simplest form of the GUP. In order to deform the WDW equation we used the relations between a set of coordinates that obeys the GUP and one that obeys the standard HUP. We used approximations up to second order in the momenta as well as in the GUP parameter. We found that there are two possible ways to obtain the WDW equation modified by the GUP regarding an exponential term writen in terms of a coordinates times its corresponding momentum. The first possibility interprets such term as an scaling on the wave functional as done in \cite{GUP}, this option was useful to look for analytical solutions in the UV. On the other hand, there is a second option that consists on expanding as a power series such term, this was useful to provide a general modified WDW equation. 
	
	We presented the general form of the modified WDW equation. However in order to obtain analytical solutions we used the limiting cases considered before. In the IR limit we found consistency with the result obtained in GR in \cite{GUP}. Furthermore, we considered a particular UV limit obtained after a proper choice of the parameters as in the standard case. We also employed the UV limit in the form $\beta'\gg1$ and obtain an adequate approximation. Then we were able to solve analytically the WDW equation in terms of the Whittaker functions. With this solution we look for oscillatory behaviour. We encountered that contrary to the standard case $\nu$ was not constrained to be a purely imaginary number, in order to obtain the correct behaviour however we find that it can not take positive real values. We found that the oscillatory behaviour depends on the correct choice of the parameters $\nu$, $\gamma$ and $\lambda$ and it is not a generic feature since it can be spoiled easily. We presented plots of  $|\Psi|^2$ in Figures \ref{WFL1.5NC}, \ref{WFLVNCF}	and \ref{WF3D} by taking different values of $\gamma$ and $\lambda$. We found that the oscillatory behaviour is present for small values of $\beta'$, it then decreases until it eventually stops, thus since we make an approximation for big $\beta'$, we need to find a region where the approximation is valid and the oscillatory behaviour is present.  We also encounter that  $|\Psi|^2$ decreases inversely with $\gamma$ or $\lambda$. In addition we found that $\lambda=1$ represented an special case since for this case the oscillatory behaviour does not stop as shown in  Figure \ref{WFL1NI}. This is a particular feature of the GUP deformation since for the standard case the behaviour found is the same for all values of $\lambda$.
	
	Finally, let us discuss a possible way to expand the present work. The GUP proposal naturally implies a non-commutativity for the coordinates on the minisuperspace. However, we can also pursue a deformation of the WDW equation (\ref{WDWO}) through the standard non-commutative approach employing a $\theta$ parameter as in \cite{Garcia-Compean:2001jxk}. This deformation will be different that the one studied in this chapter but it will also be relevant at high energies, thus it may be relevant in the same line of the GUP deformation. Some	solutions of the noncommutative deformation of (\ref{WDWOGR}) have been studied previously \cite{Garcia-Compean:2001jxk}. Thus it will be interesting to pursue a general deformation of such equation as the general analysis presented in this chapter.

\chapter{Lorentzian vacuum transitions}\label{CH-Transitions}

In the canonical approach to quantum gravity, the quantum transitions between different configurations have been one of the main topics to study. For example, the tunnelling from nothing scenario described in section \ref{S-CQG} is a particular example of this kind of processes. In order to correctly study such transitions we need to employ the quantum features of gravity since there is not a classical analog. One particular example of this kind that will be of interest of us in the present chapter are vacuum transitions in the presence of gravity. There has been great interest over the years to obtain a correct quantum description of such process. Euclidean methods to study these processes were developed originally by Sidney Coleman and collaborators for field theory \cite{ColemanFT1,ColemanFT2}  and then, they were extended to include gravity by Coleman and De Luccia \cite{ColemanDeLuccia} using a path integral approach. Their results were later generalized by Parke \cite{Parke:1982pm}. This process is described by the nucleation of true vacuum bubbles and the subsequent evolution is obtained after making an analytic continuation to the Lorentzian description. It is found that only open universes can arise as the end result. 

On the other hand, an alternative procedure to compute these transitions without using Euclidean methods, employing instead the hamiltonian approach which is explicitly Lorentzian, was developed  by Fischler, Morgan and Polchinski (FMP) in \cite{FMP1,FMP2}. This method employs the formalism of the Wheeler-DeWitt equation presented in section \ref{S-CQG}. The vacuum is implemented through a cosmological constant which is interpreted as the vacuum energy and the transition are carried out through a bubble nucleation \cite{deAlwis:2019dkc}. This procedure was later generalized in \cite{deAlwis:2019dkc,Cespedes:2020xpn}, it was used to compute tunnelling probabilities of transitions between two de Sitter universes as well as transitions from a Minkowski to a de Sitter space.  In this kind of models it was shown that the transitions between closed universes are allowed, in contrast to the result obtained by using the Euclidean methods. It was conjectured that the Euclidean result could be only a limitation of relying on an analytic continuation. However, in these works there were considered only transitions between universes with a different cosmological constant. A generalization to consider explicitly a scalar field with the vacuum states described by the minima of its potential was carried out in \cite{Cespedes:2020xpn}. In order to solve the WDW equation a WKB approximation was used. Although the method can be defined on the general superspace, in order to compute the transition probability  explicitly, they used a closed FLRW metric. For this metric they obtained the same results as the one obtained by using the Euclidean methods. However, since the minisuperspace approximation was used, in this approach a description of bubble nucleation is not present, then it is conjectured that this approach can describe a generalization of the tunnelling from nothing scenario described in subsection \ref{SS-WDWBC}. That is, it can be used to describe the creation of universes by the process of a vacuum transition. Furthermore, a proposal to leave the minisuperspace approximation was studied in \cite{Oshita:2021aux}.

In this chapter we will present the full analytical study of the vacuum transition probabilities between two minima of a scalar field potential in various contexts:

First of all, in section \ref{S-TransitionsGeneralMethod} we present a general method to compute such transition probabilities by solving the Wheeler-DeWitt equation in a semiclassical approach, that is employing a WKB approximation. This method will be valid for any model on the superspace whose Hamiltonian constraint is composed of terms of squared as well as linear momenta. In this way we generalize the treatment developed in  \cite{Cespedes:2020xpn}. We remark that this method only assumes a generic form of the Hamiltonian constraint, thus it will be applicable not only to an isotropic universe in GR as studied before. As we will see, this method will be general enough to be applied to any metric and any gravitational theory that leads to the appropriate form for the WDW equation, at least in principle.

In section \ref{S-TransitionsGR} we will apply the method to compute the transition probabilities in General Relativity. Firstly we will consider a closed FLRW metric in subsection \ref{S-FLRW} and show that our procedure gives the same result as in \cite{Cespedes:2020xpn} but now only as a particular solution of the general method. Next, we obtain the transition probabilities for the flat FLRW  metric in subsection \ref{S-FLRWP}. Moreover, since in principle, we can apply our method to any metric, we will be interested in studying the transition probabilities of homogeneous but anisotropic metrics. Vacuum transitions for these types of universes have also been studied previously. For example, the transitions between two values of the cosmological constant considering only gravity, as well as considering a scalar field using the Bianchi IX metric were studied in \cite{DelCampo}. Moreover, these transitions were studied with Euclidean methods for a Kantowski-Sachs metric in \cite{JensenRuback}. Furthermore, the tunnelling from nothing scenario was studied for the Bianchi IX metric in \cite{Folomeev}. In addition, the transitions between two minima of a scalar field potential using the Bianchi I metric were studied in \cite{Mansouri}. However, in this section we will present the study of such transition probabilities using the Lorentzian formalism developed in the previous section without using any Euclidean method. We will carry out this procedure for the Kantowski-Sachs metric in subsection \ref{S-KS}, the Bianchi III metric in subsection \ref{S-B3} and the biaxial Bianchi IX metric in subsection \ref{S-B9}. We remark that the isotropic limit of the Bianchi III metric is found to be the flat FLRW metric. Thus, the study of the Bianchi III metric allows us to find what is the effect of anisotropy in the transition probabilities explicitly.   

As we remarked in section \ref{S-HLGravity}, the Ho\v{r}ava-Lifshitz theory of gravity has a better UV behaviour than GR, then this theory is more adequate to describe the quantum effects of gravity. Thus, we expect that the study of the vacuum transition  using this theory can provide more information about the quantum features of gravity. For this reason, in section \ref{S-TransitionsHL} we will apply the general method to the HL theory. We will particularly focus on the closed FLRW metric. We will derive the corresponding WDW equation in subsection \ref{S-HL}. However, for this theory we will study two types of scalar fields. First, since the anisotropic scaling of spacetime variables is the property defining the HL theory as we remarked in subsection \ref{SS-HLMatter}, we will consider a scalar field which is allowed to depend on all spacetime coordinates. In subsection \ref{S-Trans} we will obtain the transition probability in this case. Then, we will consider also a scalar field which only depends on the time variable, as usual on cosmological studies, in subsection  \ref{S-TT}. In both cases the infrared and ultraviolet limits will be discussed for comparison with the results obtained in General Relativity. For the inhomogeneous case in which the scalar field depends on all spacetime variables we will obtain that the behaviour in the infrared limit will be similar to the one encountered in GR. However, the behaviour found in the ultraviolet limit  will be completely opposite. This result will allow us to obtain a plausible resolution of the initial singularity. On the other hand, for the homogeneous scalar field which depends only on the time variable, the behaviour will coincide with that of GR in both limits, although in the intermediate region the probability will be slightly altered.  

On the other hand, in chapter \ref{CH-GUPWDW} we presented a method to incorporate a GUP in the variables of minisuperspace to the WDW equation. It was found that the modification of the WDW equation is not trivial but an equation that takes into consideration the GUP can be found starting from the WDW equation with the standard commutators. Using this procedure, in section \ref{S-TransitionsGUP} we will apply our general method to study the effect of the GUP in the vacuum transition probabilities, in this case the variables on minisuperspace will be the gravitational degrees of freedom and the scalar field.  We will consider a closed FLRW metric with positive curvature in subsection \ref{S-FLRWClosed}, an FLRW metric with zero curvature in subsection \ref{S-FLRWF}. Furthermore we will also study the Bianchi III metric in subsection \ref{S-Bianchi} in order to incorporate the anisotropic effect as well. In all cases we compare the results obtained with the GUP with the results obtained in the standard HUP scenario in section \ref{S-TransitionsGR}. We will find for all metrics that the effect of considering a GUP is to enhance the probability for small values of the scale factor, but it produces a faster decay so when the corresponding scale factor is big enough the probability is reduced. Finally, in subsection \ref{GUP-EUP} we will briefly discuss other extensions of the uncertainty principle that could lead to interesting results as well as the possible applicability of the general method. 

The results of this chapter were originally presented in  \cite{Garcia-Compean:2021syl,Garcia-Compean:2021vcy,Garcia-Compean:2022ysy}.

\section{General Method}\label{S-TransitionsGeneralMethod}
In this section we will introduce a general method to compute the transition probabilities between two minima of a scalar field potential. This will be achieved after obtaining a solution, using a WKB semiclassical approximation, to a generic form of WDW equation that is described by quadratic and linear momentum terms. This approach was originally proposed in \cite{Cespedes:2020xpn} for a Hamiltonian described only by quadratic terms in the momenta. Therefore, the method introduced in the present section represents its generalization. 

Let us consider a generic form of the Hamiltonian constraint written in the context of the ADM formalism of subsection \ref{SS-ADM} in the form
\begin{equation}\label{GeneralHamiltonian}
	\tcboxmath[colback=red!10!white,colframe=red]{
	H=\frac{1}{2}G^{MN}(\Phi)\pi_{M}\pi_{N}+W^{M}(\Phi)\pi_{M}+f(\Phi)\simeq 0 ,}
\end{equation}
where $\Phi^M$ denotes the variables on superspace which are the degrees of freedom of the three-metric, as well as all the matter field variables considered  (collectively they will all be denoted just by $\Phi$), their corresponding canonical conjugate momenta are respectively $\pi_{M}$. The superspace in general may be infinite dimensional, furthermore in this case it is equipped with the inverse metric $G^{MN}(\Phi)$. We have defined a vector in superspace with entries $W^{M}(\Phi)$ that can depend in general on all the fields. Finally, $f(\Phi)$ is a scalar function that will be composed of the scalar field potential, as well as any other term that could arise in the quantization procedure that does not depend on any of the momenta. The Wheeler-De Witt equation will be obtained after doing a canonical quantization procedure to this generic Hamiltonian constraint,  in the coordinates representation we will promote the momenta to operators as $\pi_{M}\to-i\hbar\frac{\delta}{\delta\Phi^M}$ in (\ref{GeneralHamiltonian}). Then, we obtain
\begin{equation}\label{WDWEqGeneral}
	H\Psi(\Phi)=\left[-\frac{\hbar^2}{2}G^{MN}(\Phi)\frac{\delta}{\delta\Phi^M}\frac{\delta}{\delta\Phi^N}-i\hbar W^M(\Phi)\frac{\delta}{\delta\Phi^M}+f(\Phi)\right]\Psi(\Phi)\simeq0 ,
\end{equation}
where $\Psi(\Phi)$ is the wave functional. We will now consider a semiclassical approximation to the WDW equation of the WKB type, that is we propose the ansatz
\begin{equation}\label{WKBAnsatz}
	\Psi(\Phi)=\exp\left\{\frac{i}{\hbar}S[\Phi]\right\} ,
\end{equation}
where $S=S[\Phi]$ is a function that can be expanded in $\hbar$ as is usual in the WKB procedure, that is
\begin{equation}\label{ActionExpansionh}
	S[\Phi]=S_{0}[\Phi]+\hbar S_{1}[\Phi]+\mathcal{O}({\hbar}^2) ,
\end{equation}
where $S_{0}$ is the classical action and $S_{i}$ with $i\geq1$ will correspond to quantum correction terms. We will see in the following sections that employing this semiclassical approximation and keeping only the first term will be sufficient to study the vacuum transitions probabilities obtaining relevant results. Substituting this approximation back in (\ref{WDWEqGeneral}) we obtain an equation for every order in $\hbar$. For the first two powers of $\hbar$ we obtain the set of equations
\begin{equation}\label{WDWFirstH}
	\frac{G^{MN}(\Phi)}{2}\frac{\delta S_{0}}{\delta\Phi^M}\frac{\delta S_{0}}{\delta\Phi^N}+W^M(\Phi)\frac{\delta S_{0}}{\delta\Phi^M}+f(\Phi)\simeq0 ,
\end{equation}
\begin{equation}\label{WDWSecondH}
	-\frac{i}{2}G^{MN}(\Phi)\frac{\delta^2S_{0}}{\delta\Phi^M\delta\Phi^N}+G^{MN}(\Phi)\frac{\delta S_{0}}{\delta\Phi^M}\frac{\delta S_{1}}{\delta\Phi^N}+W^{M}(\Phi)\frac{\delta S_{1}}{\delta\Phi^M}\simeq 0 .
\end{equation}
For the remaining of this chapter, we will focus only at first order in the semiclassical expansion, that is only up to first order in $\hbar$. Then all the information of the WDW equation will be contained in (\ref{WDWFirstH}). Let us note that if we choose a different ordering in the quantization procedure leading to the WDW equation (\ref{WDWEqGeneral}), the modifications will appear at second order in $\hbar$, therefore at the level of approximation that we will employ, we can ignore any related issues. Next, we will define a set of integral curves in superspace parametrized by a parameter $s$ in the following form
\begin{equation}\label{IntegralCurves}
	C(s)\frac{d\Phi^M}{ds}=G^{MN}(\Phi)\frac{\delta S_{0}}{\delta\Phi^N} ,
\end{equation}
where $C(s)$ is a function to be determined. We have the freedom to incorporate this function in (\ref{IntegralCurves}) so we can take into account the information coming from the WDW equation. From (\ref{WDWFirstH}) we obtain
\begin{equation}\label{System1}
	G_{MN}(\Phi)\frac{d\Phi^M}{ds}\frac{d\Phi^N}{ds}=-\frac{2}{C(s)}\left[\frac{f(\Phi)}{C(s)}+W_{P}(\Phi)\frac{d\Phi^P}{ds}\right] ,
\end{equation}
where $G_{MN}$ is the inverse of $G^{MN}$ and we lower and raise indices with this metric, that is $W_{P}=G_{LP}W^{L}$. Using (\ref{WDWFirstH}) we can also obtain for the classical action 
\begin{equation}\label{ClassicalAction}
	S_{0}=-2\int_{X}\int_{s}ds\left[\frac{f(\Phi)}{C(s)}+W_{N}(\Phi)\frac{d\Phi^N}{ds}\right] ,
\end{equation}
where $X$ is the spatial slice being considered. We note that equations (\ref{IntegralCurves}) and (\ref{System1}) actually describe a system of equations for the variables $\frac{d\Phi^M}{ds}$ and $C(s)$. Furthermore we have the same number of equations as variables, then in principle we have all the information we need to solve this system. With the solutions, we can use (\ref{ClassicalAction}) to obtain the classical action, then we can obtain the correct form for the wave functional (\ref{WKBAnsatz}), in this way we can obtain a solution to the WDW equation at first order in $\hbar$ regardless of the number of fields in superspace.

Let us simplify the system of equations by assuming that all the fields depend only on the time variable. In this case, the variational derivative of the classical action can be explicitly computed, then (\ref{IntegralCurves}) takes the form
\begin{multline}\label{System2}
	\frac{d\Phi^M}{ds}=-\frac{2{\rm Vol}(X)}{C^2(s)}G^{MN}\frac{\partial f}{\partial\Phi^N}-\frac{2{\rm Vol}(X)}{C(s)}\left[\left(G^{MN}G_{LP}\frac{\partial W^{P}}{\partial\Phi^N}-\frac{\partial W^M}{\partial\Phi^L}\right)\frac{d\Phi^L}{ds}\right. \\ \left. +G^{MN}W^{P}\left(\frac{\partial G_{LP}}{\partial\Phi^N}-\frac{\partial G_{NP}}{\partial\Phi^L}\right)\frac{d\Phi^L}{ds}\right] ,
\end{multline}
where  $\rm Vol(X)$ represents the volume of the spatial slice.

If we consider the  particular simple case in which $W^{M}=0$, we can obtain a general solution to the system which is written as
\begin{equation}\label{CsGeneral}
	C^2(s)=-\frac{2{\rm Vol}^2(X)}{f[\Phi]}G^{MN}\frac{\partial f}{\partial\Phi^M}\frac{\partial f}{\partial\Phi^N} ,
\end{equation}
\begin{equation}\label{DerivSolGen}
	\frac{d\Phi^M}{ds}=\frac{f[\Phi]}{{\rm Vol}(X)}\frac{G^{MN}\frac{\partial f}{\partial\Phi^N}}{G^{LO}\frac{\partial f}{\partial\Phi^L}\frac{\partial f}{\partial\Phi^{O}}} .
\end{equation}
Furthermore, from this general solution (\ref{DerivSolGen}) we can see that the fields on superspace are not independent, they must fulfill the following system of differential equations
\begin{equation}\label{FieldsRelations}
	\frac{d\Phi^M}{d\Phi^N}=\frac{G^{ML}\frac{\partial f}{\partial\Phi^L}}{G^{NP}\frac{\partial f}{\partial\Phi^{P}}} ,
\end{equation}
which is valid for every value of $M$ and $N$ such that $d\Phi^{M,N}\neq0$. Therefore, in this particular case the independent degrees of freedom are reduced and then we will need less fields for the computation of the classical action. This is similar to the procedure used in \cite{Folomeev} in which the degrees of freedom are also reduced in order to compute the classical action. However, in the general case in which $W^{M}\neq0$,  we cannot longer obtain a general solution, since the system of equations is far more difficult, thus for this scenario we will only solve the system in special cases of interest.

As we pointed oud in subsection \ref{SS-WDWI}, the interpretation of the solutions of the WDW equation is still an ongoing debate. One of the possible interpretations proposed is that the squared ratio of two solutions represent a relative probability of obtaining a certain universe configuration with respect to the other. With this framework in mind, in \cite{Cespedes:2020xpn} the transition probability between two minima of the scalar field potential was proposed to be obtained by
\begin{equation}\label{DefProb}
	P(A\to B)=|e^{-\Gamma}|^2=\bigg|\frac{\Psi(\varphi^I_{0},\phi_{B};\varphi^I_{f},\phi_{A})}{\Psi(\varphi^I_{0},\phi_{A};\varphi^I_{f},\phi_{A})}\bigg|^2,
\end{equation}
where $\phi_{A}$ and $\phi_{B}$ are the values that the scalar field takes in each minimum. In the rest of this chapter we will always consider transitions from a false minimum to a true one, where the potential will have the false minimum at $\phi_{A}$ and the true one at $\phi_{B}$. Furthermore, in this expression $\Psi(\varphi^I_{0},\phi_{B};\varphi^M_{f},\phi_{A})$ represents the path in superspace in which the scalar field varies from $\phi_{B}$ to $\phi_{A}$ and all the others fields evolve from an initial value $\varphi^I_{0}$ to a final value $\varphi^I_{f}$ (the notation here indicates that the scalar field is denoted by $\phi$ and  all the other fields in superspace are denoted by $\varphi^{I}$). On the other hand, $\Psi(\varphi^I_{0},\phi_{A};\varphi^M_{f},\phi_{A})$ represents the path in which the scalar field remains constant in the false minimum whereas all the other fields evolve. We will consider that both paths are described by the parameter $s$ defined in (\ref{IntegralCurves}) taking values in the interval $[0,s_{M}]$. Thus, considering the WKB expansion proposed previously, the transition probability will be given in general by
\begin{equation}\label{Probability}
	\tcboxmath[colback=red!10!white,colframe=red]{
	P(A\to B)= \exp(-2\Re[\Gamma]) ,}
\end{equation}
where 
\begin{equation}\label{DefGamma}
	\tcboxmath[colback=red!10!white,colframe=red]{
	\pm\Gamma=\frac{i}{\hbar}\left[S_{0}(\varphi^I_{0},\phi_{B};\varphi^I_{f},\phi_{A})-S_{0}(\varphi^I_{0},\phi_{A};\varphi^I_{f},\phi_{A})\right] .}
\end{equation}
The sign ambiguity arrives from the fact that the general solution in (\ref{DefProb}) will consist of a linear superposition of two exponentials with diferent signs. However, we can only keep the dominant term in both cases that leads to an overall sign ambiguity.

It is worth mentioning that this method only assumes a Hamiltonian constraint written in the general form of (\ref{GeneralHamiltonian}). Therefore, it is not tied to any theory of gravitation and it can be applied, in principle, to any model of superspace that fulfils this condition. Furthermore, the case when there is only quadratic terms in the Hamiltonian constraint, that is, when $W^{M}=0$, is a natural general form that will be used to study the transitions in General Relativity with different metrics,  as well as in the Ho\v{r}ava-Lifshitz theory. Moreover, the linear terms in momenta will be used to study a deformation of the WDW equation to take into account a GUP in the variables of the minisuperspace. It will be interesting to keep exploring further the applicability of the method to more scenarios.

Let us also remark that in this thesis we will only consider the first term in the WKB expansion (\ref{ActionExpansionh}), this will be sufficient to capture some features of the microscopic spacetime. Moreover, the quantum correction terms arriving from higher orders in $\hbar$ can be incorporated into this method in the following way: The general method presented allows us to compute, in principle, $S_{0}$ for any model. Thus, we note from (\ref{WDWSecondH}) that once we have obtained $S_{0}$, we can obtain the next term in the expansion $S_{1}$ by solving this equation. Then, the first correction term can be incorporated in the transition probability by adding $S_{1}$ to both solutions in (\ref{DefGamma}). We note that for every higher order term in $\hbar$ that we want to take into account, a new equation will be obtained from the WDW equation that defines the correction term $S_{i}$ in terms of all the others $S_{j}$ with $0\leq j<i$. Thus, in principle we can obtain enough number of equations to solve for every quantum correction term $S_{i}$ that we want to incorporate. Therefore, the method presented is applicable not only to different gravitational theories and any model in the superspace in a semiclassical approximation, we have enough information to compute the wave functional, and thus the effect on the transition probabilities, up to any quantum correction term that we desire.

\section{Transitions in General Relativity}\label{S-TransitionsGR}
Let us begin in this section by applying the method to study the transition probabilities in General Relativity. First of all, we will study the FLRW metrics, in particular in the case with positive curvature we will obtain consistent results with the ones obtained previously on \cite{Cespedes:2020xpn} and using Euclidean methods in \cite{Parke:1982pm}. Furthermore, we will use homogeneous but anisotropic metrics to study the effect of considering anisotropy in the transition probabilities as well.

\subsection{Transitions for a FLRW metric with positive curvature}
\label{S-FLRW}

We start by studying the simplest model on minisuperspace, that is the homogeneous and isotropic model described by a FLRW metric with positive curvature. The transition probability for this metric was already obtained using the Lorentzian formalism in \cite{Cespedes:2020xpn}, however here we will derive it as a particular application of the general method, in addition we will point out a subtlety not considered in that work. This metric can be written as the first form of (\ref{DefFLRW}), however, considering for more generality a lapse function we can write the metric as
	 \begin{equation}\label{MetricFLRWP}
		ds^2=-N^2(t)dt^2+a^2(t)\left(dr^2+\sin^2\theta d\Omega^2_{2}\right) ,
	\end{equation}
In the following we will consider a homogeneous scalar field that only depends the time variable $\phi=\phi(t)$, then the Lagrangian for this system takes the form
\begin{equation}\label{FLRWLagr}
	\mathcal{L}=3N(t)a(t)-\frac{3a(t)\dot{a}(t)^2}{N(t)}+\frac{a^3(t)\dot{\phi}^2}{2N(t)}-N(t)a^3(t)V(\phi) ,
\end{equation}
where as usual $\dot{a}(t)=\frac{da}{dt}$ and $V(\phi)$ is the scalar field potential. Following the standard procedure, the canonical momenta are defined by
\begin{equation}\label{FLREMomenta}
	\pi_{N}=0 , \hspace{1cm} \pi_{a}=-\frac{6a\dot{a}}{N} , \hspace{1cm} \pi_{\phi}=\frac{a^3\dot{\phi}}{N} ,
\end{equation}
which lead to the Hamiltonian constraint
\begin{equation}\label{FLREHam}
	H=N\left[\frac{\pi^2_{\phi}}{2a^3}-\frac{\pi^2_{a}}{12a}-3a+a^3V\right]\simeq
	0.
\end{equation}
As we expected, the canonical momentum with respect to $N$ vanishes. Thus we can ignore the prefactor and focus only on the term inside brackets in the last expression as the full Hamiltonian constraint. Therefore we see that we have obtained a Hamiltonian constraint of the same form as (\ref{GeneralHamiltonian}). Furthermore, the variables in minisuperspace in this case are $\{\Phi^M\}=\{a,\phi\}$, while the non-vanishing elements of the metric are
\begin{equation}\label{FLREMetricSuper}
	G^{aa}=-\frac{1}{6a} , \hspace{1cm} G^{\phi\phi}=\frac{1}{a^3} ,
\end{equation}
furthermore, we also have in this case
\begin{equation}\label{FLRWf}
	f(a,\phi)=-3a+a^3V(\phi) .
\end{equation}
and $\left\{W^{M}\right\}=0.$ We note that for this metric the volume of the spatial slice will be given by
\begin{equation}\label{FLRWVolume}
	{\rm Vol}(X)=\int_{\phi=0}^{2\pi}\int_{\theta=0}^{\pi}\int_{r=0}^{\pi}\sin^2r\sin\theta drd\theta
	d\phi=2\pi^2,
\end{equation}
which is evidently finite.

In order to proceed with the application of the general method we need to choose and appropriate  parameter $s$ so that we can correctly describe the process. We will choose in this case the parameter such that for the interval $[0,\bar{s}-\delta s]$, where $s=0$ is the initial value, the field remains close to its value at the true minimum $\phi_{B}$, and for
the interval $[\bar{s}+\delta s,s_{M}]$ the field remains very close to its value at the false minimum $\phi_{A}$, that is, we choose the parameter $s$ such that
\begin{equation}\label{ChooseS}
	\phi(s) \approx
	\begin{cases}
		\phi_{B} , & 0<s<\bar{s}-\delta s,\\
		\phi_{A} , & \bar{s}+\delta s<s<s_{M} .
	\end{cases}
\end{equation}
In this way we are implementing a configuration where there is a bubble of true minimum in the background of a false minimum with a thin wall. Since we cannot describe such bubble in the minisuperspace approach, this is the only way to implement such configuration. Therefore, from (\ref{ClassicalAction}) we obtain for the classical action
\begin{multline}\label{FLRWAccionComplA}
	S_{0}(a_{0},\phi_{B};a_{m},\phi_{A})=-4\pi^2\int_{0}^{s_{M}}\frac{ds}{C(s)}\left[-3a+a^3V(\phi)\right] \\
	=-4\pi^2\bigg\{\int_{0}^{\bar{s}-\delta s}\frac{ds}{C(s)}\left[-3a+a^3V_{B}\right] +\int_{\bar{s}-\delta s}^{\bar{s}+\delta s}\frac{ds}{C(s)}\left[-3a+a^3V(\phi)\right] \\ +\int_{\bar{s}+\delta s}^{s_{M}}\frac{ds}{C(s)}\left[-3a+a^3V_{A}\right]\bigg\} ,
\end{multline}
where $V_{B}=V(\phi=\phi_{B})$ and $V_{A}=V(\phi=\phi_{A})$. For the second path we also obtain
\begin{equation}\label{FLRWAccionComplB}
	S_{0}(a_{0},\phi_{A};a_{m},\phi_{A})=-4\pi^2\int_{0}^{s_{M}}\frac{ds}{C(s)}\left[-3a+a^3V_{A}\right] .
\end{equation}
Substituting these two expressions back in (\ref{DefGamma}) we obtain
\begin{multline}\label{FLRWGammaA}
	\pm\Gamma=-\frac{4\pi^2i}{\hbar}\int_{0}^{\bar{s}-\delta s}\frac{ds}{C(s)}\left[-3a+a^3V_{B}\right]+\frac{4\pi^2i}{\hbar}\int_{0}^{\bar{s}-\delta s}\frac{ds}{C(s)}\left[-3a+a^3V_{A}\right] \\ -\frac{4\pi^2i}{\hbar}\int_{\bar{s}-\delta s}^{\bar{s}+\delta s}\frac{ds}{C(s)}a^3\left[V(\phi)-V_{A}\right].
\end{multline}
The first two integrals can be computed by studying the general solution of the system of equations (\ref{System1}) and (\ref{System2}) for this metric, when the scalar field is a constant. Such general
solutions (\ref{CsGeneral}) and (\ref{DerivSolGen}) leads in this case to
\begin{equation}\label{FLRWSolutions}
		C^{2}(s)=4\pi^{4}\frac{\left(aV_{A,B}-\frac{1}{a}\right)^{2}}{\frac{a^{2}}{3}V_{A,B}-1} ,
\end{equation}
\begin{equation}\label{FLRWSolutions2}
		\frac{da}{ds}=\frac{1}{2\pi^{2}}\frac{\frac{a^{2}}{3}V_{A,B}-1}{aV_{A,B}-\frac{1}{a}} ,
\end{equation}
where in both expressions $V_{A,B}$ represents either $V_{A}$ or $V_{B}$. We perform a change of variables in the integral from $s$ to $a$ in the form $ds=\left(\frac{da}{ds}\right)^{-1}da$, we then obtain
\begin{multline}\label{FLRWIntegralsF0}
		-4\pi^2\int \frac{ds}{C(s)}\left[-3a+a^3V_{A,B}\right]=\pm12\pi^{2}i\int a\left(\sqrt{1-\frac{a^{2}}{3}V_{A,B}}\right) da \\ = \pm \frac{12\pi^2i}{V_{A,B}}\left[1-\frac{V_{A,B}}{3}a^2\right]^{3/2} .
\end{multline}
We can write for the scale factor $a(\bar{s}\pm\delta s)=\bar{a}\pm\delta a$, then we obtain
\begin{multline}\label{FLRWIntegralsF1E}
	-4\pi^2i\int_{0}^{\bar{s}-\delta s}\frac{ds}{C(s)}\left[-3a+a^3V_{A,B}\right]=\mp \frac{12\pi^2}{V_{A,B}}\left\{\left[1-\frac{V_{A,B}}{3}(\bar{a}-\delta a)^2\right]^{3/2}\right. \\ \left. -\left[1-\frac{V_{A,B}}{3}a_{0}^2\right]^{3/2}\right\} .
\end{multline}
We note that in the limit $a_{0}\to0$ this expression is well behaved. Therefore, even if we are working at first order in a semiclassical approximation, we can safely choose the origin to be at $a_{0}=a(s=0)=0$. Thus we finally obtain for the first two  integrals in (\ref{FLRWGammaA})
\begin{equation}\label{FLRWIntegralsF1}
	-4\pi^2i\int_{0}^{\bar{s}-\delta s}\frac{ds}{C(s)}\left[-3a+a^3V_{A,B}\right]=\mp \frac{12\pi^2}{V_{A,B}}\left\{\left[1-\frac{V_{A,B}}{3}(\bar{a}-\delta a)^2\right]^{3/2}-1\right\} .
\end{equation}
The third term in (\ref{FLRWGammaA}) can be treated as follows: by analogy with the computation performed in the Euclidean approach, we can define a tension parameter $T$ as the contribution to $\Gamma$ coming from the portion of the path in which the scalar field can vary in the following way
\begin{equation}\label{FLRWTension}
	2\pi^2\bar{a}^3T=-4\pi^2i\int_{\bar{s}-\delta s}^{\bar{s}+\delta s}\frac{ds}{C(s)}a^3\left[V(\phi)-V_{A}\right]
	.
\end{equation}
Then, substituting these two results back into (\ref{FLRWGammaA}) we obtain that the logarithm of the transition probability (that is $\Gamma$) for this metric is given by
\begin{multline}\label{FLRWGamma}
\pm\Gamma=\mp
\frac{12\pi^2}{\hbar}\left\{\frac{1}{V_{B}}\left[\left(1-\frac{V_{B}}{3}(\bar{a}-\delta
a)^2\right)^{3/2}-1\right]-\frac{1}{V_{A}}\left[\left(1-\frac{V_{A}}{3}(\bar{a}-\delta
a)^2\right)^{3/2}-1\right]\right\}
\\
	+\frac{2\pi^2}{\hbar}\bar{a}^3T .
\end{multline}
We note that our result has two sign ambiguities $\pm$. However, they are independent to each other, since they arrive from different reasons. The sign ambiguity in the left appeared in (\ref{DefGamma}) and as was explained before it comes from the fact that the general solution of the WKB type is a superposition of two exponentials with different signs. On the other hand, the sign ambiguity on the right hand side appears because as it was shown in (\ref{CsGeneral}) , the system of equations gives a solution for $C^2(s)$, but in order to compute the classical action we need $C(s)$. In practice we have the freedom to choose both signs in an appropriate way to obtain a well defined probability\footnote{This argument will be valid throughout this chapter for any equations involving two sign ambiguities.}.

Furthermore, this transition probability was derived in \cite{Cespedes:2020xpn} using the same Lorentzian formalism. However, they derived this result by proposing $C^2(s)=-1$. We note that such proposal is inconsistent with the general system of equations (\ref{System1}) and (\ref{System2}), actually the general solution of this system of equations provides an unique value for such function given by (\ref{FLRWSolutions}). Nevertheless the solution found in \cite{Cespedes:2020xpn} gives the correct expression for the classical action (\ref{FLRWIntegralsF0}), and thus the correct solution for the transition probability.

The transition probability obtained can be compared to the Euclidean result when we apply the thin wall approximation, that is considering $\delta s\to0$, therefore we obtain in this limit
\begin{equation}\label{FLRWGammaCDL}
	\tcboxmath[colback=red!10!white,colframe=red]{
		\begin{split}
	\pm\Gamma=\mp \frac{12\pi^2}{\hbar}\left\{\frac{1}{V_{B}}\left[\left(1-\frac{V_{B}}{3}\bar{a}^2\right)^{3/2}-1\right]-\frac{1}{V_{A}}\left[\left(1-\frac{V_{A}}{3}\bar{a}^2\right)^{3/2}-1\right] \right\}\\ +\frac{2\pi^2}{\hbar}\bar{a}^3T .
	\end{split}}
\end{equation}
We note that the transition probability is described by two parameters, $\bar{a}$ and $T$. However, in the Euclidean formalism it is proposed that both parameters are related by requiring that the tension is obtained when $\Gamma$ acquires a maximum value with respect to $\bar{a}$. Therefore, the condition to find an extremum of the latter expression with respect to $\bar{a}$ is $\frac{\partial\Gamma}{\partial\bar{a}}=0$, which in the case leads to
\begin{equation}\label{FLRWExtrem}
	\frac{1}{\bar{a}^2}=\frac{V_{B}}{3}+\left(\frac{\Delta V}{3T}+\frac{T}{4}\right)^2=\frac{V_{A}}{3}+\left(\frac{\Delta V}{3T}-\frac{T}{4}\right)^2 ,
\end{equation}
where  $\Delta V=V_{A}-V_{B}$. Considering such value for $\bar{a}$ in (\ref{FLRWGammaCDL}), and choosing the plus sign in the right, we finally obtain
\begin{equation}\label{FLRWGammaF}
	\pm\Gamma=\frac{12\pi^2}{\hbar}\left\{\frac{\left[4(V_{A}-V_{B})^2+3T^2(V_{A}+V_{B})\right]\bar{a}}{12TV_{A}V_{B}}+\frac{1}{V_{A}}-\frac{1}{V_{B}}\right\} .
\end{equation}
These results coincide with the Euclidean approach for this case presented originally in \cite{Parke:1982pm}. Furthermore, we note that the above analysis is valid for all non-zero values of $V_{A}$ and $V_{B}$, there is not a constriction regarding their signs. Moreover, for the case in which one of the potential is zero, we can still apply the same procedure, the difference arrives in the integral (\ref{FLRWIntegralsF0}), since it will provide a dependence of the form $\bar{a}^2$, therefore the following equations must be changed in the corresponding manner.

Furthermore, we note that our final result for the logarithm of the transition probability given by (\ref{FLRWGammaF}) is still written in terms of two factors just to obtain the same expression as in the Euclidean approach. However, in order to obtain such solution  we relate both parameters in (\ref{FLRWExtrem}). Therefore, we must be able to write effectively the logarithm of the probability in terms of just one of the parameters. Solving for the tension term in (\ref{FLRWExtrem}) we can write
\begin{equation}\label{FLRWTensionA}
	T=\pm 2\left(\sqrt{\frac{1}{\bar{a}^2}-\frac{V_{A}}{3}}-\sqrt{\frac{1}{\bar{a}^2}-\frac{V_{B}}{3}}\right) .
\end{equation}
Then choosing the plus sign in the right hand side of (\ref{FLRWGammaCDL}) we finally obtain in terms of just one parameter the transition probability
\begin{multline}\label{FLRWGammaCDLF}
	\pm\Gamma= \frac{12\pi^2}{\hbar}\left\{\frac{1}{V_{B}}\left[\left(1-\frac{V_{B}}{3}\bar{a}^2\right)^{3/2}-1\right]-\frac{1}{V_{A}}\left[\left(1-\frac{V_{A}}{3}\bar{a}^2\right)^{3/2}-1\right]\right. \\ \left. -\frac{\bar{a}^2}{3}\left(\sqrt{1-\frac{V_{A}}{3}\bar{a}^2}-\sqrt{1-\frac{V_{B}}{3}\bar{a}^2}\right)\right\} .
\end{multline}
It can be shown that the right hand side is always positive, therefore in order to have a well defined probability, we also choose the plus sign in the left hand side as well. Furthermore, we note from (\ref{FLRWTensionA}) that $\lim_{\bar{a}\to0}T=0$, therefore from the latter expression we obtain that the logarithm of the transition probability in the limit $\bar{a}\to0$ is
well behaved, giving a vanishing value. Therefore, the transition probability obeys $\lim_{\bar{a}\to0}P(A\to B)=1$. Thus, even if we have employed a limited approach by keeping only the first term in the semiclassical expansion and used General Relativity as the gravitational theory, we can still obtain a well defined transition probability in the ultraviolet limit. Furthermore, we note that choosing the appropriate signs in (\ref{FLRWGammaCDLF}) we obtain that the transition probability decreases as $\bar{a}$ increases. Thus, the maximum value is encountered in $\bar{a}=0$, then the probability falls to zero.

Let us discuss a particular complication regarding this extremizing procedure, we note from (\ref{FLRWTensionA}) that the tension will only be well defined if the terms inside the square roots are positive. This can be achieved if the potential minima are negative. However, if the potential minima are positive, this expression implies an upper bound for the scale factor. Therefore, since we require that $T$ must be a real number, in this case the above expressions are not valid for all values of $\bar{a}$. This complication is not fatal for our solutions (it is actually present in the Euclidean approach as well), since if the extremizing procedure fails in some region, we should simply abandon this condition and consider both parameters $\bar{a}$ and $T$ as independent. When we apply this procedure to new scenarios (that have not been considered previously) in the following subsections, we will encounter that in general this extramizing procedure will be troublesome, thus in the majority of the cases we will consider the tension term in the transition probability as an independent parameter.

The transition probability in this subsection was obtained by choosing a path in field space parametrized by $s$ in the form of (\ref{ChooseS}). We showed that this choice was useful to obtain consistency with the Euclidean approach. However, we have the freedom to choose such parameter in different forms. For instance, we could define the $s$ parameter as the distance in field space along the trajectories, that is we can propose
\begin{equation}\label{ParameterSGeneral}
	ds^2=\int_{X}G_{MN}d\Phi^M\Phi^N .
\end{equation}
Following the same procedure as before, the classical action (\ref{ClassicalAction}) in this case is written as
\begin{equation}\label{FLRWClassicalActionGeneral}
	S_{0}=\pm 2\pi\int_{a_{0},\phi_{B}}^{a_{m},\phi_{A}}\frac{\left(a^3V(\phi)-3a\right)^{3/2}}{\sqrt{\frac{3}{2a}\left(a^2V(\phi)-1\right)^2-a^3(V'(\phi))^2}}\sqrt{a^3d\phi^2-6ada^2} ,
\end{equation}
where as usual $V'(\phi)=\frac{dV}{d\phi}$. We note from the Hamiltonian constraint (\ref{FLREHam}) that the kinetic part of the WDW equation is proportional to the same for of the term, when instead of differentials they are  derivatives with respect to time, that is we obtain
\begin{equation}\label{FLRWKinteicGeneral}
	a^3\dot{\phi}^2-6a\dot{a}^2=2N^2\left[\frac{\pi^2_{\phi}}{2a^3}-\frac{\pi^2_{a}}{12a}\right]=2N^2\left[3a-a^3V(\phi)\right] ,
\end{equation}
Furthermore, from the definition of the transition probability (\ref{Probability}) we note that the imaginary part of the classical action is actually the term defining the transition probability. Therefore, if the integrand in (\ref{FLRWClassicalActionGeneral}) is real, we note that it is possible to obtain an imaginary contribution to the classical action when the momenta $\pi_{A}$ and $\pi_{\phi}$ take values such that the last expression is negative. This is in general possible only because the metric in superspace is not definite positive, therefore this allows us to find a classical path for the transition. The reason for the non-positiveness of the metric in superspace can be related  to the attractive nature of gravity \cite{Kiefer:2004xyv}, thus the existence for the transition probabilities that we are studying is closely related to the quantum behaviour of gravity.

Finally, before we end up this case let us discuss a subtle issue that will be relevant in the following subsections. We note that the original Hamiltonian constraint was written with a lapse function as a global multiplicative factor in (\ref{FLREHam}), since the lapse function is non-dynamical we ignored this factor and consider only the terms within square brackets. However, if we factorize any other term that corresponds to dynamical variables, the Hamiltonian constraint will be modified. At the classical level it will be equivalent to the form without the factorization. However, at the quantum level the WDW equation obtained will be different, and in general we will not be able to relate both equations on a simple form. This ambiguity on the correct choice for the Hamiltonian constraint is equivalent to the ordering ambiguity appearing when the quantization procedure is performed for the squared momentum. Thus, we have the freedom to factorize any term in order to simplify the computation of the transition probabilities.

Let us illustrate this point with a particular example that will be relevant in section \ref{S-TransitionsGUP}. Let us write the scale factor as $a(t)=e^{A(t)}$. Then, considering an adequate factorization of the Hamiltonian constraint we obtain
\begin{equation}\label{FLRWHC}
	H=Ne^{-3A}\left[-\frac{\pi^2_{A}}{12}+\frac{\pi^2_{\phi}}{2}+e^{6A}V(\phi)-3e^{4A}\right]\simeq0 ,
\end{equation}
where the canonical momenta are in this case
\begin{equation}\label{FLRWMomenta}
	\pi_{\phi}=\frac{e^{3A}}{N}\dot{\phi} , \hspace{1cm} \pi_{A}=-\frac{6e^{3A}}{N}\dot{A} .
\end{equation}
We will consider the Hamiltonian constraint to be defined by the terms inside the squared brackets. Thus, we note that in this case the fields in superspace are $\{A,\phi\}$, the inverse metric is still diagonal but it now takes the form $G^{AA}=-\frac{1}{6}$, $G^{\phi\phi}=1$, furthermore now $f(A,\phi)=e^{6A}V(\phi)-3e^{4A}$ and we still have $W^{M}=0$. Following the same procedure as before we will obtain $\Gamma$ in terms of two integrals that can be solved employing the general solutions, furthermore we will also obtain a third term regarding the region where the scalar field is allowed to vary. Thus, we treat such term in the same way, that is by defining a tension term, however in this case we propose the definition
\begin{equation}\label{FLRWTensionDef}
	2\pi^2e^{6\bar{A}}T=-4\pi^2i\int_{\bar{s}-\delta s}^{\bar{s}+\delta s}\frac{ds}{C(s)}e^{6A}\left[V(\phi)-V_{A}\right] .
\end{equation}
Therefore, in the thin wall limit we obtain
\begin{equation}\label{FLRWTRP}
	\pm\Gamma=\pm12\pi^2\bigg\{\frac{1}{V_{B}}\bigg[\left(1-\frac{V_{B}}{3}e^{2\bar{A}}\right)^{3/2}-1\bigg]-\frac{1}{V_{A}}\bigg[\left(1-\frac{V_{A}}{3}e^{2\bar{A}}\right)^{3/2}-1\bigg]\bigg\}+2\pi^2e^{6\bar{A}}T .
\end{equation}
We note that this result differs from the one obtained without the factorization in (\ref{FLRWGammaCDL}) only on the tension term, since the power of the scale factor for this term is modified.

Following the procedure, we relate $\bar{A}$ and $T$ by looking for an extremum with respect to $\bar{A}$. This condition leads to
\begin{equation}\label{FLRWTensionOri}
	T=\mp e^{-4\bar{A}}\left[\sqrt{1-\frac{V_{A}}{3}e^{2\bar{A}}}-\sqrt{1-\frac{V_{B}}{3}e^{2\bar{A}}}\right] ,
\end{equation}
which is similar to (\ref{FLRWTensionA}) with an extra term of $e^{-3\bar{A}}/2$. In the same way the tension term will only be well defined if the terms inside the square roots are positive which can create an upper bound for the scale factor and correspondingly to the validity of this procedure. Finally the transition probability will be written in terms of just one parameter in the form
\begin{multline}\label{FLRWTRPF}
	\pm\Gamma=\pm12\pi^2\left\{\frac{1}{V_{B}}\left[\left(1-\frac{V_{B}}{3}e^{2\bar{A}}\right)^{3/2}-1\right]-\frac{1}{V_{A}}\left[\left(1-\frac{V_{A}}{3}e^{2\bar{A}}\right)^{3/2}-1\right]\right. \\ \left. -\frac{e^{2\bar{A}}}{6}\left[\sqrt{1-\frac{V_{A}}{3}}e^{2\bar{A}}-\sqrt{1-\frac{V_{B}}{3}}e^{2\bar{A}}\right]\right\} .
\end{multline}

This result differs from (\ref{FLRWGammaCDLF}) only by a constant factor regarding the last term. However, as we have seen, in order to obtain an appropriate transition probability that coincides with the Euclidean result in (\ref{FLRWGammaCDLF}) the correct choices of sign are the plus sign in the right as well as a plus sign in the left. Nonetheless, since for this result (\ref{FLRWTRPF}) the second term is smaller, it is not always possible to obtain a positive quantity in the right hand side if we choose the plus sign, therefore in this case in order to have a well defined probability we have to choose the minus sign in the right hand side and the plus sign in the left.	Thus,  by factorizing a term in the Hamiltonian constraint we obtained a different solution for the transition probability that can not be related in a simple form to the one obtained if we do not perform such factorization. Therefore, the particular choice for the classical Hamiltonian constraint is relevant and it implies a concrete transition probability.

\subsection{Transitions for a flat FLRW metric}
\label{S-FLRWP}

Let us study now the transitions probability using a flat FLRW metric. In cartesian coordinates this metric is given by (\ref{DefFlatFLRW}), however considering a lapse function as well we can write it as
\begin{equation}\label{FLRWPMetric}
	ds^2=-N(t)dt^2+a^2(t)\left[dx^2+dy^2+dz^2\right] .
\end{equation}
We will consider the homogeneous case where the scale factor and the scalar field depend only on the time variable $t$ as usual. Then the Lagrangian takes the form
\begin{equation}\label{FLRWPLagr}
	\mathcal{L}=-\frac{3a(t)\dot{a}(t)^2}{N(t)}+\frac{a^3(t)\dot{\phi}^2}{2N(t)}-N(t)a^3(t)V(\phi) .
\end{equation}
The canonical momenta are the defined by (\ref{FLREMomenta}) as in the latter case, then the Hamiltonian constraint leads to
\begin{equation}\label{FLRWPHam}
	H=N\left[\frac{\pi^2_{\phi}}{2a^3}-\frac{\pi^2_{a}}{12a}+a^3V\right]\approx 0 	.
\end{equation}
As in the previous case we can identify the terms within brackets as the Hamiltonian constraint and apply the same procedure as before. Although this can be done without issues, it will be useful for comparison with the anisotropic metrics to factorize the Hamiltonian as follows
\begin{equation}\label{FLRWPHamT}
	H=\frac{N}{a^3}\left[\frac{\pi^2_{\phi}}{2}-\frac{a^2\pi^2_{a}}{12}+a^6V\right]\approx 0 ,
\end{equation}
and then focus on the terms inside the brackets. Then, we have obtained the general form (\ref{GeneralHamiltonian}), with the  coordinates of minisuperspace $\{\Phi^M\}=\{a,\phi\}$ and metric
\begin{equation}\label{FLRWPMetricSuper}
	G^{aa}=-\frac{a^2}{6} , \hspace{1cm} G^{\phi\phi}=1 ,
\end{equation}
furthermore the function $f$ reads in this case
\begin{equation}\label{FLRWPf}
	f(a,\phi)=a^6V(\phi) ,
\end{equation}
and $\{W^{M}\}=0$. We remark that the spatial volume for this metric is given by
\begin{equation}\label{FLRWPVolume}
	{\rm Vol}(X)=\int\int\int dxdydz ,
\end{equation}
which may diverge because the spatial slice described by this metric is not compact. We can obtain a finite volume if we restrict the variables to a  finite interval. However, we are considering a homogeneous case  where both fields depend only on time. Therefore, we can consider an appropriately compactified spatial section to obtain a finite volume which will lead to a regularized form for the probability. The transition probability obtained in this form will be also well defined and finite, thus it will be useful to describe the process in the non-compact scenario. In any case, the following procedure will be valid without any change, thus in the following we will treat this term as a constant\footnote{This argument will hold for any volume arriving from a non-compact spatial slice for the the rest of this chapter.}.

Considering the  $s$ parameter as in (\ref{ChooseS}), we obtain for the classical actions in (\ref{ClassicalAction}) the solutions
\begin{multline}\label{FLRWPAccionComplA}
	S_{0}(a_{0},\phi_{B};a_{m},\phi_{A})=-2{\rm
		Vol}(X)\int_{0}^{s_{M}}\frac{ds}{C(s)}\left[a^6V(\phi)\right]=-2{\rm
		Vol}(X)\left[\int_{0}^{\bar{s}-\delta
		s}\frac{ds}{C(s)}\left[a^6V_{B}\right]\right. \\ \left. +
	\int_{\bar{s}-\delta s}^{\bar{s}+\delta
		s}\frac{ds}{C(s)}\left[a^6V(\phi)\right]+\int_{\bar{s}+\delta
		s}^{s_{M}}\frac{ds}{C(s)}\left[a^6V_{A}\right]\right] ,
\end{multline}
\begin{equation}\label{FLRWPAccionComplB}
	S_{0}(a_{0},\phi_{A};a_{m},\phi_{A})=-2{\rm
		Vol}(X)\int_{0}^{s_{M}}\frac{ds}{C(s)}\left[a^6V_{A}\right] .
\end{equation}
Substituting these results into (\ref{DefGamma}) we obtain 
\begin{multline}\label{FLRWPGammaA}
	\pm\Gamma=-\frac{2{\rm Vol}(X)i}{\hbar}\int_{0}^{\bar{s}-\delta
		s}\frac{ds}{C(s)}\left[a^6V_{B}\right]+\frac{2{\rm
			Vol}(X)i}{\hbar}\int_{0}^{\bar{s}-\delta
		s}\frac{ds}{C(s)}\left[a^6V_{A}\right]\\-\frac{2{\rm
			Vol}(X)i}{\hbar}\int_{\bar{s}-\delta s}^{\bar{s}+\delta
		s}\frac{ds}{C(s)}a^6\left[V(\phi)-V_{A}\right].
\end{multline}

We proceed as before, that is we use the general solution (\ref{CsGeneral}) and (\ref{DerivSolGen})  to compute the first two integrals in (\ref{FLRWPGammaA}), which in this case take the form 
\begin{equation}\label{FLRWPSolutions}
		C^{2}(s)=12{\rm Vol}^2(X)a^6V_{A,B} ,
\end{equation}
\begin{equation}\label{FLRWPSolutions2}
		\frac{da}{ds}=\frac{1}{{\rm Vol}(X)}\frac{a}{6}.
\end{equation}
Changing the integration variable to $a$ we then can perform the integral to obtain
\begin{equation}\label{FLRWPIntegralsF0}
		-2{\rm Vol}(X)\int \frac{ds}{C(s)}\left[a^6V_{A,B}\right]=\pm2{\rm Vol}(X)\sqrt{\frac{V_{A,B}}{3}}a^3.
\end{equation}
Since this expression is regular in zero, we can safely choose $a_{0}=0$ once again, then we get
\begin{equation}\label{FLRWPIntegralsF0Aux}
	\begin{split}
		-2{\rm Vol}(X)\int_{0}^{\bar{s}-\delta s} \frac{ds}{C(s)}\left[a^6V_{A,B}\right]=\pm2{\rm Vol}(X)\sqrt{\frac{V_{A,B}}{3}}(\bar{a}-\delta a)^3.
	\end{split}
\end{equation}
For the third term we also define a tension parameter $T$ as the contribution coming from the portion of the path in which the scalar field varies in the following way
\begin{equation}\label{FLRWPTension}
	{\rm Vol}(X)\bar{a}^6T=-2{\rm Vol}(X)i\int_{\bar{s}-\delta s}^{\bar{s}+\delta s}\frac{ds}{C(s)}a^6\left[V(\phi)-V_{A}\right]	.
\end{equation}
Using these results in (\ref{FLRWPGammaA}) and considering the thin wall limit ($\delta s\to0$), we obtain
\begin{equation}\label{FLRWPGamma}
	\tcboxmath[colback=red!10!white,colframe=red]{
	\pm\Gamma=\pm\frac{2i{\rm Vol}(X)}{\sqrt{3}\hbar}\left(\sqrt{V_{B}}-\sqrt{V_{A}}\right)\bar{a}^3+\frac{{\rm Vol}(X)}{\hbar}\bar{a}^6T .}
\end{equation}
We now try to relate both parameter as in the Euclidean approach, that is by imposing the condition $\frac{\partial\Gamma}{\partial\bar{a}}=0$, which leads to
\begin{equation}\label{FLRWPMinimumConditions}
	\bar{a}^3 =
	\begin{cases}
		0 ,\\
		\mp\frac{i}{\sqrt{3}T}\left(\sqrt{V_{B}}-\sqrt{{V_{A}}}\right) .
	\end{cases}
\end{equation}
The first option of (\ref{FLRWPMinimumConditions}) implies the vanishing of (\ref{FLRWPGamma}), thus it is not relevant. For the second option we note that $\bar{a}$ must be real, then this expression is only valid when both $V_{A}$ and $V_{B}$ are negative, in which case we obtain
\begin{equation}\label{FLRWPMinimumValue}
	\bar{a}^3=\pm\frac{1}{\sqrt{3}T}\left(\sqrt{|V_{B}|}-\sqrt{|V_{A}|}\right) ,
\end{equation}
thus the transition probability (\ref{FLRWPGamma}) is expressed finally as
\begin{equation}\label{FLRWPGammaF}
	\pm2{\rm Re}[\Gamma]=-\frac{2{\rm Vol}(X)}{3\hbar T}\left(\sqrt{|V_{B}|}-\sqrt{|V_{A}|}\right)^2 .
\end{equation}

Taking in general non-negative values of $V_{A,B}$ the transition probability will be described by two independent parameters in the form
\begin{equation}\label{FLRWPGammaGen}
	\pm2{\rm Re}[\Gamma]=\mp\frac{4{\rm Vol}(X)}{\sqrt{3}\hbar}\Im\left[\left(\sqrt{V_{B}}-\sqrt{V_{A}}\right)\right]\bar{a}^3+\frac{2{\rm Vol}(X)}{\hbar}\bar{a}^6T ,
\end{equation}
From this expression we note that for a transition between two dS spaces values (that is for positive values of $V_{A}$ and $V_{B}$), the probability is described only by the tension term.

We can see that in any of the possible cases, that is, regarding $T$ as an independent parameter or finding a relation with $\bar{a}$ by the extremizing procedure, whenever possible, leading to (\ref{FLRWPMinimumValue}), the logarithm of the transition probability (\ref{FLRWPGammaGen}) is well behaved in the ultraviolet limit $\bar{a}\to0$ giving once again a vanishing value at this point.

Finally, the transition probability for this metric (\ref{FLRWPGamma}) can be written with the change of variables on the scale factor $a(t)=e^{A(t)}$,  as
\begin{equation}\label{FLRWFProb}	
	\pm\Gamma=\pm\frac{2{\rm Vol}(X)i}{\sqrt{3}}\left(\sqrt{V_{B}}-\sqrt{V_{A}}\right)e^{3\bar{A}}+{\rm Vol}(X)e^{6\bar{A}}T .
\end{equation}
This form will be important in section \ref{S-TransitionsGUP}.

\subsection{Transitions for a Kantowski-Sachs metric}\label{S-KS}
So far we have only considered the simplest models on minisuperspace employing metrics that describe homogeneous and isotropic universes in GR. Let us move on to the study of homogeneous but anisotropic metrics. We will begin by considering the Kantowski-Sachs metric, which can be written as (\ref{KSMetric}) with $0\leq\theta\leq\pi$ and $0\leq\psi\leq2\pi$. For simplicity we define $\zeta(t)=e^{\sqrt{3}\beta(t)}$ and $\sigma(t)=e^{-\sqrt{3}\Omega(t)}$, then the metric is written in the form
\begin{equation}\label{MetricKS}
	ds^2=-N^2(t)dt^2+\zeta^2(t)dr^2+\frac{\sigma^2(t)}{\zeta^{2}(t)}\left[d\theta^2+\sin^2\theta d\psi^2\right] .
\end{equation}
Considering the corresponding homogeneous scalar field $\phi(t)$, the Lagrangian is given by
\begin{equation}\label{KSLagrangian}
	\mathcal{L}=\frac{\sigma^2(t)}{\zeta^3(t)N(t)}\dot{\zeta}^2-\frac{\dot{\sigma}^2}{\zeta(t)N(t)}+N(t)\zeta(t)+\left[\frac{\dot{\phi}^2}{2N(t)}-N(t)V(\phi)\right]\frac{\sigma^2(t)}{\zeta(t)} .
\end{equation}
In these case the corresponding canonical momenta are given by
\begin{equation}\label{KSMomenta}
	\pi_{N}=0 , \hspace{1cm} \pi_{\zeta}=\frac{2\sigma^2}{N\zeta^3}\dot{\zeta} , \hspace{1cm} \pi_{\sigma}=-\frac{2}{N\zeta}\dot{\sigma} , \hspace{1cm} \pi_{\phi}=\frac{\sigma^2}{N\zeta}\dot{\phi}
\end{equation}
and the Hamiltonian constraint takes the form
\begin{equation}\label{KSHamilt}
	H=N\left[\frac{\zeta^3}{4\sigma^2}\pi^2_{\zeta}-\frac{\zeta}{4}\pi^2_{\sigma}+\frac{\zeta}{2\sigma^2}\pi^2_{\phi}+\frac{\sigma^2}{\zeta}V(\phi)-\zeta\right]\approx 0 .
\end{equation}
In this case we do not factorize any term besides the lapse function, then we have the  general form (\ref{GeneralHamiltonian})  with coordinates  $\{\Phi^M\}=\{\zeta,\sigma,\phi\}$, and metric
\begin{equation}\label{KSMetricSup}
	G^{\zeta\zeta}=\frac{\zeta^{3}}{2\sigma^2} , \hspace{0.5cm} G^{\sigma\sigma}=-\frac{\zeta}{2} , \hspace{0.5cm} G^{\phi\phi}=\frac{\zeta}{\sigma^2} ,
\end{equation}
the function $f$ is written  as
\begin{equation}\label{KSf}
	f(\zeta,\sigma,\phi)=\frac{\sigma^2}{\zeta}V(\phi)-\zeta ,
\end{equation}
and $\{W^{M}\}=0$. Furthermore, since the metric describes a non-compact spatial slice the volume of $X$ takes the form
\begin{equation}\label{KSVOlume}
	{\rm Vol}(X)=\int_{r}\int_{\theta=0}^{\pi}\int_{\psi=0}^{2\pi}\sin\theta drd\theta d\psi=4\pi\int dr ,
\end{equation}
which can be finite by taking a finite interval on $r$.

Choosing $s$ as in (\ref{ChooseS}), we obtain the classical action (\ref{ClassicalAction}) on each path  given by
\begin{multline}\label{KSClassicalAction01}
	S_{0}(\zeta_{0},\sigma_{0},\phi_{B};\zeta_{m},\sigma_{m},\phi_{A})=-2{\rm Vol}(X)\bigg\{\int_{0}^{\bar{s}-\delta
		s}\frac{ds}{C(s)}\left[\frac{\sigma^2}{\zeta}V_{B}-\zeta\right]
	\\
	+\int_{\bar{s}-\delta s}^{\bar{s}+\delta s}\frac{ds}{C(s)}\left[\frac{\sigma^2}{\zeta}V(\phi)-\zeta\right] + \int_{\bar{s}+\delta s}^{s_{M}}\frac{ds}{C(s)}\left[\frac{\sigma^2}{\zeta}V_{A}-\zeta\right]\bigg\} ,
\end{multline}
\begin{equation}\label{KSCLassicalAction02}
	S_{0}(\zeta_{0},\sigma_{0},\phi_{A};\zeta_{m},\sigma_{m},\phi_{A})=-2{\rm Vol}(X)\int_{0}^{s_{M}}\frac{ds}{C(s)}\left[\frac{\sigma^2}{\zeta}V_{A}-\zeta\right] ,
\end{equation}
then, from (\ref{DefGamma}) it follows
\begin{multline}\label{KSGammaDef}
	\pm\Gamma=-\frac{2{\rm Vol}(X)i}{\hbar}\int_{0}^{\bar{s}-\delta s}\frac{ds}{C(s)}\left[\frac{\sigma^2}{\zeta}V_{B}-\zeta\right]+\frac{2{\rm Vol}(X)i}{\hbar}\int_{0}^{\bar{s}-\delta s}\frac{ds}{C(s)}\left[\frac{\sigma^2}{\zeta}V_{A}-\zeta\right] \\ -\frac{2{\rm Vol}(X)i}{\hbar}\int_{\bar{s}-\delta s}^{\bar{s}+\delta s}\frac{ds}{C(s)}\frac{\sigma^2}{\zeta}\left[V(\phi)-V_{A}\right] .
\end{multline}
Once again, to evaluate the first two integrals we employ the general solutions (\ref{CsGeneral}) and (\ref{DerivSolGen}), which in this case take the form
\begin{equation}\label{KSSolutionSystem}
		C^2(s)={\rm Vol}^2(X)\left(\frac{\zeta^2}{\sigma^2}+3V_{A,B}\right) ,
\end{equation}
\begin{equation}\label{KSSolutionSystem2}
	\frac{d\zeta}{ds}=\frac{1}{{\rm Vol}(X)}\frac{\zeta\left(\frac{\zeta^2}{\sigma^2}+V_{A,B}\right)}{\frac{\zeta^2}{\sigma^2}+3V_{A,B}} ,
		\hspace{1cm} \frac{d\sigma}{ds}=\frac{1}{{\rm Vol}(X)}\frac{2\sigma V_{A,B}}{\frac{\zeta^2}{\sigma^2}+3V_{A,B}} .
\end{equation}
We now assume that $V_{A}$ and $V_{B}$ are different from zero, then (\ref{FieldsRelations}) is written in this case as
\begin{equation}\label{KSRelationFields}
	\frac{1}{\zeta\left(\frac{\zeta^2}{\sigma^2}+V_{A,B}\right)}d\zeta=\frac{1}{2\sigma V_{A,B}}d\sigma ,
\end{equation}
which leads to the differential equation
\begin{equation}
	2\sigma V_{A,B}d\zeta-\zeta\left(\frac{\zeta^2}{\sigma^2}+V_{A,B}\right)d\sigma=0 .
\end{equation}
Solving this equation we obtain that the variables on minisuperspace are related by
\begin{equation}\label{KSRelationiFieldsComplete}
	\zeta^2=\frac{\sigma^2V_{A,B}}{1-c\sigma} ,
\end{equation}
where $c$ is an integration constant. We note that by definition $\zeta$ and $\sigma$ are definite positive, therefore the following condition must hold
\begin{equation}\label{KSConditConst}
	\frac{V_{A,B}}{1-c\sigma}>0 ,
\end{equation}
which establish a constraint of the $c$ constant depending on the values of $V_{A,B}$ and $\sigma$. Now that we have related these two fields, we can compute the first two integrals in (\ref{KSGammaDef}) in terms of one function only. We choose $\sigma$, then performing a change of variables $ds=\left(\frac{d\sigma}{ds}\right)^{-1}d\sigma$ the integral leads to
\begin{equation}\label{KSIntegralAu0}
	-2{\rm Vol}(X)\int\frac{ds}{C(s)}\left[\frac{\sigma^2}{\zeta}V_{A,B}-\zeta\right]=\pm {\rm Vol}(X)c_{A,B}F_{KS}[c_{A,B},\sigma] ,
\end{equation}
where we have defined the function
\begin{equation}\label{FKS}
F_{KS}[c,x]=\int\frac{\sqrt{4-3cx}}{1-cx}xdx=-\frac{2}{9c^2}\left[\sqrt{4-3cx}(5+3cx)-9\arctanh\left(\sqrt{4-3cx}\right)\right] .
\end{equation}
The third term in (\ref{KSGammaDef}) can be treated as in the FLRW cases. That is, analogously to (\ref{FLRWTension}), we can define a tension term as
\begin{equation}\label{KSTensionDef}
	{\rm Vol}(X)\frac{\bar{\sigma}^2}{\bar{\zeta}}T=-2{\rm Vol}(X)i\int_{\bar{s}-\delta s}^{\bar{s}+\delta s}\frac{ds}{C(s)}\frac{\sigma^2}{\zeta}\left[V(\phi)-V_{A}\right] ,
\end{equation}
However, we have shown that in the regions where the scalar field is a constant $\zeta$ and $\sigma$ are related by (\ref{KSRelationiFieldsComplete}). Therefore employing the $s$ parametrization we obtain
\begin{equation}\label{RelationFieldsGeneral}
	\zeta=
	\begin{cases}
		\sigma\sqrt{\frac{V_{B}}{1-c_{B}\sigma}} , & 0<s<\bar{s}-\delta s,\\
		\zeta(\sigma) , & \bar{s}-\delta s<s<\bar{s}+\delta s,\\
		\sigma\sqrt{\frac{V_{A}}{1-c_{AA}\sigma}} , & \bar{s}+\delta s<s<s_{M} ,
	\end{cases}
\end{equation}
where $c_{AA}$ is an integration constant that must satisfy the condition (\ref{KSConditConst}), however it is in general different from $c_{A}$ in (\ref{KSIntegralAu0}) since this constant appears on a different path. Furthermore, the value of $\zeta(\sigma)$ in the intermediate region can be found by solving the system of equations when the scalar field is allowed to vary. However, this process needs an explicit particular form of the potential. Furthermore, since we are interested in the thin wall limit  $\delta s\to0$, the above expression simplifies to
\begin{equation}\label{RelationFieldsGeneralTWL}
	\zeta=
	\begin{cases}
		\sigma\sqrt{\frac{V_{B}}{1-c_{B}\sigma}} , & 0<s<\bar{s},\\
		\sigma\sqrt{\frac{V_{A}}{1-c_{AA}\sigma}} , & \bar{s}<s<s_{M} ,
	\end{cases}
\end{equation}
thus we can forget about the intermediate region. Then imposing continuity on $\zeta$ we obtain
\begin{equation}\label{KSRelationsTWL0}
	\bar{\zeta}=\bar{\sigma}\sqrt{\frac{V_{B}}{1-c_{B}\bar{\sigma}}}=\bar{\sigma}\sqrt{\frac{V_{A}}{1-c_{AA}\bar{\sigma}}}=\frac{V_{B}-V_{A}}{\sqrt{(c_{B}-c_{AA})(c_{B}V_{A}-c_{AA}V_{B})}} ,
\end{equation}
and
\begin{equation}\label{KSRelationTWL1}
	\bar{\sigma}=\frac{V_{A}-V_{B}}{c_{B}V_{A}-c_{AA}V_{B}} .
\end{equation}
Substituting  (\ref{KSTensionDef}) and  (\ref{KSIntegralAu0}) back into (\ref{KSGammaDef}) and taking into account the above consideration, we finally obtain in the thin wall limit
\begin{equation}\label{KSGammaOK}
	\tcboxmath[colback=red!10!white,colframe=red]{
		\begin{split}
	\pm\Gamma=\pm \frac{{\rm Vol}(X)i}{\hbar}\left[c_{B}F_{KS}[c_{B},\sigma]\bigg\rvert^{\bar{\sigma}}_{\sigma_{0}}-c_{A}F_{KS}[c_{A},\sigma]\right. & \left. \bigg\rvert^{\bar{\sigma}}_{\sigma_{0}}\right]  \\ &+\frac{{\rm Vol}(X)}{\hbar}\frac{\bar{\sigma}T}{\sqrt{V_{B}}}\sqrt{1-c_{B}\bar{\sigma}}	.
	\end{split}}
\end{equation}
Therefore, the transition probability is described by two variables, $\bar{\sigma}$ and $T$, in complete analogy to the FLRW cases. We also note that the expressions found allow us to safely
choose $\sigma_{0}=0$ since the above expression is well behaved in such limit. 

If we want to relate the two parameters by an extremizing procedure, that is to find $\bar{\sigma}$ such that $\frac{\partial \zeta}{\partial \bar{\sigma}}=0$, we are lead to the condition 
\begin{equation}\label{KSCoditionExt}
	\frac{2-3c_{B}\bar{\sigma}}{2\sqrt{V_{B}(1-c_{B}\bar{\sigma})}}T=\mp i\bar{\sigma}\left[\frac{c_{B}\sqrt{4-3c_{B}\bar{\sigma}}}{1-c_{B}\bar{\sigma}}-\frac{c_{A}\sqrt{4-3c_{A}\bar{\sigma}}}{1-c_{A}\bar{\sigma}}\right] 	.
\end{equation}
We note that for two positive values of the potentials, this condition leads inevitably to $T=0$, which is not a realistic scenario. Thus we will consider once again the two parameters as independent.

Since all the constants of integrations are real numbers, we finally obtain
\begin{multline}\label{KSGammaOKF}
	\pm2{\rm Re}[\Gamma]=\frac{2{\rm Vol}(X)}{\hbar}\left\{\mp \left[c_{B}\Im[F_{KS}[c_{B},\sigma]]\bigg\rvert^{\bar{\sigma}}_{\sigma_{0}}-c_{A}\Im[F_{KS}[c_{A},\sigma]]\bigg\rvert^{\bar{\sigma}}_{\sigma_{0}}\right]\right. \\ \left.+\frac{\bar{\sigma}T}{\sqrt{V_{B}}}\sqrt{1-c_{B}\bar{\sigma}}\right\}	.
\end{multline}
Let us remark that this result is valid for any values of the potentials different from zero. The only difference between choosing positive or negative values for the potential minima are the constraints on the integration constants so the condition (\ref{KSConditConst}) is satisfied in each case. We have also obtained that if we regard $T$ as an independent parameter for positive values of the potentials, or related to $\bar{\sigma}$ as in (\ref{KSCoditionExt}), the limit $\bar{\sigma}\to\sigma_{0}=0$ is well behaved, it actually leads in both  case to a unit value for the probability, therefore we can safely study the ultraviolet limit, and the general behaviour that the probability starts at its maximum and then decays to zero with varying $\bar{\sigma}$ as in the FLRW cases is present in this scenario as well.

For a vanishing value of the any of the potential minima, the general solution (\ref{CsGeneral}) and (\ref{DerivSolGen}) leads to
\begin{equation}\label{KSSolutionGenC0}
		C^2(s)={\rm Vol}^2(X)\frac{\zeta^2}{\bar{\sigma}^2} ,
\end{equation}
\begin{equation}\label{KSSolutionGenC02}
		\frac{d \zeta}{ds}=\frac{\zeta}{{\rm Vol}(X)} ,
		\hspace{1cm} \frac{d\sigma}{ds}=0 , \hspace{0.3cm} \therefore \hspace{0.3cm} \sigma=\bar{\sigma},
\end{equation}
then we obtain for this case
\begin{equation}\label{KSIntegralAu0C0}
	-2{\rm Vol}(X)\int\frac{ds}{C(s)}\left[\frac{\sigma^2}{\zeta}V-\zeta\right]\bigg\rvert_{V=0}=\pm 2{\rm Vol}(X)\bar{\sigma}\ln\zeta .
\end{equation}
Let us consider first that the zero minimum is at the false one, that is $V_{A}=0$, then $V_{B}<0$. In this case the path on superspace in which the system tunnels from the false minimum to the true one is described as follows; $\sigma$ starts with one specific value and evolves until it reaches a constant $\bar{\sigma}$ then it remains there. For this case we obtain
\begin{equation}\label{KSGammaOKC01}
	\pm\Gamma=\pm \frac{{\rm Vol}(X)i}{\hbar}\left[c_{B}F_{KS}[c_{B},\sigma]\bigg\rvert^{\bar{\sigma}}_{\sigma_{0}}-2\bar{\sigma}\ln{\zeta}\bigg\rvert^{\bar{\sigma}\sqrt{\frac{V_{B}}{1-c_{B}\bar{\sigma}}}}_{\sigma_{0}\sqrt{\frac{V_{B}}{1-c_{B}\sigma_{0}}}}\right]+\frac{{\rm Vol}(X)}{\hbar}\frac{\bar{\sigma}T}{\sqrt{V_{B}}}\sqrt{1-c_{B}\bar{\sigma}} ,
\end{equation}
we note that in this case we cannot longer choose $\sigma_{0}=0$ since there is a divergence in the logarithm, thus we loose the ability to explore the UV region. Furthermore, the extremizing condition to relate both parameters leads to the equation
\begin{multline}\label{KsMinimumConditionC01}
\frac{2-3c_{\sigma}\bar{\sigma}}{2\sqrt{V_{B}(1-c_{\sigma}\bar{\sigma})}}T=\mp i \bigg\{
	-2\bigg[\ln\left(\bar{\sigma}\sqrt{\frac{V_{B}}{1-c_{\sigma}\bar{\sigma}}}\right)-\ln\left(\sigma_{0}\sqrt{\frac{V_{B}}{1-c_{\sigma}\sigma_{0}}}\right)
	+1-\frac{c_{\sigma}\bar{\sigma}}{2(1-c_{\sigma}\bar{\sigma})}\bigg] \\+\frac{\sqrt{4-3c_{\sigma}\bar{\sigma}}}{1-c_{\sigma}\bar{\sigma}}c_{\sigma}\bar{\sigma}  \bigg\} .
\end{multline}

On the other hand, if the zero is at the true minimum, that is $V_{B}=0$, we then have $V_{A}>0$. In this case the probability is described in terms of $\zeta$ instead of $\sigma$, we obtain
\begin{equation}\label{KSRelationC02}
	\bar{\sigma}=\frac{\bar{\zeta}}{2V_{A}}\left[\sqrt{c_{A}^2\bar{\zeta}^2+4V_{A}}-c_{A}\bar{\zeta}\right] ,
\end{equation}
and consequently
\begin{multline}\label{KSGammaOKC02}
	\pm\Gamma=\pm \frac{{\rm Vol}(X)i}{\hbar}\bigg\{\frac{\bar{\zeta}}{V_{A}}\left[\sqrt{c_{A}^2\bar{\zeta}^2+4V_{A}}
	-c_{A}\bar{\zeta}\right]\ln\zeta\bigg\rvert^{\bar{\zeta}}_{\zeta_{0}}\\
	-c_{A}F_{KS}\left[c_{A},\frac{\zeta}{2V_{A}}\left(\sqrt{c_{A}^2\zeta^2+4V_{A}}-c_{A}\zeta\right)\right]\bigg\rvert^{\bar{\zeta}}_{\zeta_{0}}\bigg\}\\+\frac{{\rm Vol}(X)}{\hbar}\frac{\bar{\zeta} T}{4V_{A}^2}\left(\sqrt{c_{A}^2\bar{\zeta}^2+4V_{A}}-c_{A}\bar{\zeta}\right)^2 ,
\end{multline}
we  note that once again we loose the ability to explore the UV region since the limit $\zeta_{0}\to0$ causes a divergence. The condition for this expression to have an extremum with respect to $\bar{\zeta}$ leads to
\begin{multline}\label{KSConditionExtremumC02}
	\frac{T}{4V_{A}^2}\left[6c_{A}^2\bar{\zeta}^2+4V_{A}-\frac{2c_{A}\bar{\zeta}}{\bar{d}}\left(3c^2_{A}\bar{\zeta}^2+8V_{A}\right)\right] \\
	=\mp i\left[-\frac{c_{A}}{V_{A}}\left(\frac{c^2_{A}\bar{\zeta}^2+2V_{A}}{\bar{d}}-c_{A}\bar{\zeta}\right)\frac{\sqrt{4-\frac{3c_{A}}{2V_{A}}\bar{\zeta}\bar{b}}}{1-\frac{c_{A}}{2V_{A}}\bar{\zeta}\bar{b}} +\frac{\bar{b}}{V_{A}} \right. \\ \left. +\frac{2}{V_{A}}\left(\frac{c^2_{A}\bar{\zeta}^2+2V_{A}}{\bar{d}}-c_{A}\bar{\zeta}\right)\left(\ln\bar{\zeta}-\ln\zeta_{0}\right) \right] ,
\end{multline}
where
\begin{equation}\label{KSConditionExtremumC02Aux}
	d=\sqrt{c^2_{A}\bar{\zeta}^2+4V_{A}} , \hspace{0.5cm} \bar{b}=\bar{d}-c_{A}\bar{\zeta} .
\end{equation}
Therefore, the procedure allows us to consider the case with vanishing potential minima, but the results obtained are restricted, making it impossible to explore the UV region.

As we mentioned in the final part of subsection \ref{S-FLRW}, we can employ a general definition of parameter $s$ as the distance along the trajectories in the form of (\ref{ParameterSGeneral}). In this case, doing the same procedure we obtain for the classical action 
\begin{equation}\label{KSClassicalActionGeneral}
	S_{0}=\pm2\sqrt{{\rm Vol}(X)}\int_{\zeta_{0},\sigma_{0},\phi_{B}}^{\zeta_{m},\sigma_{m},\phi_{A}}\frac{\sigma^2\left(V(\phi)-\frac{\zeta^2}{\sigma^2}\right)^{3/2}}{\zeta\sqrt{\left(3V(\phi)+\frac{\zeta^2}{\sigma^2}\right)\left(V(\phi)-\frac{\zeta^2}{\sigma^2}\right)-2(V'(\phi))^2}}ds ,
\end{equation}
where
\begin{equation}\label{KSdsGeneral}
	ds=\sqrt{\frac{2\sigma^2}{\zeta^3}d\zeta^2-\frac{2}{\zeta}d\sigma^2+\frac{\sigma^2}{\zeta}d\phi^2} ,
\end{equation}
however from the Hamiltonian constraint (\ref{KSHamilt}) it follows that the $ds$ term can be written as
\begin{equation}\label{KSRelationDiffGeneral}
	\frac{2\sigma^2}{\zeta^3}\dot{\zeta}^2-\frac{2}{\zeta}\dot{\sigma}^2+\frac{\sigma^2}{\zeta}\dot{\phi}^2=2N^2\left[\frac{\zeta^3}{4\sigma^2}\pi^2_{\zeta}-\frac{\zeta}{4}\pi^2_{\sigma}+\frac{\zeta}{2\sigma^2}\pi^2_{\phi}\right]=2N^2\left[\zeta-\frac{\sigma^2}{\zeta}V(\phi)\right] .
\end{equation}
Thus if the integrand in (\ref{KSClassicalActionGeneral}) is real, we can have a transition probability simply by obtaining a negative kinetic term in the WDW equation. This is possible because the metric on superspace is not positive definite, allowing again for a classical path in consistency with the FLRW cases. However, in this case we actually have three terms, thus there are more possibilities to realize such paths. Since the only negative term is in the momentum $\pi_{\sigma}$, it may seem from the first equality that we could have transitions even if the scalar field is not considered. However, if we neglect the scalar field from (\ref{KSClassicalActionGeneral}) we obtain
\begin{equation}\label{KSClassicalActionGeneralNS}
	S_{0}\bigg\rvert_{\phi=0,V=0}=\mp2\sqrt{{\rm Vol}(X)}\int_{\zeta_{0},\sigma_{0}}^{\zeta_{m},\sigma_{m}}\sigma\sqrt{\frac{2\sigma^2}{\zeta^3}d\zeta^2-\frac{2}{\zeta}d\sigma^2} ,
\end{equation}
but from (\ref{KSRelationDiffGeneral}) we obtain that the last term cannot be negative, since it only depends on $\zeta$. Therefore, if we do not consider the scalar field, we have no contributions for the transition probability as we expected.

\subsection{Transitions for a Bianchi III metric}\label{S-B3} 
The transitions described for the Kantowski-Sachs metric were illustrative to show how to apply the general procedure to anisotropic models. However, the isotropic limit of such metric is not easily related to the FLRW scenario. Therefore, let us consider an anisotropic metric whose isotropy limit can be related to one of the FLRW metric, this will be helpful because with this metric we will be able to explicitly obtain the effects of anisotropy. Then, let us study the transitions for the Bianchi III metric that presents this behaviour. The explicit form for the metric can be written as
\cite{Ryan:1972Hm}
\begin{equation}\label{MetricB3}
	ds^{2}=-N^{2}(t)dt^{2}+\mathbb{A}^{2}dx^{2}+\mathbb{B}^{2}(t)e^{-2\alpha x}dy^{2}+\mathcal{C}^{2}(t)dz^{2} ,
\end{equation}
where $\alpha\neq0$ is a constant important to the anisotropy behaviour. Let us note that in the limit $\mathbb{A}(t)=\mathbb{B}(t)=\mathcal{C}(t)$ and $\alpha\to0$, the above expression results  in the FLRW flat metric (\ref{FLRWPMetric}), thus we have an easy connection between both metrics. We once again consider a homogeneous scalar field, then the Lagrangian is found to be
\begin{multline}\label{B3Lagrangian}
	\mathcal{L}=-\frac{1}{N(t)}\left(\dot{\mathbb{A}}(t)\dot{\mathbb{B}}(t)\mathcal{C}(t)+\dot{\mathbb{A}}(t)\dot{\mathcal{C}}(t)\mathbb{B}(t)+\dot{\mathbb{B}}(t)\dot{\mathcal{C}}(t)\mathbb{A}(t)\right)-\frac{\alpha^{2}}{\mathbb{A}(t)}N(t)\mathbb{B}(t)\mathcal{C}(t)\\ +\frac{\dot{\phi}^{2}}{2N(t)}\mathbb{A}(t)\mathbb{B}(t)\mathcal{C}(t)-V(\phi)N(t)\mathbb{A}(t)\mathbb{B}(t)\mathcal{C}(t) .
\end{multline}
Then, the corresponding canonical momenta take the form
\begin{equation}\label{B3CanoncialMomenta}
	\begin{split}
		\pi_{N}=0 , \hspace{0.5cm} \pi_{\mathbb{A}}=-\frac{1}{N}(\dot{\mathbb{B}}\mathcal{C}+\dot{\mathcal{C}}\mathbb{B}) , \hspace{0.5cm} &\pi_{\mathbb{B}}=-\frac{1}{N}(\dot{\mathbb{A}}\mathcal{C}+\dot{\mathcal{C}}\mathbb{A}) , \\
		\pi_{\mathcal{C}}=-\frac{1}{N}(\dot{\mathbb{A}}\mathbb{B}+\dot{\mathbb{B}}\mathbb{A}) , \hspace{0.5cm} &\pi_{\phi}=\frac{\dot{\phi}}{N}\mathbb{A}\mathbb{B}\mathcal{C}
	\end{split}
\end{equation}
and the Hamiltonian constraint is written as
\begin{multline}\label{B3Ham0}
	H=N\left[\frac{\mathbb{A}}{4\mathbb{B}\mathcal{C}}\pi_{\mathbb{A}}^{2}+\frac{\mathbb{B}}{4\mathbb{A}\mathcal{C}}\pi_{\mathbb{B}}^{2}+\frac{\mathcal{C}}{4\mathbb{A}\mathbb{B}}\pi_{\mathcal{C}}^{2}-\frac{1}{2\mathcal{C}}\pi_{\mathbb{A}}\pi_{\mathbb{B}}-\frac{1}{2\mathbb{B}}\pi_{\mathbb{A}}\pi_{\mathcal{C}}-\frac{1}{2\mathbb{A}}\pi_{\mathbb{B}}\pi_{\mathcal{C}} \right. \\ \left. +\frac{\pi_{\phi}^{2}}{2\mathbb{A}\mathbb{B}\mathcal{C}}+\frac{\alpha^{2}\mathbb{B}\mathcal{C}}{\mathbb{A}}+V(\phi)\mathbb{A}\mathbb{B}\mathcal{C}\right] .
\end{multline}
However, in this case it will be more useful to consider the following factorization
\begin{multline}\label{B3Ham}
	H= \frac{N}{\mathbb{A}\mathbb{B}\mathcal{C}}\bigg\{\frac{\mathbb{A}^2}{4}\pi^2_{\mathbb{A}}-\frac{\mathbb{A}\mathbb{B}}{2}\pi_{\mathbb{A}}\pi_{\mathbb{B}}-\frac{\mathbb{A}\mathcal{C}}{2}\pi_{\mathbb{A}}\pi_{\mathcal{C}}+\frac{\mathbb{B}^2}{4}\pi^2_{\mathbb{B}}-\frac{\mathbb{B}\mathcal{C}}{2}\pi_{\mathbb{B}}\pi_{\mathcal{C}}+\frac{\mathcal{C}^2}{4}\pi^2_{\mathcal{C}}+\frac{1}{2}\pi^2_{\phi}  \\  + \mathbb{B}^2C^2\alpha^2+\mathbb{A}^2\mathbb{B}^2\mathcal{C}^2V(\phi) \bigg\} ,
\end{multline}
then, we focus on the terms within brackets, we note that this form reduces consistently in the isotropy limit to the form considered in the FLRW flat case (\ref{FLRWPHamT}). We then obtain the same general form as in (\ref{GeneralHamiltonian}) with coordinates $\{\Phi^M\}=\{\mathbb{A},B,\mathcal{C},\phi\}$. Moreover the metric in this case takes the non-diagonal form
\begin{equation}\label{B3MetricSuper}
	(G^ {MN})= \frac{1}{2}\left( \begin{array}{cccc}
		\mathbb{A}^2 & -\mathbb{A}\mathbb{B} & -\mathbb{A}\mathcal{C} & 0 \\
		-\mathbb{A}\mathbb{B} & \mathbb{B}^2 & -\mathbb{B}\mathcal{C} & 0 \\
		-\mathbb{A}\mathcal{C} & -\mathbb{B}\mathcal{C} & \mathcal{C}^2 & 0  \\
		0 & 0  & 0 & 2   \end{array} \right) ,
\end{equation}
and the function $f$ is written as
\begin{equation}\label{B3f}
	f(\mathbb{A},\mathbb{B},\mathcal{C},\phi)=\mathbb{B}^2\mathcal{C}^2\left(\alpha^2+\mathbb{A}^2V(\phi)\right) ,
\end{equation}
and $\{W^{M}\}=0$. Furthermore, the spatial slice volume in this case is written as
\begin{equation}\label{B3Volume}
	{\rm Vol}(X)=\int\int\int e^{-\alpha x}dxdyxdz ,
\end{equation}
which is finite only if we restrict to a finite interval in $y$ and $z$ since the spatial slice is non-compact. 

Using the parameter $s$ defined by (\ref{ChooseS}) we obtain for the two classical actions
\begin{multline}\label{B3ClassActi01}
	S_{0}(\mathbb{A}_{0},\mathbb{B}_{0},\mathcal{C}_{0},\phi_{\mathbb{B}};\mathbb{A}_{m},\mathbb{B}_{m},\mathcal{C}_{m},\phi_{A})=-2{\rm Vol}(X)\left[\int_{0}^{\bar{s}-\delta s}\frac{ds}{C(s)}\left[\mathbb{B}^2\mathcal{C}^2\alpha^2+\mathbb{A}^2\mathbb{B}^2\mathcal{C}^2V_{B}\right] \right. \\ \left. +\int_{\bar{s}-\delta s}^{\bar{s}+\delta s}\frac{ds}{C(s)}\left[\mathbb{B}^2\mathcal{C}^2\alpha^2+\mathbb{A}^2\mathbb{B}^2\mathcal{C}^2V(\phi)\right]  +\int_{\bar{s}+\delta s}^{s_{M}}\frac{ds}{C(s)}\left[\mathbb{B}^2\mathcal{C}^2\alpha^2+\mathbb{A}^2\mathbb{B}^2\mathcal{C}^2V_{A}\right]\right] ,
\end{multline}
\begin{equation}\label{B3ClassActi02}
	S_{0}(\mathbb{A}_{0},\mathbb{B}_{0},\mathcal{C}_{0},\phi_{A};\mathbb{A}_{m},\mathbb{B}_{m},\mathcal{C}_{m},\phi_{A})=-2{\rm Vol}(X)\int_{0}^{s_{M}}\frac{ds}{C(s)}\left[\mathbb{B}^2\mathcal{C}^2\alpha^2 +\mathbb{A}^2\mathbb{B}^2\mathcal{C}^2V_{A}\right]  ,
\end{equation}
then
\begin{multline}\label{B3GammaDef}
	\pm\Gamma=-\frac{2{\rm Vol}(X)i}{\hbar}\int_{0}^{\bar{s}-\delta s}\frac{ds}{C(s)}\mathbb{B}^2\mathcal{C}^2\left[\alpha^2+\mathbb{A}^2V_{B}\right] \\ +\frac{2{\rm Vol}(X)i}{\hbar}\int_{0}^{\bar{s}-\delta s}\frac{ds}{C(s)}\mathbb{B}^2\mathcal{C}^2\left[\alpha^2+\mathbb{A}^2V_{A}\right] \\ -\frac{2{\rm Vol}(X)i}{\hbar}\int_{\bar{s}-\delta s}^{\bar{s}+\delta s}\frac{ds}{C(s)}\mathbb{A}^2\mathbb{B}^2\mathcal{C}^2\left[V(\phi)-V_{A}\right] .
\end{multline}

Using the same procedure as in the previous examples, we obtain that when the scalar field is constant and the potential minima are different from zero, the general solution is
\begin{equation}\label{B3SolutionGen}
		C^2(s)=4{\rm Vol}^2(X)\mathbb{A}^2\mathbb{B}^2\mathcal{C}^2V_{A,B}\frac{4\alpha^2+3\mathbb{A}^2V_{A,\mathbb{B}}}{\alpha^2+\mathbb{A}^2V_{A,B}} ,
\end{equation}
\begin{equation}\label{B3SolutionGen2}
		\frac{d\mathbb{A}}{ds}=\frac{1}{{\rm Vol}(X)}\frac{1}{2\mathbb{A}V_{A,B}}\frac{(2\alpha^2+\mathbb{A}^2V_{A,B})(\alpha^2+\mathbb{A}^2V_{A,B})}{4\alpha^2+3\mathbb{A}^2V_{A,B}} ,
\end{equation}
\begin{equation}\label{B3SolutionGen3}
		\frac{d\mathbb{B}}{ds}=\frac{1}{{\rm Vol}(X)}\frac{\mathbb{B}}{2}\frac{\alpha^2+\mathbb{A}^2V_{A,B}}{4\alpha^2+3\mathbb{A}^2V_{A,B}} , \hspace{1cm}
		\frac{d\mathcal{C}}{ds}=\frac{1}{{\rm Vol}(X)}\frac{\mathcal{C}}{2}\frac{\alpha^2+\mathbb{A}^2V_{A,B}}{4\alpha^2+3\mathbb{A}^2V_{A,B}} .
\end{equation}
In addition, the coordinates are related by
\begin{equation}\label{B3RelField}
	\frac{1}{\mathcal{C}}\frac{d\mathcal{C}}{ds}=\frac{1}{\mathbb{B}}\frac{d\mathbb{B}}{ds}=\frac{\mathbb{A}V_{A,B}}{2\alpha^2+\mathbb{A}^2V_{A,B}}\frac{d\mathbb{A}}{ds} ,
\end{equation}
which leads to
\begin{equation}\label{B3FieldRelAux}
	\mathbb{B}=b_{0}\mathcal{C}=c_{0}\sqrt{\frac{2\alpha^2}{V_{A,B}}+\mathbb{A}^2} ,
\end{equation}
where $\mathbb{B}_{0}$ and $c_{0}$ are integration constants. Since these constants appear only as global multiplicative terms to each coordinate, we can in general choose a unit value for all of them by redefinitions on the spatial variables. Thus, in the two regions of constant scalar field in (\ref{B3GammaDef}) we have 
\begin{equation}\label{B3FieldRel}
	\mathbb{B}=\mathcal{C}=\sqrt{\frac{2\alpha^2}{V_{A,B}}+\mathbb{A}^2} .
\end{equation}
We notice from this relation that the limit $\alpha\to0$ implies $\mathbb{A}=\mathbb{B}=\mathcal{C}$. Thus at the semiclassical level the isotropy limit that allows to recover the flat FLRW metric is simply $\alpha\to0$. Employing these relations we can express the first two integrals in (\ref{B3GammaDef}) in terms of any of the coordinates. Let us choose $\mathbb{A}$, then by performing a change of variables in the form $ds=\left(\frac{d\mathbb{A}}{ds}\right)^{-1}d\mathbb{A}$, the integral leads to
\begin{equation}\label{B3ClassiActionResult}
	-2{\rm Vol}(X)\int\frac{ds}{C(s)}\mathbb{B}^2\mathcal{C}^2\left[\alpha^2+\mathbb{A}^2V_{A,B}\right]=\pm2{\rm Vol}(X)\sqrt{V_{A,B}}F_{III}[V_{A,B},\mathbb{A}] ,
\end{equation}
where we have defined
\begin{multline}\label{F3}
	F_{III}[V,x]=\int\sqrt{\left(3x^2+\frac{4\alpha^2}{V}\right)\left(x^2+\frac{\alpha^2}{V}\right)}dx\\ =\frac{1}{9\sqrt{\frac{4\alpha^4}{V^2}+\frac{7\alpha^2}{V}x^2+3x^4}}\left\{\frac{12\alpha^4}{V^2}x+\frac{21\alpha^2}{V}x^3+9x^5 \right. \\ \left. + \frac{2\alpha^3i}{V^{5/2}}\sqrt{x^2V+\alpha^2}\sqrt{3Vx^2+4\alpha^2}\left[F\left(i\arcsin h\left(\sqrt{\frac{Vx}{\alpha^2}}\right),\frac{3}{4}\right)\right.\right. \\ \left.\left. -7E\left(i\arcsin
	h\left(\sqrt{\frac{Vx}{\alpha^2}}\right),\frac{3}{4}\right)\right]\right\} ,
\end{multline}
where $F(x,m)$ and $E(x,m)$ denote the elliptic integrals of first and second kind respectively. Note that in the isotropy limit $\alpha\to0$ the integral in (\ref{B3ClassiActionResult}) simplifies to the form encountered in the flat FLRW case (\ref{FLRWPIntegralsF0}) as expected. We will deal with the third term in (\ref{B3GammaDef}) as in the other cases, that is by defining a tension term valid in the thin wall limit. In this case we propose such term as
\begin{equation}\label{B3TensionDef}
	{\rm Vol}(X)\bar{\mathbb{A}}^2\bar{\mathbb{B}}^2\bar{\mathcal{C}}^2T=-2{\rm Vol}(X)i\int_{\bar{s}-\delta s}^{\bar{s}+\delta s}\frac{ds}{C(s)}\mathbb{A}^2\mathbb{B}^2\mathcal{C}^2\left[V(\phi)-V_{A}\right] .
\end{equation}
Furthermore from (\ref{B3FieldRel}), we obtain in the thin wall limit the relations
\begin{equation}\label{B3RelationsCases}
	\mathbb{B}=\mathcal{C}=
	\begin{cases}
		\sqrt{\frac{2\alpha^2}{V_{B}}+\mathbb{A}^2} , & 0<s<\bar{s} \\
		b\sqrt{\frac{2\alpha^2}{V_{A}}+\mathbb{A}^2} , & \bar{s}<s<s_{M} ,
	\end{cases}
\end{equation}
where $b$ is a constant of integration that cannot be absorbed by redefinition of the space variables since it arrives in the same path. Therefore, imposing continuity of $\mathbb{B}$ and $\mathcal{C}$ in $\bar{s}$ we obtain the relations
\begin{equation}\label{B3Continuity1}
	\bar{\mathbb{B}}=\bar{\mathcal{C}}=\sqrt{\frac{2\alpha^2}{V_{B}}+\bar{\mathbb{A}}^2}=b\sqrt{\frac{2\alpha^2}{V_{A}}+\bar{\mathbb{A}}^2}=\sqrt{2}\alpha b\sqrt{\left(\frac{1}{b^2-1}\right)\left(\frac{1}{V_{B}}-\frac{1}{V_{A}}\right)} ,
\end{equation}
and
\begin{equation}\label{B3Continuity2}
	\bar{\mathbb{A}}^2=\frac{2\alpha^2}{b^2-1}\left(\frac{1}{V_{B}}-\frac{b^2}{V_{A}}\right) .
\end{equation}

Thus, we finally obtain the transition probability in the thin wall limit
\begin{equation}\label{B3GammaResult}
	\tcboxmath[colback=red!10!white,colframe=red]{
	\begin{split}
	\pm\Gamma=\pm\frac{2{\rm Vol}(X)i}{\hbar}\left[\sqrt{V_{B}}F_{III}[V_{B},\mathbb{A}]\bigg\rvert^{\bar{\mathbb{A}}}_{\mathbb{A}_{0}}- \sqrt{V_{A}}\right. & \left.F_{III}[V_{A},\mathbb{A}]\bigg\rvert^{\bar{\mathbb{A}}}_{\mathbb{A}_{0}}\right]\\ +&\frac{{\rm Vol}(X)}{\hbar}\bar{\mathbb{A}}^2\left(\frac{2\alpha^2}{V_{B}}+\bar{\mathbb{A}}^2\right)^2T  .
	\end{split}}
\end{equation}

Therefore the transition probability for the Bianchi III metric is also described by two parameters, one coming from the metric, we have chosen $\bar{\mathbb{A}}$, and the tension term $T$. Furthermore, we note that $\lim_{x\to0}F_{III}[V,x]=0$ for any value of $V$, therefore in all cases we can choose $\mathbb{A}_{0}=0$. We remark that this expression is valid for positive potential minima, since the relations (\ref{B3FieldRel}) must hold, and the functions $\mathbb{B}$ and $\mathcal{C}$ are real.

Thus, we obtain for positive potential minima $V_{A,B}$  
\begin{multline}\label{B3GammaResultF}
	\pm2{\rm Re}[\Gamma]=\mp\frac{4{\rm Vol}(X)}{\hbar}\left\{\Im\left[\sqrt{V_{B}}F_{III}[\mathbb{A},V_{B}]\right]\bigg\rvert^{\bar{\mathbb{A}}}_{0}-\Im\left[\sqrt{V_{A}}F_{III}[\mathbb{A},V_{A}]\bigg\rvert^{\bar{\mathbb{A}}}_{0}\right]\right\} \\+\frac{2{\rm Vol}(X)}{\hbar}\bar{\mathbb{A}}^2\left(\frac{2\alpha^2}{V_{B}}+\bar{\mathbb{A}}^2\right)^2T  .
\end{multline}
As we remarked earlier, the  isotropy limit $\alpha\to0$ of this result gives consistency with the flat FLRW metric (\ref{FLRWPGamma}). 

Following the extremizing procedure to relate the two parameters describing the probability we obtain that $\frac{\partial\Gamma}{\partial\bar{A}}=0$ leads to
\begin{multline}\label{B3MinCondit}
T\bar{\mathbb{A}}\left(3\bar{\mathbb{A}}^4+\frac{8\alpha^2}{V_{B}}\bar{\mathbb{A}}^2+\frac{4\alpha^4}{V^2_{B}}\right)=\mp i\bigg[\sqrt{\left(3\bar{\mathbb{A}}^2V_{B}+4\alpha^2\right)\left(\bar{\mathbb{A}}^2+\frac{\alpha^2}{V_{B}}\right)}
\\	-\sqrt{\left(3\bar{\mathbb{A}}^2V_{A }+4\alpha^2\right)\left(\bar{\mathbb{A}}^2+\frac{\alpha^2}{V_{A}}\right)}\bigg].
\end{multline}
However, for positive values for the potentials the solution is given by $T=0$. Thus, we must consider each term as independent in this case as well. Furthermore, for positive values of the potential, we can study the UV region where $\bar{A}\to0$ in (\ref{B3GammaResultF}). It leads to a vanishing result for $\Gamma$. Thus, the general form of the probability is the same as for the other metrics. 

If we want to consider a vanishing value for one ot the potential minima, we obtain that the system of equations (\ref{System1}) and (\ref{System2}) for a constant scalar field is inconsistent. Therefore, the results found for this metric are only valid when $V_{A,B}\neq 0$. We could try another factorization of the Hamiltonian in order to obtain a well defined result, but we fail to obtain a particular factorization that could lead to a solvable system for the vanishing of the potential minima at the same time that gives an integral in (\ref{B3ClassiActionResult}) that is computable. Thus, we do not consider this case any more.

\begin{figure}[h!]
	\centering
	\includegraphics[width=0.7\textwidth]{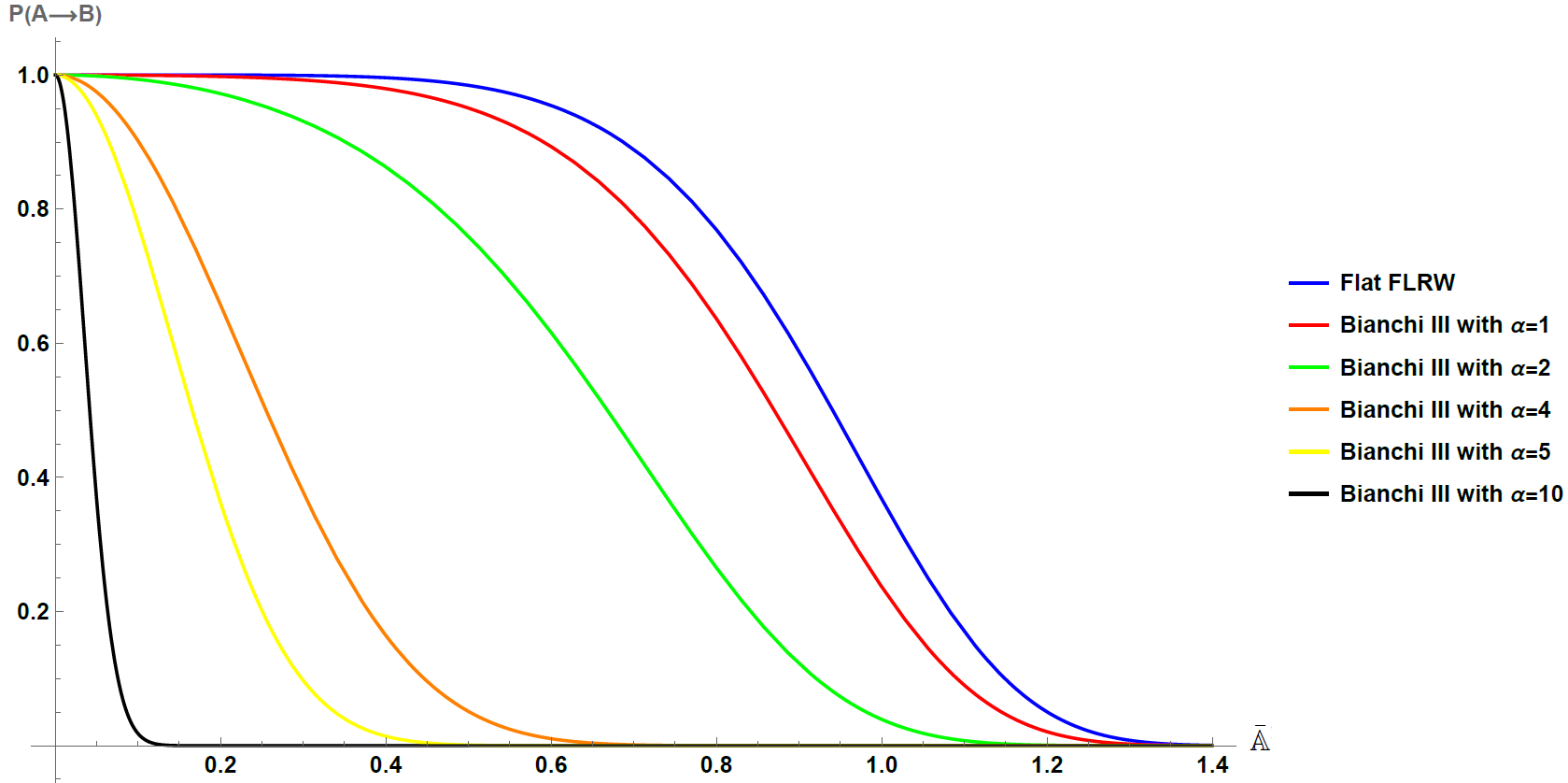}
	\caption{Transition probability in units such that $\frac{2{\rm Vol}(X)}{\hbar}=1$, with $V_{A}=100$, $V_{B}=10$ and $T=1$ for the Bianchi III metric with $\alpha=10$ (black line), $\alpha=5$ (yellow line), $\alpha=4$ (orange line), $\alpha=2$ (green line), $\alpha=1$ (red line) and $\alpha=0$  corresponding to the flat FLRW result (blue line). The contribution comes only for the tension term in (\ref{B3GammaResultF}).} \label{FigTransProbN}
\end{figure}

\begin{figure}[h!]
	\centering
	\includegraphics[width=0.7\textwidth]{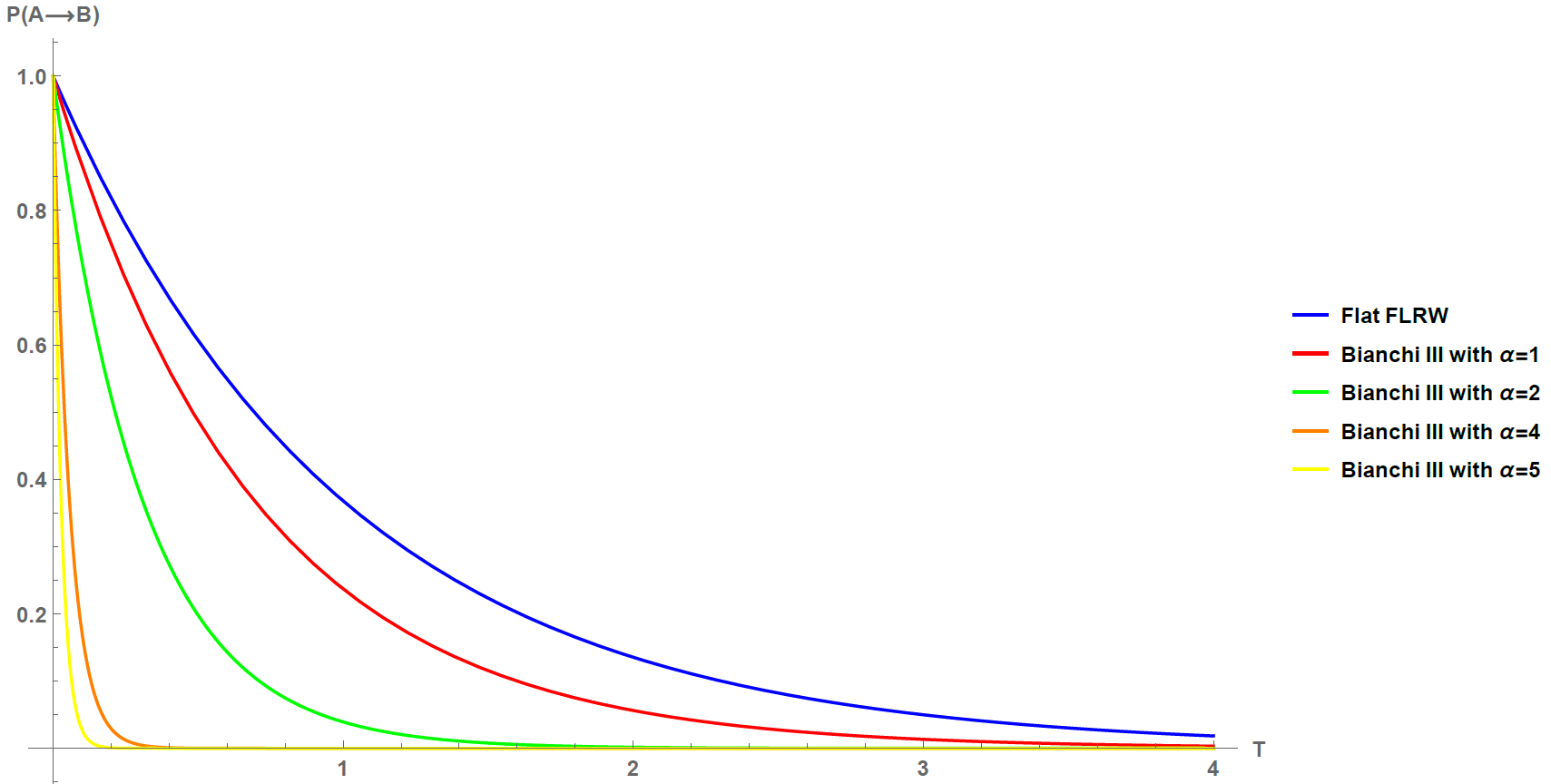}
	\caption{Transition probability in units such that $\frac{2{\rm Vol}(X)}{\hbar}=1$, with $V_{A}=100$, $V_{B}=10$ and $\bar{\mathbb{A}}=1$ for
		the Bianchi III metric with $\alpha=5$ (yellow line), $\alpha=4$ (orange line),
		$\alpha=2$ (green line), $\alpha=1$ (red line) and $\alpha=0$ corresponding to the
		flat FLRW  result (blue line).} \label{FigTransProbS}
\end{figure}

We have obtained analytical solutions for the transition probabilities in this metric, and we have obtained that the effect of anisotropy is described by the appearance of $\alpha$ in a complicated form. Thus, let us plot our solutions in on order to visualize what is the consequence of anisotropy. We choose the minus sign both in the left and in the right of (\ref{B3GammaResultF}) in order to obtain a well defined probability. We present plots for the probability obtained for this metric in (\ref{B3GammaResultF}) for different values of $\alpha$ (different levels of anisotropy) as well as the isotropy limit $\alpha\to0$ that gives the flat FLRW result (\ref{FLRWPGammaGen}).  We choose units such that $\frac{2{\rm Vol}(X)}{\hbar}=1$, as well as positive values for the potential minima $V_{B}=10$, $V_{A}=100$. Since the probability is described by two parameters we present to plots. First, in Figure \ref{FigTransProbN} we set $T=1$ and plot the probability versus $\bar{\mathbb{A}}$, or equivalently $\bar{a}$.  For the FLRW metric (\ref{FLRWPGammaGen}) and for the Bianchi III metric (\ref{B3GammaResultF}) the contribution comes only from the tension term. Furthermore, in Figure \ref{FigTransProbS} we also show the plot of the probability versus the tension term and choose $\bar{\mathbb{A}}=1$. For both figures we conclude that the effect of anisotropy is to reduce the transition probability and make it fall faster as $\alpha^2$ increases. 

\begin{figure}[h!]
	\centering
	\includegraphics[width=0.7\textwidth]{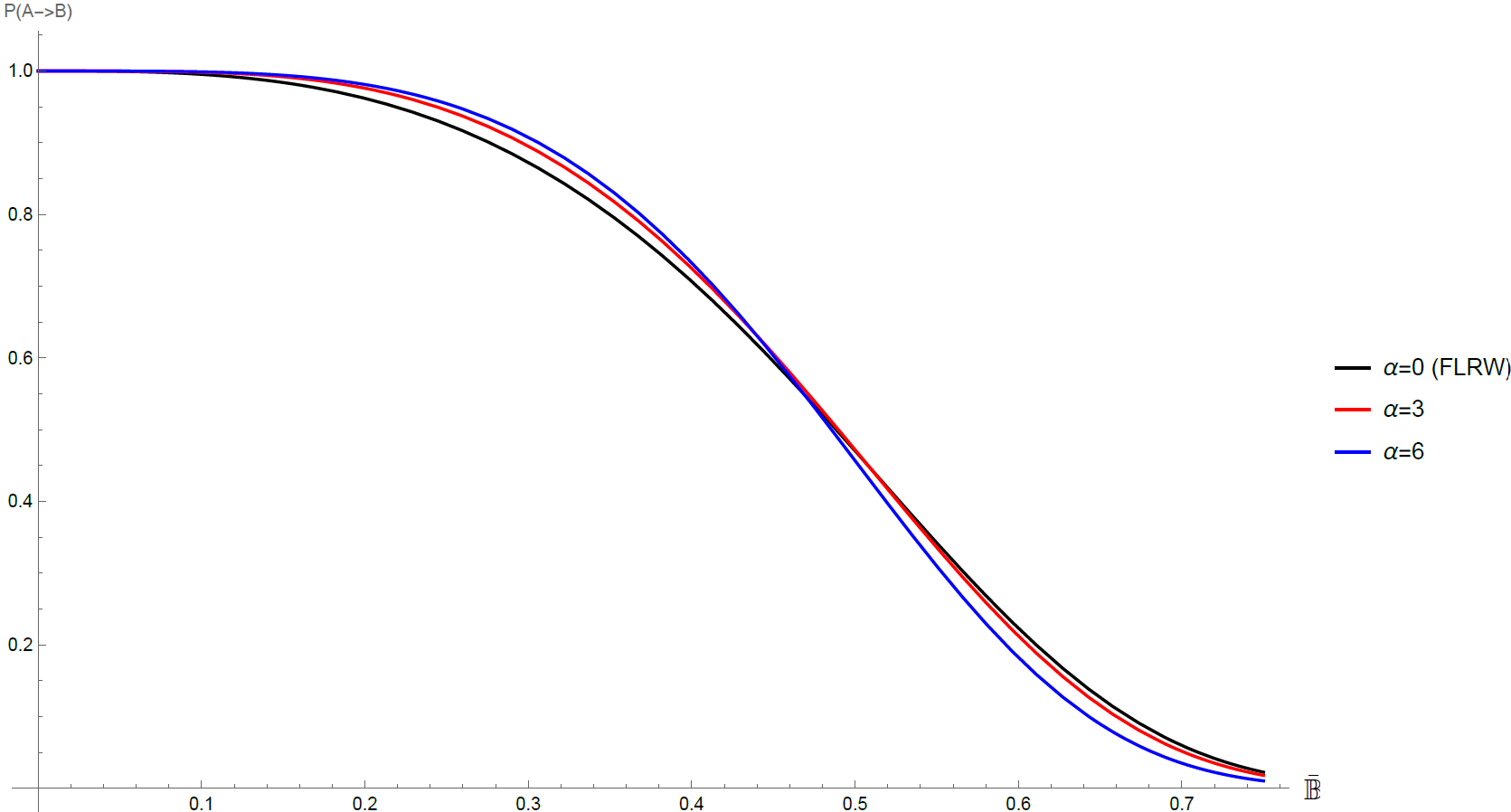}
	\caption{Transition probability in units such that $\frac{2{\rm Vol}(X)}{\hbar}=1$ for the Bianchi III metric with  $\alpha=6$ (Blue curve), $\alpha=3$ (Red curve) and the isotropy limit with leads to the FLRW flat result when  $\alpha=0$ (Black curve) choosing $V_{A}=-100$, $V_{B}=-200$ and $T=10$.}
	\label{F-BianchiCom}
\end{figure}

Let us remark that for this metric we can write the transition probability in terms of any of the two different variables, that is $\mathbb{A}$ or $\mathbb{B}$, we note that they are not independent and thus the probability is equivalent with the two variables, however we will see that each variable is useful for a particular scenario. We can easily see from (\ref{B3FieldRel}) that for positive values of the potential minima, we can take $\mathbb{A}_{0}=0$ without any problem, thus the transition probability is well defined for all values of $\bar{\mathbb{A}}$. However, for negative values of the potential minima, this expression implies that $\mathbb{A}_{0}\neq0$ in order to have a correctly defined form for $\mathbb{B}$. Therefore, for negative values of the potential minima, the probabilities written in terms of $D$ does not allow to explore the UV region. In contrast, from  (\ref{B3FieldRel}) wee see that in this case we can indeed choose $\mathbb{B}_{0}=0$. Thus if we express instead the probability in terms of $\mathbb{B}$, we will obtain a consistent result valid for all values of $\bar{\mathbb{B}}$ and we can explore the UV region. Therefore, writing the transition probability in terms of one of the two variables is a matter of convenience depending on the signs of the potential minima that we want to consider. Therefore, we can also study the effect of anisotropy with negative potential minima, but now employing only the parameter $\bar{\mathbb{B}}$. Choosing $\mathbb{B}_{0}=0$ and making the corresponding change of variables, we obtain that (\ref{B3GammaResult}) is written as

\begin{multline}\label{B3GammaResultB}
	\pm\Gamma=\pm\frac{2{\rm Vol}(X)i}{\hbar}\left[\sqrt{V_{B}}G_{III}[V_{B},\mathbb{B}]\bigg\rvert^{\bar{\mathbb{B}}}_{\mathbb{B}_{0}}-\sqrt{V_{A}}G_{III}[V_{A},\mathbb{B}]\bigg\rvert^{\bar{\mathbb{B}}}_{\mathbb{B}_{0}}\right]\\ +\frac{{\rm Vol}(X)}{\hbar}\bar{\mathbb{B}}^4\left(\bar{\mathbb{B}}^2-\frac{2\alpha^2}{V_{B}}\right)T  ,
\end{multline}
where we have defined the function
	\begin{equation}\label{G3}
		G_{III}[V,x]=\int\sqrt{\frac{\left(3x^2-\frac{2\alpha^2}{V}\right)\left(x^2-\frac{\alpha^2}{V}\right)}{\left(x^2-\frac{2\alpha^2}{V}\right)}}xdx ,
	\end{equation}
which can also be written in terms of elliptic integrals. Studying the effect of anisotropy on the tension term we note that as $\alpha^2$ increases, this term will also increase because the potential minima are negative, the same behaviour as the one encountered in (\ref{B3GammaResult}). However, contrary to what is encountered with positive potentials, in this case the first term will contribute to the probability. Regarding the sign ambiguity we find that choosing the plus sign in the right hand side of (\ref{B3GammaResultB}) we obtain the same behaviour with the variable $A$, that is, as $\alpha^2$ increases the probability decreases. However for this choice of signs there will be a region near $\bar{\mathbb{B}}\to0$ where the isotropy limit $\alpha\to0$ would lead to an ill defined probability, since it will produce values  greater than $1$. Therefore, in order to obtain well defined probabilities for all values of $\alpha$, we choose the minus sign. This choice leads to a scenario where the tension term has the same behaviour as in the latter case, but the first term has the opposite behaviour. We plot the probability (\ref{B3GammaResultB}) in Figure \ref{F-BianchiCom} for different values of $\alpha$, choosing unit in which $2{\rm Vol}(X)=1$, we also choose $V_{A}=-100$, $V_{B}=-200$ and $T=10$. We note that for small values of $\bar{\mathbb{B}}$ the first term is dominant and makes the probability  increase with $\alpha$, however the tension term grows with $\bar{\mathbb{B}}$,  eventually this term becomes dominant and we recover the same behaviour as the one encountered with the positive potentials. On the other hand, we can easily see from (\ref{B3GammaResultB}) that if we vary the tension the behaviour will be the same as the one presented in Figure \ref{FigTransProbS}.

Pursuing a more general analysis we can choose  $s$  as in (\ref{ParameterSGeneral}). In this case, the classical action is given by
\begin{equation}\label{B3ActionClassicalGeneral}
	S_{0}=\pm\sqrt{2{\rm Vol}(X)}\int_{\mathbb{A}_{0},\mathbb{B}_{0},\mathcal{C}_{0},\phi_{B}}^{\mathbb{A}_{m},\mathbb{B}_{m},\mathcal{C}_{m},\phi_{A}}\frac{\mathbb{B}\mathcal{C}\left(\alpha^2+\mathbb{A}^2V(\phi)\right)^{3/2}}{\mathbb{A}\sqrt{2V(\phi)\left(4\alpha^2+3\mathbb{A}^2V(\phi)\right)-\mathbb{A}^2(V'(\phi))^2}} ds ,
\end{equation}
where
\begin{equation}\label{B3dsGeneral}
	ds=\sqrt{-\frac{2}{\mathbb{A}\mathbb{B}}d\mathbb{A}d\mathbb{B}-\frac{2}{\mathbb{A}\mathcal{C}}d\mathbb{A}d\mathcal{C}-\frac{2}{\mathbb{B}\mathcal{C}}d\mathbb{B}d\mathcal{C}+d\phi^2} .
\end{equation}
From the Hamiltonian constraint (\ref{B3Ham}), we obtain a relation between the kinetic term in the WDW equation and the $ds$ term in the form
\begin{multline}\label{B3RelationsKinetic}
	-\frac{2}{\mathbb{A}\mathbb{B}}\dot{\mathbb{A}}\dot{\mathbb{B}}-\frac{2}{\mathbb{A}\mathcal{C}}\dot{\mathbb{A}}\dot{\mathcal{C}}-\frac{2}{\mathbb{B}\mathcal{C}}\dot{\mathbb{B}}\dot{\mathcal{C}}+\dot{\phi}^2\\=\frac{2N^2}{\mathbb{A}^2\mathbb{B}^2\mathcal{C}^2}\left[\frac{\mathbb{A}^2}{4}\pi^2_{\mathbb{A}}-\frac{\mathbb{A}\mathbb{B}}{2}\pi_{\mathbb{A}}\pi_{B}-\frac{\mathbb{A}\mathcal{C}}{2}\pi_{\mathbb{A}}\pi_{\mathcal{C}}+\frac{\mathbb{B}^2}{4}\pi^2_{\mathbb{B}}-\frac{\mathbb{B}\mathcal{C}}{2}\pi_{\mathbb{B}}\pi_{\mathcal{C}}+\frac{\mathcal{C}^2}{4}\pi^2_{\mathcal{C}}+\frac{1}{2}\pi^2_{\phi}\right] \\ =-\frac{2N^2}{\mathbb{A}^2}\left(\alpha^2+\mathbb{A}^2V(\phi)\right) .
\end{multline}
Therefore, when the integrand in (\ref{B3ActionClassicalGeneral}) is real, we can obtain a transition probability simply by considering a negative kinetic term in the WDW equation, which can be
achieved by having correct values for the momenta, or just by having a positive potential as the second equality of the last expression shows. If we want to explore a system without a scalar field, the expression for the classical action does not have a well defined behaviour, this happens because, as we have seen previously, the system of differential equations in this case is inconsistent.

Finally, let us write the results of this chapter using a change of variables that will be helpful in section \ref{S-TransitionsGUP}. We define
\begin{equation}\label{B3VariablesDef}
	\mathbb{A}(t)=e^{p(t)} , \hspace{0.5cm} \mathbb{B}(t)=e^{q(t)} , \hspace{0.5cm} \mathcal{C}(t)=e^{u(t)} ,
\end{equation}
then, the Hamiltonian constraint with the proper factorization takes the form
\begin{multline}\label{B3HC}
	H= \frac{N}{e^{p+q+u}}\bigg\{\frac{\pi^2_{p}}{4}+\frac{\pi^2_{q}}{4}+\frac{\pi^2_{u}}{4}-\frac{\pi_{p}\pi_{q}}{2}-\frac{\pi_{p}\pi_{u}}{2}-\frac{\pi_{q}\pi_{u}}{2}+\frac{\pi^2_{\phi}}{2} \\  +  e^{2(q+u)}\left[\alpha^2+e^{2p}V(\phi)\right] \bigg\}\simeq 0 ,
\end{multline}
where the canonical momenta are given by
\begin{equation}\label{B3DefMomenta}
	\pi_{p}=-\frac{\dot{q}+\dot{u}}{N}e^{p+q+u} , \hspace{0.5cm} \pi_{q}=-\frac{\dot{p}+\dot{u}}{N}e^{p+q+u} , \hspace{0.5cm} \pi_{u}=-\frac{\dot{p}+\dot{q}}{N}e^{p+q+u} , \hspace{0.5cm} 
	\pi_{\phi}=\frac{\dot{\phi}}{N}e^{p+q+u} .
\end{equation}

In this case, the relation between the coordinates on minisuperspace in the region where the scalar field is constant (\ref{B3FieldRel}) is written as
\begin{equation}\label{B3RelationsC}
	q=u=\ln\sqrt{\frac{2\alpha^2}{V}+e^{2p}} ,
\end{equation}
Therefore, we can write the transition probability (\ref{B3GammaResult}) for positive values of the potential minima in terms of the $p$ variable as
\begin{multline}\label{B3ProbComp} 
	\pm\Gamma=\pm2{\rm Vol}(X)i\left[\sqrt{V_{B}}\int_{-\infty}^{\bar{p}}e^{p}\sqrt{\left(e^{2p}+\frac{\alpha^2}{V_{B}}\right)\left(3e^{2p}+\frac{4\alpha^2}{V_{B}}\right)}dp \right. \\ \left. -\sqrt{V_{A}}\int_{-\infty}^{\bar{p}}e^{p}\sqrt{\left(e^{2p}+\frac{\alpha^2}{V_{A}}\right)\left(3e^{2p}+\frac{4\alpha^2}{V_{A}}\right)}dp\right]+{\rm Vol}(X)e^{2\bar{p}}\left(e^{2\bar{p}}+\frac{2\alpha^2}{V_{B}}\right)^2T ,
\end{multline}
Moreover, for negative values of the potential minima we can write the probability (\ref{B3GammaResultB}) in terms of the $q$ variable as
\begin{multline}\label{B3Prob}
	\pm\Gamma=\pm2{\rm Vol}(X)i\left[\sqrt{V_{B}}\int_{q_{0}}^{\bar{q}}e^{2q}\sqrt{\frac{\left(e^{2q}-\frac{\alpha^2}{V_{B}}\right)\left(3e^{2q}-\frac{2\alpha^2}{V_{B}}\right)}{e^{2q}-\frac{2\alpha^2}{V_{B}}}}dq \right. \\ \left. -\sqrt{V_{A}}\int_{q_{0}}^{\bar{q}}e^{2q}\sqrt{\frac{\left(e^{2q}-\frac{\alpha^2}{V_{A}}\right)\left(3e^{2q}-\frac{2\alpha^2}{V_{A}}\right)}{e^{2q}-\frac{2\alpha^2}{V_{A}}}}dq\right]+{\rm Vol}(X)e^{4\bar{q}}\left(e^{2\bar{q}}-\frac{2\alpha^2}{V_{B}}\right)T .
\end{multline}

\subsection{Transitions for a Bianchi IX metric}\label{S-B9}
The last metric that we will consider in this section will be the biaxial Bianchi IX metric \cite{Ryan:1972Hm}, which is an anisotropic metric that describes a compact spatial slice. Such metric is given by 
\begin{equation}\label{B9MetricDefOr}
	ds^2=-\frac{N^2(t)}{Q(t)}dr^2+\frac{P(t)}{4}\left(\sigma^2_{1}+\sigma^2_{2}\right)+\frac{Q(t)}{4}\sigma^2_{3} ,
\end{equation}
where
\begin{equation}\label{B9MetricDefOrA}
		\sigma_{1}=\sin\psi d\theta-\cos\psi\sin\theta d\omega ,
		\hspace{1cm} \sigma_{2}=\cos\psi d\theta+\sin\psi\sin\theta d\omega , 
\end{equation}
\begin{equation}
	\sigma_{3}=-\left(d\psi+\cos\theta d\omega\right) ,
\end{equation}
and $0\leq\psi\leq4\pi$, $0\leq\theta\leq\pi$ and $0\leq\omega\leq2\pi$.  Considering the above definitions we can expand the metric to take the form
\begin{equation}\label{B9Metric}
	ds^2=-\frac{N^2(t)}{Q}dt^2+\frac{Q(t)}{4}d\psi^2+\frac{Q(t)}{2}\cos\theta d\psi d\omega+\frac{P(t)}{4}d\theta^2+\frac{1}{4}\left(P(t)\sin^2\theta+Q(t)\cos^2\theta\right)d\omega^2 .
\end{equation}
Considering a homogeneous scalar field the Lagrangian takes the form
\begin{equation}\label{B9LagrangianDef}
	\mathcal{L}=\frac{1}{16}\left[-\frac{Q}{2NP}\dot{P}^2-\frac{\dot{Q}\dot{P}}{N}-\frac{2Q}{P}N+8N\right]+\frac{P}{8}\left[\frac{Q\dot{\phi}^2}{2N}-NV\right].
\end{equation}
The canonical momenta derived from this expression are
\begin{equation}
	\pi_{N}=0 , \hspace{0.5cm} \pi_{P}=-\frac{1}{16N}\left(\frac{Q}{P}\dot{P}+\dot{Q}\right) , \hspace{0.5cm} \pi_{Q}=-\frac{\dot{P}}{16N} , \hspace{0.5cm}
	\pi_{\phi}=\frac{PQ}{8N}\dot{\phi}.
\end{equation}
Thus the Hamiltonian constraint in this case is
\begin{equation}\label{B9HamiltConstra}
	H=N\left[\frac{8Q}{P}\pi^2_{Q}-16\pi_{P}\pi_{Q}+\frac{4}{PQ}\pi^2_{\phi}+\frac{1}{8}\left(\frac{Q}{P}+PV(\phi)-4\right)\right] \approx 0 .
\end{equation}
We will not factorize any other term in this case, then we only focus on the expression inside brackets, therefore we have the general form (\ref{GeneralHamiltonian}) with the coordinates $\{\Phi^M\}=\{Q,P,\phi\}$, the metric is given by
\begin{equation}\label{B9MetricSuper}
	G^{QQ}=\frac{16Q}{P} , \hspace{0.5cm} G^{PQ}=-16 , \hspace{0.5cm} G^{\phi\phi}=\frac{8}{PQ} ,
\end{equation}
the function $f$ is 
\begin{equation}\label{B9fDef}
	f(P,Q,\phi)=\frac{1}{8}\left(\frac{Q}{P}+PV(\phi)-4\right) ,
\end{equation}
and $\{W^M\}=0$. For this metric the spatial volume is given by
\begin{equation}\label{B9Volume}
	{\rm Vol}(X)=\int_{\omega=0}^{2\pi}\int_{\theta=0}^{\pi}\int_{\psi=0}^{4\pi}\sin\theta d\psi d\theta d\omega = 16\pi^2 ,
\end{equation}
which is finite since the spatial slice is compact.

Taking $s$ as in the previous cases, we obtain for the two classical actions
\begin{multline}\label{B9ClassicalActionAu1}
	S_{0}(P_{0},Q_{0},\phi_{B};P_{m},Q_{m},\phi_{A})=-4\pi^2\bigg\{\int_{0}^{\bar{s}-\delta
		s}\frac{ds}{C(s)}\left[\frac{Q}{P}+PV_{B}-4\right] \\
	+\int_{\bar{s}-\delta s}^{\bar{s}+\delta s}\frac{ds}{C(s)}\left[\frac{Q}{P}+PV(\phi)-4\right] + \int_{\bar{s}+\delta s}^{s_{M}}\frac{ds}{C(s)}\left[\frac{Q}{P}+PV_{A}-4\right]\bigg\} ,
\end{multline}
\begin{equation}\label{B9ClassicalActionAu2}
	S_{0}(P_{0},Q_{0},\phi_{A};P_{m},Q_{m},\phi_{A})=-4\pi^2\int_{0}^{s_{M}}\frac{ds}{C(s)}\left[\frac{Q}{P}+PV_{A}-4\right].
\end{equation}
We then obtain
\begin{multline}\label{B9GammaDefO}
	\pm\Gamma=-\frac{4\pi^2i}{\hbar}\int_{0}^{\bar{s}-\delta s}\frac{ds}{C(s)}\left[\frac{Q}{P}+PV_{B}-4\right]+\frac{4\pi^2i}{\hbar}\int_{0}^{\bar{s}-\delta s}\frac{ds}{C(s)}\left[\frac{Q}{P}+PV_{A}-4\right] \\ -\frac{4\pi^2i}{\hbar}\int_{\bar{s}-\delta s}^{\bar{s}+\delta s}\frac{ds}{C(s)}P\left[V(\phi)-V_{A}\right] .
\end{multline}

For a constant scalar field the general solution gives
\begin{equation}\label{B9SolSyst}
		C^2(s)=-1024\pi^4\frac{3Q-2V_{A,B}P^2}{P^2\left(Q+P^2V_{A,B}-4P\right)} ,
\end{equation}
\begin{equation}
	\frac{dQ}{ds}=\frac{1}{16\pi^2}\frac{(2Q-V_{A,B}P^2)(Q+P^2V_{A,B}-4P)}{3Q-2V_{A,B}P^2} , \hspace{1cm} \frac{dP}{ds}=-\frac{1}{16\pi^2}\frac{P(Q+P^2V_{A,B}-4P)}{3Q-2V_{A,B}P^2} ,
\end{equation}
which imply the relation
\begin{equation}\label{B9RelationFieldsOr}
	\frac{1}{2Q-V_{A,B}P^2}\frac{dQ}{ds}=-\frac{1}{P}\frac{dP}{ds} .
\end{equation}
This expressions leads to the differential equation
\begin{equation}\label{B9RelationFieldsDE}
	PdQ+(2Q-V_{A,B}P^2)dP=0 ,
\end{equation}
where the general solution is
\begin{equation}\label{B0RelationFieldsT}
	Q=\frac{c}{P^2}+\frac{V_{A,B}}{4}P^2 ,
\end{equation}
with $c$ an integration constant. By definition $P$ and $Q$ are positive functions, furthermore we  impose the condition $\lim_{P\to0}Q=0$ in order to have an spatial singularity at the beginning as standard in cosmological solutions, thus we see for this expression that we can only consider positive values for the potential minima and we have to neglect the integration constant, thus we obtain the relation
\begin{equation}\label{B9RelationFields}
	Q=\frac{V_{A,B}}{4}P^2 .
\end{equation}
With this expression, we can compute the first two terms in (\ref{B9GammaDefO}) as integrals on any of the two variables, we choose  $P$ , then we obtain
\begin{equation}\label{B9ClassACtion}
	-4\pi^2\int\frac{ds}{C(s)}\left[\frac{Q}{P}+PV_{A,B}-4\right]=\pm2\pi^{2}\sqrt{5V_{A,B}}F_{IX}[V_{A,B},P] ,
\end{equation}
where we have defined
\begin{multline}\label{F9}
	F_{IX}[V,x]= \int\sqrt{x\left(\frac{5V}{16}x-1\right)}dx \\ =\frac{1}{4\left(\frac{5V}{16}\right)^{3/2}\sqrt{x\left(\frac{5V}{16}x-1\right)}}
	\left[\sqrt{\frac{5V}{16}}x\left(1-\frac{15V}{16}x+\frac{25V^2}{128}x^2\right) \right. \\ \left.-\sqrt{x\left(\frac{5V}{16}x-1\right)}\ln\left[\frac{5V}{16}\sqrt{x}+\sqrt{\frac{5V}{16}}\sqrt{\frac{5V}{16}x-1}\right]\right] .
\end{multline}

Analogously as it was done in all the previous cases, we propose for this metric the tension term
\begin{equation}\label{B9TensionDef}
	16\pi^2\bar{P}T=-4\pi^2i\int_{\bar{s}-\delta s}^{\bar{s}+\delta s}\frac{ds}{C(s)}P\left[V(\phi)-V_{A}\right] .
\end{equation}
Therefore, in the thin wall limit we obtain the logarithm of the transition probability
\begin{equation}\label{B9GammaRes1}
	\tcboxmath[colback=red!10!white,colframe=red]{
	\pm\Gamma=\pm\frac{2\pi^2i\sqrt{5}}{\hbar}\bigg\{\sqrt{V_{B}}F_{IX}[V_{B},P]\bigg\rvert^{\bar{P}}_{P_{0}}-\sqrt{V_{A}}F_{IX}[V_{A},P]\bigg\rvert^{\bar{P}}_{P_{0}}\bigg\}+\frac{16\pi^2}{\hbar}\bar{P}T ,}
\end{equation}
which is again expressed in terms of only two independent variables, namely $\bar{P}$ and $T$. Furthermore, this expression allows us to explore the UV limit by choosing safely $P_{0}=0$. Therefore we finally obtain
\begin{equation}\label{B9GammaRes}
	\pm2{\rm Re}[\Gamma]=\mp\frac{4\pi^2\sqrt{5}}{\hbar}\bigg\{\Im\left[\sqrt{V_{B}}F_{IX}[V_{B},P]\right]\bigg\rvert^{\bar{P}}_{0}-\Im\left[\sqrt{V_{A}}F_{IX}[V_{A},P]\right]\bigg\rvert^{\bar{P}}_{0}\bigg\}+\frac{32\pi^2}{\hbar}\bar{P}T .
\end{equation}
Seeking an extremum for this expression leads to the equation
\begin{equation}\label{B9CondtionMinimumGamma}
	8T=\mp \sqrt{5}i\left[\sqrt{V_{B}\bar{P}\left(\frac{5V_{B}}{16}\bar{P}-1\right)}-\sqrt{V_{A}\bar{P}\left(\frac{5V_{A}}{16}\bar{P}-1\right)}\right] ,
\end{equation}
which puts constrictions on $\bar{P}$, thus we can consider both terms as independent as usual. We remark that the results obtained for this metric is only valid for positive potential minima. If we want to consider negative values for any of the $V_{A,B}$, the integration constant cannot be set to zero, then we will describe a cosmology without an initial singularity. Furthermore, we obtain that $\lim_{\bar{P}\to0}\Gamma=0$, then the transition probability near the primordial singularity is regular and it acquires its maximum value at $\bar{P}=0$.

If we consider that one of the potential  minima is zero, the general solution for a constant scalar field leads to
\begin{equation}\label{B9SolSystC0}
		C^2(s)=-1024\pi^4\frac{3Q}{P^2\left(Q-4P\right)} ,
\end{equation}\label{B9SolSystC01}
	\begin{equation} 
		\frac{dQ}{ds}=\frac{1}{24\pi^2}(Q-4P) , \hspace{1cm} \frac{dP}{ds}=-\frac{1}{48\pi^2}\frac{P(Q-4P)}{Q} ,
\end{equation}
then, the coordinates are related by
\begin{equation}\label{B9RelationFieldsC00}
	\frac{1}{Q}\frac{dQ}{ds}=-\frac{2}{P}\frac{dP}{ds} .
\end{equation}
Solving this equation and absorbing a multiplicative constant by redefinition of the spatial variables we obtain
\begin{equation}\label{B9RelationFieldsC0}
	Q=\frac{1}{P^2} .
\end{equation}
Thus, we get
\begin{equation}\label{B9ClaasicActionC0}
	-4\pi^2\int\frac{ds}{C(s)}\left[\frac{Q}{P}+PV-4\right]\bigg\rvert_{V=0}=\pm2\sqrt{3}\pi^{2}F^{0}_{IX}(P) ,
\end{equation}
where we defined
\begin{multline}\label{F90}
	F^0_{IX}(x)= \int\frac{\sqrt{4x^3-1}}{x^3}dx \\
	=-\frac{\sqrt{4x^3-1}}{2x^2} +\frac{i2^{1/3}3^{3/4}}{\sqrt{4x^3-1}}\sqrt{(-1)^{5/6}(2^{2/3}x-1)}\sqrt{1+2^{2/3}x+2^{4/3}x^2}
	\\
	\times F\left(\arcsin\left(\frac{\sqrt{-(-1)^{5/6}-i2^{2/3}x}}{3^{1/4}}\right),(-1)^{1/3}\right) .
\end{multline}
Since we cannot choose negative values for the potential, the only possible case is $V_{B}=0$ and $V_{A}>0$. Then, we obtain in the thin wall limit
\begin{equation}\label{B9GammaRes1C01}
	\pm\Gamma=\pm\frac{2\pi^2i}{\hbar}\bigg\{\sqrt{3}F^{0}_{IX}(P)\bigg\rvert^{\bar{P}}_{P_{0}}-\sqrt{5V_{A}}F_{IX}[V_{A},P]\bigg\rvert^{\bar{P}}_{P_{0}}\bigg\}+\frac{16\pi^2}{\hbar}\bar{P}T
	.
\end{equation}
We note that $F^{0}_{IX}(P)$ is divergent in the limit $P\to0$, thus we loose the ability to explore the UV region since $P_{0}\neq0$. Therefore, we finally obtain
\begin{equation}\label{B9GammaRes1C01F}
	\pm2{\rm Re}[\Gamma]=\mp\frac{4\pi^2}{\hbar}\bigg\{\sqrt{3}\Im[F^{0}_{IX}(P)]\bigg\rvert^{\bar{P}}_{P_{0}}-\sqrt{5V_{A}}\Im[F_{IX}[V_{A},P]]\bigg\rvert^{\bar{P}}_{P_{0}}\bigg\}+\frac{32\pi^2}{\hbar}\bar{P}T .
\end{equation}
If we look for an extremum of the above expression we obtain the condition
\begin{equation}\label{B9ConditionExtremumC01}
	8T=\mp i \left[\frac{\sqrt{3(4\bar{P}^3-1)}}{\bar{P}^3}-\sqrt{5V_{A}\bar{P}\left(\frac{5V_{A}}{16}\bar{P}-1\right)}\right]
	.
\end{equation}
Since $T$ is real, the above condition leads to upper bounds for $\bar{P}$ in general, therefore we will not have a correctly defined tension for all values of $\bar{P}$.

Finally, in a more general setup we choose $s$ as in (\ref{ParameterSGeneral}). Then, the classical actions takes the form
\begin{equation}\label{B9ActionClassicalGeneral}
	S_{0}=\pm\frac{\pi}{2}\int_{Q_{0},P_{0},\phi_{B}}^{Q_{m},P_{m},\phi_{A}} \frac{\left(Q+P^2V(\phi)-4P\right)^{3/2}}{\sqrt{2P^2V(\phi)-3Q-\frac{P^4}{2Q}(V'(\phi))^2}} ds ,
\end{equation}
where
\begin{equation}\label{B9dsGeneral}
	ds=\sqrt{-\frac{1}{8}dQdP-\frac{Q}{16P}dP^2+\frac{PQ}{8}d\phi^2} .
\end{equation}
From the Hamiltonian constraint  (\ref{B9HamiltConstra}) we obtain
\begin{equation}\label{B9RelationsKinetic}
	-\frac{\dot{Q}\dot{P}}{8}-\frac{Q}{16P}\dot{P}^2+\frac{PQ}{8}\dot{\phi}^2=2N^2\left[\frac{8Q}{P}\pi^2_{Q}-16\pi_{P}\pi_{Q}+\frac{4}{QP}\pi^2_{\phi}\right]=-\frac{N^2}{4}\left(\frac{Q}{P}+pV(\phi)-4\right)
\end{equation}
Therefore, if the integrand in (\ref{B9ActionClassicalGeneral}) is real, we can have transition probabilities simply with a negative kinetic term in the WDW equation. It may seem from the first equality above that if we remove the scalar field, we could have a negative term contributing to the probability. However, if we remove the scalar field, the action is reduced to
\begin{equation}\label{B9ActionClassicalGeneralNS}
	S_{0}\bigg\rvert_{\phi=0,V=0}=\pm\frac{\pi}{2}\int_{Q_{0},P_{0}}^{Q_{m},P_{m}}(Q-4P)\sqrt{\frac{4P-Q}{3Q}}\sqrt{-\frac{1}{8}dQdP-\frac{Q}{16P}dP^2} 
\end{equation}
On the other hand, we obtain from  (\ref{B9RelationsKinetic}) that
\begin{equation}\label{B9RelationsKineticNS}
	-\frac{\dot{Q}\dot{P}}{8}-\frac{Q}{16P}\dot{P}^2=\frac{N^2}{4}\left(\frac{4P-Q}{P}\right) .
\end{equation}
Thus, from this result we show that it is not possible to obtain an imaginary term from the classical action, thus if there is not a scalar field, there is not a transition.

\section{Transitions in Ho\v{r}ava-Lifshitz gravity}\label{S-TransitionsHL}
Now that we have applied the general method to minisuperspace models derived from General Relativity with different forms of the metric, let us move on to consider the Ho\v{r}ava-Lifshitz gravity theory. As we remarked earlier, we will be able to apply the method since the only requirement is that the Hamiltonian constraint is written in the general form (\ref{GeneralHamiltonian}). Furthermore, we note that in the last section we were able to explore the UV limits (for example in the FLRW metric such limit was $a\to0$) because the expressions found are well behaved functions in the corresponding limit. However, since HL gravity represents a better description than GR in the UV, we expect that we can obtain more information from the transition probabilities, in particular in the UV regime, by employing this theory.

\subsection{WDW equation for Ho\v{r}ava-Lifshitz gravity coupled to a scalar field}\label{S-HL}
Let us begin by obtaining the WDW equation in Ho\v{r}ava-Lifshitz theory considering the coupling to an inhomogeneous scalar field that depends on the time variable as well as the spatial variables. Although in cosmological studies it is more common that the field depends only on the time variable, we will allow the scalar field to depend also on the spatial variables in this case since the anisotropic scaling of both sets of variables is the key ingredient defining the HL theories. This type of dependence has been explored in a cosmological scenario previously by studying the perturbations derived from a scalar field in HL theories in \cite{Mukohyama:2009gg}. However, for completeness and in line with standard cosmological models, we will also study the homogeneous case where the field will only depend on the time coordinate in subsection \ref{S-TT}.

We will consider the most general action in projectable HL gravity without a cosmological constant and without detailed balance given by (\ref{ActionHL}). We will also consider the closed FLRW metric with positive curvature (\ref{MetricFLRWP}). For this metric the action takes the form
\begin{multline}\label{FLRWActionGrav}
	S_{\rm HL}=2\pi^2\int dt N\left[-\frac{3(3\lambda-1)M^2_{p}a}{2N^2}\dot{a}^2+3M^2_{p}a-\frac{6}{a}(3g_{2}+g_{3})\right. \\ \left. -\frac{12}{a^3M^2_{p}}(9g_{4}+3g_{5}+g_{6})\right] ,
\end{multline}
Where the volume of the spatial slice is given by ${\rm Vol}(X)=2\pi^2$ and it appears as a global constant. Furthermore, we couple the inhomogeneous scalar field $\phi(t,x^{i})$ using the action (\ref{ActionMatter}). We note from (\ref{DefFLRW}) that the three-metric in this case is simply the metric of the three-sphere times the scale factor, then we can employ the spherical harmonic functions defined in ${\bf S}^{3}$  \cite{Lindblom:2017maa,Sandberg:1978} in order to make expansions of the scalar functions involved. Such harmonic functions are defined  as eigenfunctions of the Laplacian in the form
\begin{equation}\label{SpHarDeg}
	\Delta Y_{nlm}(x^{i})=-\frac{n(n+2)}{a^2}Y_{nlm}(x^{i}),
\end{equation}
where $n$ is an integer. They obey the orthonormality condition
\begin{equation}\label{OrtoCondiO}
	\frac{1}{a^3}\int \sqrt{h}Y_{nlm}(x^{i})Y^{*}_{n'l'm'}(x^{i})d^3x=\delta_{nn'}\delta_{ll'}\delta_{mm'} .
\end{equation}
These harmonic functions form a complete basis in this space, thus we can expand any scalar function defined on the three-sphere as
\begin{equation}\label{ExpansionScalarFunc}
	f(x^{i})=\sum_{n=0}^{\infty}\sum_{l=0}^{n}\sum_{m=-l}^{l}\alpha_{nlm}Y_{nlm}(x^{i})=\sum_{\{n,l,m\}}\alpha_{nlm}Y_{nlm}(x^{i}) .
\end{equation}
Therefore the scalar field can be expanded as
\begin{equation}\label{ExpansionScalarField}
	\phi(t,x^{i})=\sum_{\{n,l,m\}}\phi_{nlm}(t)Y_{nlm}(x^{i}) ,
\end{equation}
where the fields $\phi_{nlm}(t)$ are real functions that depend only on the time variable. Furthermore, we can also make an expansion of the scalar field potential as
\begin{equation}\label{ExpansiconScalarPotential}
	V(\phi)=\sum_{\{n,l,m\}}V_{nlm}(t)Y_{nlm}(x^{i}) ,
\end{equation}
where the functions $V_{nlm}(t)$ will depend on all the functions $\phi_{nlm}(t)$ in general. Substituting (\ref{ExpansionScalarField}) and (\ref{ExpansiconScalarPotential}) back into the action for the scalar field (\ref{ActionMatter}) we obtain 
\begin{multline}\label{ActionMatterF}
	S_{m}=\sum_{\{n,l,m\}}\frac{1}{2}\int dt \left\{ \frac{3\lambda-1}{2N}a^3\dot{\phi}_{nlm}^2-Na\left[c_{1}\beta_{n}+\frac{c_{2}\beta^2_{n}}{a^2}+\frac{c_{3}\beta^3_{n}}{a^4}\right]\phi_{nlm}^2\right. \\ \left. -Na^3\gamma_{nlm}V_{nlm} \right\} ,
\end{multline}
where $\beta_{n}=n(n+2)$ and
\begin{equation}\label{DefGammasAux}
	\gamma_{nlm}=\int_{r=0}^{\pi}\int_{\psi=0}^{2\pi}\int_{\theta=0}^{\pi}\sin^2r\sin\theta Y_{nlm}(r,\theta,\psi)drd\psi d\theta ,
\end{equation}
are constants. Finally, considering both actions (\ref{FLRWActionGrav}) and (\ref{ActionMatterF}), the full Lagrangian describing HL gravity coupled to an inhomogeneous scalar field is given by
\begin{multline}\label{LagrangianFull}
	\mathcal{L}=2\pi^2N\left[-\frac{3M^2_{p}\dot{a}^2a}{2N^2}(3\lambda-1)+3M^2_{p}a-\frac{6}{a}(3g_{2}+g_{3})-\frac{12}{a^3M^2_{p}}(9g_{4}+3g_{5}+g_{6})\right] \\ +\sum_{\{n,l,m\}}\left\{ \frac{3\lambda-1}{4N}a^3\dot{\phi}_{nlm}^2-\frac{Na}{2}\left[c_{1}\beta_{n}+\frac{c_{2}\beta^2_{n}}{a^2}+\frac{c_{3}\beta^3_{n}}{a^4}\right]\phi_{nlm}^2-\frac{Na^3}{2}\gamma_{nlm}V_{nlm} \right\} .
\end{multline}
The superspace in this case is defined by the fields  $\left\{a,\phi_{nlm}\right\}$, which are in principle infinite. Their corresponding canonical momenta are
\begin{equation}\label{CanonicalMomenta}
	\pi_{a}=-\frac{6\pi^2M^2_{p}(3\lambda-1)}{N}a\dot{a} , \hspace{1cm} \pi_{\phi_{nlm}}=\frac{3\lambda-1}{2N}a^3\dot{\phi}_{nlm} ,
\end{equation}
and, as usual $\pi_{N}=0$. Therefore, the most general form of the  Hamiltonian constraint is
\begin{multline}\label{HamiltonianConstraint}
	H=N\left\{-\frac{\pi^2_{a}}{12\pi^2M^2_{p}(3\lambda-1)a}+\sum_{\{n,l,m\}}\frac{\pi^2_{\phi_{nlm}}}{(3\lambda-1)a^3}\right. \\ \left. +2\pi^2\left[-3M^2_{p}a+\frac{6}{a}(3g_{2}+g_{3})+\frac{12}{M^2_{p}a^3}(9g_{4}+3g_{5}+g_{6})\right]\right.\\\left.+\frac{1}{2}\sum_{\{n,l,m\}}\left[\left(c_{1}\beta_{n}+\frac{c_{2}}{a^2}\beta^2_{n}+\frac{c_{3}}{a^4}\beta^3_{n}\right)a\phi^2_{nlm}+a^3\gamma_{nlm}V_{nlm}\right]\right\} \simeq 0 .
\end{multline}

\subsection{Transitions for an inhomogeneous scalar field}\label{S-Trans}
Now that we have obtained the most general form of the Hamiltonian constraint with an inhomogeneous scalar field, let us study the transition probabilities. We note that the form of the Hamiltonian constraint (\ref{HamiltonianConstraint}) has the general form  (\ref{GeneralHamiltonian}) with the coordinates on  superspace $\left\{a,\phi_{nlm}\right\}$, which in this case will be in general infinite dimensional. Furthermore, we note the metric on superspace is given by
\begin{equation}
	(G^{MN})={\rm diag}\left(-\frac{1}{6\pi^2M^2_{p}(3\lambda-1)a},\frac{2}{(3\lambda-1)a^3}\mathbf{1}_{nlm}\right) ,
\end{equation}
where $\mathbf{1}_{nlm}$ denotes a vector with the same length to all the possible allowed values that the set $\{n,l,m\}$ can have and with $1$ in all its entries. In this case we also obtain
\begin{multline}
	f(a,\phi_{nlm},V_{nlm})=2\pi^2\left[-3M^2_{p}a+\frac{6}{a}(3g_{2}+g_{3})+\frac{12}{M^2_{p}a^3}(9g_{4}+3g_{5}+g_{6})\right]\\+\frac{1}{2}\sum_{\{n,l,m\}}\left[\left(c_{1}\beta_{n}+\frac{c_{2}}{a^2}\beta^2_{n}+\frac{c_{3}}{a^4}\beta^3_{n}\right)a\phi^2_{nlm}+a^3\gamma_{nlm}V_{nlm}\right] ,
\end{multline}
and $\{W^M\}=0$. Therefore, even if the Hamiltonian constraint has a more complicated form, the general procedure to study the transition probabilities described in section \ref{S-TransitionsGeneralMethod} is applicable to the HL gravity theory as well. In order to correctly apply the method we make some assumptions. We consider that all fields $V_{nlm}$ appearing in the expansion of the potential (\ref{ExpansiconScalarPotential}) are well defined functions with the same minima, that is they all have a false minimum at $\phi^{A}_{nlm}$ and a true minimum at $\phi^{B}_{nlm}$. Therefore, we assume that the two minima of the general homogeneous scalar field $\phi(t,x^{i})$ arrives only from its time dependence. With these assumptions the method proceed in the same form as in all the cases in GR.

Let us choose $s$ as usual in the form of (\ref{ChooseS}). However, since we employed the expansion (\ref{ExpansionScalarField}), and the spherical harmonics are an orthonormal set, (\ref{ChooseS}).
implies in this case
\begin{equation}\label{ChooseSField}
	\phi_{nlm}(s) \approx
	\begin{cases}
		\phi_{nlm}^{B} , & 0<s<\bar{s}-\delta s,\\
		\phi_{nlm}^{A} , & \bar{s}+\delta s<s<s_{M} ,
	\end{cases}
\end{equation}
in an analogous manner we obtain for the potentials
\begin{equation}\label{ChooseSPotential}
	V_{nlm}(s) \approx
	\begin{cases}
		V_{nlm}^{B} , & 0<s<\bar{s}-\delta s,\\
		V_{nlm}^{A} , & \bar{s}+\delta s<s<s_{M} .
	\end{cases}
\end{equation}
Therefore, the classical actions in the form of (\ref{ClassicalAction}) will be given in general for this case
\begin{multline}\label{ClassicalAction1}
	S_{0}\left(a_{0},\phi^{B}_{nlm};a_{m},\phi^{A}_{nlm}\right)=-4\pi^2\left[\int_{0}^{\bar{s}-\delta s}\frac{ds}{C(s)}f\big\rvert_{\phi_{nlm}=\phi^{B}_{nlm}}+\int_{\bar{s}-\delta s}^{\bar{s}+\delta s}\frac{ds}{C(s)}f\right. \\ \left. +\int_{\bar{s}+\delta s}^{s_{M}}\frac{ds}{C(s)}f\big\rvert_{\phi_{nlm}=\phi^{A}_{nlm}}\right] ,
\end{multline}
and
\begin{equation}\label{ClassicalAction2}
	S_{0}\left(a_{0},\phi^{A}_{nlm};a_{m},\phi^{A}_{nlm}\right)=-4\pi^2\int_{0}^{s_{M}}\frac{ds}{C(s)}f\big\rvert_{\phi_{nlm}=\phi^{A}_{nlm}}	.
\end{equation}
Therefore, the logarithm of the transition probability (\ref{DefGamma}) takes the form
\begin{multline}\label{GammaF1}
	\pm\Gamma=\frac{i}{\hbar}\left[-4\pi^2\int_{0}^{\bar{s}-\delta s}\frac{ds}{C(s)}f\big\rvert_{\phi_{nlm}=\phi^{B}_{nlm}}+4\pi^2\int_{0}^{\bar{s}-\delta s}\frac{ds}{C(s)}f\big\rvert_{\phi_{nlm}=\phi^{A}_{nlm}}\right. \\
	\left.-4\pi^2\int_{\bar{s}-\delta s}^{\bar{s}+\delta s}\frac{ds}{C(s)}\left\{\frac{1}{2}\sum_{\{n,l,m\}}\left[a\left(c_{1}\beta_{n}+\frac{c_{2}}{a^2}\beta^2_{n}+\frac{c_{3}}{a^4}\beta^3_{n}\right)\left(\phi^2_{nlm}-(\phi^{A}_{nlm})^2\right)\right.\right.\right. \\
	\left. \left.\left. +a^3\gamma_{nlm}(V_{nlm}-V^{A}_{nlm})\right]\right\}\right] .
\end{multline}

Let us discuss firstly the terms that arrive in the region where the scalar field can vary. We note that the last term in (\ref{GammaF1}) can be written in the following form
\begin{equation}\label{NewPotentials0}
	-4\pi^2i\int_{\bar{s}-\delta s}^{\bar{s}+\delta s}\frac{ds}{C(s)}\left[\frac{1}{2}\sum_{\{n,l,m\}}a^3\gamma_{n,l,m}(V_{n,l,m}-V^{A}_{nlm})\right]=-4\pi^2i\int_{\bar{s}-\delta s}^{\bar{s}+\delta s}\frac{ds}{C(s)}a^3\left[V_{0}-V_{0}^{A}\right] ,
\end{equation}
with a potential function defined by
\begin{equation}\label{DefPot0}
	V_{0}=\frac{1}{2}\sum_{\{n,l,m\}}\gamma_{nlm}V_{nlm} .
\end{equation}
We note that (\ref{NewPotentials0}) is of the same general form as the term considered in (\ref{FLRWTension}) in GR. Therefore we can also propose a tension term for this integral in the form
\begin{equation}\label{DefTension0}
	2\pi^2\bar{a}^3T_{0}=-4\pi^2i\int_{\bar{s}-\delta s}^{\bar{s}+\delta s}\frac{ds}{C(s)}a^3\left[V_{0}-V_{0}^{A}\right] .
\end{equation}
Moreover, we note that the term that contains $c_{1}$ in (\ref{GammaF1}) can be written as 
\begin{multline}\label{NewPotential1}
	-4\pi^2i\int_{\bar{s}-\delta s}^{\bar{s}+\delta s}\frac{ds}{C(s)}\left[\frac{1}{2}\sum_{\{n,l,m\}}ac_{1}\beta_{n}\left(\phi^2_{nlm}-(\phi^{A}_{nlm})^2\right)\right]\\=-4\pi^2i\int_{\bar{s}-\delta s}^{\bar{s}+\delta s}\frac{ds}{C(s)}c_{1}a\left[V_{1}-V_{1}^{A}\right] ,
\end{multline}
with
\begin{equation}\label{DefPotential1}
	V_{1}=\frac{1}{2}\sum_{\{n,l,m\}}\beta_{n}\phi_{nlm}^2 .
\end{equation}
We note that this function does not have a minimum in the scalar fields where $V_{0}$ does. However, it is a well defined function of the scalar fields, therefore we can interpret $V_{1}$ as a new effective potential which has the form of a mass term for the fields. Therefore, applying the same logic used so far regarding these types of integrals, we can define a new contribution for the tension term in the form
\begin{equation}\label{DefTension1}
	2\pi^2c_{1}\bar{a}T_{1}=-4\pi^2i\int_{\bar{s}-\delta s}^{\bar{s}+\delta s}\frac{ds}{C(s)}c_{1}a\left[V_{1}-V_{1}^{A}\right] .
\end{equation}
In a similar manner, for the two remaining terms we define
\begin{equation}\label{DefPotentials23}
	V_{2}=\frac{1}{2}\sum_{\{n,l,m\}}\beta_{n}^2\phi_{nlm}^2 , \hspace{1cm}
	V_{3}=\frac{1}{2}\sum_{\{n,l,m\}}\beta_{n}^3\phi_{nlm}^2 ,
\end{equation}
which lead to two new contributions to the tension terms as
\begin{equation}\label{DefTension2}
	2\pi^2\frac{c_{2}}{\bar{a}}T_{2}=-4\pi^2i\int_{\bar{s}-\delta s}^{\bar{s}+\delta s}\frac{ds}{C(s)}\frac{c_{2}}{a}\left[V_{2}-V_{2}^{A}\right]
	,
\end{equation}
\begin{equation}\label{DefTension3}
	2\pi^2\frac{c_{3}}{\bar{a}^{3}}T_{3}=-4\pi^2i\int_{\bar{s}-\delta s}^{\bar{s}+\delta s}\frac{ds}{C(s)}\frac{c_{3}}{a^3}\left[V_{3}-V_{3}^{A}\right] .
\end{equation}

On the other hand, as it was done in GR, we can employ the general solutions (\ref{CsGeneral}) and (\ref{DerivSolGen}) in order to compute the first two integrals in (\ref{GammaF1}), since they are defined in the region  where all scalar fields are constants, then making a change of coordinates as $ds=\left(\frac{da}{ds}\right)^{-1}da$ we obtain
\begin{multline}\label{IntegralConst}
	-4\pi^2\int_{0}^{\bar{s}-\delta s}\frac{ds}{C(s)}f\big\rvert_{\phi_{nlm}=\phi^{A,B}_{nlm}} \\ =\pm 4\pi^3M_{p}\sqrt{3(3\lambda-1)}\int_{a_{0}}^{\bar{a}-\delta a}\sqrt{-\alpha^{A,B}_{1}a^2+\alpha^{A,B}_{2}+\frac{\alpha^{A,B}_{3}}{a^2}+V^{A,B}_{0}a^4}da ,
\end{multline}
where
\begin{equation}\label{DefConstIntegralO}
		\alpha^{A.B}_{1}=6\pi^2M^2_{p}-c_{1}V^{A,B}_{1}  ,
\end{equation}
\begin{equation}
	 \alpha^{A,B}_{2}=12\pi^2(3g_{2}+g_{3})+c_{2}V^{A,B}_{2} ,
\end{equation}
\begin{equation}
	\alpha^{A,B}_{3}=\frac{24\pi^2}{M^2_{p}}(9g_{4}+3g_{5}+g_{6})+c_{3}V^{A,B}_{3} .
\end{equation}
Therefore, considering all the definitions (\ref{DefTension0}), (\ref{DefTension1}), (\ref{DefTension2}), (\ref{DefTension3}) and taking into account (\ref{IntegralConst}), we obtain from (\ref{GammaF1}) that the logarithm of the transition probability in this scenario takes the form
\begin{multline}\label{GammaF2}
	\pm\Gamma=\pm\frac{ 4\pi^3M_{p}\sqrt{3(3\lambda-1)}}{\hbar}\left[\int_{a_{0}}^{\bar{a}-\delta a}F\left(\alpha^{B}_{1},\alpha^{B}_{2},\alpha^{B}_{3},V^{B}_{0},a\right)da \right. \\ \left. -\int_{a_{0}}^{\bar{a}-\delta a}F\left(\alpha^{A}_{1},\alpha^{A}_{2},\alpha^{A}_{3},V^{A}_{0},a\right)da \right]+\frac{2\pi^2}{\hbar}\left[\bar{a}^3T_{0}+c_{1}\bar{a}T_{1}+\frac{c_{2}}{\bar{a}}T_{2}+\frac{c_{3}}{\bar{a}^3}T_{3}\right] .
\end{multline}
where we have defined the function
\begin{equation}\label{FDef}
	F(a,b,c,e,x)=\sqrt{a x^2-b-\frac{c}{x^2}-e x^4} .
\end{equation}

Let us discuss a particular issue regarding the choice of units. As we described in subsection \ref{SS-HLInfraredLimit}, the IR limit of HL gravity for an FLRW metric is achieved in the limit $\lambda\to1$ and $a\gg1$, it should lead to GR with an extra degree of freedom, apart from instability issues. However, let us note that the kinetic term for the scale factor in (\ref{LagrangianFull}) is
\begin{equation}
	-2\pi^2\left[\frac{3M^2_{p}a\dot{a}^2}{2N}(3\lambda-1)\right] .
\end{equation}
On the other hand, in the GR case (\ref{FLRWLagr}) the kinetic term was written as
\begin{equation}
	-\frac{3a\dot{a}^2}{N} ,
\end{equation}
because the spatial volume given by $2\pi^2$, is a global multiplicative factor to the full Lagrangian in that case, and thus it was ignored. Therefore since we want to compare directly the result obtained in this case with the one obtained in GR,  we need to get the same kinetic term in both cases in the limit $\lambda\to1$, thus we choose units such that $2\pi^2M^2_{p}=1$ in this case.

Then, the logarithm of the transition probability is finally written in the form
\begin{equation}\label{GammaFF}
	\tcboxmath[colback=red!10!white,colframe=red]{
	\begin{split}
	\pm\Gamma&=\pm\frac{ 2\pi^2\sqrt{6(3\lambda-1)}}{\hbar}\left[\int_{a_{0}}^{\bar{a}-\delta a}F\left(\alpha^{B}_{1},\alpha^{B}_{2},\alpha^{B}_{3},V^{B}_{0},a\right)da  \right. \\ \left.-\right.&\left. \int_{a_{0}}^{\bar{a}-\delta a}F\left(\alpha^{A}_{1},\alpha^{A}_{2},\alpha^{A}_{3},V^{A}_{0},a\right)da
	\right]+\frac{2\pi^2}{\hbar}\left[\bar{a}^3T_{0}+c_{1}\bar{a}T_{1}+\frac{c_{2}}{\bar{a}}T_{2}+\frac{c_{3}}{\bar{a}^3}T_{3}\right]
 	\end{split}
 	}
\end{equation}
with
\begin{equation}\label{DefConstIntegral}
		\alpha^{A.B}_{1}=3-c_{1}V^{A,B}_{1}  ,
\end{equation}
\begin{equation}
	\alpha^{A,B}_{2}=12\pi^2(3g_{2}+g_{3})+c_{2}V^{A,B}_{2} , 
\end{equation}
\begin{equation}
	 \alpha^{A,B}_{3}=48\pi^3(9g_{4}+3g_{5}+g_{6})+c_{3}V^{A,B}_{3} .
\end{equation}
In contrast to all the results obtained for General Relativity in section \ref{S-TransitionsGR}, we see that this transition probability for HL gravity is described by five parameters. By looking for an extremum with respect to $\bar{a}$ we can at most reduce the number of parameters by one. We also note that the integrals involved cannot be done explicitly for any values of the constants $\alpha_{i}$. Nonetheless, the result is valid for any value of the potential minima and it is not necessary to consider the different modes contributing to the expansion in (\ref{ExpansionScalarField}) separately.

As we stated in subsection \ref{S-FLRW} the choice of $s$ as in (\ref{ChooseS}) is useful to obtain exact solutions for the transition probabilities that leads to consistent results with the Euclidean approach. However, as we saw in all the cases considered in the last section, we can also choose $s$  as the distance in field space. This choice allows us to show that we can have classical transitions just because the metric in superspace for the WDW equation considered here coming from the Hamiltonian constraint (\ref{HamiltonianConstraint}) is non-positive definite as well. Therefore we obtain consistency with all the results found in section \ref{S-TransitionsGR} for GR.

Now that we have a concrete expression for the transition probability for HL gravity with an inhomogeneous scalar field in general, let us consider its infrared and ultraviolet limiting behaviour. With the IR limit we can directly compare to the result found in GR. On the other hand, the UV limit will allow us to highlight the contributions for high energies that are expected to change the result since it is in this region where HL gravity improves over GR.

Implementing the IR limit in the form $\lambda\to1$ and $a\gg1$, we obtain from (\ref{GammaFF}) that
\begin{multline}\label{GammaIR}
	\pm\Gamma_{IR}=\mp \frac{4\pi^2}{\hbar}\sqrt{\frac{1}{3}}\left[\frac{(\alpha^{B}_{1})^{3/2}}{V^{B}_{0}}\left(1-\frac{V^{B}_{0}}{\alpha^{B}_{1}}a^2\right)^{3/2}\bigg\rvert_{a_{0}}^{\bar{a}-\delta a}-\frac{(\alpha^{A}_{1})^{3/2}}{V^{A}_{0}}\left(1-\frac{V^{A}_{0}}{\alpha^{A}_{1}}a^2\right)^{3/2}\bigg\rvert_{a_{0}}^{\bar{a}-\delta a}\right]\\ +\frac{2\pi^2}{\hbar}\left[\bar{a}^3T_{0}+c_{1}\bar{a}T_{1}\right] .
\end{multline}
Thus, the infrared results provides an expression similar to the GR result (\ref{FLRWGamma}), we encounter  an extra degree of freedom arriving from the $c_{1}$ term in the action for the scalar field (\ref{ActionMatterF}) as we anticipated. Let us choose for the moment $c_{1}=0$, then the last result simplifies to 
\begin{equation}\label{GannaIRF}
	\pm\Gamma_{IR}=\mp \frac{12\pi^2}{\hbar}\left[\frac{1}{V^{B}_{0}}\left(1-\frac{V^{B}_{0}}{3}a^2\right)^{3/2}\bigg\rvert_{a_{0}}^{\bar{a}-\delta a}-\frac{1}{V^{A}_{0}}\left(1-\frac{V^{A}_{0}}{3}a^2\right)^{3/2}\bigg\rvert_{a_{0}}^{\bar{a}-\delta a}\right] +\frac{2\pi^2}{\hbar}\bar{a}^3T_{0},
\end{equation}
which is the same result as the one obtained for GR in (\ref{FLRWGamma}). However, for consistency in the approximation of the integral here we must impose that $a_{0}\gg1$, therefore it cannot be chosen to be zero. Thus, the difference between this result and the GR one are only constants. We also remark that the potential appearing in this expression is not the original potential, it is an effective potential appearing after integration of the harmonic functions (\ref{DefPot0}). Implementing the thin wall limit $\delta a\to0$ and looking for an extremum with respect to $\bar{a}$ we obtain
\begin{equation}\label{FLRWTensionAHL}
	T_{0}=\pm 2\left(\sqrt{\frac{1}{\bar{a}^2}-\frac{V^{A}_{0}}{3}}-\sqrt{\frac{1}{\bar{a}^2}-\frac{V^{B}_{0}}{3}}\right) .
\end{equation}
Then, choosing the plus sign in the right of (\ref{GannaIRF})  side we finally obtain in the infrared
\begin{multline}\label{GammaIRCGRF} 
	\pm\Gamma_{IR}= \frac{12\pi^2}{\hbar}\left\{\frac{1}{V^{B}_{0}}\left[\left(1-\frac{V^{B}_{0}}{3}\bar{a}^2\right)^{3/2}-\left(1-\frac{V^{B}_{0}}{3}a_{0}^2\right)^{3/2}\right]\right. \\ \left. -\frac{1}{V^{A}_{0}}\left[\left(1-\frac{V^{A}_{0}}{3}\bar{a}^2\right)^{3/2}-\left(1-\frac{V^{A}_{0}}{3}a_{0}^2\right)^{3/2}\right]-\frac{\bar{a}^2}{3}\left(\sqrt{1-\frac{V^{A}_{0}}{3}\bar{a}^2}-\sqrt{1-\frac{V^{B}_{0}}{3}\bar{a}^2}\right)\right\} .
\end{multline}
Therefore, in the IR limit the transition probability is finally described in terms of just one parameter (since $a_{0}$ is just a constant) as in General Relativity.

We can see easily see from the GR result (\ref{FLRWGammaCDL}) that choosing the plus sign on the right hand side of (\ref{FLRWGammaF}) and the plus sign on the left, as was done in the last section, implies $T>0$ since $V_{A}>V_{B}$. The same is true regarding $T_{0}$. Thus in the following we will always choose positive values for $T_{0}$, not only in the infrared limit.

Let us study now the ultraviolet limit, this is implemented through $a\ll1$. In this case, the logarithm of the transition probability (\ref{GammaFF}) takes the form
\begin{equation}\label{GammaUV}
	\pm\Gamma_{UV}=\pm\frac{ 2\pi^2\sqrt{6(3\lambda-1)}}{\hbar}\left[\left(\sqrt{-\alpha^{B}_{3}}-\sqrt{-\alpha^{A}_{3}}\right)\ln a \bigg\rvert_{a_{0}}^{\bar{a}-\delta a}
	\right]+\frac{2\pi^2}{\hbar}\left[\frac{c_{2}}{\bar{a}}T_{2}+\frac{c_{3}}{\bar{a}^3}T_{3}\right].
\end{equation}
Let us remark that since all the $g_{n}$ constants are positive, the $V_{3}$ function defined in (\ref{DefPotentials23}) is also positive definite for both minima, then the first term is purely imaginary. Thus, it does not contribute to the transition probability and we can ignore it. Then, we finally obtain in this limit
\begin{equation}\label{GammaUF}
	\pm\Gamma_{UV}=\frac{2\pi^2}{\hbar}\left[\frac{c_{2}}{\bar{a}}T_{2}+\frac{c_{3}}{\bar{a}^3}T_{3}\right] .
\end{equation}
We note that this expression does not have an extremum with respect to $\bar{a}$. Thus, in this case the probability is necessarily described by three independent parameters. For consistency with the GR result we choose the plus sign in the left hand side of the latter equation. Thus in order to have a well defined probability we need the overall sign of the right hand side to be positive. Then, we are going to consider that both tension terms $T_{2}$ and $T_{3}$ take always positive values.

We remark from the GR result (\ref{FLRWGammaCDL}), or (\ref{FLRWGammaCDLF}), that we obtained $P(A\to B)\to1$ in the UV limit $\bar{a}\to0$. However for the HL theory result (\ref{GammaUF}) we find that  $P(A\to B)\to0$ when $\bar{a}\to0$, then the probability increases with $\bar{a}$. Therefore the result found in this case exhibits the opposite behaviour than in GR.

Now that we have presented the study in the two limits of interest analytically, let us proceed to compare the full result for the transition probability for all values of $\bar{a}$ (\ref{GammaFF}) and compare it to the GR result. Employing the correct choice of signs mentioned earlier we obtain in thin wall limit
\begin{multline}\label{GammaFFF}
	\Gamma=-\frac{ 2\pi^2\sqrt{6(3\lambda-1)}}{\hbar}\left[\int_{a_{0}}^{\bar{a}}F\left(\alpha^{B}_{1},\alpha^{B}_{2},\alpha^{B}_{3},V^{B}_{0},a\right)da \right. \\ \left. -\int_{a_{0}}^{\bar{a}}F\left(\alpha^{A}_{1},\alpha^{A}_{2},\alpha^{A}_{3},V^{A}_{0},a\right)da \right]+\frac{2\pi^2}{\hbar}\left[\bar{a}^3T_{0}+c_{1}\bar{a}T_{1}+\frac{c_{2}}{\bar{a}}T_{2}+\frac{c_{3}}{\bar{a}^3}T_{3}\right] .
\end{multline}
If we want to look for an extremum of (\ref{GammaFFF}) we will obtain an expression involving the tension terms and the functions $F(\alpha^{A,B}_{1},\alpha^{A,B}_{2},\alpha^{A,B}_{3},V^{A,B},\bar{a})$. However, these functions are defined on terms of square roots, therefore in order to have real values for the tensions terms we will need that
\begin{equation}\label{CondF}
	\alpha^{A,B}_{1}\bar{a}^2-\alpha^{A,B}_{2}-\frac{\alpha^{A,B}_{3}}{\bar{a}^2}-V^{A,B}_{0}\bar{a}^4\geq0 .
\end{equation}
In the best scenario, this expression implies only a lower bound on $\bar{a}$ coming from the $\alpha_{3}$ term. In the other cases, it can imply a lower and an upper bound for $\bar{a}$, making the tension well defined only on a specific interval of $\bar{a}$. However, it could also be the case that such constraint cannot be satisfied for any value of $\bar{a}$, in that case (\ref{GammaFFF}) would not have an extremum. In any case, we see that the extremizing procedure depends completely on the many parameters of the theory and it always forbids exploration into the UV region.  Thus, we will avoid all these difficulties by considering all the parameters involved as independent. Then we will compare the GR result described by its two parameters (\ref{FLRWGammaCDL}) choosing the correct signs used in that section, with the HL result (\ref{GammaFFF}). We are going to evaluate numerically the integrals of the HL result, furthermore, by analyzing the limiting procedures we found that we must choose  $T$, $T_{0}$, $T_{2}$ and $T_{3}$ to be positive. In order to obtain a well defined probability, we will also choose positive values for $T_{1}$. Thus, we will consider positive values for all the tension terms involved.

\begin{figure}[h]
	\centering
	\includegraphics[width=0.7\textwidth]{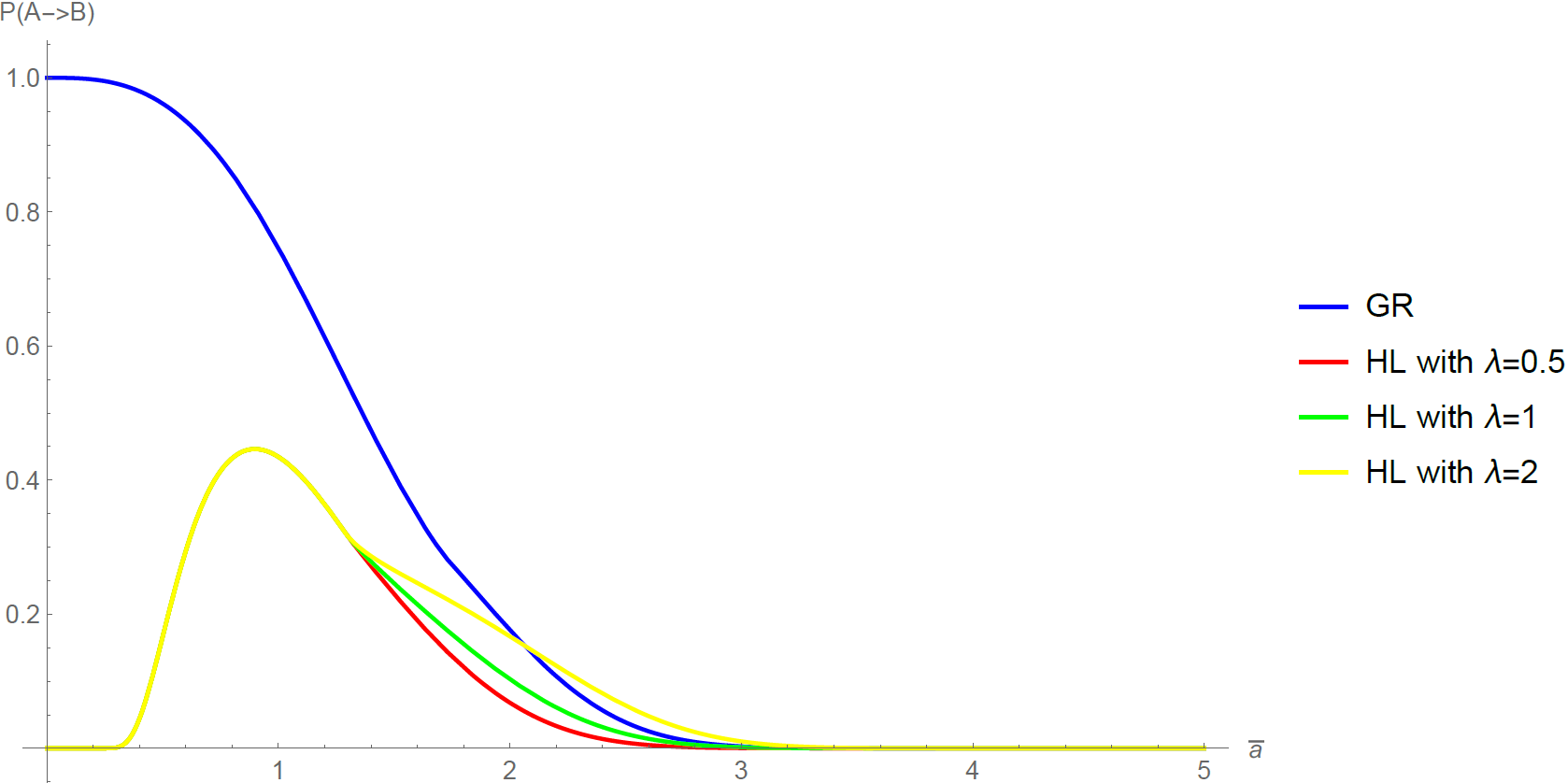}
	\caption{Transition probability for HL gravity in units where $\frac{24\pi^2}{\hbar}=1$, with $V_{A}=V^{A}_{0}=1$, $V_{B}=V^{B}_{0}=0.1$, $\alpha^{A}_{1}=\alpha^{A}_{2}=\alpha^{A}_{3}=5$,  $\alpha^{B}_{1}=\alpha^{B}_{2}=\alpha^{B}_{3}=4$, $T=T_{0}=2$, $c_{1}T_{1}=c_{2}T_{2}=c_{3}T_{3}=1$, for GR (blue line) and HL with an inhomogeneous scalar field depending on all spacetime variables with $\lambda=0.5$ (red line), $\lambda=1$ (green line) and $\lambda=2$ (yellow line). For HL we choose $a_{0}=0.000001$ but the same form is expected for $a_{0}=0$.}
	\label{PlotGeneral}
\end{figure}

In this way, in Figure \ref{PlotGeneral} we show a plot of the transition probabilities from the two theories. We employ units where $\frac{24\pi^2}{\hbar}=1$. For the GR result (blue line) we choose
$V_{A}=1$, $V_{B}=0.1$ and $T=2$. Since the first term in (\ref{FLRWGammaCDL}) is negative for this choice, we need to choose big enough values for $T$ to obtain a well defined probability. For the HL result we choose $V^{A}_{0}=1$, $V^{B}_{0}=0.1$, $\alpha^{A}_{1}=\alpha^{A}_{2}=\alpha^{A}_{3}=5$, $\alpha^{B}_{1}=\alpha^{B}_{2}=\alpha^{B}_{3}=4$, $T_{0}=2$ and $c_{1}T_{1}=c_{2}T_{2}=c_{3}T_{3}=1$, we show plots of the probability for three different values of $\lambda$ to see how the variation of this parameter affects the behaviour. In order to evaluate numerically the integral we choose $a_{0}=0.000001$, but this is only a computational limitation, since as we know from the analytical UV analysis presented before, there is not an issue with the choice $a_{0}=0$, then the general form will be unaltered. In this case, we also obtain that the first term in (\ref{GammaFFF}) is negative and increases with $\lambda$, therefore, we also need to make sure that the tension terms  are big enough to have a well defined probability. The behaviour encountered with the analytical treatment of the limits is portrayed in the figure. In the IR region the probability falls in the same manner as the GR result. In this limit we note that the first term contributes to the probability only when $\bar{a}$ is big enough, the behaviour derived from this term implies that $\lambda$ only affects the curve in the IR region and the probability increases with $\lambda$, because this term have the opposite sign that the tension terms. On the other hand, in the UV region the parameter $\lambda$ has no impact at all, all the contributions arrive by the tension terms. As explained before, we find that the probability goes to $0$ in the limit $\bar{a}\to0$, furthermore in the UV region the probability increases with $\bar{a}$. We note that this behaviour comes from the extra terms in the action for the scalar field (\ref{ActionMatter}), that is from the  $c_{2}$ and $c_{3}$ terms. The probability grows only to a maximum point, then for some small value of $\bar{a}$ the behaviour changes, then the probability decreases with $\bar{a}$ and leads to the IR behaviour. The  general behaviour just described will be valid regardless of the values that we assign to the parameters. We only have to propose appropriate values such that the tension terms dominates, and we can have a well defined probability. However, specific points as the maximum height or the point in which both plots match is completely determined by the choice of parameters, thus we cannot say something about them in general.

Let us remark that the relevant terms that led us in HL gravity to an opposite behaviour than in GR  appear in the action for the scalar field when the field depends on the spatial variables. Thus, for a homogeneous field we expect that these results will be modified. In the next subsection we are going to study such scenario.

\subsection*{Physical interpretation}
Before we end up this subsection let us discuss some phenomenological aspects regarding our results. As we discussed earlier, in the Euclidean approach to the vacuum transitions formulated in \cite{ColemanDeLuccia}, the process is described by the nucleation of true vacuum bubbles in a background defined by the false vacuum, then they describe the corresponding expansion of the bubble. This could lead to phenomenological predictions regarding this kind of phase transitions occurring at some point in the evolution of the universe. However employing the Hamiltonian formalism to study the transitions between two minima of a scalar field potential is limited, since we have to use a minisuperspace approximation as explained in \cite{Cespedes:2020xpn}. Within this formalism we do not have a notion of bubble nucleation, we only compare two three-metric configurations and interpret its ratio as a transition probability. Therefore it is believed that this approach  is describing a generalization of the tunnelling from nothing scenario. That is, the transition probabilities obtained can be interpreted as probability distributions of creating universes from a tunnelling event between two minima of the scalar potential. We will consider this interpretation seriously. Then, the scale factor $\bar{a}$ appearing in the transition probabilities with FLRW metrics would correspond to the value that the scale factor of the created universe would have at the time of creation (its corresponding 'size'). Then, the results of last section imply that in General Relativity we always expect that the creation of the universe starts with a singularity since all the probability distributions exhibit their maximum value at vanishing scale factor. On the other hand, we have obtained for HL gravity with an inhomogeneous scalar field that the universe would be created with an small scalar factor, but different from zero.  In this way  we would avoid the singularity at the beginning of the universe. This of course would have relevant phenomenological consequences in the physics of the early universe and its corresponding evolution. Therefore, even if we have used semiclassical approximations to study the transition probabilities we have been able to explore concrete examples that deals with important subjects such as the initial singularity, in fact we have obtained a toy model where we can avoid such singularity.

\subsection{Transitions for a homogeneous scalar field}\label{S-TT} 
As we stated earlier,  in cosmological scenarios it is more common to study a homogeneous scalar field that depends only on the time variable as it was done for example in \cite{Kiritsis:2009sh,Tavakoli:2021kyc,Tawfik:2016dvd}. Let us study then the vacuum transition probabilities for this type of scalar field as well. 

For a homogeneous field, the scalar field action (\ref{ActionMatter}) simplifies to
\begin{equation}\label{AcionMatterTF}
	S_{m}=2\pi^2\int dt a^3(t)\left[\frac{3\lambda-1}{4N^2}\dot{\phi}^2-NV(\phi)\right] ,
\end{equation}
where we have redefined the scalar field of (\ref{DefinitionF}) in the form $\frac{V}{2}\to V$ in order to obtain consistency with the usual scalar potential in the action. In this case the spatial volume will provide a global $2\pi^2$ to both the gravitation as well as the scalar field actions, thus we can ignore it. Then the Lagrangian takes the form
\begin{multline}\label{LagrangianFullT}
	\mathcal{L}=N\left[-\frac{3M^2_{p}\dot{a}^2a}{2N^2}(3\lambda-1)+3M^2_{p}a-\frac{6}{a}(3g_{2}+g_{3})-\frac{12}{a^3M^2_{p}}(9g_{4}+3g_{5}+g_{6})\right] \\ +a^3\left[\frac{3\lambda-1}{4N}\dot{\phi}^2-NV\right]  .
\end{multline}
Thus, this case is far simpler because as in GR the only coordinates on superspace are $a$ and $\phi$, their canonical momenta are written as
\begin{equation}\label{CanonicalMomentaT}
	\pi_{a}=-\frac{3(3\lambda-1)M^2_{p}}{N}a\dot{a} , \hspace{1cm} \pi_{\phi}=\frac{(3\lambda-1)a^3}{2N}\dot{\phi} ,
\end{equation}
and the Hamiltonian constraint takes the form
\begin{multline}\label{HamiltonianT}
	H=N\left[\frac{\pi^2_{\phi}}{a^3(3\lambda-1)}-\frac{\pi^2_{a}}{6(3\lambda-1)M^2_{p}a}-3M^2_{p}a+\frac{6}{a}(3g_{2}+g_{3})\right. \\ \left. +\frac{12}{M^2_{p}a^3}(9g_{4}+3g_{5}+g_{6})+a^3V(\phi)\right]\simeq 0 .
\end{multline}
Then we have the general form of (\ref{GeneralHamiltonian}), with inverse metric
\begin{equation}\label{MetricSuperspaceT}
	G^{\phi\phi}=\frac{2}{(3\lambda-1)a^3} , \hspace{0.5cm} G^{aa}=-\frac{1}{3(3\lambda-1)M^2_{p}a} ,
\end{equation}
\begin{equation}\label{DefinitionfT}
	f(a,\phi)=-3M^2_{p}a+\frac{6}{a}(3g_{2}+g_{3})+\frac{12}{M^2_{p}a^3}(9g_{4}+3g_{5}+g_{6})+a^3V(\phi),
\end{equation}
and $\{W^M\}=0$. We choose the parameter $s$ as usual in (\ref{ChooseS}), then following the same procedure as in the previous cases and choosing $M_{p}=1$ (as in the GR case) we obtain in the thin wall limit  
\begin{equation}\label{GammaT}
	\tcboxmath[colback=red!10!white,colframe=red,right=-0.02cm,left=-0.02cm]{
	\begin{split}
	\pm\Gamma=\pm\frac{2\pi^2\sqrt{6(3\lambda-1)}}{\hbar}\left[\int_{a_{0}}^{\bar{a}}F(3,\bar{\alpha}_{2},\bar{\alpha}_{3},V_{B},a)da-\int_{a_{0}}^{\bar{a}}F(3,\bar{\alpha}_{2},\bar{\alpha}_{3},V_{A},a)da\right]& \\ +\frac{2\pi^2}{\hbar}&\bar{a}^3T ,
	\end{split}}
\end{equation}
where
\begin{equation}\label{DefAlphasT}
	\bar{\alpha}_{2}=6(3g_{2}+g_{3}) , \hspace{0.5cm} \bar{\alpha}_{3}=12(9g_{4}+3g_{5}+g_{6}) ,
\end{equation}
and the function $F$ is the same that was defined in (\ref{FDef}).

Now that we have computed the transition probability in general, let us study the limiting cases. We implement the IR limit through $\lambda\to1$ and $a\gg1$, in this case we obtain the same result as in (\ref{GannaIRF}) with the same subtlety about $a_{0}$. On the other hand, the UV limit is described by $a\ll1$. In this limit we obtain $\Gamma\to0$ as $\bar{a}\to0$. Thus, the limiting behaviour coincides with GR in both cases.  In fact, we can look for an extremum of (\ref{GammaT}) with respect to $\bar{a}$ which leads to
\begin{equation}\label{TensionT}
	T=\pm\frac{\sqrt{6(3\lambda-1)}}{3\bar{a}^2}\left[F(3,\bar{\alpha}_{2},\bar{\alpha}_{3},V_{A},\bar{a})-F(3,\bar{\alpha}_{2},\bar{\alpha}_{3},V_{B},\bar{a})\right] .
\end{equation}
Then, from (\ref{GammaT}) we obtain finally
\begin{multline}\label{GammaTF}
	\pm 2\Re[\Gamma]=\pm\frac{4\pi^2\sqrt{6(3\lambda-1)}}{\hbar}\Re\left[\int_{a_{0}}^{\bar{a}}F(3,\bar{\alpha}_{2},\bar{\alpha}_{3},V_{B},a)da-\int_{a_{0}}^{\bar{a}}F(3,\bar{\alpha}_{2},\bar{\alpha}_{3},V_{A},a)da\right. \\ \left. +\frac{\bar{a}}{3}\left\{F(3,\bar{\alpha}_{2},\bar{\alpha}_{3},V_{A},\bar{a})-F(3,\bar{\alpha}_{2},\bar{\alpha}_{3},V_{B},\bar{a})\right\}\right]
	.
\end{multline}
Thus, the transition probability can also be written in terms of just one parameter as in the GR result. Therefore, the only difference between GR and HL in this case is that the functions involved in the integrals change, acquiring two more terms in the square root before integration, then the integral cannot be computed in a general form, furthermore there is an overall  global factor depending on $\lambda$ in (\ref{GammaTF}). However, the qualitative behaviour in both the IR and UV limit is unaltered.

\begin{figure}[h]
	\centering
	\includegraphics[width=0.7\textwidth]{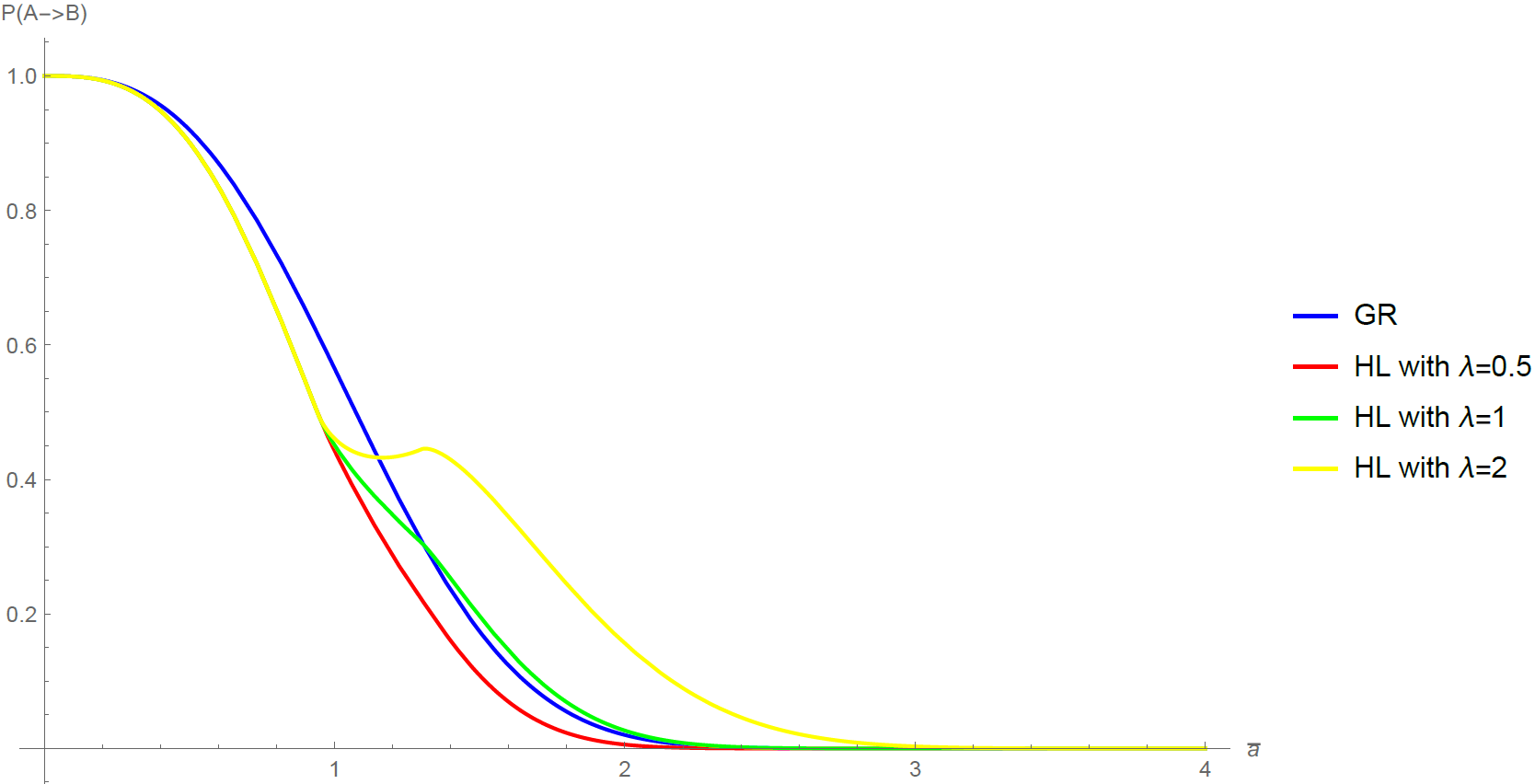}
	\caption{Transition probability for HL gravity in units where $\frac{24\pi^2}{\hbar}=1$, with $V_{A}=-1$, $V_{B}=-10$, $\bar{\alpha}_{2}=\bar{\alpha}_{3}=5$ and $T=5$, for GR (blue line) and HL with a homogeneous scalar field depending only on the time variable with $\lambda=0.5$ (red line), $\lambda=1$ (green line) and $\lambda=2$ (yellow line). For HL we choose $a_{0}=0.000001$ but the same form is expected for $a_{0}=0$.}
	\label{PlotTime}
\end{figure}

Let us compare now the result of this section with the one obtained in GR in all the regimes. We note that although it is true that we can write the probability in terms of only one parameter, the extremizing procedure leading to (\ref{TensionT}) give constrictions for the validity of (\ref{GammaTF}). In particular, we never have access to the UV region where $\bar{a}$ is small. Therefore, we are going to consider the  result depending on the two parameters (\ref{GammaT}) and consider both as independent, we will also choose the minus sign in the right hand side and the plus sign in the left for consistency with the GR result. We will compare such result with the one obtained in  GR  (\ref{FLRWGammaCDL}). In both cases we will take positive values for the tension $T$ big enough so the probability is well defined  and we will compute the integrals numerically. In Figure \ref{PlotTime} we show plots of both results. We choose units where $\frac{24\pi^2}{\hbar}=1$, with $V_{A}=-1$, $V_{B}=-10$, $\bar{\alpha}_{2}=\bar{\alpha}_{3}=5$ and $T=5$. For the HL result we show plots for three values of $\lambda$ and choose $a_{0}=0.000001$ only for a numerical limitation, but the same form is expected for $a_{0}=0$ as shown previously. This figure shows the limiting behaviour that we described earlier, that is, in the IR and in the UV limits all curves behave in the same way, the general form is unaltered. It is in the middle region where their behaviour is modified. In particular, in the same form than in the previous case we find that  the contribution of $\lambda$  appears only when $\bar{a}$ is big enough and it implies the same behaviour as before, that is the probability increases with $\lambda$. Finally, in the IR region all probabilities fall as in the GR case.

\section{Transitions with a Generalized Uncertainty Principle}\label{S-TransitionsGUP}
Until now, we have studied the transition probabilities in two gravitational theories, that is, in General Relativity in section \ref{S-TransitionsGR} and in Ho\v{r}ava-Lifshitz gravity in section \ref{S-TransitionsHL}. However, let us remark that in all the cases that have been studied so far, the Hamiltonian constraints have been of the form (\ref{GeneralHamiltonian}) with only quadratic terms that is in the special case in which $W^{M}=0$, since as we anticipated the form with only quadratic terms is a generic one. However, we will now use the procedure presented in chapter \ref{CH-GUPWDW} to incorporate a GUP in the variables of the minisuperspace to the WDW equation. As we saw in that chapter the WDW equation will be modified in a non-trivial form, in particular linear terms in the momenta will appear. Thus, in order to compute the transition probabilities for such deformed equation the full general method will be needed. We will use in this section the flat and closed FLRW metrics firstly, but we will also use the Bianchi III metric in order to study the effects of the GUP as well as anisotropy at the same time.

	\subsection{Transitions for a closed FLRW metric}
	\label{S-FLRWClosed}	
	Let us begin by considering the FLRW metric with positive curvature. We will employ the Hamiltonian constraint written in terms of the $A(t)$ variable as in (\ref{FLRWHC}). We will consider the simplest form of the GUP generalized to more variables in the form of (\ref{RelationGUP}). In order to simplify notation we will denote the prime coordinates employed in chapter \ref{CH-GUPWDW} with a subscript $g$, that is we will consider\footnote{In this section we will use $\hbar=1$ from the start.}
	\begin{equation}\label{RelationGUPTr}
		[\Phi^{M}_{g},\pi_{N}]=i\delta_{MN}\left(1+\gamma^2\mathcal{P}^2\right) .
	\end{equation}
	In this case the total momentum is given by
	\begin{equation}\label{FLRWP2}
		\mathcal{P}^2=-\frac{\pi^2_{A}}{12}+\frac{\pi^2_{\phi}}{2} .
	\end{equation}	
	As usual, the coordinates $\Phi^{M}$ obey the standard commutation relations, that is $[\Phi^M,\pi_{N}]=i\delta_{MN}$. As explained in chapter \ref{CH-GUPWDW} in order to obtain a WDW equation with this set up we start by considering that (\ref{FLRWHC}) is valid written in terms of coordinates with the $g$ subscript and then each term in that expression has to be rewritten in terms of the coordinates without subscript by using the relations between both sets of variables. Considering only up to second order in $\gamma$ and in momenta we obtain
	\begin{equation}\label{FLRWGUOTRAux}
		-3e^{4A_{g}}\Psi(A_{g},\phi_{g})\simeq -3e^{4\left(1+\frac{4\gamma^2}{9}\right)A}e^{4\gamma^2A\mathcal{P}^2}e^{\frac{4\gamma^2}{3}iA\pi_{A}}\Psi\left(A,\phi\right) .
	\end{equation}	
	As we explained in chapter \ref{CH-GUPWDW} there are two ways to proceed. In this chapter we will always use the second form, where both exponentials depending on the momenta are expanded as power series, then we obtain
	\begin{equation}\label{FLRWExpansion1}
		-3e^{4A_{g}}\Psi(A_{g},\phi_{g})\simeq -3e^{4\left(1+\frac{4\gamma^2}{9}\right)A}\left(1+\frac{4}{3}i\gamma^2A\pi_{A}+4\gamma^2A\mathcal{P}^2\right)\Psi\left(A,\phi\right) 
	\end{equation}	
	and in the same way using a power series expansion on the scalar field potential, we obtain up to second order in $\gamma$ 
	\begin{equation}\label{FLRWExpansion2}
		e^{6A_{g}}V(\phi_{g})\Psi(A_{g},\phi_{g})\simeq e^{6(1+\gamma^2)A}V(\phi)\left(1+3i\gamma^2A\pi_{A}+6\gamma^2A\mathcal{P}^2\right)\Psi\left(A,\phi\right) .
	\end{equation}
	Therefore the modified WDW equation takes the form
	\begin{multline}\label{FLRWGUPHC}
		\left\{\left[1+6\gamma^2e^{6(1+\gamma^2)A}V(\phi)A-12\gamma^2e^{4\left(1+\frac{4}{9}\gamma^2\right)A}A\right]\left(-\frac{\pi^2_{A}}{12}+\frac{\pi^2_{\phi}}{2}\right) \right. \\ \left. +\left[3i\gamma^2e^{6(1+\gamma^2)A}AV(\phi)-4i\gamma^2e^{4\left(1+\frac{4\gamma^2}{9}\right)A}A\right]\pi_{A}  \right. \\ \left. +e^{6(1+\gamma^2)A}V(\phi)-3e^{4\left(1+\frac{4\gamma^2}{9}\right)A}\right\}\Psi\left(A,\phi\right)\simeq0 .
	\end{multline}
	We note that we still have the same general form of (\ref{GeneralHamiltonian}) but in this case
	\begin{equation}\label{FLRWGUPMetric}
		\{G^{MN}\}=\left(1+6\gamma^2e^{6(1+\gamma^2)A}V(\phi)A-12\gamma^2e^{4\left(1+\frac{4\gamma^2}{9}\right)A}A\right){\rm diag}\left(-\frac{1}{6},1\right), 
	\end{equation}
	\begin{equation}\label{FLRWGUPW}
		\{W^{M}\}=\left(3i\gamma^2e^{6(1+\gamma^2)A}AV(\phi)-4i\gamma^2e^{4\left(1+\frac{4\gamma^2}{9}\right)A}A,0\right),
	\end{equation}
	\begin{equation}\label{FLRWGUPf}
		f(A,\phi)=e^{6(1+\gamma^2)A}V(\phi)-3e^{4\left(1+\frac{4\gamma^2}{9}\right)A} .
	\end{equation}
	
	The modified WDW equation by the GUP that we have obtained has been changed in a number of ways. It still has the same general form but all the components have been modified, in addition there are now extra term of linear momenta. We note that  (\ref{FLRWGUPHC}) simplifies to the WDW equation obtained after quantizing (\ref{FLRWHC}) with standard ordering in the limit $\gamma\to 0$ as expected. Therefore this deformed WDW equation  represents a non trivial deformation of the standard case, since it is defined by the $\gamma$ parameter which has the physical significance of measuring the deviation from the HUP. In contrast, if we simply consider a different ordering in (\ref{FLRWHC}) we can also obtain a linear term depending on the factor ordering. However such term  has few physical meaning, furthermore all the other terms will be unaltered. However as we discuss in section \ref{S-TransitionsGeneralMethod} the ordering ambiguity will not be present at the semiclasical level of approximation we have employed. Thus, considering a deformation defined by the GUP is a more relevant scenario to consider and its effects will be present even at the first order in the semiclassical approximation.
	
	Choosing $s$ as usual in (\ref{ChooseS}) and following the same procedure we obtain
	\begin{multline}\label{FLRWGUPGamma}
		\pm\Gamma=-4\pi^2i\left[\int_{0}^{\bar{s}-\delta s} \left(\frac{f}{C(s)}+W_{M}\frac{d\Phi^M}{ds}\right)\bigg\rvert_{\phi=\phi_{B}}ds-\int_{0}^{\bar{s}-\delta s}\left(\frac{f}{C(s)}+W_{M}\frac{d\Phi^M}{ds}\right)\bigg\rvert_{\phi=\phi_{A}}ds \right] \\ +2\pi^2e^{6(1+\gamma^2)\bar{A}}T ,
	\end{multline}
	where the corresponding tension term is defined as
	\begin{equation}\label{FLRWGUPTensionDef}
		2\pi^2e^{6(1+\gamma^2)\bar{A}}T=-4\pi^2i\int_{\bar{s}-\delta s}^{\bar{s}+\delta s}\frac{ds}{C(s)}e^{6(1+\gamma^2)A}\left[V(\phi)-V_{A}\right] .
	\end{equation}
	Once again, the first two integrals in (\ref{FLRWGUPGamma}) can be computed by solving the general system of equations (\ref{System1}) and (\ref{System2})  when the scalar field is constant. However we did not find a general solution in this scenario. Nonetheless, for this simple metric the system simplifies and we can obtain a solution in the form
	\begin{equation}\label{FLRWGUPAux}
		C(s)=\frac{2\pi^2}{f}\frac{\partial f}{\partial A}\left[W^{A}\pm\sqrt{(W^{A})^2-2fG^{AA}}\right] ,
	\end{equation}
	\begin{equation}\label{FLRWGUPAux2}
		\frac{dA}{ds}=-\frac{G^{AA}f^2}{\pi^2\frac{\partial f}{\partial A}\left[W^{A}\pm\sqrt{(W^{A})^2-2fG^{AA}}\right]^2} ,
	\end{equation}
	where the sign ambiguity appears as as a result of $C(s)$ as it happened in all the other scenarios we have considered. We make a change of variables in the form $dA=\left(\frac{dA}{ds}\right)ds$,   which leads in this case to
	\begin{equation}\label{FLRWGUPAction}
		\int \left(\frac{f}{C(s)}+W_{M}\frac{d\Phi^M}{ds}\right)ds=\frac{1}{2}\int\frac{W^{A}}{G^{AA}}dA\pm\int\frac{\sqrt{(W^{A})^2-2fG^{AA}}}{2G^{AA}}dA .
	\end{equation}
	We note that $W^{A}$ has an overall dependence on $\gamma^2$, therefore we can neglect its squared value with respect to $2fG^{AA}$, thus the second integral simplifies. Following the procedure we finally obtain up to second order in $\gamma$ and using the thin wall limit 
	\begin{equation}\label{FLRWGUPProb}
		\tcboxmath[colback=red!10!white,colframe=red,left=-0.01cm,right=-0.01cm]{
			\begin{split}
		\pm\Gamma\simeq\pm12\pi^2&\left[\int_{-\infty}^{\bar{A}}\sqrt{F(A,V_{B})}dA-\int_{-\infty}^{\bar{A}}\sqrt{F(A,V_{A})}dA\right] \\ &-12\pi^2\gamma^2\left[\int_{-\infty}^{\bar{A}}G(A.V_{B}) dA-\int_{-\infty}^{\bar{A}}G(A,V_{A})dA\right] + 2\pi^2e^{6(1+\gamma^2)A}T ,
		\end{split}}
	\end{equation}
	where we have defined the functions
	\begin{equation}\label{FLRWGUPDefF}
		F(A,V)=\frac{e^{4\left(1+\frac{4\gamma^2}{9}\right)A}-\frac{V}{3}e^{6(1+\gamma^2)A}}{1+6\gamma^2e^{6(1+\gamma^2)A}VA-12\gamma^2e^{4\left(1+\frac{4\gamma^2}{9}\right)A}A} ,
	\end{equation}
	\begin{equation}\label{FLRWGupDefG}
		G(A,V)=\frac{3Ve^{6(1+\gamma^2)A}-4e^{4\left( 1+\frac{4\gamma^2}{9}\right)A}}{1+6\gamma^2e^{6(1+\gamma^2)A}VA-12\gamma^2e^{4\left(1+\frac{4\gamma^2}{9}\right)A}A} A ,
	\end{equation}
	and we have chosen the limits of integration in order to explore the UV limit since the functions are well behaved. We note that for some values of the potentials the integrals of the $F$ functions could result in imaginary values which will not contribute to the transition probability. However the integrals of the $G$ function are always real. Thus they will always contribute to the probabilities. Therefore, the dependence of the probabilities with the GUP parameter  $\gamma$ will always be present. Furthermore, we can see that this result simplifies to the standard case (\ref{FLRWTRP}) consistently in the limit $\gamma\to0$ as expected. The probability found is described in terms of two parameters as in the standard case. Following the extremizing procedure to try to relate them we are led to
	\begin{equation}\label{FLRWGUPTension}
		T=\frac{e^{-6(1+\gamma^2)\bar{A}}}{1+\gamma^2}\left[\gamma^2\left(G(\bar{A},V_{B})-G(\bar{A},V_{A})\right)\mp\left(\sqrt{F(\bar{A},V_{B})}-\sqrt{F(\bar{A},V_{A})}\right)\right] ,
	\end{equation}
	which can only be fulfilled under the condition $F(\bar{A},V_{A,B})\geq0$. However when $\bar{A}\gg1$ we obtain
	\begin{equation}
		F(A,V_{A,B})\to-\frac{1}{18\gamma^2\bar{A}} ,
	\end{equation}
	therefore, the tension term  (\ref{FLRWGUPTension}) will only be defined until an an upper bound for $\bar{A}$, regardless of the value of the potential minima. Thus we are going to consider both parameters in (\ref{FLRWGUPProb}) as independent as usual at this point.  Furthermore, we note that both functions (\ref{FLRWGUPDefF}), (\ref{FLRWGupDefG}) have a singular value in which the denominator goes to zero for some $A$. Therefore when studying the transition probabilities we will avoid such values so the integrals can be performed numerically.
	
\begin{figure}[h!]
		\centering
		\includegraphics[width=0.7\textwidth]{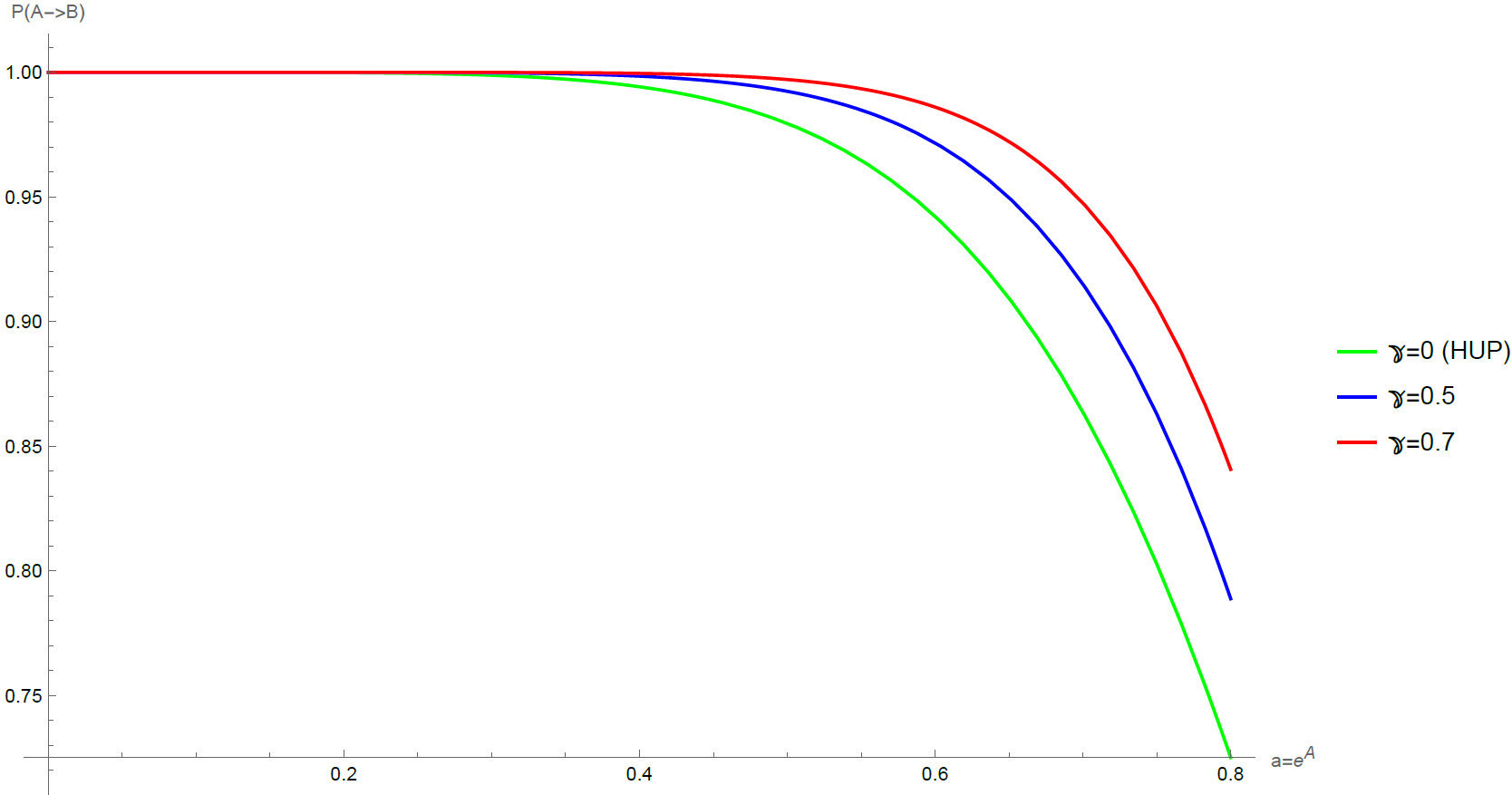}
		\caption{Transition probability for the FLRW closed metric in the UV region with HUP or $\gamma=0$ (Green curve), and considering a GUP with $\gamma=0.5$ (Blue curve) and $\gamma=0.7$ (Red curve) for $V_{A}=-1$, $V_{B}=-2$ and $T=7$.  }
		\label{F-FLRW}
	\end{figure}

\begin{figure}[h!]
		\centering
		\includegraphics[width=0.7\textwidth]{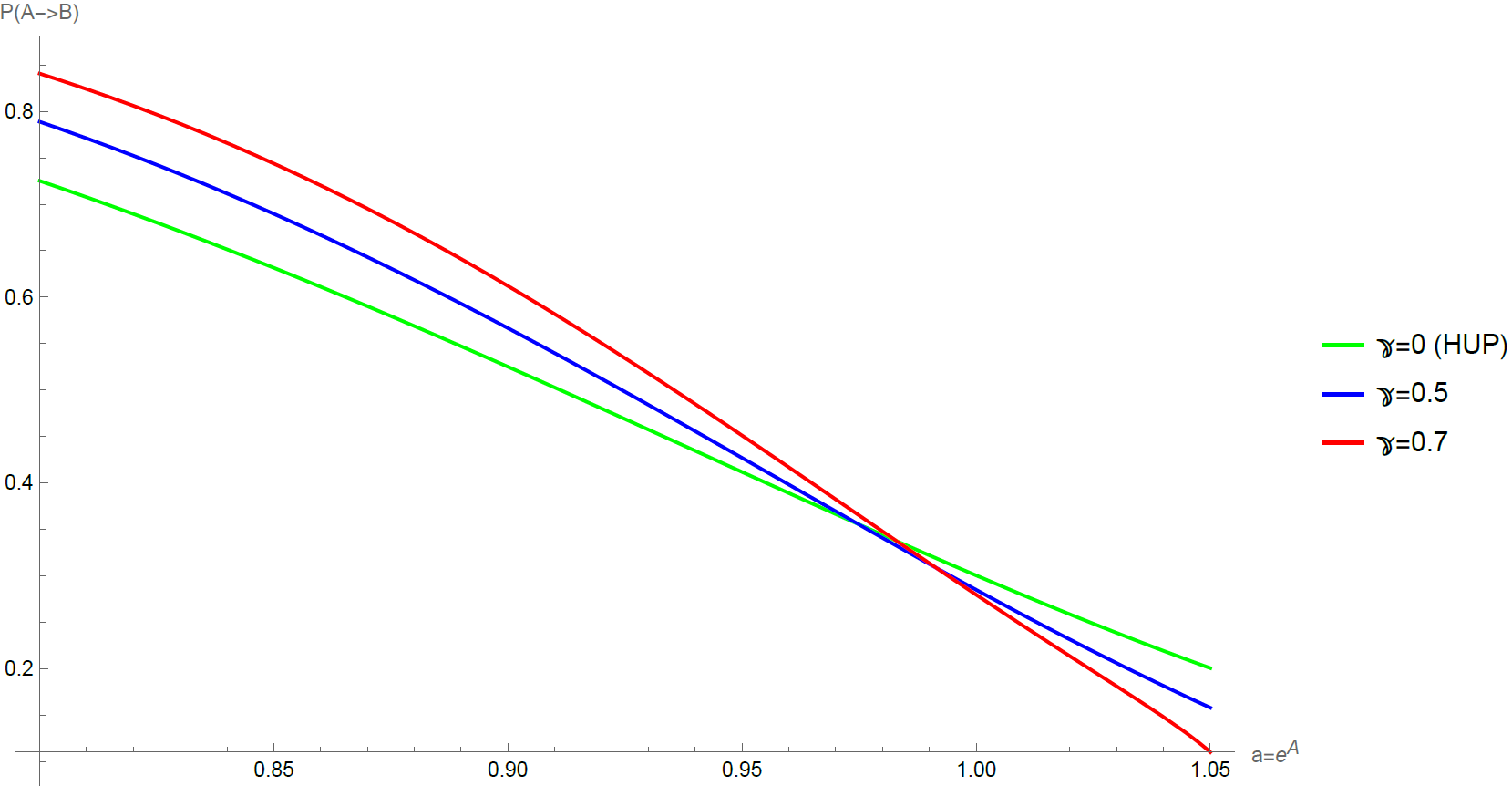}
		\caption{Transition probability for the FLRW closed metric with HUP or $\gamma=0$ (Green curve), and considering a GUP with $\gamma=0.5$ (Blue curve) and $\gamma=0.7$ (Red curve) for $V_{A}=-1$, $V_{B}=-2$ and $T=7$ focusing on the region where the change of behaviour occurs. }
		\label{F-FLRWExtra}
	\end{figure}

	The correct choice of signs for the standard case was discussed at the end of subsection \ref{S-FLRW}. For the GUP result (\ref{FLRWGUPProb}) the correct choice is the plus sign in the right (which corresponds to the minus sign in the standard case, thus we are considering the same sign in both expressions). We then can compare both results. By studying limiting cases we find that the general form of the probability is unaltered, that is it always starts at $1$ in the UV limit $\bar{a}\to0$ and falls to $0$ as the scale factor increases. However, the effect of the GUP is that the probability in the UV region (small scale factor) is increased with the GUP parameter $\gamma$, however it also decreases faster, thus in the IR region (big scale factor) the probability is smaller than the GR case. Employing the interpretation of the transition probabilities as probability distributions of creating universes as explained in section \ref{S-TransitionsHL}, this behaviour exhibits consistency with the idea of a minimal measurable value for the scale factor, since it implies that the  probability of creating a universe with small non-zero value of the scale factor increases proportional to the GUP parameter. However, this effect is subtle because $\gamma$ is a small number and we have done approximation up to second order. Thus, in order to visualize correctly this behaviour we show two plots. In Figure \ref{F-FLRW} we present a plot of the UV region and in Figure \ref{F-FLRWExtra} we show a plot in the intermediate region where the change in the behaviour of the probabilities is clearly appreciated. We note that the point where the behaviour is changed is not the same for all the curves, it depends on $\gamma$ . Therefore, there is always a small intermediate region where the probability has neither an increasing nor decreasing behaviour with $\gamma$. In both plots we choose $V_{A}=-1$, $V_{B}=-2$, $T=7$, furthermore we ignore the overall  factor $2(12\pi^2)$ in all the expressions and plot for different values of $\gamma$. We remark that we can perform the numerical integrations since the region where a singularity could appear in the $F$ or $G$ functions is encountered for bigger values of the scale factor where the probability is very small. 
	
	
	\subsection{Transitions for a flat FLRW metric}
	\label{S-FLRWF}
	
	Let us consider now a FLRW metric with vanishing spatial curvature. We will consider the procedure in terms of $A(t)$, in the standard case the transition probability is given by (\ref{FLRWFProb}).  Let us assume now that the coordinates in minisuperspace obey a GUP of the form (\ref{RelationGUPTr}) and proceed as in the previous case. Using (\ref{FLRWExpansion2}) we obtain the modified WDW equation 
	\begin{multline}\label{FLRWGUPFHC}
		\left\{\left[1+6\gamma^2e^{6(1+\gamma^2)A}V(\phi)A\right]\left(-\frac{\pi^2_{A}}{12}+\frac{\pi^2_{\phi}}{2}\right) +3i\gamma^2e^{6(1+\gamma^2)A}AV(\phi)\pi_{A}  \right. \\ \left. +e^{6(1+\gamma^2)A}V(\phi)\right\}\Psi\left(A,\phi\right)\simeq0 .
	\end{multline}
	We see that we have the general form of (\ref{GeneralHamiltonian}) with
	\begin{equation}\label{FLRWGUPFMetric}
		\{G^{MN}\}=\left(1+6\gamma^2e^{6(1+\gamma^2)A}V(\phi)A\right){\rm diag}\left(-\frac{1}{6},1\right), 
	\end{equation}
	\begin{equation}\label{FLRWGUPFW}
		\{W^{M}\}=\left(3i\gamma^2e^{6(1+\gamma^2)A}AV(\phi),0\right) , \hspace{0.5cm}
		f(A,\phi)=e^{6(1+\gamma^2)A}V(\phi) .
	\end{equation}
	Considering $s$ as usual by (\ref{ChooseS}) and following the same procedure as always we obtain
	\begin{multline}\label{FLRWFGUPGamma}
		\pm\Gamma=-2i{\rm Vol}(X)\left[\int_{0}^{\bar{s}-\delta s} \left(\frac{f}{C(s)}+W_{M}\frac{d\Phi^M}{ds}\right)\bigg\rvert_{\phi=\phi_{B}}ds\right. \\ \left. -\int_{0}^{\bar{s}-\delta s}\left(\frac{f}{C(s)}+W_{M}\frac{d\Phi^M}{ds}\right)\bigg\rvert_{\phi=\phi_{A}}ds \right] +\frac{{\rm Vol}(X)}{\hbar}e^{6(1+\gamma^2)\bar{A}}T .
	\end{multline}
	The corresponding system of equations can be solved in the same form as before leading to (\ref{FLRWGUPAction}), then we  note that for this metric we can also neglect $(W^{A})^2$  with respect to $2fG^{AA}$. in this way we finally obtain in the thin wall limit and up to second order in $\gamma$
	\begin{multline}\label{FLRWGUPFProb}
		\tcboxmath[colback=red!10!white,colframe=red,left=-0.01cm,right=-0.01cm]{
		\begin{split}
		\pm&\Gamma\simeq\pm2\sqrt{3}i{\rm Vol}(X)\left[\int_{-\infty}^{\bar{A}}\sqrt{F_{f}(A,V_{B})}dA-\int_{-\infty}^{\bar{A}}\sqrt{F_{f}(A,V_{A})}dA\right] \\ &-18{\rm Vol}(X)\gamma^2\left[\int_{-\infty}^{\bar{A}}G_{f}(A.V_{B}) dA-\int_{-\infty}^{\bar{A}}G_{f}(A,V_{A})dA\right] + {\rm Vol}(X)e^{6(1+\gamma^2)\bar{A}}T ,
		\end{split}}
	\end{multline}
	where 
	\begin{equation}\label{FLRWGUPFDefFG}
		F_{f}(A,V)=\frac{Ve^{6(1+\gamma^2)A}}{1+6\gamma^2e^{6(1+\gamma^2)A}VA} , \hspace{0.5cm}
		G_{f}(A,V)=AF_{f}(A,V) .
	\end{equation}
	The transition probability is described once again by two parameters. We will consider that they are independent since pursuing an extremizing procedure will lead to similar problems as for the closed FLRW metric. Furthermore, in this result we also have that the second term in (\ref{FLRWGUPFProb}) always contributes to the probability, whereas the first term could result in imaginary values in some cases that will be irrelevant for the probability, for example when both potential minima are positive. Finally,  in the limit $\gamma\to0$ this results reduces consistently to the standard result  (\ref{FLRWFProb}). Let us remark that for negative potential minima, the $F_{f}$ and $G_{f}$ functions can have a singular point where the denominator vanishes. Thus, the probabilities will be well defined in this case for regions that avoid these points. Furthermore, the correct choice to obtain well defined probabilities in this case is the minus sign in the right hand side and the plus sign on the left. This result exhibits the same behaviour as the FLRW closed metric, that is, at first the probability increases with $\gamma$, but it decreases faster and when the scale factor is big enough the behaviour is flipped. However, the general behaviour for all the cases is the same, they all start at $1$ in the UV limit and falls to $0$ as in the GR case.

	\subsection{Transitions for a Bianchi III metric} 	\label{S-Bianchi}
	Finally, in order to study the effect of anisotropy at the same time that the GUP we will consider the Bianchi III metric, in particular we will use the $p(t),q(t),u(t)$ functions in which the probability is written in the standard case as (\ref{B3ProbComp}) or (\ref{B3Prob}).
	 
	We now consider the GUP (\ref{RelationGUPTr}), in this case the total momentum is given by
	\begin{equation}\label{B3P2}
		\mathcal{P}^2=\frac{\pi^2_{p}}{4}+\frac{\pi^2_{q}}{4}+\frac{\pi^2_{u}}{4}-\frac{\pi_{p}\pi_{q}}{2}-\frac{\pi_{p}\pi_{u}}{2}-\frac{\pi_{q}\pi_{u}}{2}+\frac{\pi^2_{\phi}}{2} .
	\end{equation}	
	Then, using the relation between both sets of variables in the coordinate representation (\ref{RelationsCoordinates}) and the Zassenhaus formula (\ref{ZSFormula})  we obtain
	\begin{equation}\label{B3Expansion1}
		e^{2(q_{g}+u_{g})}\simeq e^{2(q+g)}\left[1+2i\gamma^2(q+u)\pi_{p}+2\gamma^2(q+u)\mathcal{P}^2\right],
	\end{equation}
	\begin{multline}\label{B3Expansion2}
		e^{2(p_{g}+q_{g}+u_{g})}V(\phi_{g})\simeq e^{2(1+\gamma^2)(p+q+u)}V(\phi)\left[1+i\gamma^2(p+q+u)(\pi_{p}+\pi_{q}+\pi_{u})\right. \\ \left. +2\gamma^2(p+q+u)\mathcal{P}^2\right] .
	\end{multline}
	Then we obtain the modified WDW equation 
	\begin{multline}\label{B3GUPHC}
		\left\{\tilde{L}[p,q,u,V(\phi)]\left[\frac{\pi^2_{p}}{4}+\frac{\pi^2_{q}}{4}+\frac{\pi^2_{u}}{4}-\frac{\pi_{p}\pi_{q}}{2} -\frac{\pi_{p}\pi_{u}}{2}-\frac{\pi_{q}\pi_{u}}{2}+\frac{\pi^2_{\phi}}{2}\right]\right. \\ \left. +L_{1}[p,q,u,V(\phi)](\pi_{q}+\pi_{u}) +\left(L_{1}[p,q,u,V(\phi)]+L_{2}[p,q,u]\right)\pi_{p} \right. \\ \left. +\alpha^2e^{2(q+u)}+e^{2(1+\gamma^2)(p+q+u)}V(\phi)\right\}\Psi(p,q,u,\phi)\simeq 0,
	\end{multline}
	where we have defined
	\begin{equation}\label{B3GUPHDef}
		\tilde{L}[p,q,u,V(\phi)]=1+2\gamma^2\alpha^2e^{2(q+u)}(q+u)+2\gamma^2e^{2(1+\gamma^2)(p+q+u)}(p+q+u)V(\phi) ,
	\end{equation}
	\begin{equation}\label{B3GUPL1Def}
		L_{1}[p,q,u,V(\phi)]=i\gamma^2e^{2(1+\gamma^2)(p+q+u)}(p+q+u)V(\phi) ,
	\end{equation}
	\begin{equation}\label{B3GUPL2Def}
		L_{2}[p,q,u]=2i\gamma^2\alpha^2e^{2(q+u)}(q+u) .
	\end{equation}
	We note that we have obtained a more difficult equation but it still has the same general form as in (\ref{GeneralHamiltonian}) identifying
	\begin{equation}\label{B3GUPMetricD}
		\{G^ {MN}\}= \frac{\tilde{L}[p,q,u,V(\phi)]}{2} \left( \begin{array}{cccc}
			1 & -1 & -1 & 0 \\
			-1 & 1 & -1 & 0 \\
			-1 & -1 & 1 & 0  \\
			0 & 0  & 0 & 2   \end{array} \right) ,
	\end{equation}
	\begin{equation}\label{B3GUPWDef}
		\{W^{M}\}=L_{1}[p,q,u,V(\phi)](1,1,1,0)+L_{2}[p,q,u](1,0,0,0) ,
	\end{equation}
	\begin{equation}\label{B3GUPfDef}
		f(p,q,u,\phi)=\alpha^2e^{2(q+u)}+e^{2(1+\gamma^2)(p+q+u)}V(\phi) .
	\end{equation}
	Choosing $s$ as usual in (\ref{ChooseS}) and following the same procedure  we obtain
	\begin{multline}\label{B3GUPGamma}
		\pm\Gamma=-2{\rm Vol}(X)i\left[\int_{0}^{\bar{s}-\delta s} \left(\frac{f}{C(s)}+W_{M}\frac{d\Phi^M}{ds}\right)\bigg\rvert_{\phi=\phi_{B}}ds\right. \\ \left. -\int_{0}^{\bar{s}-\delta s}\left(\frac{f}{C(s)}+W_{M}\frac{d\Phi^M}{ds}\right)\bigg\rvert_{\phi=\phi_{A}}ds \right]+\frac{{\rm Vol}(X)}{\hbar}e^{2(1+\gamma^2)(\bar{p}+\bar{q}+\bar{u})}T ,
	\end{multline}
	where the tension term is defined as usual by
	\begin{equation}\label{B3GUPTensionDeg}
		{\rm Vol}(X)e^{2(1+\gamma^2)(\bar{p}+\bar{q}+\bar{u})}T=-2{\rm Vol}(X)i\int_{\bar{s}-\delta s}^{\bar{s}+\delta s}\frac{V(\phi)-V_{A}}{C(s)}e^{2(1+\gamma^2)(p+q+u)}ds .
	\end{equation}
	The integrals in (\ref{B3GUPGamma}) can be computed by solving the system of equations (\ref{System1}) and (\ref{System2}) when the scalar field is constant. In this case such system is written as
	\begin{multline}\label{B3Systempo}
		\frac{dp}{ds}=\frac{2{\rm Vol}(X)}{C^2(s)}\tilde{L}(p,q,u,V)\left[(1+\gamma^2)e^{2(1+\gamma^2)(p+q+u)}V+2\alpha^2e^{2(q+u)}\right] \\ -\frac{{\rm Vol}(X)}{C(s)}\left[\frac{dp}{ds}+\frac{1}{2}\frac{dq}{ds}+\frac{1}{2}\frac{du}{ds}\right]\frac{8i\gamma^4\alpha^2e^{2\left[q+u+(1+\gamma^2)(p+q+u)\right]}V[2\gamma^2(p+q+u)(q+u)-p]}{\tilde{L}[p,q,u,V]} ,
	\end{multline}
	\begin{multline}\label{B3Systemqo}
		\frac{dq}{ds}=\frac{2{\rm Vol}(X)}{C^2(s)}\tilde{L}[p,q,u,V](1+\gamma^2)e^{2(1+\gamma^2)(p+q+u)}V \\ +\frac{2{\rm Vol}(X)}{C(s)}\left[\frac{dq}{ds}+\frac{du}{ds}\right]\frac{4i\gamma^4\alpha^2e^{2\left[q+u+(1+\gamma^2)(p+q+u)\right]}V[2\gamma^2(p+q+u)(q+u)-p]}{\tilde{L}[p,q,u,V]} ,
	\end{multline}
	\begin{multline}\label{B3Systemuo}
		\frac{du}{ds}=\frac{2{\rm Vol}(X)}{C^2(s)}\tilde{L}[p,q,u,V](1+\gamma^2)e^{2(1+\gamma^2)(p+q+u)}V \\ +\frac{2{\rm Vol}(X)}{C(s)}\left[\frac{dq}{ds}+\frac{du}{ds}\right]\frac{4i\gamma^4\alpha^2e^{2\left[q+u+(1+\gamma^2)(p+q+u)\right]}V[2\gamma^2(p+q+u)(q+u)-p]}{\tilde{L}[p,q,u,V]} ,
	\end{multline}
	where $V$ represents wither $V_{A}$ or $V_{B}$ depending on which path is considered. It is hard to pursue a general solution for this system, however in the three equations the second term has an overall dependence of $\gamma^4$, in contrast the first term only contains up to second order in $\gamma$. Thus, in the three equations we can safely ignore the second term with respect to the first, then the system of equations simplifies and takes the form
	\begin{equation}\label{B3Systemp}
		\frac{dp}{ds}\simeq\frac{2{\rm Vol}(X)}{C^2(s)}\tilde{L}[p,q,u,V]\left[(1+\gamma^2)e^{2(1+\gamma^2)(p+q+u)}V+2\alpha^2e^{2(q+u)}\right]  ,
	\end{equation}
	\begin{equation}\label{B3Systemq}
		\frac{dq}{ds}\simeq\frac{2{\rm Vol}(X)}{C^2(s)}\tilde{L}[p,q,u,V](1+\gamma^2)e^{2(1+\gamma^2)(p+q+u)}V ,
	\end{equation}
	\begin{equation}\label{B3Systemu}
		\frac{du}{ds}\simeq\frac{2{\rm Vol}(X)}{C^2(s)}\tilde{L}[p,q,u,V](1+\gamma^2)e^{2(1+\gamma^2)(p+q+u)}V  .
	\end{equation}
	From this equation we find that the coordinates of minisuperspace  are related to each other, in this case it follows
	\begin{multline}\label{B3GUPRelationsC}
		\frac{1}{(1+\gamma^2)e^{2(1+\gamma^2)(p+q+u)}V+2\alpha^2e^{2(q+u)}}\frac{dp}{ds}=\frac{1}{(1+\gamma^2)e^{2(1+\gamma^2)(p+q+u)}V}\frac{dq}{ds}\\=\frac{1}{(1+\gamma^2)e^{2(1+\gamma^2)(p+q+u)}V}\frac{du}{ds} ,
	\end{multline}
	that leads to the solution
	\begin{equation}\label{B3GUPRelationsF}
		u=q , \hspace{0.5cm} p=\ln\left[\left(e^{2(1+\gamma^2)q}-\frac{2\alpha^2}{V(1+3\gamma^2)}e^{-4\gamma^2q}\right)^{\frac{1}{2(1+\gamma^2)}}\right] ,
	\end{equation}
	where the integration constants have been absorbed by a proper redefinition of the coordinates. Thus, we obtain from (\ref{System1}) a quadratic equation for $C(s)$ that leads to the solution
	\begin{equation}\label{B3GUPSolCA}
		C(s)=\frac{M_{1}}{2f}\pm\frac{\sqrt{M_{1}^2+4M_{2}f}}{2f} ,
	\end{equation}
	where
	\begin{multline}\label{B3GUPDefM1}
		M_{1}=4{\rm Vol}(X)i\gamma^2e^{2(1+\gamma^2)(p+q+u)}V\left[3(1+\gamma^2)e^{2(1+\gamma^2)(p+q+u)}(p+q+u)V\right. \\ \left. +2\alpha^2e^{2(q+u)}\left[p+q+u+(1+\gamma^2)(q+u)\right]\right] ,
	\end{multline}
	\begin{multline}\label{B3GUPDefM2}
		M_{2}=4{\rm Vol}^2(X)\tilde{L}[p,q,u,V](1+\gamma^2)e^{2(1+\gamma^2)(p+q+u)}V\left[3(1+\gamma^2)e^{2(1+\gamma^2)(p+q+u)}V\right. \\ \left. +4\alpha^2e^{2(q+u)}\right].
	\end{multline}
	Moreover,  the  $M_{1}^2$ term has a global factor of $\gamma^4$ , therefore we can  neglect it with respect to $4M_{2}f$ once again. Thus we obtain
	\begin{equation}\label{B3GUPSolC}
		C(s)\simeq\frac{M_{1}}{2f}\pm\sqrt{\frac{M_{2}}{f}} ,
	\end{equation}
	Therefore employing an approximating up to second order in $\gamma$ we obtain for a constant scalar field 
	\begin{equation}\label{B3GupAction}
		\int \left(\frac{f}{C(s)}+W_{M}\frac{d\Phi^M}{ds}\right)ds\simeq\pm\int \sqrt{F_{III}(q,V)}dq -\frac{i\gamma^2}{1+\gamma^{2}}\int G_{III}(q,V)dq , 
	\end{equation}
	where 
	\begin{equation}\label{B3GUPDefF}
		F_{III}(q,V)=\frac{\left[3(1+\gamma^2)Ve^{6(1+\gamma^2)q}-\frac{2\alpha^2(1-3\gamma^2)}{1+3\gamma^2}e^{4q}\right]\left[Ve^{6(1+\gamma^2)q}-\frac{1-3\gamma^2}{1+3\gamma^2}\alpha^2e^{4q}\right]}{\tilde{L}[q,V]\left[V(1+\gamma^2)e^{6(1+\gamma^2)q}-\frac{2\alpha^2(1+\gamma^2)}{1+3\gamma^2}e^{4q}\right]} ,
	\end{equation}
	\begin{equation}\label{B3GUPDefG}
		G_{III}(q,V)=\frac{3V(1+\gamma^2)(2q+p(q))e^{6(1+\gamma^2)q}+\frac{4\alpha^2}{1+3\gamma^2}\left[(3\gamma^4+4\gamma^2-1)q-p(q)\right]e^{4q}}{\tilde{L}[q,V]} .
	\end{equation}
	In this result we have written $p(q)$ in this way to specify that $p$ is related to $q$ as stated in  (\ref{B3GUPRelationsF}), furthermore $\tilde{L}[q,V]$ means the function defined in (\ref{B3GUPHDef}) with those relations considered, which leads to
	\begin{equation}\label{B3HFunctionF}
		\tilde{L}[q,V]=1+2\gamma^2V(2q+p(q))e^{6(1+\gamma^2)q}-\frac{4\gamma^2\alpha^2}{1+3\gamma^2}\left[(1-3\gamma^2)q+p(q)\right]e^{4q} .
	\end{equation}
	
	We can also write the tension in (\ref{B3GUPGamma}) in term of just $\bar{q}$ and the potential minima after using the thin wall limit, and considering the relations in (\ref{B3GUPRelationsF}) to impose continuity in $q$ as it was done in the standard case. In this way we finally obtain 
	\begin{equation}\label{B3GUPProb}
		\tcboxmath[colback=red!10!white,colframe=red,left=-0.01cm,right=-0.01cm]{
		\begin{split}
		\pm\Gamma\simeq\pm2{\rm Vol}(X)i&\left[\int_{q_{0}}^{\bar{q}}\sqrt{F_{III}(q,V_{B})}dq-\int_{q_{0}}^{\bar{b}}\sqrt{F_{III}(q,V_{A})}dq\right]\\&-\frac{2{\rm Vol}(X)\gamma^2}{(1+\gamma^2)}\left[\int_{q_{0}}^{\bar{q}}G_{III}(q,V_{B})dq-\int_{q_{0}}^{\bar{q}}G_{III}(q,V_{A})db\right]\\&\hspace{3cm}+{\rm Vol}(X)\left[e^{6(1+\gamma^2)\bar{q}}-\frac{2\alpha^2}{V_{B}(1+3\gamma^2)}e^{4\bar{q}}\right]T ,
		\end{split}}
	\end{equation}
	where in each integral $p$ must be related to $q$ as in (\ref{B3GUPRelationsF}) with the appropriate value of the potential. 
	
	We note that as it happened in the case without a GUP in subsection \ref{S-B3}, the value of $q_{0}$ can only be chosen on the origin at $-\infty$ when the potential minima take negative values. In principle the case with positive values should be correctly describe by the $p$ function. However in this case it is very difficult to express the integral in terms of $p$ , thus we will only consider negative potential minima. In this case we also obtain a similar behaviour as before, that is the transition probability is described by two parameters, which will be considered as independent to avoid difficulties. Furthermore, once again the first term can be ignored if the integrals leads to a real value since only the overall real part of this expression will contribute to the probability, whereas the second term will always contribute since it will take real values.
	
	This result reduces to standard result with a HUP (\ref{B3Prob}) in the limit $\gamma\to0$  as expected, Furthermore, it leads to the result for the  flat FLRW metric modified by the GUP (\ref{FLRWGUPFProb}) in the isotropy limit $\alpha\to0$ as expected. Thus this result fulfils the two appropriate limits and provides a transition probability that has a more complicated form and should be evaluated numerically, but is still described by the same number of parameters as before.
	
	The correct choice of signs in this case is the minus sign on the right hand side and the plus sign on the left. Varying the $\gamma$ parameter we obtain a consistent behaviour in agreement with the earlier result for the FLRW metric, i.e. we find that the effect of the GUP is to increase the probability at first but make it fall faster so eventually the roles are interchanged and the probability decreases with $\gamma$. On the other hand, varying the anisotropy parameter $\alpha$ we obtain a consistent behaviour with the one encountered in the HUP case and shown in  Figure \ref{F-BianchiCom}. Interestingly we have found that the variation of both parameters, $\gamma$ (measuring the GUP) and  $\alpha$ (measuring the anisotropy) leads to a very similar behaviour, although the parameters are independent to each other. Moreover, for all cases the probabilities obey the same limits as found with the HUP and with isotropy, that is they all go to $1$ when $q\to-\infty$ and as $\mathbb{B}$ increases they fall to zero.

	\subsection{Other possible extensions}
	\label{GUP-EUP}
	
	To end this section let us remark that we have applied the most general version of our method to incorporate the effects of a GUP in the transition probabilities. However, the applicability of the method relies only on  a general Hamiltonian constraint of the correct form. Thus, it would be interesting to study more scenarios that deform the WDW equation in such a way that we can still apply the method. Let us briefly discuss some of these kind of scenarios.
	
	In this section we considered a GUP in the form of (\ref{RelationGUPTr}) that is a direct generalization of the simplest form of the GUP. Furthermore, we worked in the coordinate representation where we promoted the momenta to derivative operators. Moreover, both sets of coordinates can be related as (\ref{RelationsCoordinates}). Thus, we have effectively considered a Hamiltonian constraint of the form 
	\begin{equation}
		H\bigg(\Phi^{N},i(1+\gamma^2\mathcal{P}^2){\delta \over \delta \Phi^{N}}\bigg) \Psi(\Phi)= 0.
	\end{equation}
	 
	 On the other hand as we discussed in section \ref{S-GUP} when the quantization procedure is applied to an (anti)-de Sitter background, the commutator relations can modify the uncertainty principle by a term which depends on a cosmological constant $\Lambda$, this is called the EUP \cite{Park:2007az,Mignemi:2009ji,Gine:2020izd}. Using these principle in the variables of superspace we would have
	\begin{equation}
		[\Phi^{M},(\pi_{N})_{g}]=i\delta_{MN}\left(1+\lambda^2\mathcal{Q}^2\right)
	\end{equation}
	where $\mathcal{Q}^2= q^\ell q_\ell$ will take into account the metric on superspace and $\lambda$ will be the EUP parameter. Thus we can seek for the dual representation where the momenta $\pi_{M}$ will act as coordinates of superspace and the original coordinates will be promoted to differential operators, in this way the Hamiltonian constraint would take the form
	\begin{equation}
		H\bigg(-i(1+\lambda^2 \mathcal{Q}^2){\delta \over \delta \pi_N},\pi_N\bigg) \psi(\Pi)= 0.
	\end{equation}
	where $\Pi$ denotes collectively all the momenta. Therefore, employing the relations in this case between both sets of momenta and making approximations uo to second order in $\lambda$ we would obtain a modified WDW equation in the momentum representation that would contain relevant information of the EUP. Thus, it would be interesting to consider some models in quantum cosmology employing the momentum representation. In  particular if we can get a Hamiltonian constraint wit the general form of (\ref{GeneralHamiltonian}) in the momentum representation, our method will still be applicable to this context. It is worth mentioning that if we want to pursue this path the interpretation of the EUP in the variables of minisuperspace has a different meaning that the original interpretation of quantum mechanics in (anti)-de Sitter space \cite{Park:2007az,Mignemi:2009ji,Gine:2020izd}. This issue also arrives with the GUP as well. However, the relation between both descriptions can be a very interesting subject to study.

	Finally, we can consider the effects of the GUP and EUP at the same time. In this situation the commutator relations will be given by
	\begin{equation}
		[\Phi^{M}_{g},(\pi_{N})_{g}]=i\delta_{MN}\left(1+\gamma^2\mathcal{P}^2+\lambda^2\mathcal{Q}^2\right) .
	\end{equation}
	We would need two transformations to relate both sets of coordinates and momenta with the subscript in terms of the ones without the subscript that obey the standard commutator relations. One possible candidate to carry out such description would be the {\it phase space representation} \cite{GoTorres1,GoTorres2,Curtright:2011vw}, in which the WDW equation would take the generic form
	\begin{equation}
		H \bigg(-i(1+\lambda^2\mathcal{Q}^2){\delta \over \delta \Phi^{M}},i(1+\gamma^2 \mathcal{P}^2){\delta \over \delta \pi_{M}}\bigg) \Psi(\Phi,\Pi)= 0.
	\end{equation}
	It would be interesting to see if this proposal is viable to study quantum cosmological  models in the superspace using the WDW equation in the phase space representation. Furthermore, it would be interesting to explore if our general method, or a variation of it can be used to compute the corresponding transition probabilities that take into account both effects. 

	\section{Discussion}\label{S-DTransitions}	
	Finally, let us summarize and discuss the results presented in this chapter.
	
	We studied the transition probabilities between two minima of a scalar field potential using the Lorentzian formalism of the Wheleer-De Witt equation. We proposed a general method to compute such probabilities analytically by solving the WDW equation with a WKB semiclassical approximation. The method only assumes a general form of the Hamiltonian constraint consisting of squared and linear terms in the momenta that is a generic form. The method generalized the ideas proposed originally in \cite{Cespedes:2020xpn} by considering the linear terms and by treating the equations obtained as a system of equations as well. In this way the system led to an unique solution in the case of only quadratic momentum terms in the Hamiltonian. Let us point out that we presented the explicit computation of the transition probabilities in different scenarios by staying at first order in $\hbar$ in the semiclassical approximation. However, the method also provides the necessary equations to incorporate up to any order of quantum corrections to the probabilities. Thus, this procedure can in principle be applied to any model in the superspace with the correct form of the Hamiltonian constraint, and it can provide the transition probabilities up to any order in the WKB approximation.
	
	We obtained the transition probabilities in General Relativity for the homogeneous isotropic metric FLRW with both positive and zero curvature, as well as for the anisotropic metrics: Kantowski-Sachs, Bianchi III and biaxial Bianchi IX. For the FLRW metric with positive curvature, we reproduced the same result obtained in \cite{Cespedes:2020xpn}, with the advantage that the general method produces an unique possible result (up to sign ambiguities). For all the metrics considered, we found that the probabilities are described by two parameters, namely, one degree of freedom coming from the metric and one tension term.  For the anisotropic metrics the gravitational degrees of freedom were related as a consequence of the semiclasssical approximation, such that we have only one independent variable at the end, then we could write the probability in terms of any of those degrees of freedom. Moreover, we showed that in order to obtain a well defined probability we should consider both parameters as independent, since the extremizing procedure based on the Euclidean approach is troublesome. Furthermore, the Bianchi III metric allowed us to study explicitly the effect of the anisotropy. We found that the $\mathbb{A}$ variable was useful to study the probabilities for positive values of the potential minima, then we showed in Figures \ref{FigTransProbN} and \ref{FigTransProbS} that the effect of anisotropy is to reduce the probability and produce a faster decay. This result coincides with the general behaviour described in \cite{Mansouri} for the Bianchi I model using a different approach. On the other hand, taking negative values of the potential minima we needed to employ the $\mathbb{B}$ variable, then the effect of anisotropy was shown in Figure \ref{F-BianchiCom}. We showed that for negative values of the potential the probability increases with the anisotropy for small values of $\bar{\mathbb{B}}$ but when the corresponding scale factor is big enough the behaviour is flipped and the smaller probability is found with the biggest anisotropy. Let us remark that in all cases we are able to explore the ultraviolet limit since the probabilities were well behaved and represented by regular functions	in such limit. For all cases we found that in the UV limit the probability leads to $1$, thus it always have its maximum value for the initial singularity.
	
	We then applied the method to study the probabilities in the Ho\v{r}ava-Lifshitz theory using the FLRW closed metric. We considered two types of scalar fields. Firstly, an inhomogeneous scalar field which depends on all spacetime variables in order to explore the differences with General Relativity that are present for such inhomogeneous scalar fields in the action. Then, we also studied a homogeneous scalar field that depends only on the time variable as customary in cosmology. The transition probability for the inhomogeneous scalar field was found to be described by five parameters that should be considered as independent to avoid issues. We examined analytically the IR and UV limits of this expression. The IR limit was similar to the GR result with an extra degree of freedom as expected, However the UV limit showed a completely different behaviour, since it lead to a vanishing value of the probability for vanishing scale factor. We showed the overall behaviour in Figure \ref{PlotGeneral}. We found that the $\lambda$ parameter is relevant only in the IR region where the probability increases with this parameter, the UV region is described exclusively by the tension terms. On the other hand, for the homogeneous scalar field we obtained a result described only by two parameters as in GR. Furthermore, the IR and UV limits coincide with the ones on GR. Thus, the only difference appeared in the intermediate region and a dependence on $\lambda$ in the IR that behaves as in the inhomogeneous case. We presented such behaviour in Figure \ref{PlotTime}. It is interesting to note that using HL theory instead of GR we were able to explore the UV region for both cases. In particular, although we only used up to first order in the semiclassical expansion, we were able to obtain a relevant result for the inhomogeneous case.	
	
	Finally, we used the most general form of the method to study the transition probabilities when the variables of the minisuperspace obey a GUP. In order to modify the WDW equations obtained using the standard commutators we used the procedure presented in chapter \ref{CH-GUPWDW}. In order to simplify the GUP analysis a change of variables was carried out in all the metrics considered, as we all an appropriate factorization in each case. We studied the closed and flat FLRW metrics, as well as the Bianchi III metric in order to incorporate anisotropy as well and consider in all cases up to second order terms in the GUP parameter and in the momenta. For all the metrics used, it was found that the effect of considering a  GUP is that the probability increases at first but it decays faster and then it decreases with the parameter $\gamma$. We presented a plot of  this behaviour for the closed FLRW metric in Figures \ref{F-FLRW} and \ref{F-FLRWExtra}. Furthermore, for the Bianchi III metric we also were able to relate the degrees of freedom coming from the metric using approximations up to second order in $\gamma$, then again all the probabilities were described by two parameters that should be considered as independent. The result or the Bianchi III case has a very complicated form but it fulfils the two expected limits. That is, in the isotropy limit $\alpha\to 0$ it reduces to the flat FLRW result and in the standard HUP limit $\gamma\to0$ it reduces to the Bianchi III with standard commutators result. By varying both parameters, $\alpha$ and $\lambda$ we encountered similar behaviour as explained previously, although both parameters are independent. Finally, we described some other possible ways to deform the WDW equation that could lead to scenarios where the general method, or variations from it, can also be applicable. 
	
	Let us point out that one of the differences encountered when studying this type of transition probabilities with the Lorentzian formalism, instead of relying on the analytic continuation used in the Euclidean formalism,  was that the end result of such transition respects the closeness of the spatial universe as was shown in \cite{Cespedes:2020xpn}. Contrary to the prediction obtained using the Euclidean formalism. However, in the present chapter we were limited by a minisuperspace approximation, in particular, we do not have an explicit description of the bubble, thus we could not explore the end result of the transition. However, we obtained transitions between metrics with the same sign of the spatial curvature in the three different scenarios. Thus, we expect that debate landscape/swampland corresponding to open/closed universes \cite{Freivogel:2005vv,Kleban:2012ph,Hartle:2013oda,Hawking:2017wrd} can be extended to the anisotropic context in General Relativity, but also to the Ho\v{r}ava-Lifshitz theory of gravity as well as within the context of a GUP.
	
	Furthermore, given the minisuperspace limitation, in \cite{Cespedes:2020xpn} it was proposed that we can interpret the expressions computed as probabilities of creating universes by vacuum transitions, that is a generalization of the tunnelling from nothing scenario. In that sense the value of the parameter coming to the metric will correspond to the size of the universe at the moment of creation. Thus, we see that in General Relativity for all the metrics studied the most probable scenario is to have a singularity as the origin for the universe. However, for the HL gravity with an inhomogeneous scalar field we obtained that the most probable value for the creation of the universe is small but different than zero. In this way HL gravity predicts that if a universe is created with this mechanism, it can start without a singularity. Furthermore, for the last case considered, we found that the probability of creating a universe with a non-zero small size is enhanced when a GUP is considered compared to the standard case which indicates that the idea of a minimal measurable length derived from the GUP is present in the transition probability even if the GUP is taken in the minisuperspace with the gravitational degrees of freedom as well as the scalar field. Thus, even if we restricted ourselves to first order in the semiclassical approximations, we obtained in the transition probabilities  relevant information of the UV region by applying the method to scenarios that change the UV behaviour. Therefore, even with this limitation, the computation of the probabilities have been proven to be very insightful.
	
	Let us remark that as we explained before, the addition of quantum correction terms to the transition probabilities can in principle be done up to any desired order. Thus, we can expand the result of this chapter by considering more terms in the WKB expansion. We expect that by doing so we can obtain important implications to the UV region in particular. Furthermore, it will be interesting to keep exploring more models of the minisuperspace, or deformations of the WDW equation that lead to the correct form of the Hamiltonian constraint such that the general procedure can still be applicable, in this way keep exploring the transition probabilities for even more scenarios. Finally, it would also be interesting to explore a possible way to left behind the minisuperspace limitation so we can explore more properties regarding the end result of the transition which in this case would correspond to the initial conditions of the universe. 

\chapter{Final Remarks}\label{Ch-FinalRemarks}

In the present thesis we have explored different classical and quantum cosmological scenarios and we have obtained relevant results. The main idea was to use different ingredients that have been proposed in the literature that try to describe gravity at a quantum level with different levels of rigour, and use them in cosmological scenarios in order to obtain insightful information about the phenomenological aspects that we can expect. With this general idea in mind we have presented several interesting results.

First of all, in chapter \ref{CH-GUPInflation} we presented an analysis of the implications of a GUP to the classical description of inflation driven by a scalar field by using a classical limit. Instead of using a particular form of the GUP, we proposed a general deformation of the Heisenberg algebra and we restricted its form by looking for the appropriate cosmological solutions. In this way, we obtained that inflation can still be obtained by an approximation procedure, as in the standard case, but now the constrictions on the potential can be less restrictive. Furthermore, the most interesting result was that there is always an analytical solution describing inflation driven by a scalar field when the GUP is restricted to take an specific form. We showed that such solution can be seen as a generalization of the cosmological constant solution of the standard case. However, it can also be seen as a solution arriving exclusively because of the GUP. In this way, we showed that such solution has the property that when the GUP can be neglected, the scalar field vanishes and the only remaining ingredient can be a cosmological constant or nothing at all. Therefore, even if the scenarios that we are presented are at the level of a toy model, and the form of the GUP obtained has not been explored before, we have obtained explicit analytical solutions where the scalar field and the inflationary period of expansion are described exclusively by the presence of the GUP. Thus, our result supports the idea that a modification of the standard quantum theory, inspired presumably by quantum gravity should provide answers to the origin of inflation.

Then, in chapter \ref{CH-HLFRSwampland} we explored the compatibility of the dS conjecture with the Ho\v{r}ava-Lifshitz $F(\bar{R})$ theories. We showed that using the flat FLRW metric, commonly used in inflationary scenarios, the conjecture can provide constraints on the parameters of the theory. It is in this sense that we can study the compatibility of the conjectures with the theory itself. We showed that standard $f(R)$ theories can be compatible with the conjecture in a very restrictive way. On the other, the $F(\bar{R})$ HL theories are compatible by restricting the parameters of the theory. We showed that for a power law form of the scale factor, the conjecture is consistent in the UV region of the theories. Furthermore, for a constant Hubble parameter, where GR and standard $f(R)$ theories are not compatible with the conjecture, we showed that these theories are in fact compatible. Let us point out that this study was particularly interesting because we know that the HL theory of gravity, as well as its generalizations have a better UV description than GR. On the other hand, the swampland conjectures are an attempt to capture features of effective theories that can be completed to a full theory of quantum gravity. Thus, the agreement of the conjectures with these theories can be interpreted in two ways: If we assume that the conjectures are in fact true, we have obtained a proof that these theories with the corresponding regions for the parameters are in fact describing a UV behaviour, and thus there should be a way to describe this theory as an effective theory derived from some configuration of a string theory. However, since the UV behaviour of the parameters is well known without use of the conjecture, we can interpret our results as an statement that the dS conjecture may possibly be right, since in one case it is in agreement with the UV behaviour and in the other it can be fulfilled only in these type of theories. Furthermore, since we have applied the conjecture in a scenario where the Lorentz invariance is  broken and obtain consistent results, we can also support the idea that the conjecture may be applicable to more scenarios than originally thought.

We point out that in these two first chapters we have explored classical implications of ideas that are derived from a scenario of quantum gravity, such as the GUP and the swampland conjectures. On the remaining of the thesis, we studied explicitly a proposal for a quantum gravity by exploring the Wheeler-De Witt equation. In chapter \ref{CH-GUPWDW} we studied the implications of considering a GUP in the variables of the minisuperspace to the canonical quantum version of HL gravity. We showed how at the quantum level the different versions of the HL theory can be related, only by an appropriate choice of constants. Furthermore, in the infrared region we obtained consistent results with GR. The UV region on the other hand, was studied and analytical solutions were presented. We showed that oscillatory behaviour can still be present after deforming the WDW equation. Let us remark that the canonical quantization approach of the WDW equation is believed to be limited in nature, and it is thought that it is equivalent to an s-wave approximation to the full quantum theory of gravity. In this way, studying such equation with the HL theory of gravity and with the GUP represents a form in which we expect to improve the ability of such proposal to describe gravity at a quantum level, since we are adding to the formalism an expected quantum gravitational behaviour in the GUP and we are applying the formalism to a theory with a better UV description. It is thus important to study the UV behaviour, and it is for this reason that finding analytical solutions that describe oscillatory behaviour can provide relevant information.

Finally, in chapter \ref{CH-Transitions} we presented a thorough analysis of Lorentzian vacuum transitions. We focused on this processes because it is quantum in nature and thus it needs a proper approach to quantum gravity in order to obtain the transition probabilities. We presented a general method using a semiclassical approximation and used it in different scenarios. We remark that this method only relies on a generic form of the hamiltonian constraint and thus, it is in principle applicable to any model in the minisuperspace, furthermore, it can also provide quantum correction terms up to any desired order in the semiclassical expansion. We discussed that the probabilities obtained can be interpreted as the probability of creating universes through a vacuum transition with a given size. With this interpretation we showed that in GR the most probable scenario is that the universe starts at a singularity in all cases. Furthermore, the effect of  anisotropy does not change the general form of the probabilities, it only increases or decreases their value depending on the sign of the potential minima. However, when studying these transitions in the HL gravity theory we encountered that for an inhomogeneous scalar field, the UV region (that is, for small scale factors) is entirely modified, giving in this way a prediction for a universe to be created at small values of the scale factor but different than zero, thus avoiding the initial singularity. Finally, the transitions were studied when the WDW equation is deformed through the incorporation of a GUP in the variables of minisuperspace.  In this way we obtained that although the general behaviour is unaltered, the probability of creating universes with a non-zero size increases with the GUP parameter. We remark that even if we only used the first order on the semiclassical approximation we have obtained results that allow us to explore the UV region near the initial singularity, thus even if the WDW approach is limited, it can in fact be used to explore quantum cosmological features even at the semiclassical level used. The HL result is particularly interesting because we believe this result supports a previously proposed idea that HL theory can be related to a form of the GUP. Furthermore, if the GUP is a general feature of quantum gravity, we can expect that the transition probabilities do not lead to a a prediction of a singularity, therefore the HL supports this idea. Moreover, since the introduction of the GUP allows the probability of creating a universe with a non-zero size, we believe this supports the idea that the probabilities can indeed be interpreted in this way, since a minimum value for the scale factor does have the expected behaviour in the probabilities even after approximations. Furthermore, the vacuum transitions for different values of the cosmological constant have been recently studied further on various situations in \cite{Pasquarella:2022ibb,Cespedes:2023jdk}. Thus the study of vacuum transitions employing Lorentzian methods is at present a very relevant subject where many new scenarios can be studied. For example, we point out that the method presented in this work can be explored further, for different theories of modified gravity or in other scenarios that lead to a deformation of the WDW equation in an adequate form.  Moreover, in principle, we can also study the quantum corrections to the transition probabilities with this method. These corrections are being studied at the moment and will be presented in the future. Therefore this study has shown that indeed, as it is generally believed, a quantum theory of gravity is capable of giving information about the initial singularity, in particular, we have obtained a fairly simple toy model in which such singularity is avoided.

In summary, all the results presented have provided information of keys aspects that are generally hoped to be resolved when a full theory of quantum gravity is finally obtained. Even if we have been limited in our studies, for example incorporating the GUP as an extra ingredient, or using semiclassical approximations to the WDW equation for the transition probabilities, we have been able to explore in relatively simple models important subjects such as: the origin of inflation, the validity of the swampland conjectures, the limiting nature of the WDW equation and the possible avoidance of the initial singularity. There is of course much to do in order to provide definite answers to all of these subjects. However, we expect that the results presented in this work provide a solid step in many directions that can be used to get a little bit closer to such answers.

\end{document}